\documentclass[twoside]{article}

\pdfoutput=1

\usepackage{coffee2}
\usepackage{times}
\usepackage{amsmath,amssymb,mathrsfs}
\usepackage{graphicx}
\usepackage{feynmp}
\usepackage{url}
\usepackage{hyperref}
\usepackage{longtable}
\usepackage{soul}
\makeatletter
\@ifundefined{pdfoutput}{}{\DeclareGraphicsRule{*}{mps}{*}{}}
\makeatother

\usepackage[bottom]{footmisc}

\numberwithin{equation}{section}
\raggedright

\newcommand{\arxiv}[1]{\href{http://arxiv.org/abs/#1}{arXiv:#1}}

\newcommand{\msu}{m_{\tilde{u}}}
\newcommand{\mmed}{m_\text{med}}
\newcommand{\eqx}[1]{\stackrel{\text{Eq.\eqref{#1}}}{=}}

\newcommand{\adec}{a_\text{dec}}
\newcommand{\xdec}{x_\text{dec}}
\newcommand{\Tdec}{T_\text{dec}}
\newcommand{\geff}{g_\text{eff}}
\newcommand{\mpl}{M_\text{Pl}}

\newcommand\one{\leavevmode\hbox{\small1\normalsize\kern-.33em1}}
\newcommand{\p}{\partial}

\newcommand{\met}{\slashchar{E}_T}

\newcommand{\prol}{\mathbb{P}_L}
\newcommand{\pror}{\mathbb{P}_R}

\newcommand{\lag}{\mathscr{L}}
\newcommand{\lumi}{\mathcal{L}}
\newcommand{\ope}{\mathcal{O}}

\newcommand{\mat}{\mathcal{M}}
\newcommand{\qqquad}{\qquad \qquad}
\newcommand{\qqqquad}{\qquad \qquad \qquad}

\newcommand{\really}{\stackrel{!}{=}}

\providecommand{\nni}{\tilde{\chi}_i^0}
\providecommand{\nnj}{\tilde{\chi}_j^0}
\providecommand{\nne}{\tilde{\chi}_1^0}
\providecommand{\nnz}{\tilde{\chi}_2^0}

\providecommand{\cpe}{\tilde{\chi}_1^+}

\providecommand{\cme}{\tilde{\chi}_1^-}

\providecommand{\cpmj}{\tilde{\chi}_j^\pm}
\providecommand{\cpme}{\tilde{\chi}_1^\pm}

\newcommand{\ev}{\text{eV}}
\newcommand{\mev}{\text{MeV}}
\newcommand{\gev}{\text{GeV}}
\newcommand{\tev}{\text{TeV}}
\newcommand{\fb}{\text{fb}}

\newcommand{\pb}{\text{pb}}
\newcommand{\br}{\text{BR}}

\newcommand{\ifb}{\text{fb}^{-1}}

\def\slashchar#1{\setbox0=\hbox{$#1$}           
   \dimen0=\wd0                                 
   \setbox1=\hbox{/} \dimen1=\wd1               
   \ifdim\dimen0>\dimen1                        
      \rlap{\hbox to \dimen0{\hfil/\hfil}}      
      #1                                        
   \else                                        
      \rlap{\hbox to \dimen1{\hfil$#1$\hfil}}   
      /                                         
   \fi}

\def\ie{{\sl i.e.} \,}
\DeclareMathOperator{\tr}{Tr}

\usepackage{makeidx}
\makeindex
\begin{document}

\title{Yet Another Introduction to Dark Matter \\[1mm]
       The Particle Physics Approach}

\author{Martin Bauer$^{1}$ and Tilman Plehn$^{2}$ \\[7mm]
        $^1$ Institute for Particle Physics Phenomenology, Durham University \\
        $^2$ Institut f\"ur Theoretische Physik, Universit\"at Heidelberg}

\date{\today}

\maketitle
\thispagestyle{empty}

\begin{abstract}
Dark matter is, arguably, the most widely discussed topic in
contemporary particle physics.  Written in the language of particle
physics and quantum field theory, these notes focus on a set of
standard calculations needed to understand different dark matter
candidates. After introducing some general features of such dark
matter agents, we introduce a set of established models which guide us
through four experimental aspects: the dark matter relic density
extracted from the cosmic microwave background, indirect detection
including the Fermi galactic center excess, direct detection, and
collider searches.\footnote{ continuously updated under
  \url{www.thphys.uni-heidelberg.de/\~plehn}.  The original
  publication will be available at \url{www.springer.com}} \\ \coffeeA
\end{abstract}

\newpage

\tableofcontents 

\newpage

\begin{fmffile}{feynman}
\section*{Foreword}

As expected, this set of lecture notes is based on a course on dark
matter at Heidelberg University. The course is co-taught by a theorist and an
experimentalist, and these notes cover the theory half.
Because there exist a large number of text books
and of lecture notes on the general topic of dark matter, the obvious
question is why we bothered collecting these notes. The first answer
is: because this was the best way for us to learn the
topic. Collecting and reproducing a basic set of interesting
calculations and arguments is the way to learn physics. The only
difference between student and faculty is that the latter get to
perform their learning curve in front of an audience. The second
answer is that we wanted to specifically organize material on weakly
interacting dark matter candidates with the focus on four key
measurements \\[2mm]

1. current relic density; \\[1mm]
2. indirect searches; \\[1mm]
3. direct searches; \\[1mm]
4. LHC searches.\\[2mm]

All of those aspects can be understood using the language of
theoretical particle physics. This implies that we will mostly talk
about particle relics in terms of quantum field theory and not about
astrophysics, nuclear physics, or general relativity. Similarly, we
try to avoid arguments based on thermodynamics, with the exception of
some quantum statistics and the Boltzmann equation. With this in mind,
these notes include material for at least 20 times 90 minutes of
lectures for master-level students, preparing them for using the many
excellent black-box tools which are available in the field. As
indicated by the coffee stains, these notes only make sense if people
print them out and go through the formulas one by one. This way any
reader is bound to find a lot of typos, and we would be grateful if
they could send us an email with them.

\vspace*{-5mm} \coffeeC

\newpage

\section{History of the Universe}
\label{sec:history}

When we study the history of the Universe with a focus on the matter
content of the Universe, we have to define three key parameters:
\begin{itemize}
\item[--] the \ul{Hubble constant}\index{Hubble constant} $H_0$ which describes the
  expansion of the Universe. Two objects anywhere in
  the Universe move away from each other with a velocity proportional
  to their current distance $r$. The proportionality constant is defined
  through \ul{Hubble's law}\index{Hubble law}
\begin{align}
  H_0 &:= \frac{\dot r}{r} 
 \approx 70 \; \frac{\text{km}}{\text{s}\; \text{Mpc}} \notag \\
 &= 70 \; \frac{10^5~\text{cm}}{3.1 \cdot 10^{24}~\text{cm}} \; \frac{1}{\text{s}}
  = 2.3 \cdot 10^{-18} \; 6.6 \cdot 10^{-16}~\ev
  = 1.5 \cdot 10^{-33}~\ev \; .
\label{eq:hubble_linear}
\end{align}
  Throughout these lecture notes we will use these
  high-energy units with $\hbar = c = 1$, eventually adding $k_B=1$.
  Because $H_0$ is not at all a number of order one we can replace
  $H_0$ with the dimensionless ratio
\begin{align}
  h := \frac{H_0}{100 \; \dfrac{\text{km}}{\text{s} \; \text{Mpc}}} \approx 0.7 \; .
\end{align}
  The Hubble `constant' $H_0$ is defined at the current point in
  time, unless explicitly stated otherwise.
\item[--] the \ul{cosmological constant} $\Lambda$\index{cosmological constant} , which describes
  most of the energy content of the Universe and which is defined
  through the gravitational Einstein-Hilbert action \index{Einstein-Hilbert action} 
\begin{align}
S_{EH} 
\equiv \frac{\mpl^2}{2} \; \int d^4 x \; \sqrt{-g} \; \left( R - 2 \Lambda \right) \; . 
\end{align}
   The reduced Planck mass is defined as 
\begin{align}
 \mpl = \frac{1}{\sqrt{8 \pi G}} = 2.4 \cdot 10^{27}~\ev \; . 
\label{eq:mplanck}
\end{align}
   It is most convenient to also
   combine the Hubble constant and the cosmological constant to a
   dimensionless parameter
\begin{align} 
\boxed{
 \Omega_\Lambda := \frac{\Lambda}{3H_0^2} \; .
}
\label{eq:def_omegalambda}
\end{align}
\item[--] the \ul{matter content of the Universe} which changes
  with time. As a mass density we can define it as $\rho_m$, but as
  for our other two key parameters we switch to the dimensionless
  parameters
\begin{align}
\boxed{
  \Omega_m := \frac{\rho_m}{\rho_c} 
}
\qquad \text{and} \qquad 
  \Omega_r := \frac{\rho_r}{\rho_c} \;.
\label{eq:def_omegam}
\end{align}
  The denominator $\rho_c$ is defined as the critical density
  separating an expanding from a collapsing Universe with $\Lambda =
  0$.  
  If we study the early Universe, we need
  to consider a sum of the relativistic matter or radiation content
  $\Omega_r$ and non-relativistic matter
  $\Omega_m$ alone. Today, we can also separate the non-relativistic
  baryonic matter content of the Universe. This is the matter content
  present in terms of atoms and molecules building stars, planets, and
  other astrophysical objects. The remaining matter content is
  dark matter, which we indicate with an index $\chi$
\begin{align}
  \Omega_b := \frac{\rho_b}{\rho_c} 
\qquad \Rightarrow \qquad 
  \boxed{ \Omega_\chi := \Omega_m - \Omega_b }\;.
\label{eq:def_omegab}
\end{align}

  If the critical density $\rho_c$ separates an expanding universe
  (described by $H_0$) and a collapsing universe (driven by the
  gravitational interaction $G$) we can guess that it should be given by
  something like a ratio of $H_0$ and $G$. Because the unit of the
  critical density has to be $\ev^4$ we can already guess that $\rho_c
  \sim \mpl^2 H_0^2$.

  In classical gravity we can estimate $\rho_c$ by computing the
  escape velocity of a massive particle outside a spherical piece of
  the Universe expanding according to Hubble's law. We start by
  computing the velocity $v_\text{esc}$ a massive particle has to have
  to escape a gravitational field. Classically, it is defined by equal
  kinetic energy and gravitational binding energy for a test mass $m$
  at radius $r$,
\begin{align}
\frac{m v_\text{esc}^2}{2} 
&\really \frac{GmM}{r} 
 = \frac{G m \dfrac{4 \pi r^3}{3} \rho_c}{\dfrac{v_\text{esc}}{H_0}}
 = \frac{\dfrac{m r^3}{6 \mpl^2} \rho_c}{\dfrac{v_\text{esc}}{H_0}}
\qquad \text{with} \quad 
G  = \frac{1}{8 \pi \mpl^2} \notag \\
\Leftrightarrow \qquad 
H_0^3 r^3 
\eqx{eq:hubble_linear} v_\text{esc}^3 
&\really \frac{1}{3 \mpl^2} H_0 \rho_c r^3  \notag \\
\Leftrightarrow \qquad 
\rho_c &= 3 \mpl^2 \; H_0^2
       =  (2.5 \cdot 10^{-3}~\ev)^4 \;.
\label{eq:rhoc_1}
\end{align}
  We give the numerical value based on the current Hubble expansion
  rate. For a more detailed account for the history of the Universe
  and a more solid derivation of $\rho_c$ we will resort to the theory
  of general relativity in the next section.
\end{itemize}
%

\subsection{Expanding Universe}
\label{sec:expanding}

Before we can work on dark matter as a major constituent of our
observable Universe we need to derive a few key properties based on
general relativity. For example, the Hubble constant $H_0$ as given by
the linear relation in Eq.\eqref{eq:hubble_linear} is not actually a
constant.  To describe the history of the Universe through the time
dependence of the Hubble constant $H(t)$ we start with the definition
of a line element in flat space-time,
\begin{align}
ds^2 
= dt^2 - dr^2 - r^2 d\theta^2 - r^2 \sin^2 \theta d\phi^2 
= \begin{pmatrix} dt \\ dr \\ r d\theta \\ r \sin \theta d\phi \end{pmatrix}^T
  \begin{pmatrix}  1 & 0 & 0 & 0 \\ 
                   0 &-1 & 0 & 0 \\
                   0 & 0 &-1 & 0 \\
                   0 & 0 & 0 &-1 \\
  \end{pmatrix}
  \begin{pmatrix} dt \\ dr \\ r d\theta \\ r \sin \theta d\phi \end{pmatrix} \; .
\label{eq:metric_minkowski}
\end{align} 
The diagonal matrix defines the \ul{Minkowski metric}, which we
know from special relativity or from the covariant notation of
electrodynamics. We can generalize this line element or metric to
allow for a modified space-time, introducing a \ul{scale
  factor}\index{scale factor} $a^2$ as
\begin{align}
ds^2 
&= dt^2 - \left( \frac{dr^2}{1 - \dfrac{r^2}{a^2}} + r^2 d\theta^2 + r^2 \sin^2 \theta d\phi^2 
             \right) \notag \\
&= dt^2 - a^2 \left( \frac{d \dfrac{r^2}{a^2} }{1 - \dfrac{r^2}{a^2}} + \dfrac{r^2}{a^2} d\theta^2 + \frac{r^2}{a^2} \sin^2 \theta d\phi^2 
             \right) \notag \\
&= dt^2 - a^2 \left( \frac{d r^2}{1 - r^2} + r^2 d\theta^2 + r^2 \sin^2 \theta d\phi^2 
             \right) 
\qquad \text{with} \quad \frac{r}{a} \to r \; .
\label{eq:metric_curved1}
\end{align} 
We define $a$ to have the unit length or inverse energy and the last
form of $r$ to be dimensionless.  In this derivation we
implicitly assume a positive curvature through $a^2 > 0$. However,
this does not have to be the case. We can allow for a free sign of the
scale factor by introducing the free \ul{curvature}\index{curvature} $k$, with
the possible values $k = -1,0,1$ for negatively, flat, or positively
curved space. It enters the original form in
Eq.\eqref{eq:metric_curved1} as
\begin{align}
ds^2 
= dt^2 - a^2 \left( \frac{d r^2}{1 - k r^2} + r^2 d\theta^2 + r^2 \sin^2 \theta d\phi^2 
             \right)
\; .
\label{eq:metric_curved2}
\end{align} 
%
At least for constant $a$ this looks like a metric with a modified
distance $r(t) \to r(t) a$. It is also clear that the choice $k=0$ switches
off the effect of $1/a^2$, because we can combine $a$ and $r$ to
arrive at the original Minkowski metric.  

Finally, there is really no reason to assume that the scale factor is
constant with time. In general, the history of the Universe has to
allow for a time-dependent scale factor $a(t)$, defining the line
element or metric as
\begin{align}
ds^2 
= dt^2 - a(t)^2 \left( \frac{dr^2}{1 - kr^2} + r^2 d\theta^2 + r^2 \sin^2 \theta d\phi^2 
             \right)
\; .
\label{eq:metric_time}
\end{align} 
From Eq.\eqref{eq:metric_minkowski} we can read off the 
corresponding metric including the scale factor,
\begin{align}
g_{\mu \nu} = 
\begin{pmatrix}
1 & 0 & 0 & 0 \\
0 & - \dfrac{a^2}{1 - kr^2} & 0 & 0 \\
0 & 0 & - a^2 & 0 \\
0 & 0 & 0 & - a^2  
\end{pmatrix} \; .
\label{eq:metric_gmunu}
\end{align}
Now, the time-dependent scale factor $a(t)$ indicates a motion of
objects in the Universe, $r(t) \to a(t) r(t)$. If we look at objects
with no relative motion except for the expanding Universe, we can
express Hubble's law given in Eq.\eqref{eq:hubble_linear} in terms of
\begin{align}
r'(t) = a(t) \, r 
\qquad \Leftrightarrow \qquad 
\dot r'(t) = \dot a(t) \, r
\really H(t) r'(t) = H(t) a(t) \, r
\qquad \Leftrightarrow \qquad 
\boxed{ H(t) = \frac{\dot a(t)}{a(t)} }\;.
\label{eq:scale_linear}
\end{align}
This relation reads like a linearized treatment of $a(t)$, because it depends only on the first derivative
$\dot a(t)$. However, higher derivatives of $a(t)$ appear
through a possible time dependence of the Hubble constant $H(t)$. From
the above relation we can learn another, basic aspect of cosmology: we
can describe the evolution of the universe in terms of
\begin{enumerate}
\item time $t$, which is fundamental, but hard to directly observe;
\item the Hubble constant $H(t)$ describing the expansion of the Universe;
\item the scale factor $a(t)$ entering the distance metric;
\item the temperature $T(t)$, which we will use from Section~\ref{sec:rad_matter} on.
\end{enumerate}
Which of these concepts we prefer depends on the kind of observations
we want to link. Clearly, all of them should be interchangeable. For
now we will continue with time.\bigskip

Assuming the general metric of Eq.\eqref{eq:metric_time} we can solve
\ul{Einstein's equation} \index{Einstein equation}  including the coupling to matter
\begin{align}
R_{\mu \nu}(t) - \frac{1}{2} g_{\mu \nu}(t) R(t) + \Lambda(t) g_{\mu \nu}(t) = \frac{T_{\mu \nu}(t)}{\mpl^2} \; .
\label{eq:einstein}
\end{align}
The energy-momentum tensor includes the energy
density $\rho_t = T_{00}$ and the corresponding pressure $p$.
The latter is defined as the direction-independent
contribution to the diagonal entries $T_{jj}=p_j$ of the energy-momentum
tensor.  The Ricci tensor $R_{\mu \nu}$ and Ricci scalar $R = g^{\mu
  \nu} R_{\mu \nu}$ are defined in terms of the metric; their explicit
forms are one of the main topics of a lecture on general relativity. In
terms of the scale factor the Ricci tensor reads
\begin{align}
R_{00}(t) = - \frac{3 \ddot a(t)}{a(t)}
\qquad \text{and} \qquad 
R_{ij}(t) = \delta_{ij} \left( 2 \dot a(t)^2 + a(t) \ddot a(t) \right)\;.
\end{align}
If we use the $00$ component of Einstein's equation to determine the
variable scale factor $a(t)$  \index{scale factor} , we arrive at the \ul{Friedmann
  equation} \index{Friedmann equation} 
\begin{align}
\frac{\dot a(t)^2}{a(t)^2} + \frac{k}{a(t)^2} 
&= \frac{\rho_t(t)}{3 \mpl^2} \notag \\
&:= \frac{\rho_m(t) + \rho_r(t) + \rho_\Lambda(t)}{3 \mpl^2} 
\qquad \text{with} \quad \rho_\Lambda(t) := \Lambda(t) \mpl^2 = 3 H_0^2 \mpl^2 \Omega_\Lambda(t) \; ,
\label{eq:friedmann1a}
\end{align}
with $k$ defined in Eq.\eqref{eq:metric_curved2}.  A similar, second
condition from the symmetry of the energy-momentum tensor and its
derivatives reads
\begin{align}
\frac{2 \ddot a(t)}{a(t)} + \frac{\dot a(t)^2}{a(t)^2} + \frac{k}{a(t)^2} 
&= - \frac{p(t)}{\mpl^2} \; .
\label{eq:friedmann1b}
\end{align}
If we use the quasi-linear relation Eq.\eqref{eq:scale_linear} and define
the time-dependent critical total density of the Universe following
Eq.\eqref{eq:rhoc_1}, we can write the Friedmann equation as
\begin{align}
H(t)^2 + \frac{k}{a(t)^2} 
&= \frac{ \rho_t(t)}{3 \mpl^2} \notag \\
\Leftrightarrow \qquad
1 + \frac{k}{H(t)^2 a(t)^2} 
&= \frac{\rho_t(t)}{\rho_c(t)} =: \Omega_t(t) 
\qqquad \text{with} \quad 
\boxed{ \rho_c(t) := 3H(t)^2 \mpl^2 } \; .
\label{eq:def_rho_c}
\end{align}
This is the actual definition of the \ul{critical density} \index{critical density}$\rho_c(t)$.  It
means that $k$ is determined by the time-dependent total
energy density of the Universe,
\begin{align} 
k = H(t)^2 a(t)^2 \; 
\left( \Omega_t(t) - 1 \right) \; .
\label{eq:k_vs_omegatot}
\end{align}
This expression holds at all times $t$,
including today, $t_0$.  For $\Omega_t > 1$ the curvature is
positive, $k>0$, which means that the boundaries of the Universe are
well defined. Below the critical density the curvature is negative. In
passing we note that we can identify
\begin{align}
\frac{\Lambda(t)}{3 H(t)^2} 
\eqx{eq:def_omegalambda} 
\Omega_\Lambda(t) \equiv \frac{\rho_\Lambda(t)}{\rho_c(t)}
\eqx{eq:def_rho_c} \Lambda(t) \mpl^2 \; \frac{1}{3 H(t)^2 \mpl^2} \; .
\end{align}
\bigskip

The two separate equations Eq.\eqref{eq:friedmann1a} and
Eq.\eqref{eq:friedmann1b} include not only the energy and matter
densities, but also the pressure.  Combining them we find
\begin{align}
\frac{\ddot a(t)}{a(t)} 
&\eqx{eq:friedmann1b}
 - \frac{1}{2} \; \left( \frac{\dot a(t)^2}{a(t)^2} + \frac{k}{a(t)^2} \right)
   - \frac{p(t)}{2 \mpl^2} \notag \\
&\eqx{eq:friedmann1a} - \frac{\rho_t(t)}{6 \mpl^2} - \frac{p(t)}{2 \mpl^2} 
= - \frac{\rho_t(t) + 3 p(t)}{6 \mpl^2} \; .
\label{eq:friedmann2}
\end{align}
The cosmological model based on Eq.\eqref{eq:friedmann2} is called
\ul{Friedmann--Lemaitre--Robertson--Walker model} \index{Friedmann--Lemaitre--Robertson--Walker model} or FLRW
model. In general, the relation between pressure $p$ and density
$\rho$ defines the thermodynamic equation of state
\begin{align}
\boxed{
p_j(t) = w_j \; \rho_j(t) 
} \qqquad \text{with} \qquad
w_j = 
\begin{cases}
0 \qquad & \text{non-relativistic matter} \\
1/3 \qquad & \text{relativistic radiation} \\
-1 \qquad & \text{vacuum energy}\;.
\end{cases}
\label{eq:eos}
\end{align}
It is crucial for our understanding of the matter content of the
Universe. If we can measure $w$ it will tell us what the energy or
matter density of the Universe consists of.\bigskip

Following the logic of describing the Universe in terms of the
variable scale factor $a(t)$, we can replace the quasi-linear
description in Eq.\eqref{eq:scale_linear} with a full Taylor series
for $a(t)$ around the current value $a_0$ and in terms of $H_0$. This
will allow us to see the drastic effects of the different equations of
state in Eq.\eqref{eq:eos},
\begin{align}
a(t) - a_0 
&= \dot a(t_0) \, (t - t_0)
     + \frac{1}{2} \ddot a(t_0) \, (t - t_0)^2 
     + \ope \left( (t - t_0)^3 \right) \notag \\
&\equiv a_0 H_0 \, (t - t_0) 
     - \frac{1}{2} a_0 q_0 H_0^2 \, (t - t_0)^2 
     + \ope \left( (t - t_0)^3 \right) \; ,
\label{eq:scale_taylor}
\end{align}
implicitly defining $q_0$.  The units are correct, because the Hubble
constant defined in Eq.\eqref{eq:hubble_linear} is measured in energy.
The pre-factors in the quadratic term are historic, as is the name
deceleration parameter for $q_0$.  Combined with our former results we
find for the quadratic term
\begin{align}
q_0 
&= - \frac{\ddot a(t_0)}{a_0 H_0^2}  
\eqx{eq:friedmann2} \frac{\rho_t(t_0) + 3 p(t_0)}{6 H_0^2 \mpl^2} 
\qquad   \notag \\
&= \frac{1}{6 H_0^2 \mpl^2} \left( \rho_t(t_0) + 3 \sum_j p_j(t_0) \right) 
\eqx{eq:eos} \frac{1}{2} \left( \Omega_t(t_0) + 3 \sum_j \Omega_j(t_0) w_j \right) 
\; .
\end{align}
The sum includes the three components contributing to the total energy
density of the Universe, as listed in Eq.\eqref{eq:components}.
Negative values of $w$ corresponding to a Universe dominated by its
vacuum energy can lead to negative values of $q_0$ and in turn to an
accelerated expansion beyond the linear Hubble law. This is the basis
for a fundamental feature in the evolution of the Universe,
called \ul{inflation} \index{inflation}.\bigskip

To be able to track the evolution of the Universe in terms of the
scale factor $a(t)$ rather than time, we next compute the time
dependence of $a(t)$. As a starting point, the Friedmann equation
gives us a relation between $a(t)$ and $\rho(t)$. What we need is a
relation of $\rho$ and $t$, or alternatively a second relation between
$a(t)$ and $\rho(t)$.  Because we skip as much of general relativity
as possible we leave it as an exercise to show that from the vanishing
covariant derivative of the energy-momentum tensor, which gives rise
to Eq.\eqref{eq:friedmann1b}, we can also extract the time dependence
of the energy and matter densities,
\begin{align}
\frac{d}{dt} \left( \rho_j a^3 \right) 
= - p_j \, \frac{d}{dt} a^3 \; .
\end{align}
It relates the energy inside the volume $a^3$ to the work through the
pressure $p_j$.  From this conservation law we can extract the
$a$-dependence of the energy and matter densities
\begin{alignat}{9}
&& \dot \rho_j a^3 + 3 \rho_j a^2 \dot a 
&= - 3 p_j a^2 \dot a \notag \\
\Leftrightarrow && \quad 
 \dot \rho_j + 3 \rho_j \left( 1 + w_j \right) \frac{\dot a}{a} 
&= 0 \notag \\
\Leftrightarrow && \quad 
\frac{\dot \rho_j}{\rho_j} 
&= - 3 (1 + w_j) \frac{\dot a}{a} \notag \\
\Leftrightarrow && \quad 
\log \rho_j 
&= -3(1+w_j) \log a + C  \notag \\
\Leftrightarrow && \quad 
\rho_j(a) 
&= C \; a^{-3(1+w_j)} 
\propto 
\begin{cases}
a^{-3} & \text{non-relativistic matter} \\
a^{-4} & \text{relativistic radiation} \\
\text{const} & \text{vacuum energy}\;.
\end{cases}
\label{eq:rho_vs_a}
\end{alignat}
This functional dependence is not yet what we want.  To compute the
time dependence of the scale factor $a(t)$ we use a power-law ansatz
for $a(t)$ to find
\begin{alignat}{9}
&& \frac{\ddot a(t)}{a(t)} 
 \eqx{eq:friedmann2} - \frac{1 + 3 w_j}{6 \mpl^2}  \rho_j(t) 
&\eqx{eq:rho_vs_a} - \frac{1 + 3 w_j}{6 \mpl^2} C \; a(t)^{-3(1+w_j)} \notag \\
\Leftrightarrow && \quad 
\ddot a(t) a(t)^{2+3w_j}
&= \text{const} \notag \\
\Leftrightarrow && \quad 
t^{\beta-2} \; t^{\beta (2 + 3 w_j)}  
&= \text{const}
\qqqquad \text{assuming $a \propto t^\beta$} \notag \\
\Leftrightarrow && \quad 
t^{3 \beta+ 3 w_j \beta - 2} 
&= \text{const} \equiv t^0 
\qqquad \Leftrightarrow \qqquad 
\beta = \frac{2}{3+3w_j} \; .
\label{eq:timedep_a_1} 
\end{alignat}
We can translate the result for $a(t) \propto t^\beta$ into the
time-dependent Hubble constant
\begin{align}
H(t) = \frac{\dot{a}(t)}{a(t)}
     \sim \frac{\beta \; t^{\beta-1}}{t^\beta}
     = \frac{\beta}{t} 
     = \frac{2}{3+3w_j} \; \frac{1}{t} \; .
\end{align}

The problem with these formulas is that the power-law ansatz and the
form of $H(t)$ obviously fails for the vacuum energy with $w =
-1$. For an energy density only based on vacuum energy and neglecting
any curvature, $k \equiv 0$, in the absence of matter,
Eq.\eqref{eq:scale_linear} together with the Friedmann equation
becomes
\begin{align}
H(t)^2
&= \frac{\dot a(t)^2}{a(t)^2} 
 \eqx{eq:friedmann1a} \frac{\rho_\Lambda(t) }{3 \mpl^2} 
 = \frac{\Lambda(t)}{3} \notag \\
\Leftrightarrow \qquad 
a(t) 
&= e^{H(t) t}
 = e^{\sqrt{\Lambda(t)/3} \; t} \; .
\end{align}
Combining this result and Eq.\eqref{eq:timedep_a_1}, the functional
dependence of $a(t)$ reads
\begin{align}
\boxed{
a(t) 
\sim 
\begin{cases}
t^{2/(3 + 3w_j)} = 
\begin{cases}
t^{2/3} \qquad & \text{non-relativistic matter} \\
t^{1/2} \qquad & \text{relativistic radiation} \\
\end{cases} \\[2mm]
e^{\sqrt{\Lambda(t)/3} \; t} \qqqquad \; \, \text{vacuum energy.} 
\end{cases}
}
\label{eq:components}
\end{align}
Alternatively, we can write for the Hubble parameter
\begin{align}
\boxed{
H(t) 
\sim 
\begin{cases}
\dfrac{2}{3 + 3w} \; \dfrac{1}{t} = 
\begin{cases}
\dfrac{2}{3t} \qquad & \text{non-relativistic matter} \\[2mm]
\dfrac{1}{2t} \qquad & \text{relativistic radiation} \\
\end{cases} \\[2mm]
\sqrt{\dfrac{\Lambda(t)}{3}} \qqqquad \; \, \text{vacuum energy.} 
\end{cases} }
\label{eq:components2}
\end{align}
From the above list we have now understood the relation between the
time $t$, the scale factor $a(t)$, and the Hubble constant $H(t)$.  An
interesting aspect is that for the vacuum energy case $w = -1$ the
change in the scale factor and with it the expansion of the Universe
does not follow a power law, but an \ul{exponential law},
defining an inflationary expansion.  What is missing from our list at
the beginning of this section is the temperature as the parameter
describing the evolution of the Universe. Here we need to quote a
thermodynamic result, namely that for constant entropy\footnote{This
  is the only thermodynamic result which we will (repeatedly) use in
  these notes.} \index{entropy}
\begin{align} 
\boxed{
a(T) \propto \frac{1}{T} \; .
}
\label{eq:thermodynamics1}
\end{align}
This relation is correct if the degrees of freedom describing the
energy density of the Universe does not change. The easy reference
point is $a_0 = 1$ today. We will use an improved scaling relation in
Section~\ref{sec:relic}.\bigskip

Finally, we can combine several aspects described in these notes  and
talk about \ul{distance measures} and their link to (i) the
curved space-time metric, (ii) the expansion of the Universe, and
(iii) the energy and matter densities.  We will need it to discuss the
cosmic microwave background in Section~\ref{sec:cmb}. As a first step,
we compute the apparent distance along a line of sight, defined by
$d\phi = 0 = d\theta$. This is the path of a traveling photon.  Based
on the time-dependent curved space-time metric of
Eq.\eqref{eq:metric_time} we find
\begin{align}
0 &\really ds^2 = dt^2 - a(t)^2 \frac{dr^2}{1 - kr^2}
\qquad \Leftrightarrow \qquad 
dt = a(t) \frac{dr}{\sqrt{1 - kr^2}} \; .
\end{align}
For the definition of the co-moving distance we integrate along this path,
\begin{align}
\frac{d^c}{a_0} 
&:= \int \frac{dr}{\sqrt{1 - kr^2}} 
= \int \frac{dt}{a(t)}
= \int da \; \frac{1}{\dot a(t) a(t)} \; .
\label{eq:d_comoving1}
\end{align}
The distance measure we obtain from integrating $dr$ in the presence
of the curvature $k$ is called the \ul{co-moving distance} \index{co-moving distance}. It
is the distance a photon traveling at the speed of light can
reach in a given time. We can evaluate the integrand using the
Friedmann equation, Eq.\eqref{eq:friedmann1a}, and the relation $\rho \;
a^{3(1-w)}=$~const,
\begin{align}
\dot a(t)^2 
&= a(t)^2 \; \frac{\rho_t(t)}{3 \mpl^2} - k \notag \\
&\eqx{eq:rho_vs_a}
  \frac{\rho_m(t_0) a_0^3}{3 \mpl^2 a(t)} 
  +\frac{\rho_r(t_0) a_0^4}{3 \mpl^2 a(t)^2} 
  +\frac{\rho_\Lambda a(t)^2}{3 \mpl^2}
  - k  \notag \\
&\eqx{eq:k_vs_omegatot}
   H_0^2 \left[ \Omega_m(t_0) \frac{a_0^3}{a(t)} 
 + \Omega_r(t_0) \frac{a_0^4}{a(t)^2} 
 + \Omega_\Lambda a(t)^2
 - (\Omega_t(t_0) -1 ) \, a_0^2 \right] \notag \\
\Rightarrow \qquad 
\frac{1}{\dot a(t) a(t)} 
&= \frac{1}{H_0 \left[ \Omega_m(t_0) a_0^3 a(t)
 + \Omega_r(t_0) a_0^4
 + \Omega_\Lambda a(t)^4
 - (\Omega_t(t_0) -1 ) \, a_0^2 a(t)^2 \right]^{1/2}} \;.
 \label{eq:d_comoving_int}
\end{align}
For the integral defined in Eq.\eqref{eq:d_comoving1} this gives
\begin{align}
\frac{d^c}{a_0} 
&= \frac{1}{H_0} 
 \int \frac{da}{\left[ \Omega_m(t_0) a_0^3 a(t)
 + \Omega_r(t_0) a_0^4
 + \Omega_\Lambda a(t)^4
 - (\Omega_t(t_0) -1 ) \, a_0^2 a(t)^2 \right]^{1/2}} \notag \\
&\approx \frac{1}{H_0} 
 \int \frac{da}{\left[ (\Omega_t(t_0) - \Omega_\Lambda) a_0^3 a(t)
 + \Omega_\Lambda a(t)^4
 - (\Omega_t(t_0) -1 ) \, a_0^2 a(t)^2 \right]^{1/2}} \notag \\
&= \frac{1}{H_0} 
 \int \frac{da}{\left[ 
   \Omega_t(t_0) (a_0^3 a(t) - a_0^2 a(t)^2)
 - \Omega_\Lambda ( a_0^3 a(t) - a(t)^4 )
 +  a_0^2 a(t)^2 \right]^{1/2}} \; .
\label{eq:d_comoving2}
\end{align}
Here we assume (and confirm later) that today $\Omega_r(t_0)$ can be
neglected and hence $\Omega_t(t_0) = \Omega_m(t_0) + \Omega_\Lambda$.
What is important to remember that looking back the variable scale
factor is always $a(t) < a_0$.  The integrand only depends on all mass
and energy densities describing today's Universe, as well as today's
Hubble constant. Note that the co-moving distance integrates the
effect of time passing while we move along the light cone in Minkowski
space. It would therefore be well suited for example to see which
regions of the Universe can be causally connected.\bigskip

Another distance measure based on Eq.\eqref{eq:metric_curved2} assumes
the same line of sight $d\phi = 0 = d\theta$, but also a synchronized
time at both end of the measurement, $dt = 0$. This defines a purely
geometric, instantaneous distance of two points in space,
\begin{align}
d \theta &= d \phi = dt = 0 
\qquad \Rightarrow \qquad 
ds(t) = - a(t) \frac{dr}{\sqrt{1 - kr^2}}
\quad \text{with} \quad 
k \eqx{eq:k_vs_omegatot}
H_0^2 a_0^2 (\Omega_t(t_0)-1)
 \notag \\
\Rightarrow \qquad 
d^c_A(t) &:= \int_{d}^0 ds
      = - a(t) \int_{d}^0 \frac{dr}{\sqrt{1 - kr^2}}
= \begin{cases}
\dfrac{a(t)}{\sqrt{k}} \; \arcsin ( \sqrt{k} \, d) 
\quad & k  > 0 \\[4mm]
a(t) d 
\quad & k  = 0 \\[2mm]
\dfrac{a(t)}{\sqrt{|k|}} \; \text{arcsinh} ( \sqrt{|k|} \, d)
\quad & k < 0\;.\\
\end{cases}
\label{eq:d_angular}
\end{align}
This \ul{angular diameter distance}\index{angular diameter distance} is time dependent, but
because it fixes the time at both ends we can use it for geometrical
analyses. It depends on the assumed constant distance $d$, which can
for example be identified with the co-moving distance $d \equiv d^c$. The
curvature is again expressed in terms of today's energy density and
Hubble constant.

\subsection{Radiation and matter}
\label{sec:rad_matter}

\begin{figure}[b!]
\begin{center}
\includegraphics[width=0.6\textwidth]{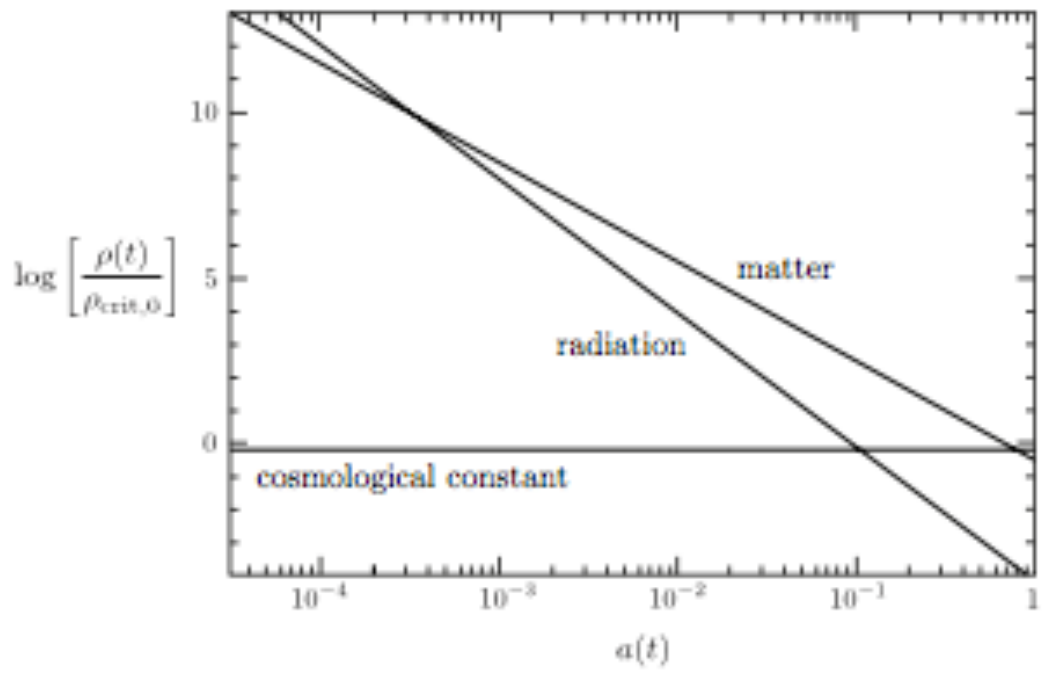}
\end{center}
\vspace*{-7mm}
\caption{Composition of our Universe as a function of the scale
  factor. Figure from Daniel Baumann's lecture notes~\cite{Baumann}.}
\label{fig:energy_vs_scale}
\end{figure}

To understand the implications of the evolution of the Universe
following Eq~\eqref{eq:rho_vs_a}, we can look at the composition of
the Universe in terms of relativistic states (radiation),
non-relativistic states (matter including dark matter), and a
cosmological constant $\Lambda$. Figure~\ref{fig:energy_vs_scale}
shows that at very large temperatures the Universe is dominated by
relativistic states.  When the variable scale factor $a$ increases,
the relativistic energy density drops like $1/a^4$. At the same time,
the non-relativistic energy density drops like $1/a^3$. This means
that as long as the relativistic energy density dominates, the
relative fraction of matter increases linear in $a$. Radiation and
matter contribute the same amount to the entire energy density around
$a_\text{eq} = 3 \cdot 10^{-4}$, a period known as \ul{matter-radiation equality}.\index{matter-radiation equality} The cosmological constant does not change,
which means eventually it will dominate. This starts happening around
now.\bigskip

We know experimentally that most of the matter content in the Universe
is not baryonic, but dark matter. To describe its production in our
expanding Universe we need to apply some basic statistical physics and
thermodynamics. We start with the observation that according to
Figure~\ref{fig:energy_vs_scale} in the early Universe neither the
curvature $k$ nor the vacuum energy $\rho_\Lambda$ play a role. This
means that the relevant terms in the \ul{Friedmann equation} \index{Friedmann equation}
Eq.\eqref{eq:friedmann1a} read
\begin{align}
\boxed{
H(t)^2 = \frac{\rho_m(t) + \rho_r(t)}{3 \mpl^2} 
} 
\qquad \Leftrightarrow \qquad 
1 = \frac{\rho_m(t) + \rho_r(t)}{\rho_c(t)} 
  = \Omega_m(t) + \Omega_r(t) \; .
\label{eq:matter_only}
\end{align}
This form will be the basis of our calculation in this section. The
main change with respect to our above discussion will be a shift to
\ul{temperature} rather than time as an evolution
variable.\bigskip

For relativistic and non-relativistic particles or radiation we can
use a unified picture in terms of their quantum fields. What we have
to distinguish are fermion and boson fields and the 
temperature $T$ relative to their respective masses $m$.  The
\ul{number of degrees of freedom} \index{degrees of freedom} are counted by a factor $g$,
for example accounting for the anti-particle, the spin, or the color
states. For example for the photon we have $g_\gamma=2$, for the
electron and positron $g_e=2$ each, and for the left-handed neutrino
$g_\nu = 1$. If we neglect the chemical potential because we assume to
be either clearly non-relativistic or clearly relativistic, and we set
$k_B=1$, we (or better \textsc{Mathematica}) find \index{number density}
\begin{align}
\label{eq:n_vs_temp}
n_\text{eq}(T)
=&  g \; 
      \int \frac{d^3 p}{(2\pi)^3} \frac{1}{e^{E/T} \pm 1}
\qqquad \; \text{for fermions/bosons} \\
=&  g \; 
      4\pi \int_m^\infty \frac{E d E}{(2\pi)^3} \frac{\sqrt{E^2 - m^2}}{e^{E/T} \pm 1}
\qquad \text{using $E^2 = p^2 + m^2$ and $p dp = E dE$} \notag \\
=& 
\begin{cases}
g \left( \dfrac{mT}{2\pi} \right)^{3/2} e^{-m/T}
\qquad &\text{non-relativistic states $T \ll m$} \notag \\
\dfrac{\zeta_3}{\pi^2} \; g T^3 
\qquad &\text{relativistic bosons $T \gg m$} \notag \\[2mm]
\dfrac{3}{4} \, \dfrac{\zeta_3}{\pi^2} \; g T^3 
\qquad &\text{relativistic fermions $T \gg m$.} \notag 
\end{cases}
\end{align}
The Riemann zeta function has the value $\zeta_3 = 1.2$.  As
expected, the quantum-statistical nature only matters once the states become
relativistic and probe the relevant energy ranges. Similarly, we can
compute the energy density in these different cases.
\begin{align}
\label{eq:rho_vs_temp}
\rho_\text{eq}(T) 
=&  g \; 
      \int \frac{d^3 p}{(2\pi)^3} \frac{E}{e^{E/T} \pm 1} 
=  g \; 
      4\pi \int_m^\infty \frac{E d E}{(2\pi)^3} \frac{E \sqrt{E^2 - m^2}}{e^{E/T} \pm 1} \\
=& 
\begin{cases}
m g \left( \dfrac{mT}{2\pi} \right)^{3/2} e^{-m/T}
\qquad &\text{non-relativistic states $T \ll m$} \notag \\
\dfrac{\pi^2}{30} \; g T^4 
\qquad &\text{relativistic bosons $T \gg m$} \notag \\[2mm]
\dfrac{7}{8} \, \dfrac{\pi^2}{30} \; g T^4 
\qquad &\text{relativistic fermions $T \gg m$.} 
\end{cases}
\end{align}
In the non-relativistic case the relative scaling of $\rho$ relative
to the number density is given by an additional factor $m \gg T$.  In
the relativistic case the additional factor is the temperature $T$,
resulting in a Stefan--Boltzmann scaling \index{Stefan--Boltzmann scaling} of the energy density, $\rho
\propto T^4$. To compute the pressure we can simply use the equation
of state, Eq.\eqref{eq:eos}, with $w = 1/3$.\bigskip

\begin{figure}[b!]
\begin{center}
\includegraphics[width=0.6\textwidth]{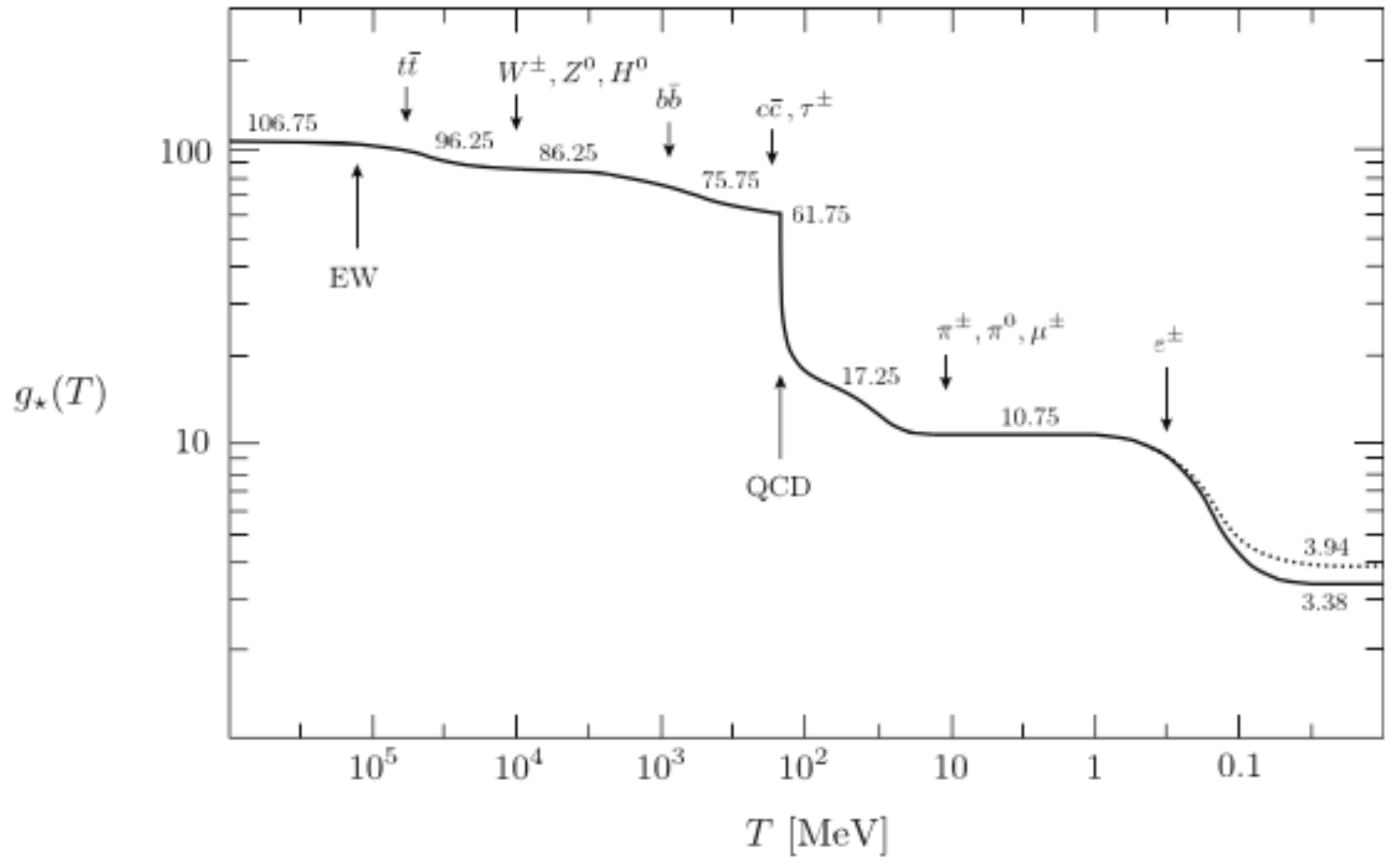}
\end{center}
\vspace*{-7mm}
\caption{Number of effective degrees of freedom $\geff$ as a function
  of the temperature, assuming the Standard Model particle
  content. Figure from Daniel Baumann's lecture notes~\cite{Baumann}.}
\label{fig:gstar_vs_t}
\end{figure}

The number of \ul{active degrees of freedom} \index{degrees of freedom}in our system
depends on the temperature.  As an example, above the electroweak
scale $v = 246$~GeV the effective number of degrees of freedom
includes all particles of the Standard Model
\begin{align}
g_\text{fermion}
&= g_\text{quark} + g_\text{lepton} + g_\text{neutrino}
 = 6 \times 3 \times 2 \times 2  + 3 \times 2 \times 2 + 3 \times 2 = 90 
\notag \\
g_\text{boson}
&= g_\text{gluon} + g_\text{weak} + g_\text{photon} + g_\text{Higgs}
 = 8 \times 2 + 3 \times 3 + 2 + 1 = 28 \; .
\label{eq:dof_sm}
\end{align}
Often, the additional factor $7/8$ for the fermions in
Eq.\eqref{eq:rho_vs_temp} is absorbed in an effective number of
degrees of freedom, implicitly defined through the unified relation
\begin{align}
\boxed{
\rho_r = \frac{\pi^2}{30} \; \geff(T) \; T^4 
} \; ,
\label{eq:def_eff_freedom1}
\end{align}
with the relativistic contribution to the matter density defined in
Eq.\eqref{eq:friedmann1a}.  Strictly speaking, this relation between the
relativistic energy density and the temperature only holds if all
states contributing to $\rho_r$ have the same temperature, \ie are in
thermal equilibrium with each other. This does not have to be the
case.  To include different states with different temperatures we
define $\geff$ as a weighted sum with the specific temperatures of
each component, namely
\begin{align}
\geff(T) 
=  \sum_\text{bosons} g_b \; \frac{T_b^4}{T^4}
 + \sum_\text{fermions} \frac{7}{8} \; g_f \; \frac{T_f^4}{T^4}\;.
\label{eq:def_eff_freedom2}
\end{align}
For the entire Standard Model particle content at equal temperatures
this gives 
\begin{align}
\geff(T > 175~\gev) 
\eqx{eq:dof_sm} 28 + \frac{7}{8} \; 90 = 106.75 \; .
\label{eq:sm_freedom}
\end{align}
When we reduce the temperature, this number of active degrees of
freedom changes whenever a particle species vanishes at the respective
threshold $T=m$. This curve is illustrated in
Figure~\ref{fig:gstar_vs_t}. For today's value we will use the value
\begin{align}
\geff(T_0) = 3.6 \; .
\end{align}
\bigskip

Finally, we can insert the relativistic matter density given in
Eq.\eqref{eq:def_eff_freedom1} into the Friedmann equation
Eq.\eqref{eq:matter_only} and find for the relativistic,
radiation-dominated case
\begin{align}
\boxed{
H(t)^2 
\eqx{eq:components2} \left( \frac{1}{2t} \right)^2 
\eqx{eq:matter_only}
\frac{\rho_r}{3 \mpl^2}
\eqx{eq:def_eff_freedom1}
\frac{1}{3 \mpl^2} \; \frac{\pi^2}{30} \; \geff(T) \; T^4
= \left(\frac{\pi \sqrt{\geff}}{\sqrt{90}} \; \frac{T^2}{\mpl} \right)^2 
}\;.
\label{eq:hubble_temp}
\end{align}
This relation is important, because it links time, temperature, and
Hubble constant as three possible \ul{scales in the evolution
  of our Universe} in the relativistic regime. The one thing we need
to check is if all relativistic relics have the same temperature.
 
\subsection{Relic photons}
\label{sec:photons}
\index{relic!photons}
Before we will eventually focus on weakly interacting massive
particles, forming the dark matter content of the Universe, it is for
many reasons instructive to understand the current photon density.  We
already know that the densities of all particles pair-produced from a
thermal bath in the early, hot Universe follows
Eq.\eqref{eq:rho_vs_temp} and hence drops rapidly with the decreasing
temperature of the expanding Universe.  This kind of behavior is
described by the Boltzmann equation, which we will study in some
detail in Section~\ref{sec:relic}.  Computing the neutrino or photon
number densities from the Boltzmann equation as a function of time or
temperature will turn out to be a serious numerical problem. An
alternative approach is to keep track of the relevant degrees of
freedom $g(T)$ and compute for example the neutrino relic density
$\rho_\nu$ from Eq.\eqref{eq:def_eff_freedom1}, all as a function of
the temperature instead of time. In this approach it is crucial to
know which particles are in equilibrium at any given point in time or
temperature, which means that we need to track the temperature of the
photon--neutrino--electron bath falling apart.\bigskip

Neutrinos, photons, and electrons maintain thermal equilibrium through
the scattering processes
\begin{align}
\bar \nu_e \; e^- \to W^* \to \bar \nu_e \; e^- 
\qquad \text{and} \qquad
e^- \; \gamma \to e^* \to e^- \; \gamma \; .
\label{eq:neutrino_scattering}
\end{align}
For low temperatures or energies $m_\nu \ll T,E \ll m_W$ the two cross
sections are approximately 
\begin{align}
\sigma_{\nu e}(T) = \frac{\pi \alpha^2 T^2}{s_w^4 m_W^4} 
\ll 
\sigma_{\gamma e}(T) = \frac{\pi \alpha^2}{m_e^2} \; .
\label{eq:neutrino_photon_electron}
\end{align}
The coupling strength $g \equiv e/\sin \theta_w \equiv e/s_w$ with
$s_w^2 \approx 1/4$ defines the weak coupling $\alpha = e^2/(4
\pi) \approx 1/137$. 
The geometric factor $\pi$ comes from the angular
integration and helps us getting to the the correct approximate numbers. 
The
photons are more strongly coupled to the electron bath, which means they
will decouple last, and in their decoupling we do not have to consider
the neutrinos anymore. The \ul{interaction rate}
\begin{align}
 \Gamma :=  \sigma \, v \, n
\end{align}
describes the probability for example of the neutrino or photon
scattering process in Eq.\eqref{eq:neutrino_scattering} to happen.  It
is a combination of the cross section, the relevant number density \index{number density} and
the velocity, measured in powers of temperature or energy, or inverse
time.  In our case, the relativistic relics move at the speed of light.
Because the Universe expands, the density of neutrinos, photons, and
charged leptons will at some point drop to a point where the processes
in Eq.\eqref{eq:neutrino_scattering} hardly occur. They will stop
maintaining the equilibrium between photons, 
neutrinos, and charged leptons roughly when the respective interaction
rate drops below the Hubble expansion. This gives us the condition
\begin{align}
\boxed{ \frac{\Gamma(\Tdec)}{H(\Tdec)} \really 1 } \; .
\label{eq:def_tdec}
\end{align}
as an implicit definition of the decoupling temperature.  

Alternatively, we can compare the \ul{mean free path} \index{mean free path}of the neutrinos or
photons, $1/(\sigma \; n)$, to the Hubble length $v/H$  to define the
point of decoupling implicitly as
\begin{align}
\frac{1}{\sigma(\Tdec) \; n(\Tdec)} \really \frac{v}{H(\Tdec)}
\qquad \Leftrightarrow \qquad
\frac{\sigma(\Tdec) \; v \; n(\Tdec)}{H(\Tdec)} \really 1 \; .
\end{align}
While the interaction rate for example for neutrino--electron
scattering is in the literature often defined using the neutrino
density $n = n_\nu$. For the mean free path we have to use the target
density, in this case the electron $n = n_e$.\bigskip

We should be able to compute the photon decoupling from the electrons
based on the above definition of $\Tdec$ and the photon--electron or
Thomson scattering rate in Eq.\eqref{eq:neutrino_photon_electron}.
The problem is, that it will turn out that at the time of photon
decoupling the electrons are no longer the relevant states. Between
temperatures of 1~MeV and the relevant eV-scale for photon decoupling,
\ul{nucleosynthesis} \index{nucleosynthesis} will have happened, and the early Universe
will be made up by atoms and photons, with a small number of free
electrons.  Based on this, we can very roughly guess the temperature
at which the Universe becomes transparent to photons from the fact
that most of the electrons are bound in hydrogen atoms. The ionization
energy of hydrogen is $13.6$~eV, which is our first guess for
$\Tdec$. On the other hand, the photon temperature will follow a
Boltzmann distribution. This means that for a given temperature
$\Tdec$ there will be a high-energy tail of photons with much larger
energies. To avoid having too many photons still ionizing the hydrogen
atoms the photon temperature should therefore come out as $\Tdec
\lesssim 13.6~\ev$.\bigskip

Going back to the defining relation in Eq.\eqref{eq:def_tdec}, we can
circumvent the problem of the unknown electron density by expressing
the density of free electrons first relative to the density of
electrons bound in mostly hydrogen, with a measured suppression factor
$n_e/n_B \approx 10^{-2}$. Moreover, we can relate the full
electron density or the baryon density $n_B$ to 
the photon density $n_\gamma$ through the measured baryon--to--photon
ratio. In combination, this gives us for the time of photon decoupling
\begin{align}
n_e(\Tdec) 
= \frac{n_e}{n_B}(\Tdec) \; n_B(\Tdec)
= \frac{n_e}{n_B}(\Tdec) \; \frac{n_B}{n_\gamma}(\Tdec) \; n_\gamma(\Tdec)
= 10^{-2} \; 10^{-10} \; \frac{2 \zeta_3 \Tdec^3}{\pi^2} \; .
\label{eq:electron_density}
\end{align}
At this point we only consider the ratio $n_B/n_\gamma \approx
10^{-10}$ a measurable quantity, its meaning will be the topic of
Section~\ref{sec:matter}. With this estimate of the \ul{relevant
  electron density} we can compute the temperature at the point of
photon decoupling.  For the Hubble constant we need the number of
active degrees of freedom in the absence of neutrinos and just
including electrons, positions, and photons
\begin{align}
\geff(\Tdec)= \frac{7}{8} \left( 2 + 2 \right) + 2 = 5.5 \; . 
\end{align}
Inserting the Hubble constant from Eq.\eqref{eq:hubble_temp} and the
cross section from Eq.\eqref{eq:neutrino_photon_electron} gives us the
condition
\begin{align}
\frac{\Gamma_\gamma}{H}
&= \frac{2 \pi \zeta_3 \alpha^2}{\pi^2} \; 10^{-12} \; \frac{T^3}{m_e^2} \; 
   \frac{\sqrt{90} \mpl}{\pi} \; \frac{1}{\sqrt{\geff(T)} T^2} && \notag \\
&= \frac{6 \sqrt{10} \, \zeta_3}{\pi^2} \;  \; 10^{-12}
   \alpha^2 \frac{1}{\sqrt{\geff(T)}} \; 
   \frac{\mpl T}{m_e^2} \really 1 
\qquad \Leftrightarrow \qquad & 
\Tdec &= 10^{12} \;
  \frac{\pi^2}{6 \sqrt{10} \, \zeta_3} \;  
         \frac{m_e^2}{\mpl} \frac{\sqrt{\geff(\Tdec)}}{\alpha^2} \notag \\
&&&\approx (0.1~...~1)~\ev \; .
\label{eq:photon_decouple1} 
\end{align}
As discussed above, to avoid having too many photons still ionizing
the hydrogen atoms, the photon temperature indeed is $\Tdec \approx
0.26~\ev < 13.6~\ev$.\bigskip

These decoupled photons form the \ul{cosmic microwave
  background} (CMB) \index{cosmic microwave background}, which will be the main topic of
Section~\ref{sec:cmb}. The main property of this photon background,
which we will need all over these notes, is its current temperature. We
can compute $T_{0,\gamma}$ from the temperature at the point of
decoupling, when we account for the expansion of the Universe between
$\Tdec$ and now. We can for example use the time evolution of the
Hubble constant $H \propto T^2$ from Eq.\eqref{eq:hubble_temp} to
compute the photon temperature today. We find the experimentally
measured value of
\begin{align}
\boxed{
T_{0,\gamma} = 2.4 \cdot 10^{-4}~\ev = 2.73~\text{K} \approx \frac{\Tdec}{1000} 
} \; .
\label{eq:cmb_temp}
\end{align}
This energy corresponds to a photon frequency around 60~GHz, which is
in the microwave range and inspires the name CMB.  We can translate
the temperature at the time of photon decoupling into the
corresponding scale factor,
\begin{align}
\adec
\eqx{eq:thermodynamics1} a_0 \frac{T_{0,\gamma}}{\Tdec}
= \frac{T_{0,\gamma}}{\Tdec}
= \frac{2.4\cdot 10^{-4}\,\text{eV}}{0.26\,\ev} \approx \frac{1}{1100} \; .
\label{eq:cmb_a}
\end{align}
From Eq.\eqref{eq:n_vs_temp} we can also compute the current density
of CMB photons,
\begin{align}
n_\gamma(T_0) 
&= \frac{2 \zeta_3}{\pi^2} \; T_{0,\gamma}^3 
= \frac{410}{\text{cm}^3} \; .
\label{eq:cmb_n}
\end{align}
%

\subsection{Cosmic microwave background}
\label{sec:cmb}

In Section~\ref{sec:photons} we have learned that at temperatures
around $0.1$~eV the thermal photons decoupled from the matter in the
Universe and have since then been streaming through the expanding
Universe. This is why their temperature has dropped to $T_0 = 2.4
\cdot 10^{-4}$~eV now. We can think of the
cosmic microwave background or CMB photons as coming from a sphere of
last scattering with the observer in the center. The photons stream
freely through the Universe, which means they come from this sphere
straight to us. 

The largest effect leading to a temperature fluctuation in the CMB
photons is that the earth moves through the photon background or any
other background at constant speed. We can subtract the corresponding
dipole correlation, because it does not tell us anything about
fundamental cosmological parameters.  The most important, fundamental
result is that after subtracting this dipole contribution the
temperature on the surface of last scattering only shows
\ul{tiny variations around $\delta T/T \lesssim
  10^{-5}$}. The entire surface, rapidly moving
away from us, should not be causally connected, so what generated such a
constant temperature? Our favorite explanation for this is a phase of
very rapid, inflationary period of expansion. This means that we
postulate a fast enough expansion of the Universe, such that the
sphere or last scattering becomes causally connected. From
Eq.\eqref{eq:components} we know that such an expansion will be driven
not by matter but by a cosmological constant. The detailed structure
of the CMB background should therefore be a direct and powerful probe
for essentially all parameters defined and discussed in
Section~\ref{sec:history}.\bigskip

The main observable which the photon background offers is their
temperature or energy --- additional information for example about the
polarization of the photons is very interesting in general, but less
important for dark matter studies. Any effect which modifies this
picture of an entirely homogeneous Universe made out of a thermal bath
of electrons, photons, neutrinos, and possibly dark matter particles,
should be visible as a modification to a constant temperature over the
sphere of last scattering. This means, we are interested in analyzing
temperature fluctuations between points on this surface.

The appropriate observables describing a sphere are the angles
$\theta$ and $\phi$. Moreover, we know that \ul{spherical
  harmonics} are a convenient set of orthogonal basis functions which
describe for example temperature variations on a sphere,
\begin{align}
\frac{\delta T(\theta,\phi)}{T_0} 
:= \frac{T(\theta,\phi) - T_0}{T_0} 
= \sum_{\ell=0}^{\infty} \sum_{m=-\ell}^\ell  
  a_{\ell m} \; Y_{\ell m}(\theta,\phi) \; .
\label{eq:spherical_def}
\end{align}
The spherical harmonics are orthonormal, which means in terms of the
integral over the full angle $d \Omega = d \phi d \cos \theta$
\begin{align}
\int d \Omega \; Y_{\ell m}(\theta,\phi) \; Y_{\ell' m'}^*(\theta,\phi) 
&= \delta_{\ell \ell'} \delta_{m m'} \notag \\
\Rightarrow \qquad 
\int d \Omega \; \frac{\delta T(\theta,\phi)}{T_0} Y_{\ell' m'}^*(\theta,\phi) 
&\eqx{eq:spherical_def}
 \sum_{\ell m}  a_{\ell m} \int d \Omega \;
 Y_{\ell m}(\theta,\phi)  \; Y_{\ell' m'}^*(\theta,\phi) \notag \\
&= \sum_{\ell m}  a_{\ell m} \;
   \delta_{\ell \ell'} \delta_{m m'} 
 = a_{\ell' m'} \; .
\label{eq:spherical_complete}
\end{align}
This is the inverse relation to Eq.\eqref{eq:spherical_def}, which
allows us to compute the set of numbers $a_{\ell m}$ from a known
temperature map $\delta T(\theta,\phi)/T_0$.\bigskip

For the function $T(\theta,\phi)$ measured over the sphere of last
scattering, we can ask the three questions which we usually ask for
distributions which we know are peaked: 
\begin{enumerate}
\item what is the peak value?
\item what is the width of the peak? 
\item what the shape of the peak?
\end{enumerate}
For the CMB we assume that we already know the peak value $T_0$ and
that there is no valuable information in the statistical distribution.
This means that we can focus on the width or the \ul{variance
  of the temperature distribution}.  Its square root defines the
standard deviation. In terms of the spherical harmonics the variance
reads
\begin{align}
\frac{1}{4 \pi} \int d \Omega \; 
\left( \frac{\delta T(\theta,\phi)}{T_0} \right)^2
&= \frac{1}{4 \pi} \int d \Omega \; 
\left[ \sum_{\ell m} \; a_{\ell m} Y_{\ell m}(\theta,\phi) \right] \;
\left[ \sum_{\ell' m'} \; a^*_{\ell' m'} Y^*_{\ell' m'}(\theta,\phi) \right] \notag \\
&\eqx{eq:spherical_complete}
\frac{1}{4 \pi}  \sum_{\ell m, \ell',m'} \; 
a_{\ell m} \; a^*_{\ell' m'} \; \delta_{\ell \ell'} \delta_{m m'} 
= \frac{1}{4 \pi} \sum_{\ell m} \; \left| a_{\ell m} \right|^2 \; .
\end{align}
We can further simplify this relation by our expectation for
the distribution of the temperature deviations. We remember for
example from quantum mechanics that for the angular
momentum the index $m$ describes the angular momentum in one specific
direction. Our analysis of the surface of last scattering, just like
the hydrogen atom without an external magnetic field, does not have
any special direction. This implies that the values of $a_{\ell m}$ do
not depend on the value of the index $m$; the sum over $m$ should 
just become a sum over $2 \ell+1$ identical terms. We therefore
define the observed \ul{power spectrum} \index{power spectrum} as the average of the
$|a_{\ell m}|^2$ over $m$,
\begin{align}
C_\ell := \frac{1}{2\ell +1} \sum_{m=-\ell}^\ell \left| a_{\ell m} \right|^2 
\qquad \Leftrightarrow \qquad 
\boxed{ \frac{1}{4 \pi} \int d \Omega \; 
\left( \frac{\delta T(\theta,\phi)}{T_0} \right)^2
= \sum_{\ell = 0}^\infty \; \frac{2 \ell +1}{4 \pi} \; C_\ell 
} \; .
\end{align}
The great simplification of this last assumption is that we now
just analyze the discrete values $C_\ell$ as a function of $\ell \ge
0$.

Note that we analyze the fluctuations averaged over the surface
of last scattering, which gives us one curve $C_\ell$ for discrete
values $\ell \ge 0$. This curve is one measurement, which means none
of its points have to perfectly agree with the theoretical
expectations. However, because of the averaging over $m$ possible
statistical fluctuation will cancel, in particular for larger values
of $\ell$, where we average over more independent orientations.\bigskip

We can compare the series in spherical harmonics
Eq.\eqref{eq:spherical_def} to a Fourier series. The latter will, for
example, analyze the frequencies contributing to a sound from a
musical instrument. The discrete series of Fourier coefficients tells
us which frequency modes contribute how strongly to the sound or
noise. The spherical harmonic do something similar, which we can
illustrate using the properties of the $Y_{\ell
  0}(\theta,\phi)$. Their explicit form in terms of the associated
Legendre polynomials $P_{\ell m}$ and the Legendre polynomials
$P_{\ell}$ is
\begin{align}
Y_{\ell m}(\theta,\phi) 
&= (-1)^m \; e^{i m \phi} 
  \sqrt{\frac{2\ell+1}{4} \frac{(\ell - m)!}{(\ell + m)!}} \;
  P_{\ell m}(\cos \theta)  \notag \\
&= (-1)^m \; e^{i m \phi} 
  \sqrt{\frac{2\ell+1}{4} \frac{(\ell - m)!}{(\ell + m)!}}  \;
  (-1)^m \left(1 - \cos^2 \theta \right)^{m/2} 
  \frac{d^m}{d (\cos \theta)^m} \;
  P_\ell(\cos \theta) \notag \\
\Rightarrow \qquad 
Y_{\ell 0}(\theta,\phi) 
&= \frac{\sqrt{2\ell+1}}{2} \; P_\ell(\cos \theta) \; .
\label{eq:legendre_def}
\end{align}
The Legendre polynomial is for example defined through
\begin{align}
P_\ell(\cos \theta) &= 
\frac{1}{2^\ell \ell!} \; 
\frac{d^\ell}{dt^\ell} \; (\cos^2 \theta -1)^\ell 
= C \cos^\ell \theta + \cdots \; ,
%
\label{eq:legendre_def2}
\end{align}
with the normalization $P_\ell(\pm1) = 1$ and $\ell$ zeros in
between. Approximately, these zeros occur at
\begin{align}
P_\ell(\cos \theta) = 0 
\qquad \Leftrightarrow \qquad
\cos \theta = \cos \left( \pi \frac{4 k -1}{4 \ell + 2} \right) 
\qquad k = 1,...,\ell \; .
\end{align}
The first zero of each mode defines an angular resolution
$\theta_\ell$ of the $\ell$th term in the hypergeometric series,
\begin{align}
\cos \left( \pi \frac{3}{4 \ell + 2} \right) 
\equiv \cos \theta_\ell
\qqquad \Leftrightarrow \qqquad
\theta_\ell \approx \frac{3 \pi}{4 \ell} \; .
\label{eq:def_cmb_angle}
\end{align}
This separation in angle can obviously be translated into a spatial
distance on the sphere of last scattering, if we know the distance of
the sphere of last scattering to us. This means that the series of
$a_{\ell m}$ or the power spectrum $C_\ell$ gives us information about
the angular distances (encoded in $\ell$) which contribute to the
temperature fluctuations $\delta T/T_0$.\bigskip

\begin{figure}[b!]
\begin{center}
\includegraphics[width=0.7\textwidth]{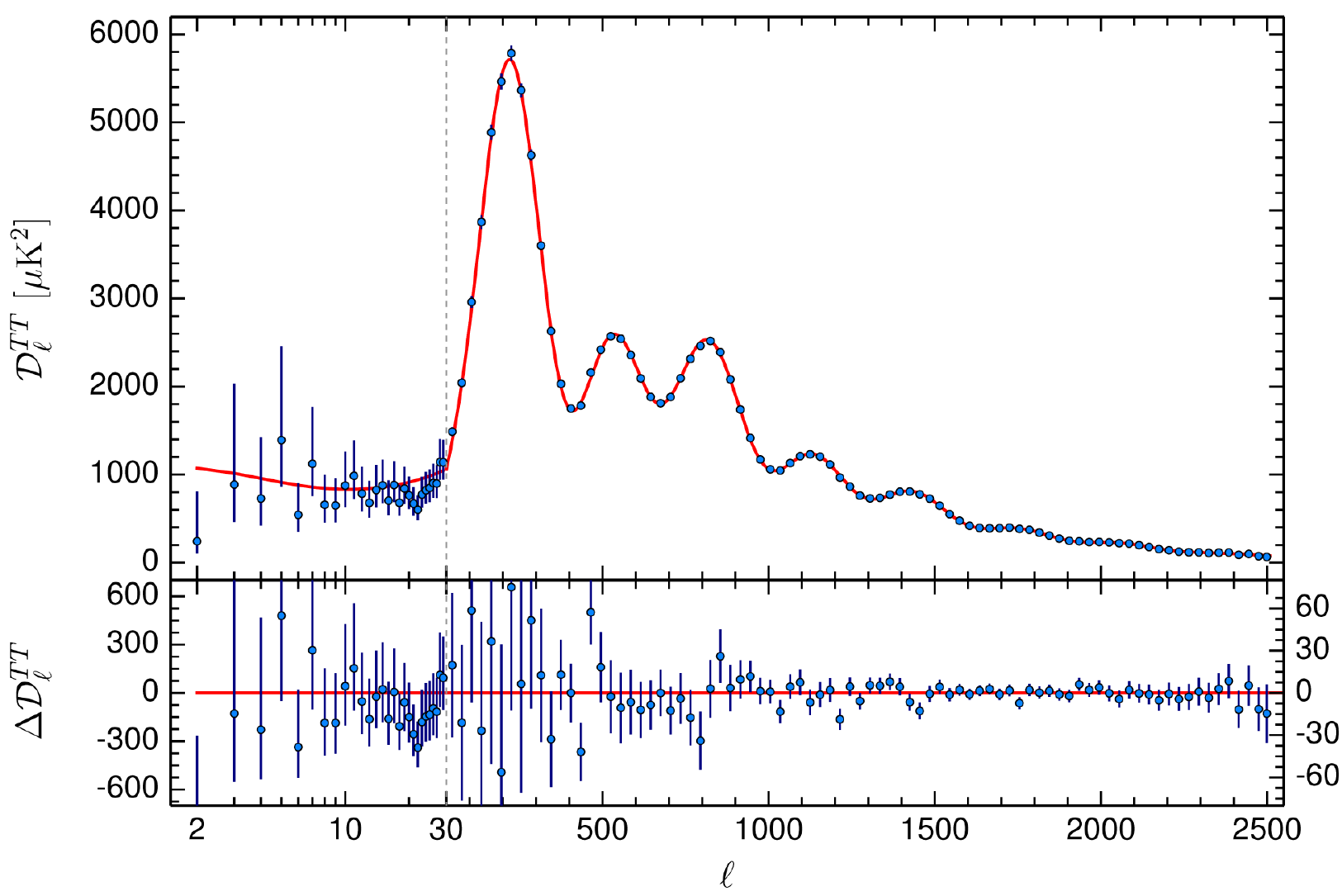}
\end{center}
\caption{Power spectrum as measured by PLANCK in 2015. Figure from
  the PLANCK collaboration~\cite{planck}.}
\label{fig:planck_cmb}
\end{figure}

Next, we need to think about how a distribution of the $C_\ell$ will
typically look.  In Figure~\ref{fig:planck_cmb} we see that the
measured power spectrum essentially consists of a set of peaks. Each
peak gives us an angular scale with a particularly large contribution
to the temperature fluctuations.  The leading physics effects
generating such temperature fluctuations are:
\begin{itemize}\index{acoustic oscillations}
\item[--] \ul{acoustic oscillations} which occur in the
  baryon--photon fluid at the time of photon decoupling. As discussed
  in Section~\ref{sec:history} the photons are initially strongly
  coupled to the still separate electrons and baryons, because the two
  components interact electromagnetically through Thomson scattering.
  Following Eq.\eqref{eq:neutrino_photon_electron} the weak
  interaction can be neglected in comparison to Thomson scattering for
  ordinary matter.  On the other hand, we can see what happens when a
  sizeable fraction of the matter in the Universe is not
  baryonic and only interacts gravitationally and possibly
  through the weak interaction. Such new, dark matter generates
  gravitational wells around regions of large matter accumulation.
  
  The baryon--photon fluid gets pulled into these gravitational
  wells. For the relativistic photon gas we can relate the pressure to
  the volume and the temperature through the thermodynamic equation of
  state $P V \propto T$. If the temperature cannot adjust rapidly
  enough, for example in an adiabatic transition, a reduced volume
  will induce an increased pressure. This photon pressure acts against
  the gravitational well. The photons moving with and against a slope
  in the gravitational potential induces a temperature fluctuation
  located around regions of dark matter concentration. Such an
  oscillation will give rise to a tower of modes with definite wave
  lengths. For a classical box-shaped potential they will be
  equi-distant, while for a smoother potential the higher modes will
  be pulled apart. Strictly speaking, we can separate the acoustic
  oscillations into a temperature effect and a Doppler shift, which
  have separate effects on the CMB power spectrum.

\item[--] the effect of general relativity on the CMB photons, not
  only related to the decoupling, but also related to the propagation
  of the streaming photons to us. In general, the so-called
  \ul{Sachs--Wolfe effect} \index{Sachs--Wolfe effect} describes this impact of gravity on
  the CMB photons. Such an effect occurs if large accumulations of
  mass or energy generate a distinctive gravitational potential which
  changes during the time the photons travel through it. This effect
  will happen before and while the photons are decoupling, but also
  during the time they are traveling towards us. From the discussion
  above it is clear that it is hard to separate the Sachs--Wolfe
  effect during photon decoupling from the other effects generating
  the acoustic oscillations.  For the streaming photons we need to
  integrate the effect over the line of sight. The
  later the photons see such a gravitational potential, the more
  likely they are to probe the cosmological constant or the
  geometrical shape of the Universe close to today.
\end{itemize}
Figure~\ref{fig:planck_cmb} confirms that the power spectrum
essentially consists of a set of peaks, \ie a set of angular scales at
which we observe a particularly strong correlation in
temperatures. They are generated through the acoustic oscillations.
Before we discuss the properties of the peaks we notice two general 
features: first, small values of $\ell$ lead to large error
bars. This is because for large angular separations there are not many
independent measurements we can do over the sphere, \ie we lose the
statistical advantage from combining measurements over the whole
sphere in one $C_\ell$ curve. Second, the peaks are washed out for
large $\ell$. This happens because our approximation that the sphere
of last scattering has negligible thickness catches up with us. If we
take into account that the sphere of last scattering has a finite
thickness, the strongly peaked structure of the power spectrum gets
washed out. Towards large values $\ell$ or small distances the
thickness effects become comparable to the spatial resolution at the
time of last scattering. This leads to an additional damping term
\begin{align} 
C_\ell \propto e^{-\ell^2/1500^2} \; ,
\end{align}
which washes out the peaks above $\ell = 1500$ and erases
all relevant information.\bigskip

Next, we can derive the position of the acoustic peaks.  Because of
the rapid expansion of the Universe, a critical angle $\theta_\ell$ in
Eq.\eqref{eq:def_cmb_angle} defines the size of patches of the sky,
which were not in causal contact during and since the time of the last
scattering. Below the corresponding $\ell$-value there will be no
correlation. It is given by two distances: the first of them is the
distance on the sphere of last scattering, which we can compute in
analogy to the co-moving distance defined in
Eq.\eqref{eq:d_comoving2}. Because the co-moving distance is best
described by an integral over the scale factor $a$, we use the value
$\adec\approx 1/1100$ from Eq.\eqref{eq:cmb_a} to integrate the
ratio of the distance to the sound velocity in the baryon--photon
fluid $c_s$ to
\begin{align}
\frac{r_s}{c_s} \eqx{eq:d_comoving1}
\int_0^{\adec}\frac{da}{a(t)\dot a(t)}\; .
\end{align}
\index{speed of sound}
For a perfect relativistic fluid the speed of sound is given by
$c_s=1/\sqrt{3}$.  This distance is called the sound horizon and
depends mostly on the matter density around the oscillating
baryon--photon fluid.  The second relevant distance is the distance
between us and the sphere of last scattering. Again, we start from the
co-moving distance $d^c$ introduced in
Eq.\eqref{eq:d_comoving1}. Following Eq.\eqref{eq:d_comoving2} it will
depend on the current energy and matter content of the universe. The
angular separation is
\begin{align}
\sin \theta_\ell = \frac{r_s(\Omega_m,\Omega_b)}{d^c} \; .
\label{eq:cmb_geometry}
\end{align}
Both $r_s(\Omega_m,\Omega_b)$ and $d^c$ are described by the same
integrand in Eq.\eqref{eq:d_comoving_int}.  It can be simplified for a
matter-dominated ($\Omega_r \ll \Omega_m$) and almost flat ($\Omega_t
\approx \Omega_m$) Universe to
\begin{align}
\left[ \Omega_m(t_0) a(t)
 + \Omega_r(t_0) 
 + \Omega_\Lambda a(t)^4
 - (\Omega_t(t_0) -1 ) \, a(t)^2 \right]^{-1/2} 
\approx \frac{1}{\sqrt{ \Omega_m(t_0)a(t)}} \; ,
\end{align}
where we also replaced $a_0 = 1$. The ratio of the two integrals then gives
\begin{align}
\sin\theta_\ell 
= \frac{\displaystyle c_s \int_0^{\adec}\frac{da}{\sqrt{a}}}
       {\displaystyle \int_{\adec}^1\frac{da}{\sqrt{a}}}
= \frac{1}{\sqrt{3}}\; \frac{\sqrt{\adec}}{1-\sqrt{\adec}}
\approx \frac{1}{55}\qquad \Rightarrow \qquad \theta_\ell\approx 1^\circ\;.
\end{align}
A more
careful calculation taking into account the reduced speed of sound and
the effects from $\Omega_r, \Omega_\Lambda$ gives a critical angle
\begin{align}
\theta_\ell \approx 0.6^\circ
\qquad \stackrel{\text{Eq.}\eqref{eq:def_cmb_angle}}{\Rightarrow} \qquad
\ell \; \Bigg|_\text{first peak} 
= \frac{4}{3\pi}  \; 0.6^\circ = 225 \; .
\end{align}
The first peak in Figure~\ref{fig:planck_cmb} corresponds to the
fundamental tone, a sound wave with a wavelength twice the size of the
horizon at decoupling. By the time of the last scattering this wave
had just compressed once.  Note that a closed or open universe predict
different result for $\theta_\ell$ following
Eq.\eqref{eq:d_angular}. The measurement of the position of the first
peak is therefore considered a measurement of the geometry of the
universe and a confirmation of its flatness.

The second peak corresponds to the sound wave which underwent one
compression and one rarefaction at the time of the last scattering and
so forth for the higher peaks. Even-numbered peaks are associated with
how far the baryon--photon fluid compresses due to the gravitational
potential, odd-numbered peaks indicate the rarefaction counter effect
of radiative pressure. If the relative baryon content in the
baryon--photon is higher, the radiation pressure decreases and the
compression peaks become higher. The relative amplitude between odd
and even peaks can therefore be used as a measure of $\Omega_b$.

Dark matter does not respond to radiation pressure, but contributes to
the gravitational wells and therefore further enhances the compression
peaks with respect to the rarefaction peaks. This makes a large third
peak a sign of a sizable dark matter component at the time of the last
scattering.

From Figure~\ref{fig:energy_vs_scale} we know that today we can
neglect $\Omega_r(t_0) \ll \Omega_m(t_0) \sim \Omega_\Lambda$.  Moreover, the
relativistic matter content is known from the accurate measurement of
the photon temperature $T_0$, giving $\Omega_r h^2$ through
Eq.\eqref{eq:omega_r_meas}.  This means that the peaks in the CMB
power spectrum will be described by: the cosmological constant defined
in Eq.\eqref{eq:def_omegalambda}, the entire matter
density defined in Eq.\eqref{eq:def_omegam}, which is dominated by the dark matter contribution, as well as by the baryonic matter
density defined in Eq.\eqref{eq:def_omegab}, and the Hubble parameter
defined in Eq.\eqref{eq:hubble_linear}. People usually choose the four
parameters
\begin{align}
\boxed{
\Omega_t(t_0), \quad \Omega_\Lambda, \quad \Omega_m(t_0) h^2, \quad \Omega_b(t_0) h^2
} \; .
\label{eq:cmb_basis}
\end{align}
Including $h^2$ in the matter densities means that we define the total
energy density $\Omega_t(t_0)$ as an independent parameter, but at the
expense of $h$ or $H_0$ now being a derived quantity,
\begin{align}
\left( \frac{H_0}{100 \dfrac{\text{km}}{\text{s Mpc}}} \right)^2 = 
h^2 = \frac{\Omega_m(t_0) h^2}{\Omega_m(t_0)}
    = \frac{\Omega_m(t_0) h^2}{\Omega_t(t_0) - \Omega_\Lambda - \Omega_r(t_0)}
    \approx \frac{\Omega_m(t_0) h^2}{\Omega_t(t_0) - \Omega_\Lambda} \; .
\label{eq:h0_vs_omegas}
\end{align}
There are other, cosmological parameters which we for example need to
determine the distance of the sphere of last scattering, but we will
not discuss them in detail. Obviously, the choice of parameter basis
is not unique, but a matter of convenience. There exist plenty of
additional parameters which affect the CMB power spectrum, but they
are not as interesting for non-relativistic dark matter
studies.\bigskip

We go through the impact of the parameters basis defined in
Eq.\eqref{eq:cmb_basis} one by one:
\begin{itemize}
\item[--] $\Omega_t$ \; affects the co-moving distance,
  Eq.\eqref{eq:d_comoving2}, such that an increase in $\Omega_t(t_0)$
  decreases $d^c$.  The same link to the curvature, $k \propto
  (\Omega_t(t_0) -1)$ as given in Eq.\eqref{eq:k_vs_omegatot}, also
  decreases $ds$, following Eq.\eqref{eq:d_angular}; this way the
  angular diameter distance $d^c_A$ is reduced.  In addition, there is
  an indirect effect through $H_0$; following
  Eq.\eqref{eq:h0_vs_omegas} an increased total energy density
  decreases $H_0$ and in turn increases $d^c$.  

  Combining all of these effects, it turns our that increasing
  $\Omega_t(t_0)$ decreases $d^c$. According to
  Eq.\eqref{eq:cmb_geometry} a smaller predicted value of $d^c$
  effectively increases the corresponding $\theta_\ell$ scale.  This
  means that the acoustic peak positions consistently appear at smaller
  $\ell$ values.
\item[--] $\Omega_\Lambda$ \; has two effects on the peak positions:
  first, $\Omega_\Lambda$ enters the formula for $d^c$ with a
  different sign, which means an increase in $\Omega_\Lambda$ also
  increases $d^c$ and with it $d^c$. At the same time, an increased
  $\Omega_\Lambda$ also increases $H_0$ and this way decreases
  $d^c$. The combined effect is that an increase in $\Omega_\Lambda$
  moves the acoustic peaks to smaller $\ell$. Because in our parameter
  basis both, $\Omega_t(t_0)$ and $\Omega_\Lambda$ have to be determined by
  the peak positions we will need to find a way to break this
  degeneracy.
\item[--] $\Omega_m h^2$ \; is dominated by dark matter and provides
  the gravitational potential for the acoustic
  oscillations. Increasing the amount of dark matter stabilizes the
  gravitational background for the baryon--photon fluid, reducing the
  height of all peaks, most visibly the first two. In addition, an
  increased dark matter density makes the gravitational potential more
  similar to a box shape, bringing the higher modes closer together.
\item[--] $\Omega_b h^2$ \; essentially only affects the height of the
  peaks. The baryons provide most of the mass of the baryon--photon
  fluid, which until now we assume to be infinitely strongly
  coupled. Effects of a changed $\Omega_b h^2$ on the CMB power
  spectrum arise when we go beyond this infinitely strong coupling.
  Moreover, an increased amount of baryonic matter increases the
  height of the odd peaks and reduces the height of the even peaks.
\end{itemize}
Separating these four effects from each other and from other
astrophysical and cosmological parameters obviously becomes easier
when we can include more and higher peaks. Historically, the WMAP \index{WMAP}
experiment lost sensitivity around the third peak. This means that its
results were typically combined with other experiments. The PLANCK
satellite clearly identified seven peaks and measures in a slight
modification to our basis in Eq.\eqref{eq:cmb_basis}~\cite{planck} \index{PLANCK}
\begin{align}
&\hspace*{-9mm}
\boxed{ \Omega_\chi h^2 = 0.1198 \pm 0.0015 } \notag \\[2mm]
\Omega_b h^2 &= 0.02225 \pm 0.00016 \notag \\
\Omega_\Lambda &= 0.6844 \pm 0.0091 \notag \\
H_0 &= 67.27 \pm 0.66  \, \frac{\text{km}}{\text{Mpc s}} \; .
\label{eq:planck_results}
\end{align}
The dark matter relic density is defined in Eq.\eqref{eq:def_omegab}.
This is the best measurement of $\Omega_\chi$ we currently have.

\subsection{Structure formation}
\label{sec:structure}
\index{structure formation}
A powerful tool to analyze the evolution of the Universe is the
distribution of structures at different length scales, from galaxies
to the largest structures. These structures are due to small
primordial inhomogeneities, tiny gravitational wells disrupting the
homogeneous and isotropic universe we have considered so far. They
have then been amplified to produce the galaxies, galaxy groups and
super-clusters we observe today. The leading theory for the origin of
these perturbations is based on quantum fluctuations of in the
inflaton field, which is responsible for the epoch of exponential
expansion of the universe. We leave the details of this idea to a
cosmology lecture, but note that quantum fluctuations behave random or
Gaussian. The evolution of these primordial seeds of over-densities
with the expansion of the universe will give us information on the
dark matter density and on dark matter properties.

We start with the evolution of a general matter density in the
Universe in the presence of a gravitational field.  As long as the
cosmic structures are small compared to the curvature of the universe
and we are not interested in the (potentially) relativistic motion of
particles we can compute the evolution of density perturbations using
Newtonian physics. The matter density $\rho$, the matter velocity
$\vec u$, and its gravitational potential $\phi$ satisfy the equations
\index{continuity equation}
\index{Euler equation}
\index{Poisson equation}
\begin{alignat}{7}
\frac{\partial \rho_m}{\partial t} &= - \nabla \cdot(\rho_m \vec u)
\qqquad &&\text{continuity equation} \label{eq:continuityeq} \\
\left(\frac{\partial }{\partial t}+\vec u\cdot\nabla\right)  \vec u&=-\frac{\nabla p}{\rho_m}-\nabla \phi 
&&\text{Euler equation} \label{eq:euler} \\
\nabla^2\phi  &=4\pi G \rho_m 
&&\text{Poisson equation} \; , \label{eq:poisson}
\end{alignat}
where $p$ denotes an additional pressure and $G=1/(8\pi \mpl^2)$ is
the gravitational coupling defined in Eq.\eqref{eq:mplanck}. \index{PLANCK} This set
of equations can be solved by a homogeneously expanding fluid
\begin{align}
\rho=\rho(t_0)\left(\frac{a_0}{a}\right)^3
\qqquad 
\vec u =\frac{\dot a}{a}\vec r =H \vec r
\qqquad 
\phi=\frac{1}{12 \mpl^2}\rho r^2 
\qqquad 
\nabla p = 0 \; .
\label{eq:stableuniv}
\end{align}
It is the Newtonian version of the matter-dominated Friedmann
model. The Euler equation turns into the second Friedmann equation,
Eq.\eqref{eq:friedmann2}, for a flat universe,
\index{Friedmann equation}
\begin{align}
\dot H \vec r + H \vec r \cdot \nabla (H \vec r)
&=-\nabla \phi =-\frac{1}{6 \mpl^2}\rho_m \vec r
\notag\\
\Leftrightarrow \qqquad 
\dot H +H^2 &= -\frac{\rho_m}{6 \mpl^2} \;.
\label{eq:2ndFriedmannH}
\end{align}
The first Friedmann equation in this approximation also follows when
we use Eq.\eqref{eq:def_rho_c} for $k \to 0$ and $\rho_t=\rho_m$,
\begin{align}
H^2=\frac{\rho_m}{3\mpl^2}\; .\label{eq:firstFMnonR}
\end{align}
We will now allow for small perturbations around the background given
in Eq.\eqref{eq:stableuniv},
\begin{alignat}{5}\label{eq:pert}
\rho(t,\vec r)&=\bar \rho(t)+\delta_\rho(t,\vec r) 
&\qqqquad 
\vec u(t,\vec r)&=H(t) \, \vec r+\vec \delta_u(t,\vec r) \notag \\
\phi(t,\vec r)&=\frac{1}{12\mpl^2}\bar \rho r^2+\delta_\phi(t,\vec r)
&\qqqquad 
p(t,\vec r)&=\bar p(t)+\delta_p(t,\vec r) \; .
\end{alignat}
The pressure and density fluctuations are linked by the speed of sound
$\delta_p = c_s^2 \delta_\rho$.  Inserting Eq.\eqref{eq:pert}, the
continuity equation becomes 
\begin{align}
0 &= \dot \rho +\nabla \cdot (\rho\, \vec u)\notag\\
&= \dot{ \bar \rho} +\dot {\delta_\rho} +\bar \rho \nabla \cdot (H\vec r+ \vec \delta_u) +  \nabla \delta_\rho\cdot (H\vec r+ \vec \delta_u) \notag\\
&= \dot{ \bar \rho} +\dot {\delta_\rho} +\bar \rho \nabla \cdot (H\vec r+ \vec \delta_u) +  \nabla \delta_\rho\cdot (H\vec r+ \vec \delta_u) \notag\\
&= \dot{ \bar \rho} +\dot {\delta_\rho} +\bar \rho \nabla \cdot H\vec r+\bar \rho \nabla\cdot \vec \delta_u +  \nabla \delta_\rho\cdot H\vec r 
  + \ope(\delta^2) \notag\\
&\eqx{eq:continuityeq}  \dot {\delta_\rho} +\bar \rho \nabla\cdot \vec \delta_u +  \nabla \delta_\rho\cdot H\vec r + \ope(\delta^2)  \; ,
\end{align}
%
where we only keep terms linear in the perturbations. In the last line
that the background fields solve the continuity equation
Eq.\eqref{eq:continuityeq}. The Euler equation for the perturbations
results in
\begin{align}
0&=\left(\frac{\partial }{\partial t}+\vec u\cdot\nabla\right)  \vec u=-\frac{\nabla p}{\rho_m}-\nabla \phi \notag\\
&=\dot H \vec r+\dot{\vec \delta}_u+ \left(H\vec r + \vec \delta_u \right) 
  \cdot \nabla \left( H\vec r +\vec \delta_u\right) +\frac{\nabla (\bar p +\delta_p)}{\bar \rho+\delta_\rho}+\nabla \left(\frac{1}{12\mpl^2}\bar \rho r^2+\delta_\phi \right)\notag\\
 &= \dot H \vec r+\dot{\vec \delta}_u+ H\vec r \cdot \nabla  H\vec r + \vec \delta_u \cdot \nabla  H\vec r+ H\vec r \cdot \nabla \vec \delta_u + \frac{\nabla \bar p}{\bar \rho}+ \frac{\nabla \delta_p}{\bar \rho}+\nabla \left(\frac{1}{12\mpl^2}\bar \rho r^2\right)+\nabla\delta_\phi + \ope (\delta^2)  \notag\\
  &\eqx{eq:euler} \dot{\vec \delta}_u + H \vec \delta_u + H\vec r \cdot \nabla \vec \delta_u +  \frac{\nabla \delta_p}{\bar \rho}+\nabla\delta_\phi + \ope (\delta^2) \; .
\end{align}
Finally, the Poisson equation for the fluctuations becomes
\begin{align}
0&=\nabla^2\phi -\frac{1}{2\mpl^2} \rho_m \notag\\
&=\nabla^2 \left(\frac{1}{12\mpl^2}\bar \rho r^2+\delta_\phi\right)-\frac{1}{2\mpl^2}  (\bar \rho+\delta_\rho)
\eqx{eq:poisson} \nabla^2 \delta_\phi - \frac{1}{2\mpl^2}  \delta_\rho \; .
\end{align}
\bigskip

In analogy with Eq.\eqref{eq:spherical_def} we define dimensionless
fluctuations in the density field at a given place $x$ and time $t$ as
\begin{align}
\delta(t, \vec x) :=\frac{\rho(t,\vec x)-\bar \rho(t)}{\bar \rho(t)}=\frac{\delta_\rho(t, \vec x)}{\bar \rho(t)}\,,
\end{align}
and further introduce co-moving coordinates 
\begin{align}
\vec x := \frac{a_0}{a}\vec r
\qqquad 
\vec v := \frac{a_0}{a}\vec u
\qqquad 
\nabla_r := \frac{a_0}{a}\nabla_x  
\qquad 
H \vec r\cdot \nabla_r+\frac{\partial}{\partial t} \rightarrow \frac{\partial}{\partial t}\; .
\end{align}
The co-moving continuity, Euler and Poisson equations then read
\begin{align}
\dot \delta + \nabla_x \vec \delta_v &= 0
\notag \\ 
\dot{\vec{\delta}}_v + 2 H \vec \delta_v &= -\left( \frac{a_0}{a}\right)^2 \nabla_x \left(c_s^2 \delta +\delta_\phi \right)
\notag \\
\nabla_x^2 \delta_\phi &= \frac{1}{2\mpl^2}  \bar \rho \left( \frac{a_0}{a} \right)^2\, \delta \; .
\label{eq:comovingeqs}
\end{align}
These three  equations can be combined into a second order differential equation for the density fluctuations $ \delta$,
\begin{align}
0
= \ddot \delta + \nabla_x \dot{\vec{\delta}}_v 
&= \ddot \delta 
 - \nabla_x \left[2H \vec \delta_v
                +\left(\frac{a_0}{a}\right)^2 \nabla_x \left(c_s^2 \delta+\delta_\phi\right)\right]\notag\\
&= \ddot \delta
 + 2H\dot \delta -\left(\frac{a_0}{a}\right)^2 c_s^2\nabla_x^2 \delta -\frac{1}{2\mpl^2}  \bar \rho \delta\; .
\label{eq:2ndorderDGLDF}
\end{align}
To solve this equation, we Fourier-transform the density fluctuation
and find the so-called \ul{Jeans equation} \index{Jeans equation}
\begin{align}
\delta (\vec x,t)=\int \frac{d^3k}{(2\pi)^3}  \hat \delta(\vec k,t)\,e^{-i\vec k \cdot \vec x}
\qquad \Rightarrow \qquad 
\boxed{\ddot{\hat\delta}+2H\dot{\hat \delta}=\hat\delta\bigg[\frac{1}{2\mpl^2}  \bar \rho -\bigg(\frac{c_sk a_0}{a}\bigg)^2\,\bigg]}\; .
\label{eq:Jeans}
\end{align}
The two competing terms in the bracket correspond to a gravitational
compression of the density fluctuation and a pressure resisting this
compression. The wave number for the homogeneous equation, where these
terms exactly cancel defines the Jeans wave length
\begin{align}
\lambda_J
=\frac{2\pi}{k} \Bigg|_\text{homogeneous}
=2\pi \frac{a_0}{a} c_s\sqrt{\frac{2\mpl^2}{\bar \rho}}\; .
\end{align}
Perturbations of this size neither grow nor get washed out by
pressure.
To get an idea what the Jeans length  \index{Jeans length} for baryons means we can compare
it to the co-moving Hubble scale,
\begin{align}
\dfrac{\lambda_J}{\dfrac{a_0}{a}H^{-1}}
= 2\pi c_s\sqrt{\frac{2\mpl^2}{\bar \rho}}\,H
\eqx{eq:firstFMnonR}
 2\pi \sqrt{\frac{2}{3}}\,c_s \; .
\end{align}
This gives us a critical fluctuation speed of sound close to the speed
of sound in relativistic matter $c_s\approx 1/\sqrt{3}$. Especially
for non-relativistic matter, $c_s\ll 1$, the Jeans length is much  
smaller than the Hubble length and our Newtonian approach is
justified.\bigskip

The Jeans equation for the evolution of a non-relativistic mass or
energy density can be solved in special regimes. First, for length
scales much smaller than the Jeans length $\lambda \ll \lambda_J$, the
Jeans equation of Eq.\eqref{eq:Jeans} becomes an equation of a damped
harmonic oscillator,
\begin{align}
\ddot{\hat\delta}+2H\dot{\hat \delta}+\bigg(\frac{c_sk a_0}{a}\bigg)^2\hat\delta=0
\qquad \Rightarrow \qquad 
\hat \delta (t)\propto e^{\pm i \omega t}
\qqquad \text{(non-relativistic, small structures)} \; ,
\label{eq:BAOsolution}
\end{align}
with $\omega = c_s k a_0/a$.  The solutions are oscillating with
decreasing amplitudes due to the Hubble friction term $2H \dot{\hat
  \delta}$. Structures with sub-Jeans lengths, $\lambda \ll
\lambda_J$, therefore do not grow, but the resulting acoustic
oscillations can be observed in the matter power spectrum today.

In the opposite regime, for structures larger than the Jeans length $\lambda
\gg \lambda_J$, the pressure term in the Jeans equation can be neglected.
The gravitational compression term can be simplified for a
matter-dominated universe with $a\propto t^{2/3}$,
Eq.\eqref{eq:components}. This gives $H =\dot a/a = 2/(3 t)$ and it
follows for the second Friedmann equation that
\begin{align}
\dot H +H^2 =-\frac{2}{9t^2}&\eqx{eq:2ndFriedmannH} - \bar \rho\frac{1}{6\mpl^2} \qquad \Rightarrow \qquad \bar \rho =\frac43\frac{\mpl^2}{t^2} \; .
\end{align}
We can use this form to simplify the Jeans equation and solve it 
\begin{alignat}{9}
\ddot{\hat\delta}+\frac{4}{3t}\dot{\hat\delta}-\frac{2}{3t^2}\hat\delta=0
\qquad \Rightarrow \qquad
\hat\delta &= A t^{2/3}+ \frac{B}{t} \notag \\
&\propto t^{2/3} \qquad \text{growing mode} \notag \\
&=: \frac{a}{a_0} \hat \delta_0 \qquad \text{using $a\propto t^{2/3}$ (non-relativistic, large structures).}
\label{eq:jeans2}
\end{alignat}
We can use this formula for the growth as a function of the scale
factor to link the density perturbations at the time of photon
decoupling to today. For this we quote that at photon decoupling we
expect $\hat \delta_\text{dec} \approx 10^{-5}$, which gives us for
today
\begin{align}
\boxed{
\hat \delta_0
= \frac{\hat \delta_\text{dec}}{\adec}
\eqx{eq:cmb_a} 1100\, \hat \delta_\text{dec}
\approx 1\% 
}\; .
\end{align}
We can compare this value with the results from numerical
\ul{N-body simulations} \index{N-body simulations}and find that those simulations prefer
much larger values $\hat \delta_0 \approx$ 1.  In other words, the
smoothness of the CMB shows that perturbations in the photon-baryon
fluid alone cannot account for the cosmic structures observed
today. One way to improve the situation is to introduce a dominant
non-relativistic matter component with a negligible pressure term,
defining the main properties of \ul{cold dark matter}.

Until now our solutions of the Jeans equation rely on the assumption
of non-relativistic matter domination.  For relativistic matter with
$a\propto t^{1/2}$ the growth of density perturbations follows a
different scaling. Following Eq.\eqref{eq:components2} we use $H =t/2$
and assume $H^2 \gg 4\pi G \bar \rho$, such that the Jeans equation
becomes
\begin{align}
\ddot{\hat \delta}+\frac{\dot {\hat\delta}}{t}=0 \qquad \Rightarrow \qquad \hat \delta =A+B\log t
\qqquad \text{(relativistic, small structures).} 
\label{eq:jeans3}
\end{align}
This growth of density perturbations is much weaker than for
non-relativistic matter. 

Finally, we have to consider relativistic density perturbations larger
than the Hubble scale, $\lambda \gg a/a_0 H$. In this case a Newtonian
treatment is no longer justified and we only quote the result of the
full calculation from general relativity, which gives a scaling
\begin{align}
\hat\delta = \left( \frac{a}{a_0} \right)^2 \hat\delta_0 
\qqquad \text{(relativistic, large structures).} 
\label{eq:growth_rel}
\end{align}
Together with Eqs.\eqref{eq:BAOsolution}, Eq.\eqref{eq:jeans2}, and
\eqref{eq:jeans3} this gives us the growth of structures as a function
of the scale parameter for non-relativistic and relativistic matter
and for small as well as large structures.  Radiation pressure in the
photon-baryon fluid prevents the growth of small baryonic structures,
but baryon-acoustic oscillations \index{acoustic oscillations} on smaller scales predicted by
Eq.\eqref{eq:BAOsolution} can be observed.  Large structures in a
relativistic, radiation-dominated universe indeed expand rapidly.
Later in the evolution of the Universe, non-relativistic structures
come close to explaining the matter density in the CMB evolving to
today as we see it in numerical simulations, but require a dominant
additional matter component.\bigskip

Similar to the variations of the cosmic microwave photon temperature
we can expand our analysis of the matter density from the central
value to its distribution with different sizes or wave numbers. To
this end we
define the \ul{matter power spectrum} \index{power spectrum} $P(k)$ in momentum space as
\begin{align}
\langle \hat \delta(\vec k)^*\hat \delta(\vec k')\rangle
=(2\pi)^3 \delta(\vec k -\vec k') P(k) \; .
\end{align}
As before, we can link $k$ to a wave length $\lambda = 2\pi/k$.  For
the scaling of the initial power spectrum the proposed relation by
Harrison and Zel'dovich is
\begin{align}
P(k) \propto k^n = \left( \frac{2\pi}{\lambda} \right)^n \; .
\label{eq:powerspec1}
\end{align}
From observations we know that $n \gg 1$ leads to an increase in
small-scale structures and as a consequence to too many black
holes. We also know that for $n\ll 1$, large structures like
super-clusters dominate over smaller structures like galaxies, again
contradicting observations. Based on this, the exponent was originally
predicted to be $n=1$, in agreement with standard inflationary
cosmology.  However, the global CMB analysis by PLANCK quoted in \index{PLANCK}
Eq.\eqref{eq:planck_results} gives 
\begin{align}
n = 0.9645 \pm 0.0049 \; .
\end{align}
%

\begin{figure}[b!]
\begin{center}
\includegraphics[width=0.45\textwidth]{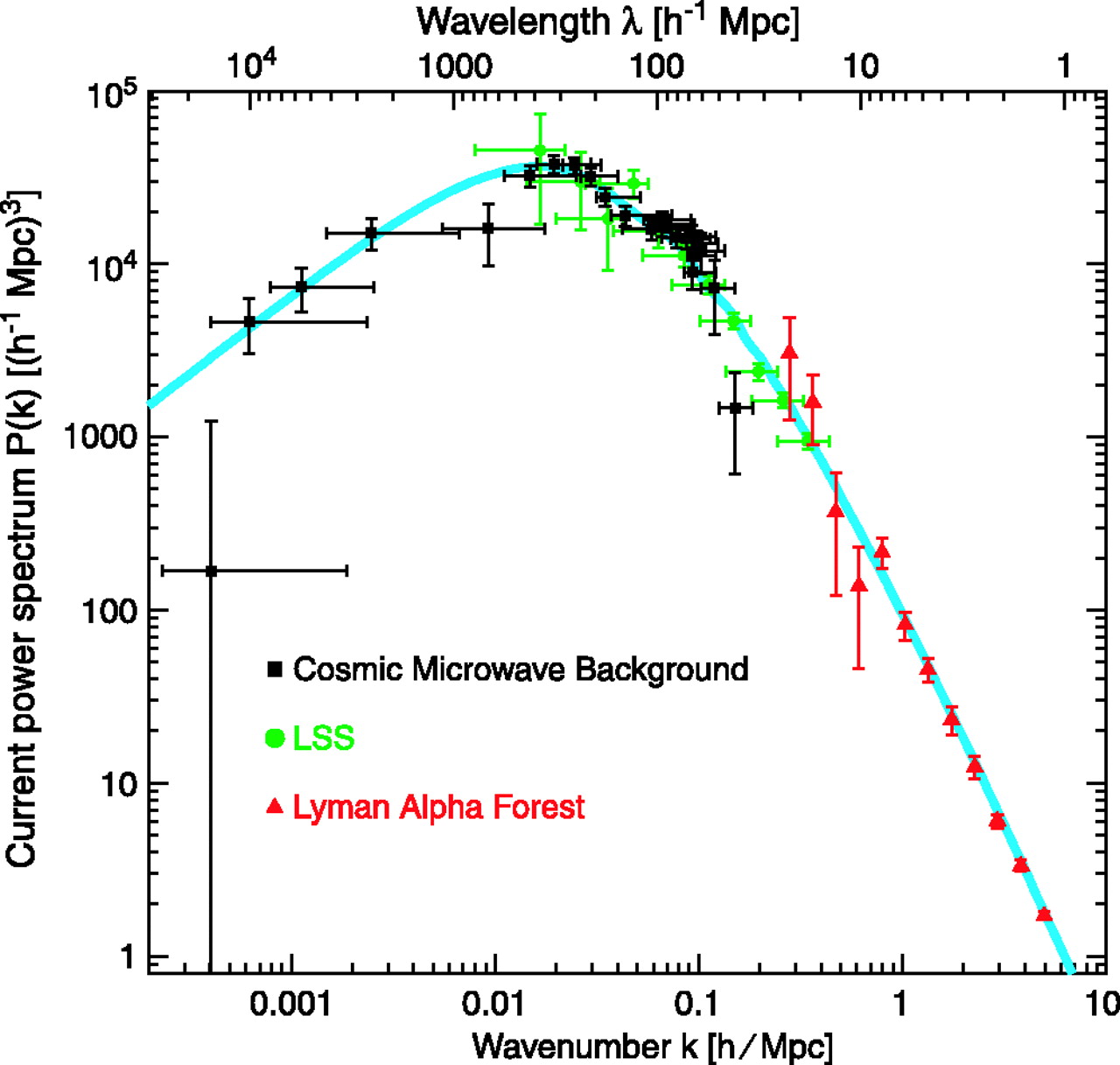} 
\end{center}
\caption{Best fit of today's matter power spectrum ($a_0=1$)
  from Max Tegmark's lecture notes~\cite{Tegmark:2002dg}.}
\label{fig:PS}
\end{figure}

We can solve this slight disagreement by considering perturbations of
different size separately.  First there are small perturbations (large
$k$), which enter the horizon of our expanding Universe during the
radiation-dominated era and hardly grow until matter-radiation
equality. Second, there are large perturbations with (small $k$),
which only enter the horizon during matter domination and never stop
growing.  This freezing of the growth before matter domination is
called the Meszaros effect. Following Eq.\eqref{eq:growth_rel} the
relative suppression in their respective growth between the entering
into the horizon and the radiation-matter equality is given by a the
correction factor relative to Eq.\eqref{eq:powerspec1} with $n=1$,
\begin{align}
P(k) 
\propto k\, \left( \frac{a_\text{enter}}{a_\text{eq}} \right)^2 \; .
\end{align}
We are interested in the wavelength of a mode that enters right at
matter-radiation equality and hence is the first mode that never stops
growing.  Assuming the usual scaling $\Omega_m/\Omega_r \propto a$ and we
first find
\begin{align}
\frac{a_\text{eq}}{a_0}
= \dfrac{\dfrac{\Omega_m(a_\text{eq})}{\Omega_r(a_\text{eq})}}{\dfrac{\Omega_m(a_0)}{\Omega_r(a_0)}}
= \frac{\Omega_r(a_0)}{\Omega_m(a_0)}
\approx 3 \cdot 10^{-4} \; ,\label{eq:OmegaRM}
\end{align}
again from PLANCK measurements. This allows us to integrate the
co-moving distance of Eq.\eqref{eq:d_comoving_int}. The lower and
upper limit of integration is $a=0$ and $a=a_\text{eq}=3\cdot
10^{-4}$, respectively. For these values of $a \ll 1$ the relativistic
matter dominates the universe, as can be seen in
Figure~\ref{fig:energy_vs_scale}. In this range the integrand of
Eq.\eqref{eq:d_comoving_int} is approximately
\begin{align}
\left[ \Omega_m(t_0) a(t)
 + \Omega_r(t_0) 
 + \Omega_\Lambda a(t)^4
 - (\Omega_t(t_0) -1 ) \, a(t)^2 \right]^{-1/2} 
\approx \frac{1}{\sqrt{ \Omega_r(t_0)}} \; .
\label{eq:matterradiationequatlity}
\end{align}
This is true even for $\Omega_\Lambda(t_0) > \Omega_m(t_0) >
\Omega_r(t_0)$ today. We can use Eq.\eqref{eq:matterradiationequatlity}
and write
\begin{align}
d^c_\text{eq} 
&\approx \int^{a_\text{eq}}_0\, da  \;
   \frac{1}{H_0\sqrt{\Omega_r(t_0)}} = \frac{a_\text{eq}}{H_0\sqrt{\Omega_r(t_0)}} \eqx{eq:OmegaRM} \frac{3\cdot 10^{-4}}{70\frac{\text{km}}{s \text{Mpc}}\sqrt{0.28 \times 3\cdot 10^{-4}}}=4.7\cdot 10^{-4} \frac{\text{Mpc }s}{\text{km}}\notag\\
&\Rightarrow \qquad 
\lambda_\text{eq}  = c \,d^c_\text{eq}\approx 3\cdot 10^5 \, \frac{\text{km}}{s} \; \times
                          4.7\cdot 10^{-4} \, \frac{\text{Mpc }s}{\text{km}}= 140\, \text{Mpc} \; . 
\label{eq:radmateqwavelenth}
\end{align}
This means that the growth of structures with a size of at least
140~Mpc never stalls, while for smaller structures the Meszaros effect
leads to a suppressed growth. The scaling of $\lambda_\text{eq}$ in
the radiation dominated era in dependence of the scale factor is given
by $\lambda_\text{eq}\propto a_\text{eq} c/H_0$. The co-moving
wavenumber is defined as $k=2\pi/\lambda$ and therefore
$k_\text{eq}\approx 0.05/\text{Mpc}$.  Using this scaling, $a\propto
1/k$, the power spectrum scales as
\begin{align}
P(k) 
\propto k\, \left( \frac{a_\text{enter}}{a_\text{eq}} \right)^2
= 
\begin{cases}k &  \qquad \text{$k < k_\text{eq}$ or $\lambda > 120$~Mpc} \\[3mm]
 \dfrac{1}{k^3} & \qquad \text{$k > k_\text{eq}$ or $\lambda < 120$~Mpc.}
\end{cases}
\label{eq:powerspectrumscaling}
\end{align}
The measurement of the power spectrum shown in Figure~\ref{fig:PS}
confirms these two regimes.\bigskip

\begin{figure}[!b]
\begin{center}
\includegraphics[width=0.8\textwidth]{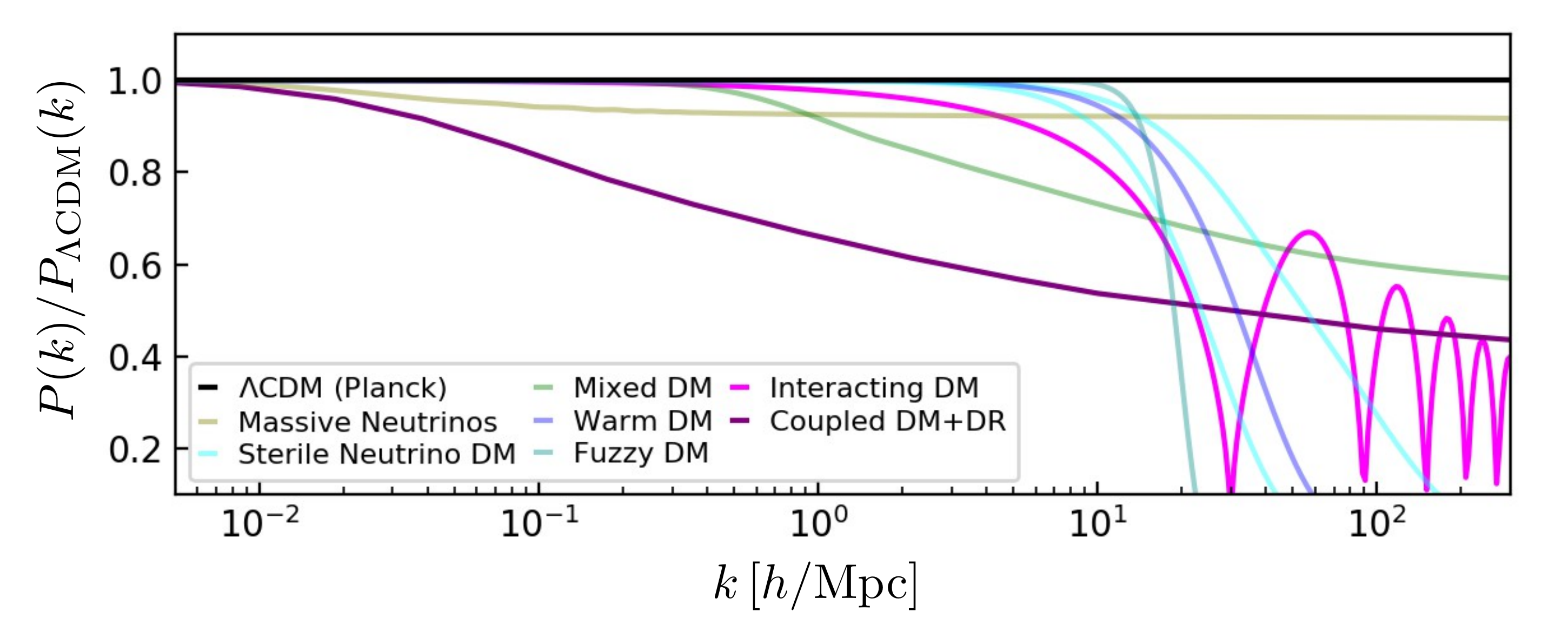} 
\end{center}
\caption{Sketch of the matter power spectrum for different dark matter
  scenarios normalized to the $\Lambda$CDM power spectrum. Figure
  from Ref.~\cite{AurelSchneider}.}
\label{fig:aurel}
\end{figure}

Even if pressure can be neglected for cold, collision-less dark matter,
its perturbations cannot collapse towards arbitrary small scales
because of the non-zero velocity dispersion. Once the velocity of dark
matter particles exceeds the escape velocity of a density
perturbation, they will stream away before they can be gravitationally
bound. This phenomenon is called \ul{free streaming} and allows
us to derive more properties of the dark matter particles from the
matter power spectrum. To this end we generalize the Jeans equation of
Eq.\eqref{eq:Jeans} to
\begin{align}
\ddot{\hat\delta}+2H\dot{\hat \delta}=\hat\delta\bigg[\frac{1}{2\mpl^2} \bar \rho -\bigg(c_s^\text{eff} \frac{ k a_0}{a}\bigg)^2\,\bigg]\; ,
\label{eq:Jeans2}
\end{align}
where in the term counteracting the gravitational attraction the speed
of sound is replaced by an effective speed of sound $c_s^\text{eff}$,\index{speed of sound}
whose precise form depends on the properties of the dark matter, We
show predictions for different dark matter particles in
Fig.~\cite{AurelSchneider}:
\begin{itemize}
\item[--] for \ul{cold dark matter} \index{dark matter!cold}with
\begin{align}
(c_s^\text{eff})^2 =  \frac{1}{m^2}\frac{\int dp\, p^2 f(p)}{\int dp\, f(p)}
\end{align}
  the speed of sound is replaced by the non-relativistic velocity
  distribution. This results in $c_s^\text{eff}\ll c_s$ and the cold
  dark matter Jeans length  \index{Jeans length} allows for halo structures as small as
  stars or planets. The dominant dark matter component in
  Figure~\ref{fig:PS} is cold collision-less dark matter and all lines
  in Figure~\ref{fig:aurel} is normalized to this the power spectrum;
\item[--] for \ul{warm dark matter} with  \index{dark matter!warm}
\begin{align}
c_s^\text{eff} = \frac{T}{m}
\end{align}
  the effective speed of sound is a function of temperature and
  mass. Warm dark matter is faster than cold dark matter and the
  effective speed of sound is larger. As a result, small structures
  are washed out as indicated by the blue line, because the free
  streaming length for warm dark matter is larger than for cold dark
  matter;
\item[--] \ul{sterile neutrinos} \index{sterile neutrinos}which we will introduce in
  Section~\ref{sec:neutrinos} feature
\begin{align}
(c_s^\text{eff})^2 = \frac{1}{m^2}\frac{\int dp\, p^2 f(p)}{\int dp\, f(p)} \; .
\end{align}
  They are a special case of warm dark matter, but the result of the
  integral depends on the velocity distribution, which is
  model-dependent. In general a suppression of small scale structures
  is expected and the resulting normalized power spectrum should end
  up between the two cyan lines;
\item[--] \ul{light, non-relativistic dark matter} \index{dark matter!light} or fuzzy dark matter \index{dark matter!fuzzy}
  which we will discuss in Section~\ref{sec:axions} gives
\begin{align}
c_s^\text{eff} = \frac{k}{m} \; .
\end{align}
  The effective speed of sound depends on $k$, leading to an even
  stronger suppression of small scale structures. The normalized power
  spectrum is shown in turquoise;
\item[--] for \ul{mixed warm and cold dark matter} \index{dark matter!mixed} with
\begin{align}
(c_s^\text{eff})^2 = \frac{T^2}{m^2} -\frac{\bar \rho}{2\mpl^2} \frac{a}{a_0 k}\frac{\hat \delta_C}{\hat \delta}
\end{align}
  the power spectrum is suppressed. Besides a temperature-dependent
  speed of sound for the warm dark matter component, a separate
  gravitational term for the cold dark matter needs to be added in the
  Jeans equation. Massive neutrinos are a special case of this
  scenario and in turn the power spectrum can be used to constrain SM
  neutrino masses;
\item[--] finally, \ul{self-interacting dark matter} \index{dark matter!self-interacting} with the
  distinctive new term
\begin{align}
(c_s^\text{eff})^2 = c_s^\text{dark} +R (\dot{\hat
    \delta}- \dot{\hat \delta}_\chi)\frac{a}{a_0 k}\frac{1}{\hat
    \delta}
\end{align}
  covers models from a dark
  force (dark radiation) to multi-component dark matter that could
  form dark atoms. Besides a potential dark sound speed, the Jeans
  equation needs to be modified by an interaction term. The effects on
  the power spectrum range from dark acoustic oscillations to a
  suppression of structures at multiple scales.
\end{itemize}

\newpage
\section{Relics}
\label{sec:relic_first}

After we understand the relic photons in the Universe, we can focus on
a set of different other relics, including the first dark matter
candidates. For those the main question is to explain the observed
value of $\Omega_\chi h^2 \approx 0.12$. Before we will eventually turn
to thermal production of massive dark matter particles, we can use a
similar approach as for the photons for relic neutrinos. Furthermore,
we will look at ways to produce dark matter during the thermal history
of the Universe, without relying on the thermal bath.

\subsection{Relic neutrinos}
\label{sec:neutrinos}

In analogy to photon decoupling, just replacing the photon--electron
scattering rate given in Eq.\eqref{eq:neutrino_photon_electron} by the
much larger neutrino--electron scattering rate, we can also compute
today's neutrino background in the Universe. At the point of
decoupling the neutrinos decouple from the electrons and photons, but
they will also lose the ability to annihilate among themselves through
the weak interaction. A well-defined density of neutrinos will
therefore \ul{freeze out} \index{freeze out} of thermal equilibrium. The two
processes
\begin{align}
\bar{\nu}_e e^- \to \bar{\nu}_e e^- 
\qqquad \text{and} \qqquad
\nu_e \bar{\nu}_e \to e^+ e^- 
\end{align}
are related through a similar scattering rate $\sigma_{\nu e}$ given
in Eq.\eqref{eq:neutrino_photon_electron}.

Because the neutrino scattering cross section is small,
we expect the neutrinos to decouple earlier. It turns out that this
happens before nucleosynthesis. This means that for the relic
neutrinos the electrons are the relevant degrees of freedom to compute
the decoupling temperature.  With the cross section given in
Eq.\eqref{eq:neutrino_photon_electron} the interaction rates for
relativistic neutrino--electron scattering \index{neutrino-electron scattering}is
\begin{align}
\Gamma_\nu
= \sigma_{\nu e} v n_\nu 
= \sigma_{\nu e} v n_e
\stackrel{\text{Eq.\eqref{eq:n_vs_temp}}}{\approx} 
  \frac{\pi \alpha^2 T^2}{s_w^4 m_W^4} \; 
  \dfrac{3\zeta_3}{4\pi^2} \; g T^3 
=\frac{3 \zeta_3}{4 \pi s_w^4} \; g \alpha^2 \;
  \frac{T^5}{m_W^4} \; .
\label{eq:neutrino_rate}
\end{align}
With only one generation of neutrinos in the initial state and a
purely left-handed coupling the number of relativistic degrees of
freedom relevant for this scattering process is $g = 1$.

Just as for the photons, we first compute the decoupling temperature.
To link the interaction rate to the Hubble constant, as given by
Eq.\eqref{eq:hubble_temp}, we need the effective number of degrees of
freedom in the thermal bath.  It now includes electrons, positrons, three
generations of neutrinos, and photons
\begin{align}
\geff(\Tdec)= \frac{7}{8} \left( 2 + 2 + 3 \times 2 \right) + 2 = 10.75 \; . 
\end{align}
With Eq.\eqref{eq:hubble_temp} and in analogy to
Eq.\eqref{eq:photon_decouple1} we find
\begin{align}
\frac{\Gamma_\nu}{H}
&= \frac{3\zeta_3 g \alpha}{4\pi^2 s_w^4} \; \frac{T^5}{m_W^4} \; 
   \frac{\sqrt{90} \mpl}{\pi} \; \frac{1}{\sqrt{\geff(T)} T^2} && \notag \\
&= \frac{9 \sqrt{10} \, \zeta_3}{4 \pi^2 s_w^4} \; 
   \alpha^2 \frac{g}{\sqrt{\geff(T)}} \; 
   \frac{\mpl T^3}{m_W^4} \really 1 
\qquad \Leftrightarrow \qquad & 
\Tdec &= 
  \left( \frac{4\pi^2 s_w^4}{9 \sqrt{10} \, \zeta_3} \;  
         \frac{m_W^4}{\mpl} \frac{\sqrt{\geff(\Tdec)}}{\alpha^2 g} \right)^{1/3} \notag \\
  &&&\approx (1~...~10)~\mev \; .
\label{eq:neutrino_decouple1} 
\end{align}
The relativistic neutrinos decouple at a temperature of a few MeV,
before nucleosynthesis.  From the full Boltzmann equation we would get
$\Tdec \approx 1$~MeV, consistent with our approximate
computation.\bigskip

Now that we know how the neutrinos and photons decouple from the
thermal bath, we follow the electron-neutrino-photon system from the
decoupling phase to today, dealing with one more relevant
effect. First, right after the neutrinos decouple around $\Tdec
\approx 1$~MeV, the electron with a mass of $m_e = 0.5$~MeV will drop
from the relevant relativistic degrees of freedom; in the following
phase the electrons will only serve as a background for the photon.
For the evolution to today we only have
\begin{align}
\geff(\Tdec~...~T_0)  = \frac{7}{8} \times 6 + 2 = 7.25
\end{align}
\index{degrees of freedom}
relativistic degrees of freedom.  The decoupling of the massive
electron adds one complication: in the full thermodynamic calculation
we need to assume that their entropy is transferred to the photons,
the only other particles still in equilibrium. We only quote the
corresponding result from the complete calculation: because the
entropy in the system should not change in this electron decoupling
process, the temperature of the photons jumps from \index{entropy}
\begin{align}
T_\gamma = T_\nu
\qquad \to \qquad 
T_\gamma = \left( \frac{11}{4} \right)^{1/3} T_\nu \; .
\label{eq:temp_jump}
\end{align}
If the neutrino and photon do not have the same temperature we can use
Eqs.\eqref{eq:def_eff_freedom1} and~\eqref{eq:def_eff_freedom2} to
obtain the combined relativistic matter density at the time of neutrino
decoupling,
\begin{align}
\rho_r (T)
&= \frac{\pi^2}{30} \; \geff(T) \; T^4 \notag \\
&= \frac{\pi^2}{30} \; 
   \left( 2 \; \frac{T_\gamma^4}{T^4} 
        + \frac{7}{8} \; 6 \; \frac{T_\nu^4}{T^4} \right) \; T^4 
\quad \Rightarrow \quad 
\rho_r (T_\gamma)
= \frac{\pi^2}{30} \; 
   \left( 2 
        + \frac{21}{4} \; \left( \frac{4}{11}  \right)^{4/3} \right) \; T_\gamma^4 
= \frac{\pi^2}{30} \; 3.4 \; T_\gamma^4 \; ,
\label{eq:neutrino_decouple2}
\end{align}
or $\geff(T) = 3.4$.  This assumes that we measure the current
temperature of the Universe through the photons. Assuming a constant
suppression of the neutrino background, its temperature and the total
relativistic energy density today are
\begin{align}
T_{0,\nu} = 1.7 \cdot 10^{-4}~\ev 
\qquad \text{and} \qquad
\rho_r (T_0)
&= \frac{\pi^2}{30} \; 3.4 \; T_{0, \gamma}^4 
 = 1.1 \; T_{0, \gamma}^4 \; .
\label{eq:nu_gamma_temp}
\end{align}
From the composition in Eq.\eqref{eq:neutrino_decouple2} we see that
the current relativistic matter density of the Universe is split
roughly $60-40$ between the photons at $T_{0,\gamma} = 2.4 \cdot
10^{-4}$~eV and the neutrinos at $T_{0,\nu} = 1.7 \cdot
10^{-4}$~eV. The normalized relativistic relic density today becomes \index{relic!neutrinos}
\begin{align}
\boxed{
\Omega_r(t_0) h^2 = \frac{\rho_r(T_0) h^2}{3 \mpl^2 H_0^2} 
= 0.54 \; \left( \frac{2.4 \cdot 10^{-4}~\ev}{2.5 \cdot 10^{-3}~\ev} \right)^4 
= 4.6 \cdot 10^{-5}
} \; .
\label{eq:omega_r_meas}
\end{align}
Note that for this result we assume that the neutrino mass never plays
a role in our calculation, which is not at all a good approximation.\bigskip

We are now in a position to answer the question whether a massive, stable fourth neutrino could explain the observed dark matter relic density. 
With a moderate
mass, this fourth neutrino decouples in a relativistic state. In that
case we can relate its number density \index{number density}to the photon temperature through
Eq.\eqref{eq:neutrino_decouple2},
\begin{alignat}{9}
n_\nu(T)
\eqx{eq:n_vs_temp} \frac{3}{4} \, \frac{2 \zeta_3}{\pi^2} \, T_\nu^3
\eqx{eq:temp_jump} \frac{6 \zeta_3}{11 \pi^2}  T_\gamma^3 \; .
\end{alignat}
With decreasing temperature a heavy neutrino will at some point become
non-relativistic. This means we use the non-relativistic relation to
compute its energy density today,
\begin{align}
\rho_\nu(T_0) &= m_\nu n_\nu(T_0) = m_\nu \frac{6 \, \zeta_3}{11 \pi^2} \; T_{0,\gamma}^3 \notag \\
\Rightarrow \qquad 
\Omega_\nu h^2 &= m_\nu \frac{6 \, \zeta_3}{11 \pi^2} \, T_{0,\gamma}^3 \; \frac{h^2}{3 \mpl^2 H_0^2}
= \frac{m_\nu}{30} \; \frac{(2.4 \cdot 10^{-4})^3}{(2.5 \cdot 10^{-3})^4} \; \frac{1}{\ev}
= \frac{m_\nu}{85~\ev} \; .
\label{eq:hot_neutrinos1}
\end{align}
For an additional, heavy neutrino to account for the observed dark matter we
need to require
\begin{align}
\Omega_\nu h^2 \really \Omega_\chi h^2 \approx 0.12
\qquad \Leftrightarrow \qquad 
m_\nu \approx 10~\ev \; .
\label{eq:hot_neutrinos2}
\end{align}
This number for \ul{hot neutrino dark matter} \index{dark matter!hot} is not
unreasonable, as long as we only consider the dark matter relic
density today. The problem appears when we study the formation of
galaxies, where it turns out that dark matter relativistic at the
point of decoupling will move too fast to stabilize the accumulation
of matter. We can look at Eq.\eqref{eq:hot_neutrinos2} another way: if
all neutrinos in the Universe add to more than this mass value, they
predict hot dark matter with a relic density more than then entire
dark matter in the Universe. This gives a stringent upper bound on the
neutrino mass scale.

\subsection{Cold light dark matter}
\label{sec:axions}

Before we introduce cold and much heavier dark matter, there is
another scenario we need to discuss.  Following
Eq.\eqref{eq:hot_neutrinos2} a new neutrino with mass around $10$~eV
could explain the observed relic density. The problem with thermal
neutrino dark matter is that it would be relativistic at the wrong
moment of the thermal history, causing serious issues with structure
formation as discussed in Section~\ref{sec:structure}. The obvious
question is if we can modify this scenario such that light dark matter
remains non-relativistic. To produce such \ul{light cold dark
  matter} \index{dark matter!light} we need a non-thermal production process. \bigskip

We consider a toy model for light cold dark matter with a
\ul{spatially homogeneous} but time-dependent complex scalar
field $\phi(t)$ with a potential $V$. For the latter, the Taylor
expansion is dominated by a quadratic mass term $m_\phi$. Based on the
invariant action with the additional determinant of the metric $g$, 
describing the expanding Universe, the Lagrangian for a single complex
scalar field reads
\begin{align}
\frac{1}{\sqrt{|g|}} \; \lag 
= (\partial^\mu \phi^*) (\partial_\mu \phi) - V(\phi)
= (\partial^\mu \phi^*) (\partial_\mu \phi) - m_\phi^2 \; \phi^* \phi \; .
\end{align}
Just as a side remark, the difference between the Lagrangians for real
and complex scalar fields is a set of factors $1/2$ in front of each
term. In our case the equation of motion for a spatially homogeneous field
$\phi(t)$ is
\begin{align}
0 
&= \partial_t \, 
   \left(\frac{\partial \lag}{\partial (\partial_t \phi^*)} \right) 
- \frac{\partial \lag}{\partial \phi^*} \notag \\
&= \partial_t \, 
   \left( \sqrt{|g|} \, \partial_t \phi \right) 
         + \sqrt{|g|} \, m_\phi^2 \phi \notag \\
&= ( \partial_t \sqrt{|g|} ) \, ( \partial_t \phi )
         + \sqrt{|g|} \, \partial_t^2 \phi 
         + \sqrt{|g|} \, m_\phi^2 \phi 
= \sqrt{|g|}
   \left[ \frac{(\partial_t \sqrt{|g|})}{\sqrt{|g|}} \; (\partial_t \phi)
         + \partial_t^2 \phi 
         + m_\phi^2 \phi \right] \; .
\end{align}
For example from Eq.\eqref{eq:metric_gmunu} we know that in flat space
$(k=0)$ the determinant of the metric is $|g| = a^6$, giving us
\begin{align}
0 
= \frac{(\partial_t a^3)}{a^3} \; (\partial_t \phi)
         + \partial_t^2 \phi 
         + m_\phi^2 \phi 
= \frac{3 \dot a}{a} \; \dot \phi
         + \ddot \phi 
         + m_\phi^2 \phi \; .
\end{align}
Using the definition of the Hubble constant in
Eq.\eqref{eq:scale_linear} we find that the expansion of the Universe
is responsible for the friction term in
\begin{align}
\boxed{ 
\ddot \phi(t) + 3 H \dot \phi(t) + m_\phi^2 \phi(t) = 0 } \; .
\label{eq:eq_axion_dm}
\end{align}
We can solve this equation for the evolving Universe, described by a
decreasing Hubble constant with increasing time or decreasing
temperature, Eq.\eqref{eq:hubble_temp}.  If for each regime we assume
a constant value of $H$ --- an approximation we need to check later
--- and find
\begin{align}
\phi(t) = e^{i \omega t} 
\qquad &\Rightarrow \qquad
\dot \phi(t) = i \omega \phi(t)
\qquad \Rightarrow \qquad
\ddot \phi(t) = - \omega^2 \phi(t) \notag \\
&\Rightarrow \qquad
- \omega^2 + 3 i H \omega + m_\phi^2 = 0 \notag \\
&\Rightarrow \qquad
\omega 
= \frac{3i}{2} H \pm \sqrt{-\frac{9}{4}H^2 + m_\phi^2} \; .
\end{align}
This functional form defines three distinct regimes in the evolution
of the Universe:

\begin{itemize}
\item[--] 
In the \ul{early Universe $H \gg m_\phi$} the two solutions are $\omega =
0$ and $\omega = 3iH$.  The scalar field value is a combination of a
constant mode and an exponentially decaying mode.
\begin{align}
\phi(t) = \phi_1 + \phi_2 \; e^{- 3 H t}
        \stackrel{\text{time evolution}}{\longrightarrow} \phi_1 \; .
\label{eq:relic_scalar1}
\end{align}
The scalar field very rapidly settles in a constant field value and
stays there.  There is no good reason to assume that this constant
value corresponds to a minimum of the potential. Due to the Hubble
friction term in Eq.\eqref{eq:eq_axion_dm}, there is simply no time
for the field to evolve towards another, minimal value. This behavior
gives the process its name, \ul{misalignment mechanism}\index{misalignment mechanism}. For
our dark matter considerations we are interested in the energy
density.  Following the virial theorem we assume that the total energy
density stored in our spatially constant field is twice the average
potential energy $V = m_\phi^2 |\phi|^2/2$.  After the rapid decay of
the exponential contribution this means
\begin{align}
\rho(t) \to m_\phi^2 \; \phi_1^2\; . 
\label{eq:relic_scalar2}
\end{align}

\item[--] A transition point in the evolution of the universe occurs when
the evolution of the field $\phi$ switches from the exponential
decay towards a constant value $\phi_1$ to an oscillation mode. If
we identify the oscillation modes of the field $\phi$ with a dark
matter degree of freedom, this point in the thermal history defines
the production of cold, light dark matter, 
\begin{align}
H_\text{prod} \approx m_\phi 
\qquad \Leftrightarrow \qquad 
\omega \approx \frac{3i}{2} H_\text{prod} \; .
\label{eq:axion_transition}
\end{align}

\item[--] 
For the \ul{late Universe $H \ll m_\phi$} we expand the complex
eigen-frequency one step further,
\begin{align}
\omega 
= \frac{3i}{2} H \pm m_\phi \sqrt{1 - \frac{9 H^2}{4 m_\phi^2}} 
\approx \frac{3i}{2} H \pm m_\phi \left( 1 - \frac{9 H^2}{8 m_\phi^2} \right) 
\approx \pm m_\phi + \frac{3i}{2} H  \; .
\end{align}
The leading time dependence of the scalar field is an oscillation. The
subleading term, suppressed by $H/m_\phi$, describes an exponentially
decreasing field amplitude,
\begin{align}
\phi(t) = \phi_3 \; e^{\pm i m_\phi t} \; e^{-3H/2 t} \; .
\label{eq:amp_late}
\end{align}
A modification of the assumed constant $H$ value changes the rapid
decay of the amplitude, but should not affect these main features.  We
can understand the physics of this late behavior when we compare it to the
variation of the scale factor for constant $H$ given in
Eq.\eqref{eq:scale_linear},
\begin{align}
H = \frac{\dot a(t)}{a(t)} 
\quad \Rightarrow \quad 
a(t) \propto e^{H t}
\quad \Rightarrow \quad 
\rho(t) 
\eqx{eq:amp_late} m_\phi^2 \; | \phi_3 |^2 \; e^{-3H t}
\propto \frac{1}{a(t)^3} 
\quad \Leftrightarrow \quad 
\boxed{ \frac{\rho(t)}{\rho_0}
\propto \frac{a_0^3}{a(t)^3} } \; .
\label{eq:scalar_scale}
\end{align}
The energy density of the scalar field in this late regime is
inversely proportional to the space volume element in the expanding
Universe.  This relation is exactly what we expect from a
\ul{non-relativistic relic} without any interaction or quantum
effects.
\end{itemize}
\bigskip

Next, we can use Eq.\eqref{eq:scalar_scale} combined with the
assumption of constant $H$ to approximately relate the dark matter
relic densities at the time of production with today $a_0 = 1$,
\begin{align}
0.12 \approx \Omega_\chi h^2 
= \frac{\rho_\chi}{\rho_c} \; h^2 
&\eqx{eq:relic_scalar2} \frac{m_\phi^2 \phi^2(t_0)}{(2.5 \cdot 10^{-3}~\ev)^4 }  \; h^2 \notag \\
&\eqx{eq:scalar_scale} \frac{m_\phi^2 \phi^2(t_\text{prod})}{(2.5 \cdot 10^{-3}~\ev)^4 } 
  \; \frac{a(t_\text{prod})^3}{a_0^3} \; h^2  \; .
\end{align}
Using our thermodynamic result $a(T) \propto 1/T$ from
Eq.\eqref{eq:thermodynamics1} and the approximate relation between the
Hubble parameter and the temperature at the time of production we find
\begin{align}
\sqrt{0.12}
= \frac{m_\phi \phi(t_\text{prod})}{(2.5 \cdot 10^{-3}~\ev)^2} 
  \; \frac{T_0^{3/2}}{T_\text{prod}^{3/2}} \; h 
\stackrel{\text{Eq}.\eqref{eq:hubble_temp}}{\approx} \frac{m_\phi \phi(t_\text{prod})}{(2.5 \cdot 10^{-3}~\ev)^2} 
  \; \frac{T_0^{3/2}}{(H_\text{prod} \mpl)^{3/4}} \; h \; .
\end{align}
Moreover, from Eq.\eqref{eq:axion_transition} we know that the Hubble
constant at the time of dark matter production is $H_\text{prod}
\sim m_\phi$.  This leads us to the relic density condition for dark
matter produced by the misalignment mechanism,
\begin{align}
0.35
&= \frac{m_\phi \phi(t_\text{prod})}{(2.5 \cdot 10^{-3}~\ev)^2} 
  \; \frac{T_0^{3/2}}{(m_\phi \mpl)^{3/4}} \; h 
\label{eq:scalar_density} \\
\Leftrightarrow \qquad 
m_\phi \phi(t_\text{prod})
&= \frac{1}{2} \; \frac{(2.5 \cdot 10^{-3}~\ev)^2}{(2.4 \cdot 10^{-4}~\ev)^{3/2}}
  \; (m_\phi \mpl)^{3/4} 
\qquad \Leftrightarrow \qquad 
\boxed{
m_\phi \phi(t_\text{prod})
\approx (m_\phi \mpl)^{3/4} \ev^{1/2} } \; . \notag 
\end{align}
This is the general relation between the mass of a cold dark matter
particle and its field value, based on the observed relic density. If
the misalignment mechanism should be responsible for today's dark
matter, inflation occuring after the field $\phi$ has picked its
non-trivial starting value will have to give us the required spatial
homogeneity. This is exactly the same argument we used for the
relic photons in Section~\ref{sec:cmb}.  We can then link today's
density to the density at an early starting point through the
evolution sketched above.

Before we illustrate this behavior with a specific model we can
briefly check when and why this dark matter candidate is
non-relativistic. If through some unspecified quantization we identify
the field oscillations of $\phi$ with dark matter particles, their
non-relativistic velocity is linked to the field value $\phi$ through
the quantum mechanical definition of the momentum operator,
\begin{align}
v = \frac{\hat{p}}{m} \propto \frac{\partial \phi}{\partial x} \; ,
\end{align}
assuming an appropriate normalization by the field value $\phi$. It
can be small, provided we find a mechanism to keep the field $\phi$
spatially constant.  What is nice about this model for cold, light
dark matter is that it requires absolutely no particle physics
calculations, no relativistic field theory, and can always be tuned to
work.\bigskip

\subsection{Axions}
\label{sec:axions2}

The best way to guarantee that a particle is massless or light is
through a symmetry in the Lagrangian of the quantum field theory. For
example, if the Lagrangian for a real spin-0 field $\phi(x) \equiv
a(x)$ is invariant under a constant shift $a(x)\to a(x)+ c$, a mass
term $m_a^2 a^2$ breaks this symmetry. Such particles, called
\ul{Nambu-Goldstone bosons}\index{Nambu-Goldstone boson}, appear in
theories with broken global symmetries.  Because most global symmetry
groups are compact or have hermitian generators and unitary
representations, the Nambu-Goldstone bosons are usually CP-odd.

We illustrate their structure using a complex scalar field
transforming under a $U(1)$ rotation, $\phi \to \phi \,e^{i a/f_a}$. A
vacuum expectation value $\langle \phi \rangle =f_a$ leads to
spontaneous breaking of the $U(1)$ symmetry, and the Nambu-Goldstone
boson $a$ will be identified with the broken generator of the
phase. If the complex scalar has couplings to chiral fermions $\psi_L$
and $\psi_R$ charged under this $U(1)$ group, the Lagrangian includes
the terms
\begin{align}
\lag \supset 
  i \overline \psi_L \gamma^\mu \partial_\mu \psi_L 
+ i \overline \psi_R \gamma^\mu \partial_\mu \psi_R 
- y\, \phi \, \overline \psi_R \psi_L+ \text{h.c.}
\end{align}
We can rewrite the Yukawa coupling such that after the rotation the
phase is absorbed in the definition of the fermion fields,
\begin{align}
y \, \phi \, \overline \psi_R  \psi_L 
\to 
y\, f_a \, \overline \psi_R  e^{ia/f_a} \,\psi_L 
\equiv y \, f_a \, \overline \psi_R' \psi'_L 
\qquad \text{with} \quad 
\psi_{R,L}' = e^{\mp ia/(2f_a)}\psi_{R,L} \; .
\end{align}
This gives us a fermion mass $m_\psi=y f_a$.  In the new basis the
kinetic terms read
\begin{align}
  i \overline \psi_L \gamma^\mu \partial_\mu \psi_L 
+ i \overline \psi_R \gamma^\mu \partial_\mu \psi_R
&= i\,\overline \psi_L' e^{-ia/(2f_a)} \gamma^\mu \partial_\mu e^{ia/(2f_a)} \psi_L'
 + i\,\overline \psi_R' e^{ia/(2f_a)} \gamma^\mu \partial_\mu e^{-ia/(2f_a)} \psi_R'
  \notag\\
&= i\,\overline \psi_L' \gamma^\mu\left( \partial_\mu +i\frac{(\partial_\mu a)}{2f_a}\right) \psi_L' 
 + i\,\overline \psi_R' \gamma^\mu \left(\partial_\mu -i\frac{(\partial_\mu a)}{2f_a}\right) \psi_R' +\mathcal{O}(f_a^{-2})\notag\,\\
&=i\,\overline \psi \gamma^\mu \partial_\mu \psi 
+ \frac{(\partial_\mu a)}{2f_a} \, \overline \psi\, \gamma^\mu \gamma_5 \,\psi  +\mathcal{O}(f_a^{-2}) \; ,
\label{eq:axcoup}
\end{align}
where in the last line we define the four-component spinor $\psi
\equiv (\psi_L', \psi_R')$. The derivative coupling and the axial
structure of the new particle $a$ are evident. Other structures arise
if the underlying symmetry is not unitary, as is the case for
space-time symmetries for which the group elements can be written as
$e^{\alpha/f}$ and a calculation analogous to Eq.\eqref{eq:axcoup}
leads to scalar couplings. The Nambu-Goldstone boson of the scale
symmetry, the dilaton, is an example of such a case.

Following Eq.\eqref{eq:axcoup}, the general shift-symmetric Lagrangian
for such a CP-odd pseudo-scalar reads
\begin{align}
\lag =
\frac{1}{2} (\partial_\mu a) \, (\partial^\mu a)
+\frac{a}{f_a}\frac{\alpha_s}{8\pi}G^a_{\mu\nu}\widetilde G^{a\,\mu\nu}
+c_\gamma\frac{a}{f_a}\frac{\alpha}{8\pi} F_{\mu\nu}\widetilde F^{\mu\nu}
+\frac{(\partial_\mu a)}{2f_a}  \sum_\psi c_\psi \; \overline \psi \gamma^\mu\gamma_5 \psi \; .
\label{eq:axlag}
\end{align}
The coupling to the Standard Model is mediated by a derivative
interactions to the (axial) current of all SM fermions.  Here $\widetilde
F_{\mu\nu}=\epsilon_{\mu\nu\rho\tau} F^{\rho\tau}/2$ and
correspondingly $\widetilde G_{\mu \nu}$ are the dual field-strength
tensors.  This setup has come to fame as a possible solution to the
so-called \ul{strong CP-problem}. In QCD, the dimension-4
operator
\begin{align}
\frac{\theta_\text{QCD}}{8\pi}G^a_{\mu\nu}\widetilde G^{a\,\mu\nu}
\label{eq:thetaangle}
\end{align}
respects the $SU(3)$ gauge symmetry, but would induce observable
CP-violation, for example a dipole moment for the neutron. Note that
it is non-trivial that this operator cannot be ignored, because it
looks like a total derivative, but due to the topological structure of
$SU(3)$, it doesn't vanish. 
The non-observation of a
neutron dipole moment sets very strong constraints on $\theta_\text{QCD} <
10^{-10}$. This almost looks like this operator shouldn't be there and
yet there is no symmetry in the Standard Model that forbids it.

Combining the gluonic operators in Eq.\eqref{eq:axlag} and
Eq.\eqref{eq:thetaangle} allows us to solve this problem
\begin{align}
\lag =
\frac{1}{2} (\partial_\mu a) \, (\partial^\mu a)
+\frac{\alpha_s}{8\pi} 
 \left(  \frac{a}{f_a} - \theta_\text{QCD} \right) \;
 G^a_{\mu\nu}\widetilde G^{a\,\mu\nu}
+c_\gamma\frac{a}{f_a}\frac{\alpha}{8\pi} F_{\mu\nu}\widetilde F^{\mu\nu}
+\frac{(\partial_\mu a)}{2f_a}  \sum_\psi c_\psi \; \overline \psi \gamma^\mu\gamma_5 \psi \; .
\end{align}
With this ansatz we can combine the $\theta$-parameter and the scalar
field, such that after quarks and gluons have formed hadrons, we can
rewrite the corresponding effective Lagrangian including the terms
\begin{align}
\lag_\text{eff} 
\supset \frac{1}{2} (\partial_\mu a) \, (\partial^\mu a) 
- \frac{1}{2} \kappa^2 \left(\theta_\text{QCD} -\frac{a}{f_a} \right)^2
- \lambda_a \left(\theta_\text{QCD} -\frac{a}{f_a} \right)^4+\mathcal{O}(f_a^{-6})\; .
\label{eq:axlag2}
\end{align}
The parameters $\kappa$ and $\lambda_a$ depend on the QCD
dynamics. This contribution provides a potential for $a$ with a 
minimum at $\langle a \rangle/f_a = \theta_\text{QCD}$.
In other words, the shift symmetry has eliminated the CP-violating
gluonic term from the theory. Because of its axial couplings to matter
fields, the field $a$ is called \ul{axion}.\index{axion}

The axion would be a bad dark matter candidate if it was truly
massless. However, the same effects that induce a potential for the
axion also induce an axion mass. Indeed, from Eq.\eqref{eq:axlag2} we
immediately see that
\begin{align}
m_a^2 \equiv 
\frac{\partial^2 V}{\partial a^2} \Bigg|_{a=f_a\theta_\text{QCD}}=\frac{\kappa^2}{f_a^2} \,,
\end{align}
This seems like a contradiction, because a mass term breaks the shift
symmetry and for a true Nambu-Goldstone boson, we expect this term to
vanish.  However, in the presence of quark masses, the transformations
in Eq.\eqref{eq:axcoup} do not leave the Lagrangian invariant under
the shift symmetry
\begin{align}
m_q \overline \psi_R \psi_L 
\to m_q \overline \psi_R' e^{2ia/f_a} \psi_L' \; .
\end{align}
Fermion masses lead to an explicit breaking of the shift symmetry and
turn the axion a pseudo Nambu-Goldstone boson, similar to the pions in
QCD. For more than one quark flavor it suffices to have a single
massless quark to recover the shift symmetry. We can determine this
mass term from a chiral Lagrangian in which the fundamental fields are
hadrons instead of quarks and find
\begin{align}
m_a^2
=\frac{m_um_d}{(m_u+m_d)^2}\frac{m_\pi^2 f_\pi^2}{f_a^2}
\approx \frac{m_\pi^2 f_\pi^2}{2 f_a^2} \; ,
\label{eq:axionmass}
\end{align}
where $f_\pi \approx m_\pi \approx 140$~MeV are the pion decay
constant and mass, respectively. This term vanishes in the limit
$m_u\to 0$ or $m_d\to 0$ as we expect from the discussion above. In
the original axion proposal, $f_a \sim v$ and therefore $m_a\sim
10$~keV. Since the couplings of the axion are also fixed by the value
of $f_a$, such a particle was excluded very fast by searches for rare
kaon decays, like for instance $K^+\to \pi^+ a$. In general, $f_a$ can
be a free parameter and the mass of the axion can be smaller and its
couplings can be weaker.

This leaves the question for which model parameters the axion makes a
good dark matter candidate. Since the value of the axion field is not
necessarily at the minimum of the potential at the time of the QCD
phase transition, the axion begins to oscillates around the minimum
and the oscillation energy density contributes to the dark matter
relic density. This is a special version of the more general
misalignment mechanism described in the previous section. We can then
employ Eq.\eqref{eq:scalar_density} and find the relation for the
observed relic density
\begin{align}
m_a a(t_\text{prod})\approx (m_a \mpl)^{3/4} \, \text{eV}^{1/2}\,.
\end{align}
The maximum field value of the oscillation mode is given by $a(t_\text{prod})\approx f_a$ and therefore 
\begin{align}
m_a f_a \stackrel{\eqref{eq:axionmass}}{\approx} m_\pi f_\pi\,\qquad 
\Rightarrow \qquad  \,\, \boxed{m_a\approx \frac{(m_\pi f_\pi)^{4/3}}{\mpl} \,\text{eV}^{-2/3}}\,.
\label{eq:axionrelic}
\end{align}
This relation holds for $m_a\approx 2 \cdot 10^{-6}$~eV, which
corresponds to $f_a\approx 10^{13}$~GeV. For heavier axions and
smaller values of $f_a$, the axion can still constitute a part of the
relic density. For example with a mass of $m_a=6 \cdot  10^{-5}$~eV and
$f_a\approx 3 \cdot 10^{11}$ GeV, axions make up one per-cent of the
observed relic density.\bigskip

\begin{figure}[b!]
\begin{center}
\includegraphics[width=0.70\textwidth]{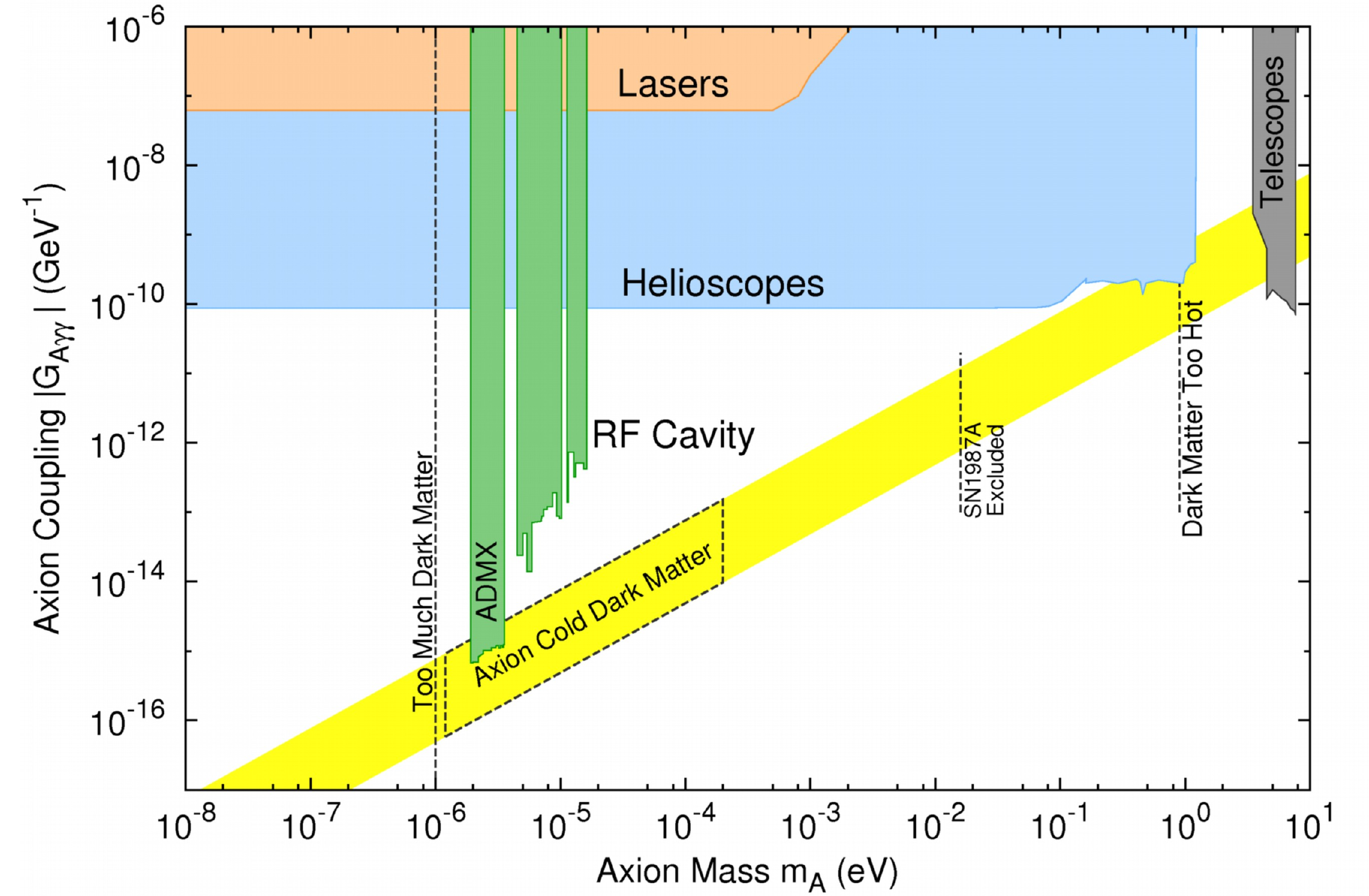}
\end{center}
\caption{Range of masses and couplings for which the axion can be a
  viable cold dark matter candidate. Figure from
  Ref.~\cite{Tanabashi:2018oca}.}
\label{fig:axionplot}
\end{figure}

Dark Matter candidates with such low masses are hard to detect and we
usually takes advantage of their couplings to photons. In
Eq.\eqref{eq:axlag}, there is no reason why the coupling $c_\gamma$
needs to be there. It it neither relevant for the strong CP problem
nor for the axion to be dark matter. However, from the perspective of
the effective theory we expect all couplings which are allowed by the
assumed symmetry structure to appear. This includes the axion coupling
to photons.  If the complete theory, the axion coupling to gluons needs
to be induced by some physics at the mass scale $f_a$. This can be
achieved by axion couplings to SM quarks, or by axion couplings to
non-SM fields that are color-charged but electrically neutral. Even in
the latter case there is a non-zero coupling induced by the axion
mixing with the SM pion after the QCD phase transition. Apart from
really fine-tuned models the axion therefore couples to photons with
an order-one coupling constant $c_\gamma$.

In Figure~\ref{fig:axionplot} the yellow band shows the range of axion
couplings to photons for which the models solve
Eq.\eqref{eq:axionmass}. The regime where the axion is a viable dark
matter candidate is dashed. It is notoriously hard to probe axion dark
matter in the parameter space in which they can constitute dark
matter. Helioscopes try to convert axions produced in the sun into
observable photons through a strong magnetic field. Haloscopes like
ADMX use the same strategy to search for axions in the dark matter
halo.

The same axion coupling to photons that we rely on for axion detection
also allows for the decay $a\to \gamma\gamma$. This looks dangerous
for a dark matter candidate, to we can estimate the corresponding
decay width,
\begin{align}
\Gamma(a\to \gamma\gamma)
&=\frac{\alpha^3}{256\,\pi^3}\frac{m_a^3}{f_a^2}| \; c_\gamma|^2\\
&= \frac{1}{137^3} \frac{1}{256\,\pi^3}\frac{(6 \cdot 10^{-6}\,\text{eV})^3}{(10^{22}\,\text{eV})^2} \; |c_\gamma|^2
\approx 1 \cdot  10^{-70} \,\text{eV} \; |c_\gamma|^2\notag
\end{align}
Assuming $c_\gamma=1$ this corresponds to a lifetime of
$\tau=1/\Gamma\approx 2 \cdot 10^{47}$ years, many orders of magnitude
larger than the age of the universe. \bigskip

While the axion is particularly interesting because it addresses the
strong CP problem and dark matter at the same time, we can drop the
relation to the CP problem and study \ul{axion-like
  particles}\index{axion-like particle} (ALPs) as dark matter. For
such a light pseudoscalar particle Eqs.\eqref{eq:axionmass}
and~\eqref{eq:axionrelic} are replaced by the more general relations
\begin{align}
m_a= \frac{\mu^2}{f_a} 
\qquad \Rightarrow \qquad 
m_a=\frac{\mu^{8/3}}{\mpl}\,\text{eV}^{-2/3}\,.
\end{align}
where $\mu$ is a mass scale not related to QCD.  In such models, the
axion-like pseudoscalar can be very light. For example for $\mu
\approx 100$~eV, the axion mass is $m_a\approx 10^{-22}$~eV. For such
a low mass and a typical velocity of $v\approx 100
\,\text{km}/\text{s}$, the de-Broglie wavelength is around $1$~Kpc,
the size of a galaxy.  This type of dark matter is called
\ul{fuzzy dark matter}\index{dark matter!fuzzy} and can inherit
the interesting properties of Bose-Einstein condensates or
super-fluids.

\subsection{Matter vs anti-matter}
\label{sec:matter}
\index{anti-matter}
Before we look at a set of relics linked to dark matter, let us follow
a famous argument fixing the conditions which allow us to live in a Universe
dominated by matter rather than anti-matter. In this section we will
largely follow Kolb \& Turner~\cite{KolbTurner}. The observational
reasoning why matter dominates the Universe goes in two steps: first,
matter and anti-matter cannot be mixed, because we do not observe
constant macroscopic annihilation; second, if we separate matter and
anti-matter we should see a boundary with constant annihilation
processes, which we do not see, either. So there cannot be too 
much anti-matter in the Universe.

The corresponding measurement is usually formulated in terms of the
observed baryons, protons and neutrons, relative to the number of photons,
\begin{align}
\frac{n_B}{n_\gamma} \approx 6 \cdot 10^{-10} \; .
\label{eq:baryon_photon}
\end{align}
The normalization to the photon density is motivated by the fact that
this ratio should be of order unity in the very early
Universe. Effects of the Universe's expansion and cooling to first
approximation cancel. Its choice is only indirectly related to the
observed number of photons and instead assumes that the photon density
as the entropy density in thermal equilibrium. As a matter of fact, we use this
number already in Eq.\eqref{eq:electron_density}. \index{entropy}\bigskip

To understand Eq.\eqref{eq:baryon_photon} we start by remembering that
in the hot universe anti-quarks and quarks or anti-baryons and baryons are pair-produced out
of a thermal bath and annihilate with each other in thermal
equilibrium. Following the same argument as for the photons, the
baryons and anti-baryons decouple from each other when the temperature
drops enough. In this scenario we can estimate the
ratio of baryon and photon densities from Eq.\eqref{eq:n_vs_temp},
assuming for example $T = 20~\mev \ll m_B= 1$~GeV
\begin{align}
\frac{n_B(T)}{n_\gamma(T)} 
= \frac{n_{\bar{B}}(T)}{n_\gamma(T)}
&= \frac{g_B \left( \dfrac{m_BT}{2\pi} \right)^{3/2} e^{-m_B/T}}
       {\dfrac{\zeta_3}{\pi^2} \; g_\gamma T^3 } \notag \\
&= \frac{g_B}{g_\gamma} \; \frac{\sqrt{\pi}}{2 \sqrt{2} \zeta_3} \left( \dfrac{m_B}{T} \right)^{3/2} e^{-m_B/T}
 = 3.5 \cdot 10^{-20}
\label{eq:antibaryon_photon}
\end{align}
The way of looking at the baryon asymmetry is that independent of the
actual anti-baryon density the density of baryons observed today is
\ul{much larger} than what we would expect from thermal
production. While we will see that for dark matter the problem is to
get their interactions just right to produce the correct freeze-out
density, for baryons the problem is to avoid their annihilation as
much as possible. \index{freeze out}

We can think of two ways to avoid such an over-annihilation in our
thermal history. First, there could be some kind of mechanism stopping
the annihilation of baryons and anti-baryons when $n_B/n_\gamma$
reaches the observed value. The problem with this solution is that we
would still have to do something with the anti-baryons, as discussed
above. 

The second solution is to assume that through the baryon annihilation
phase there exists an initially small asymmetry, such that almost all
anti-baryons annihilate while the observed baryons remain. As a rough
estimate, neglecting all degrees of freedom and differences between
fermions and bosons, we assume that in the hot thermal bath we start
with roughly as many baryons as photons.  After cooling we assume that
the anti-baryons reach their thermal density given in
Eq.\eqref{eq:antibaryon_photon}, while the baryons through some
mechanism arrive at today's density given in
Eq.\eqref{eq:baryon_photon}. The baryon vs anti-baryon asymmetry
starting at an early time then becomes
\begin{align}
\frac{n_B - n_{\bar{B}}}{n_B} 
\approx \frac{n_B}{n_\gamma} - \frac{n_{\bar{B}}}{n_\gamma} 
\stackrel{\text{cooling}}{\longrightarrow}
6 \cdot 10^{-10} - 3.5 \cdot 10^{-20} 
\approx 6 \cdot 10^{-10} \; .
\end{align}
If we do the proper calculation, the correct number for a net quark
excess in the early Universe comes out around 
\begin{align}
\boxed{ \frac{n_B - n_{\bar{B}}}{n_B} \approx 3 \cdot 10^{-8} } \; .
\label{eq:baryon_asymmetry}
\end{align}
In the early Universe we start with this very small net asymmetry
between the very large individual densities of baryons and
anti-baryons. Rather than through the freeze-out mechanism introduced for neutrinos in Section~\ref{sec:neutrinos}, the baryons
decouple when all anti-baryons are annihilated away. This mechanism
can explain the very large baryon density measured today. The question
is now how this asymmetry occurs at high temperatures.\bigskip

Unlike the rest of the lecture notes, the discussion of the matter
anti-matter asymmetry is not aimed at showing how the relic densities
of the two species are computed. Instead, we will get to the general
\ul{Sakharov conditions} \index{Sakharov conditions} which tell us what ingredients our
theory has to have to generate a net baryon excess in the early
Universe, where we naively would expect the number of baryons and
anti-baryons (or quarks and anti-quarks) to be exactly the same and in
thermal equilibrium. Let us go through these condition one by
one:\bigskip

\ul{Baryon number violation} \index{baryon number violation}--- to understand this condition
we just need to remember that we want to generate a different density
of baryons (baryon number $B=+1$) and anti-baryons (baryon number
$B=-1$) dynamically during the evolution of the Universe. We assume
that our theory is described by a Lagrangian including
finite temperature effects. If our Lagrangian is fully symmetric with
respect to exchanging baryons with anti-baryons there will be no
effect, no interaction, no scattering rate, no decay rate, nothing
that can ever distinguish between baryons and anti-baryons and hence
generate an asymmetry from a symmetric starting point. Let us assume
that we want to generate a baryon asymmetry from an interaction of
quarks and leptons with a heavy state $X$ of the kind
\begin{align}
X \to d d \qqqquad
X \to \bar{d} \ell^- \; ,
\label{eq:sakharov1}
\end{align}
where the $d$ quark carries baryon number 1/3. A scattering
process induced by these two interactions,
\begin{align}
dd \to X^* \to \bar{d} \ell^- \; ,
\label{eq:sakharov1b}
\end{align}
links an initial state with $B = 2/3$ to a final state with $B=-1/3$.
The combination $B-L$ is instead conserved. Such heavy bosons can appear in
grand unified theories.

In the Standard Model the situation is a little more
complicated: instead of the lepton number $L$ and the baryon number
$B$ individually, the combination $B-L$ is indeed an (accidental) global
symmetry of the electroweak interaction to all orders. In contrast,
the orthogonal $B+L$ is anomalous, \ie there are quantum contributions
to scattering processes which respect $B-L$ but violate $B+L$.  One can
show that non-perturbative finite-temperature \ul{sphaleron
  processes} \index{sphaleron}can generate the combined state
\begin{align}
\epsilon_{ijk}
\left( u_{L,i} d_{L,j} u_{L,k} e_L 
     + \cdots 
\right)
\end{align}
for one generation of fermions with $SU(2)_L$ indices $i,j,k$ out of
the vacuum. It violates lepton and baryon number,
\begin{align}
\Delta L = 1 \qqqquad 
\Delta B = 1 \qqqquad 
\Delta (B-L) = 0 \; .
\end{align}
The probability of these sphaleron transition to happen at zero
temperature (where they are called instanton transitions) scales like
$e^{-8\pi^2/g^2}$ with the weak $SU(2)_L$ coupling $g \approx 0.7$.
At high temperatures their rate increases significantly.  The main
effect of such interactions is that we can replace the
condition of baryon number violation with lepton number violation when
we ensure that sphaleron-induced processes transform a lepton
asymmetry into a baryon asymmetry and neither of them gets washed
out. This process is called leptogenesis rather than
baryogenesis.\bigskip

\ul{Departure from thermal equilibrium} --- in our above setup
we can see what assumptions we need to be able to generate a net
baryon asymmetry from the interactions given in
Eq.\eqref{eq:sakharov1} and the scattering given in
Eq.\eqref{eq:sakharov1b}. If we follow the reasoning for the relic
photons we reduce the temperature until the two sides of the $2 \to 2$
scattering process in Eq.\eqref{eq:sakharov1b} drop out of thermal
equilibrium. Our Universe could settles on one of the two sides of the
scattering process, \ie either with a net excess of $d$ over $\bar{d}$
particles or vice versa. The problem is that the process $\bar{d}
\bar{d} \to \bar{X}^* \to d \ell^+$ with $m_X = m_{\bar{X}}$ is protected
by CPT invariance and will compensate everything exactly.

The more promising approach are out-of-equilibrium decays of the heavy
$X$ boson. This means that a population of $X$ and $\bar{X}$ bosons
decouple from the thermal bath early and induce the baryon asymmetry
through late decays preferably into quarks or anti-quarks.  In both
cases we see that baryon number violating interactions require a
departure from thermal equilibrium to generate a net baryon asymmetry
in the evolution of the Universe.

In the absence of late-decaying particles, for example in the Standard
Model, we need to rely on another mechanism to deviate from thermal
equilibrium. The electroweak phase transition, like any phase
transition, can proceed in two ways: if the phase transition is of
first order the Higgs potential develops a non-trivial minimum while
we are sitting at the unbroken field value $\phi = 0$. At the critical
temperature the broken minimum becomes the global minimum of the
potential and we have to tunnel there. The second order phase
transition instead develops the broken minimum smoothly such that
there is never a potential barrier between the two and we can smoothly
transition into the broken minimum around the critical temperature.
For a \ul{first-order phase transition} \index{phase transition}different regions of the Universe
will switch to the broken phase at different times, starting with
expanding bubbles of broken phase regions. At the bubble surface the thermal
equilibrium will be broken, allowing for a generation of the baryon
asymmetry through the electroweak phase transition. Unfortunately, the
Standard Model Higgs mass would have had to be below $60$~GeV to allow
for this scenario.\bigskip
\index{CP violation}
\ul{C and CP violation} --- this condition appears more
indirectly. First, even if we assume that a transition of the kind
shown in Eq.\eqref{eq:sakharov1} exists we need to generate a baryon
asymmetry from these decays where the heavy state and its
anti-particle are produced from the vacuum. Charge
conjugation links particles and anti-particles, which means that C
conservation implies
\begin{align}
\Gamma (X \to d d) = \Gamma (\bar{X} \to \bar{d} \bar{d}) 
\qquad \text{and} \qquad 
\Gamma (X \to d \ell^-) = \Gamma (\bar{X} \to \bar{d} \ell^+) \; .
\label{eq:sakharov2}
\end{align}
In that case there will always be the same numbers of baryons $d$ and
anti-baryons $\bar{d}$ on average in the system. We only quote the
statement that statistical fluctuations of the baryon and anti-baryon
numbers are not large enough to explain the global asymmetry observed.

Next we assume a theory where C is violated, but CP is intact. This could
for example be the electroweak Standard Model with no CP-violating
phases in the quark and lepton mixing matrices.  For our toy model
we introduce a quark chirality $q_{L,R}$ which
violates parity P but restores CP as a symmetry. For our decay widths
transforming under C and CP this means
\begin{align}
\Gamma (X \to d_L d_L) &\ne \Gamma (\bar{X} \to \bar{d}_L \bar{d}_L) 
&\qquad &\text{(C violation)} \notag \\
\Gamma (X \to d_L d_L) &= \Gamma (\bar{X} \to \bar{d}_R \bar{d}_R) 
&&\text{(CP conservation)} \notag \\
\Gamma (X \to d_R d_R) &= \Gamma (\bar{X} \to \bar{d}_L \bar{d}_L) 
&&\text{(CP conservation)} \notag \\
\Rightarrow \qquad 
\Gamma (X \to dd) &\equiv \Gamma (X \to d_L d_L)  + \Gamma (X \to d_R d_R) \notag \\
 &= \Gamma (\bar{X} \to \bar{d}_R \bar{d}_R) + \Gamma (\bar{X} \to \bar{d}_L \bar{d}_L) = \Gamma (\bar{X} \to \bar{d} \bar{d})
\label{eq:sakharov3}
\end{align}
This means unless C and CP are both violated, there will be no baryon
asymmetry from $X$ decays to $d$ quarks.

In the above argument there is, strictly speaking, one piece missing:
if we assume that we start with the same number of $X$ and $\bar{X}$
bosons out of thermal equilibrium, once all of them have decayed to
$dd$ and $\bar{d} \bar{d}$ pairs irrespective of their chirality there
is again no asymmetry between $d$ and $\bar{d}$ quarks in the
Universe.  An asymmetry only occurs if a competing $X$ decay channel
produces a different number of baryons and allows the different
partial widths to generate a net asymmetry.  This is why we include
the second term in Eq.\eqref{eq:sakharov1}. Assuming C and CP
violation it implies
\begin{align}
\Gamma (X \to dd) &\ne \Gamma (\bar{X} \to \bar{d}\bar{d}) \notag \\
\Gamma (X \to \bar{d} \ell^-) &\ne \Gamma (\bar{X} \to d\ell^+)
\qquad \text{but} \quad 
\Gamma_\text{tot}(X) = 
\Gamma_\text{tot}(\bar{X}) 
\end{align}
because of CPT invariance, just like $m_X = m_{\bar{X}}$.

\subsection{Asymmetric dark matter}
\label{sec:asymmetric}
\index{dark matter!asymmetric}

Starting from the similarity of the measured baryon and dark matter
densities in Eq.\eqref{eq:planck_results}

\begin{align}
\frac{\Omega_\chi}{\Omega_b} 
= \frac{0.12}{0.022} = 5.5 \; ,
\label{eq:baryon_dm_asymmetry}
\end{align}
an obvious question is if we can link these two matter densities. We
know that the observed baryon density in the Universe today is not
determined by a thermal freeze-out, but by an initial small asymmetry
between the baryon and anti-baryon densities. If we assume that dark
matter is very roughly as heavy as baryons, that dark matter states
carry some kind of charge which defines dark matter anti-particles,
and that the baryon and dark matter asymmetries are linked, we can
hope to explain the observed dark matter relic density. Following the
leptogenesis example we could assume that the sphaleron transition not
only breaks $B+L$, but also some kind of dark matter number $D$. Dark
matter is then generated thermally, but the value of the relic density
is not determined by thermal freeze-out. Still, from the structure
formation constraints discussed in Section~\ref{sec:structure} we know that
the dark matter agent should not be too light.\bigskip

First, we can roughly estimate the dark matter masses this scenario
predicts. From Section~\ref{sec:matter} we know how little we
understand about the mechanism of generating the baryon asymmetry in
models structurally similar to the Standard Model. For that reason, we
start by just assuming that the particle densities of the baryons and
of dark matter trace each other through some kind of mechanism, 
\begin{align}
n_\chi(T) \approx n_B(T) \; . 
\end{align}
This will start in the relativistic regime and remain true after the
two sectors decouple from each other and both densities get diluted
through the expansion of the Universe.  For the observed densities by \index{PLANCK}
PLANCK we use the non-relativistic relation between number and energy
densities in Eq.\eqref{eq:n_vs_temp} and Eq.\eqref{eq:rho_vs_temp},
\begin{align}
\frac{\Omega_\chi}{\Omega_b} 
= \frac{\rho_\chi}{\rho_B} 
= \frac{m_\chi n_\chi}{m_B n_B} 
\approx \frac{m_\chi}{m_B} 
\qquad \Leftrightarrow \qquad 
m_\chi \approx 5.5 \, m_B \approx 5~\gev \; .
\end{align}
Corrections to this relation can arise from the mechanism linking the two
asymmetries.

Alternatively, we can assume that at the temperature $\Tdec$ at which the link
between the baryons and the dark matter decouples, the baryons are
relativistic and dark matter is non-relativistic. For the two energy
densities this means
\begin{align}
\rho_\chi(\Tdec) 
&= m_\chi n_\chi(\Tdec)  \notag \\
&\approx m_\chi n_B(\Tdec)
= m_\chi \frac{30 \zeta_3}{\pi^4} \; \frac{\rho_B(\Tdec)}{\Tdec} \notag \\
\Rightarrow \qquad
\frac{m_\chi}{\Tdec} 
&= \frac{\rho_\chi(\Tdec)}{\rho_B(\Tdec)} \frac{\pi^4}{30 \zeta_3}
 \approx 15 \; .
\end{align}
The relevant temperature is determined by the interactions between the
baryonic and the dark matter sectors. However, in general this
scenario will allow for heavy dark matter, $m_\chi \gg m_B$.\bigskip

In a second step we can analyze what kind of dark matter annihilation
rates are required in the asymmetric dark matter scenario. Very generally, we
know the decoupling condition of a dark matter particle of the thermal
bath of Standard Model states from the relativistic case.  The
mediating process can include a Standard Model fermion, $\chi f \to
\chi f$. The corresponding annihilation process for dark matter which
is not its own anti-particle is
\begin{align}
\chi \bar{\chi} \to f\bar{f} \; .
\end{align}
As long as these scattering processes are active, the dark matter
agent follows the decreasing temperature of the light Standard Model
states in an equilibrium between production out of the thermal bath
and annihilation. At some point, dark-matter \ul{freezes out} \index{freeze out}
of the thermal bath, and its density is only reduced by the expansion
of the Universe. This point of decoupling is roughly given by
Eq.\eqref{eq:neutrino_decouple1}, or
\begin{align}
n_\chi(\Tdec)
= \frac{H}{\sigma_{\chi \chi} \, v}  
= \frac{\pi \sqrt{\geff(\Tdec)}}{\sqrt{90} \mpl} \; \frac{\Tdec^2}{\sigma_{\chi \chi} \, v} \; 
\end{align}
in terms of the dark matter annihilation cross section $\sigma_{\chi \chi}$.

The special feature of asymmetric dark matter is that this relation
does not predict the dark matter density $n_\chi(\Tdec)$ leading to
the observed relic density. Instead, this annihilation has to remove
all dark matter anti-particles and almost all dark matter particles,
while the observed relic density is generated by a very small dark
matter asymmetry. If we follow the numerical example of the baryon
asymmetry given in Eq.\eqref{eq:baryon_asymmetry} this means we need a
dark matter annihilation rate which is $10^8$ times the rate necessary
to predict the observed relic density for pure freeze-out dark matter. \index{freeze out}
From the typical expressions for cross sections in
Eq.\eqref{eq:neutrino_photon_electron} we see that a massless mediator \index{mediator}
or a $t$-channel diagram in combination with light dark matter leads
to large cross sections,
\begin{align}
\sigma_{\chi \chi} \approx \frac{\pi \alpha_\chi^2}{m_\chi^2} \; ,
\end{align}
with the generic dark matter coupling $\alpha_\chi$ to a dark gauge
boson or another light mediator. For heavier dark matter we will see
in Section~\ref{sec:models_portal} how we can achieve large
annihilation rates through a $2 \to 1$ annihilation topology.

\newpage
\section{Thermal relic density}
\label{sec:relic}
\index{relic!thermal abundance}

After introducing the observed relic density of photons in
Section~\ref{sec:photons} and the observed relic density of neutrinos
in Section~\ref{sec:neutrinos} we will now compute the relic density
of a hypothetical massive, weakly interacting dark matter agent. As
for the photons and neutrinos we assume dark matter to be created
thermally, and the observed relic density to be determined by the
freeze-out combined with the following expansion of the Universe. We
will focus on masses of at least a few GeV, which guarantees that dark matter
will be non-relativistic when it decouples from thermal
equilibrium. At this point we do not have specific particles in mind,
but in Section~\ref{sec:models} we will illustrate this scenario with
a set of particle physics models.

The general theme of this section and the following
Sections~\ref{sec:indirect}-\ref{sec:coll} is the typical four-point
interaction of the dark matter agent with the Standard Model. For
illustration purposes we assume the dark matter agent to be a fermion
$\chi$ and the Standard Model interaction partner a fermion $f$:\\[10mm]

\begin{center}
\begin{fmfgraph*}(70,70)
\fmfleft{i1,i2}
\fmfright{o1,o2}
\fmf{fermion,width=0.6,lab.side=left}{i1,v1}
\fmf{fermion,width=0.6,lab.side=left}{v1,i2}
\fmf{fermion,width=0.6,lab.side=left}{o1,v1}
\fmf{fermion,width=0.6,lab.side=left}{v1,o2}
\fmflabel{$\chi$}{i1}
\fmflabel{$\chi$}{i2}
\fmflabel{$f$}{o1}
\fmflabel{$f$}{o2}
\fmfblob{.4w}{v1} 
\end{fmfgraph*}
\end{center}

Unlike for asymmetric dark matter, in this process it does not matter
if the dark matter agent has an anti-particle $\bar{\chi}$, or if is
it's own anti-particle $\chi = \bar{\chi}$.  This Feynman diagram, or
more precisely this amplitude mediates three different scattering
processes:
\begin{enumerate}
\item[--] left-to-right we can compute dark matter annihilation,
  $\chi \bar{\chi} \to f\bar{f}$, see Sections~\ref{sec:relic}-\ref{sec:indirect};
\item[--] bottom-to-top it describes dark matter scattering of
  visible matter $\chi f \to \chi f$, see Section~\ref{sec:direct};
\item[--] right-to-left it describes dark matter pair-production, $f
  \bar{f} \to \chi \bar{\chi}$, see Section~\ref{sec:coll}.
\end{enumerate}
This strong link between very different observables is what makes dark
matter so interesting for particle physicists, including the
possibility of global analyses for any model which can predict this
amplitude. Note also that we will see how different the kinematics of
the different scattering processes actually are.

\subsection{WIMP miracle}
\label{sec:miracle}
\index{WIMP}

As for the relativistic neutrinos, we will first avoid solving the
full Boltzmann equation for the number density as a function of
time. Instead, we assume that some kind of interaction keeps the dark
matter particle $\chi$ in thermal equilibrium with the Standard Model
particles and at the same time able to annihilate. At the point of
thermal decoupling the dark matter freezes out with a specific
density. As for the neutrinos, the underlying process is described by the 
matrix element for \ul{dark matter annihilation} \index{dark matter!annihilation}
\begin{align}
\chi \chi \to f\bar{f} \; .
\label{eq:dm_annihilation}
\end{align}
As in Eq.\eqref{eq:def_tdec} the interaction rate $\Gamma$
corresponding to this scattering process just compensates the
increasing scale factor at the point of decoupling,
\begin{align}
\Gamma(\Tdec) \really H(\Tdec) \; .
\label{eq:dm_equilib}
\end{align}
Assuming this interaction rate is set by electroweak interactions, for
non-relativistic dark matter agents, the temperature dependence in
Eq.\eqref{eq:neutrino_photon_electron} vanishes and gets replaced by
the dark matter mass. To allow for an $s$-channel process in
Eq.\eqref{eq:dm_annihilation} we use the $Z$-mass and $Z$-coupling in
the corresponding annihilation cross section
\begin{align}
\boxed{ \sigma_{\chi \chi}(T \ll m_\chi) = \frac{\pi \alpha^2 m_\chi^2}{c_w^4 m_Z^4} } \; .
\label{eq:wimp_ann_approx}
\end{align}
This formula combines the \ul{dark matter mass $m_\chi$} with a
\ul{weak interaction} represented by a $1/m_Z$ suppression, implicitly assuming $m_\chi \ll m_Z$. We will check this assumption later. Following
Eq.\eqref{eq:neutrino_rate} we can use the non-relativistic number
density. For the the non-relativistic decoupling we
should not assume $v=1$, as we did before.  Given the limited number
of energy scales in our description we instead estimate very roughly
\begin{align}
\frac{m_\chi}{2} v^2 = T
\qqquad \Leftrightarrow \qqquad 
v = \sqrt{ \frac{2T}{m_\chi} } \; ,
\label{eq:non_relativistic_v}
\end{align}
remembering that we need to check this later. Moreover, we set the
number of relevant degrees of freedom of the dark matter agent to
$g=2$, corresponding for example to a complex scalar or a Majorana
fermion. In that case the condition of \ul{dark matter freeze-out} \index{freeze out} is
\begin{align}
\Gamma := \sigma_{\chi \chi} \, v n_\chi 
\eqx{eq:n_vs_temp}
\sigma_{\chi \chi}
\; \sqrt{\frac{2 \Tdec}{m_\chi}} \; g \; \left( \frac{m_\chi \Tdec}{2\pi} \right)^{3/2} e^{-m_\chi/\Tdec}
&\really H
\eqx{eq:hubble_temp}
\frac{\pi}{3 \sqrt{10} \, \mpl} \, \sqrt{\geff(\Tdec)} \, \Tdec^2 \notag \\ 
\Leftrightarrow \qquad 
\sigma_{\chi \chi} \; \frac{m_\chi \Tdec^2 }{\pi^{3/2}} \; e^{-\xdec}
&= \frac{\pi}{3 \sqrt{10} \, \mpl} \, \sqrt{\geff(\Tdec)} \Tdec^2
\qquad \text{with} \quad \boxed{ x := \frac{m_\chi}{T} } \notag \\
\Leftrightarrow \qqqquad \qquad 
 e^{-\xdec}
&= \frac{\pi^{5/2}}{3 \sqrt{10}} \frac{\sqrt{\geff(\Tdec)}}{m_\chi \mpl \; \sigma_{\chi \chi}} \,  \notag \\
&= 1.8 \; \frac{\sqrt{\geff(\Tdec)}}{m_\chi \mpl \; \sigma_{\chi \chi}} \; .
\label{eq:wimp_ann_equilibrium1}
\end{align}
Note how in this calculation the explicit temperature dependence drops
out.  This means the result can be considered an equation for the
ratio $\xdec$.  If we want to include the temperature dependence of
$\geff$ we cannot solve this equation in a closed form, but we can
estimate the value of $\xdec$.  First, we can use the generic
electroweak annihilation cross section from
Eq.\eqref{eq:wimp_ann_approx} to find
\begin{align} 
 e^{-\xdec}
= \frac{\pi \sqrt{\pi}}{3 \sqrt{10} \, \alpha^2} \; \frac{c_w^4 m_Z^4}{m_\chi^3 \mpl} \; \sqrt{\geff(\Tdec)} \; .
\label{eq:wimp_ann_equilibrium2}
\end{align}
\index{degrees of freedom}
Next, we assume that most of the Standard Model particles contribute
to the active degrees of freedom.  From Eq.\eqref{eq:sm_freedom} we
know that the full number gives us $\geff = 106.75$. In the slightly
lower range $\Tdec = 5~...~80$~GeV the weak bosons and the top quark
decouple, and Eq.\eqref{eq:def_eff_freedom2} gives the slightly
reduced value
\begin{align}
\geff(\Tdec)
= \left( 8 \times 2 + 2 \right) 
+ \frac{7}{8} \left( 5 \times 3 \times 2 \times 2 
                   + 3 \times 2 \times 2 
                   + 3 \times 2 \right) 
= 18 + \frac{7}{8} \; 78 
= 86.25\;.
\label{eq:dof_wimp_dec}
\end{align}
Combining all prefactors we find the the range
\begin{align}
e^{-\xdec}
\approx 6 \cdot 10^5 \; \frac{m_Z^4}{m_\chi^3 \mpl} 
= 
\begin{cases}
2 \cdot 10^{-9} 
\qquad \; \Leftrightarrow \qquad \xdec \approx 20 \qqquad  \text{($m_\chi = 10$~GeV)} \\
6 \cdot 10^{-11} 
\qquad \Leftrightarrow \qquad \xdec \approx 23 \qqquad  \text{($m_\chi = 30$~GeV)} \\
8 \cdot 10^{-12} 
\qquad \Leftrightarrow \qquad \xdec \approx 26 \qqquad  \text{($m_\chi = 60$~GeV)\; .} 
\end{cases}
\label{eq:xdec_wimp}
\end{align}
As a benchmark we will use $m_\chi=30$ GeV with $\xdec \approx 23$ from now on.
We need to eventually check these assumptions, but because of the
leading exponential dependence we expect this result for
$\xdec$ to be insensitive to our detailed assumptions. Following
Eq.\eqref{eq:n_vs_temp} and Eq.\eqref{eq:wimp_ann_equilibrium2} the
temperature at the point of decoupling gives us the non-relativistic
number density at the point of decoupling,\index{number density}
\begin{align}
n_\chi(\Tdec)
&= g \left( \frac{m_\chi \Tdec}{2\pi} \right)^{3/2} e^{-\xdec} 
 = \frac{\pi}{3 \sqrt{20} \, \mpl} \, \sqrt{ \frac{m_\chi}{\Tdec}} \, \sqrt{\geff(\Tdec)} \, \Tdec^2 \; \frac{c_w^4 m_Z^4}{\pi \alpha^2 m_\chi^2} \notag \\
&\approx 10^3 \, \;
 \frac{m_Z^4}{\mpl} \; \left( \frac{\Tdec}{m_\chi} \right)^{3/2} 
\approx \frac{10^3}{\xdec^{3/2}} \; \frac{m_Z^4}{\mpl} \; .
\label{eq:wimp_relic_approx1}
\end{align}
From the time of non-relativistic decoupling we have to
\ul{evolve the energy density} to the current time or
temperature $T_0$. We start with the fact that once a particle has
decoupled, its number density drops like $1/a^3$, as we
can read off Eq.\eqref{eq:rho_vs_a} in the non-relativistic case,
\begin{align}
\rho_\chi(T_0) 
= m_\chi \; n_\chi(T_0)
&= m_\chi \; n_\chi(\Tdec) \; \left( \frac{a(\Tdec)}{a(T_0)} \right)^3 \;.
\label{eq:wimp_relic_approx2a}
\end{align}
To translate this dependence on the scale factor $a$ into a
temperature dependence we need to quote the same, single thermodynamic
result as in Section~\ref{sec:axions}, namely that according to
Eq.\eqref{eq:thermodynamics1} the combination $a(T) \, T$ is almost
constant. When we take into account the active degrees of freedom and
their individual temperature dependence the relation is more precisely
\begin{align}
\left( \frac{a(\Tdec) \Tdec}{a(T_0) T_0} \right)^3
= \frac{\geff(T_0)}{\geff(\Tdec)} 
\approx
\frac{3.6}{100} = \frac{1}{28} \; ,
\label{eq:thermodynamics2}
\end{align}
again for $\Tdec > 5$~GeV and depending slightly on the number of
neutrinos we take into account. We can use this result to compute the
non-relativistic energy density now
\begin{align}
\rho_\chi(T_0) 
&= m_\chi \; \left( \frac{a(\Tdec) \Tdec}{a(T_0) T_0} \right)^3 \; 
   \frac{T_0^3}{\Tdec^3} \, n_\chi(\Tdec)
= \frac{\xdec}{28} \; 
   T_0^3 \, \frac{n_\chi(\Tdec)}{\Tdec^2} 
=T_0^3 \, \frac{n_\chi(\Tdec)  \xdec^3}{28 m_\chi^2} \notag \\
&\stackrel{\text{Eq.\eqref{eq:wimp_relic_approx1}}}{\approx}
3\cdot 10^3 \; \frac{m_Z^4}{m_\chi^2 \mpl} \; T_0^3 \; .
\label{eq:wimp_relic_approx2}
\end{align}
Using this result we can compute the dimensionless dark matter density
in close analogy to the neutrino case of Eq.\eqref{eq:hot_neutrinos1},
\begin{align}
\label{eq:wimp_relic_approx3}
\Omega_\chi h^2 
&= \frac{\rho_\chi(T_0) h^2}{3 \mpl^2 H_0^2} \notag \\
&\approx
3\cdot 10^3 \; \frac{m_Z^4}{m_\chi^2 \mpl} \;
\frac{(2.4 \cdot 10^{-4})^3}{(2.5 \cdot 10^{-3})^4} \; \frac{h^2}{\ev} \\
&\approx
3\cdot 10^3 \; \frac{7\cdot 10^7}{2 \cdot 10^{18}} \; \frac{\gev^3}{m_\chi^2} \; 
\frac{1}{5} \; \frac{10^9}{\gev} 
\approx
20 \; \frac{\gev^2}{m_\chi^2} \;
\qquad \Leftrightarrow \qquad 
\boxed{
\Omega_\chi h^2 \approx
0.12 \; \left( \frac{13~\gev}{m_\chi} \right)^2 \;
} \; . \notag 
\end{align}
This outcome is usually referred to as the \ul{WIMP miracle}: \index{WIMP}
if we assume an  dark matter agent with an electroweak-scale mass and an
annihilation process mediated by the weak interaction, the predicted
relic density comes out exactly as measured.

Let us recapitulate where the WIMP mass dependence of
Eq.\eqref{eq:wimp_relic_approx3} comes from: first, the annihilation
cross section in Eq.\eqref{eq:wimp_ann_approx} is assumed to be
mediated by electroweak interactions and includes a dependence on
$m_\chi$.  Our original assumption $m_\chi \ll m_W$ is not perfectly
fulfilled, but also not completely wrong.  Second, the WIMP mass
enters the relation between the number and energy density, but some of
this dependence is absorbed into the value $\xdec = 23$, which means
that the decoupling of the non-relativistic WIMPs is supposed to
happen at a very low temperature of $\Tdec \approx m_\chi/23$. Making
things worse, some of the assumption we made in this non-relativistic
and hence multi-scale calculation are not as convincing as they were
for the simpler relativistic neutrino counterpart, so let us check
Eq.\eqref{eq:wimp_relic_approx3} with an alternative estimate. One of
the key questions we will try to answer in our alternative approach is
how the $m_\chi$-dependence of Eq.\eqref{eq:wimp_relic_approx3}
occurs.

\subsection{Boltzmann equation}
\label{sec:boltzmann}

Because the derivation for the non-relativistic dark matter agent is
at the heart of these lecture notes, we will also show how to properly
compute the current relic density of a weakly interacting, massive
dark matter agent. This calculation is based on the
\ul{Boltzmann equation}. \index{Boltzmann equation} It
describes the change of a number density $n(t)$ with time. The first
effect included in the equation is the increasing scale factor
$a(t)$. It even occurs in full equilibrium,
\begin{align}
0 
= \frac{d}{dt} \; \left[ n(t) a(t)^3 \right]
= \dot n(t) a(t)^3 + 3 n(t) a(t)^2 \dot a(t)
\qquad \Leftrightarrow \qquad 
\dot n(t) + 3 H(t) n(t) = 0 \; .
\label{eq:before_boltzmann}
\end{align}

At some point, the underlying assumption of thermal equilibrium breaks
down.  For the number of WIMPs \index{WIMP} the relevant process is not a process
which guarantees thermal equilibrium with other states, but the
explicit pair production or pair annihilation via a weakly interacting
process
\begin{align}
\chi \chi \leftrightarrow f \bar{f} \; ,
\label{eq:wimp_annihilation}
\end{align}
with any available pair of light fermions in the final state.  The
\ul{depletion rate} from the WIMP pair annihilation process in
Eq.\eqref{eq:wimp_annihilation} is given by the corresponding
$\sigma_{\chi \chi} \, v \, n_\chi^2$.  This rate describes the
probability of the WIMP annihilation process in
Eq.\eqref{eq:wimp_annihilation} to happen, given the WIMP density and
their velocity. For the relativistic relic neutrinos we could safely
assume $v=1$, while for the WIMP case we did not even make this
assumption for our previous order-of-magnitude estimate.

When we derive the Boltzmann equation from first principles it turns
out that we need to thermally average. This reflects the fact that the
WIMP number density is a global observable, integrated over the
velocity spectrum. In the non-relativistic limit the \ul{velocity} of a
particle with momentum $\vec k$ and energy $k_0$ is
\begin{align}
v_k := \frac{|\vec k|}{k_0} \approx \frac{|\vec{k}|}{m_\chi} \ll 1 \; .
\end{align}
The external momenta of the two fermions then have the form
\begin{align}
k^2 
= k_0^2 - \vec k^2 
= k_0^2 - (m_\chi v_k)^2 
\really m_\chi^2 
\qquad \Leftrightarrow \qquad 
k_0 = \sqrt{m_\chi^2 + m_\chi^2 v_k^2} 
\approx m_\chi \left( 1 + \frac{v_k^2}{2} \right) \; .
\label{eq:scatter_velocity}
\end{align}
For a $2 \to 2$ scattering process we have to distinguish the
velocities of the individual states and the relative velocity. The
energy of initial state is given by the Mandelstam variable $s = (k_1
+ k_2)^2$, in terms of the incoming momenta $k_1$ and $k_2$. \index{Mandelstam variables} These
momenta are linked to the masses of the incoming dark matter state via
$k_1^2 = k_2^2 = m_\chi^2$.  For two incoming states with the same
mass this gives us the velocity of each of the two particles as
\begin{align}
s &= (k_1 + k_2)^2 
  = 2 m_\chi^2 + 2 k_1^0 k_2^0 - 2 \vec{k}_1 \vec{k}_2
  \stackrel{\text{cms}}{=} 2 m_\chi^2 + 2 (k_1^0)^2 + 2 |\vec{k}_1|^2 \notag \\
  &= 4 m_\chi^2 + 4 |\vec{k}_1|^2 
  = 4 m_\chi^2 ( 1 + v_1^2 ) 
\qqquad \Leftrightarrow \qqquad 
v_1^2 = \frac{s - 4 m_\chi^2}{4 m_\chi^2} = \frac{s}{4 m_\chi^2} - 1\;.
\end{align}
The relative velocity of the two incoming particles in the
non-relativistic limit is instead defined as
\begin{align}
v 
= \left| \frac{\vec k_1}{k_1^0} - \frac{\vec k_2}{k_2^0} \right|
\stackrel{\text{cms}}{=} \left| \frac{\vec k_1}{k_1^0} + \frac{\vec k_1}{k_1^0} \right|
= \frac{2 |\vec k_1|}{k_1^0} 
\approx 2 v_1 
\qquad \Leftrightarrow \qquad 
m_\chi^2 v^2 
= 4 m_\chi^2 v_1^2 
= s - 4 m_\chi^2 \; . 
\label{eq:def_velocity}
\end{align}
Using the relative velocity the thermal average of $\sigma_{\chi \chi}
v$ as it for example appears in Eq.\eqref{eq:wimp_ann_equilibrium2} is
defined as
\begin{align}
\langle \sigma_{\chi \chi \to ff} \; v \rangle
&:= \dfrac{\int d^3 p_{\chi,1} d^3 p_{\chi,2} \; e^{-(E_{\chi,1}+E_{\chi,2})/ T} \; \sigma_{\chi \chi \to ff} \; v} 
         {\int d^3 p_{\chi,1} d^3 p_{\chi,2} \; e^{-(E_{\chi,1}+E_{\chi,2}) /T}} \notag \\
&= \dfrac{2 \pi^2 T \int_{4 m_\chi^2}^\infty ds \; \sqrt{s} (s - 4 m_\chi^2)K_1 \left( \dfrac{\sqrt{s}}{T} \right) \; \sigma_{\chi \chi \to ff}(s) }
         {\left( 4 \pi m_\chi^2 T \; K_2 \left( \dfrac{m_\chi}{T}\right) \right)^2} \; ,
\label{eq:def_sigma_v}
\end{align}
in terms of the modified Bessel functions of the second kind
$K_{1,2}$. Unfortunately, this form is numerically not very helpful in
the general case. The thermal averaging replaces the global value of
$\sigma_{\chi \chi} \; v$, as it gets added to the equilibrium Boltzmann equation
Eq.\eqref{eq:before_boltzmann} on the right-hand side,
\begin{align}
\boxed{
\dot n(t) + 3 H(t) n(t) 
= - \langle \sigma_{\chi \chi} \, v \rangle \; 
  \left( n(t)^2  - n_\text{eq}(t)^2 \right) 
} \; .
\label{eq:boltzmann}
\end{align}
\index{Boltzmann equation}
The time dependence of $n$ induced by the annihilation process is
proportional to $n^2$ because of the two WIMPs in the initial state of
the annihilation process. The form of the equation guarantees that for
$n = n_\text{eq}$ the only change in density occurs from the expanding
Universe.\bigskip

We can analytically solve this Boltzmann equation using a set of
approximations.  We start with a re-definition, introducing the \ul{yield} $Y$\index{yield}, we get rid of the
linear term,
\begin{alignat}{7}
&&
\frac{1}{a(t)^3} \; \frac{d}{dt} \left( n(t) a(t)^3 \right)
&= - \langle \sigma_{\chi \chi} \, v \rangle \; 
  \left( n(t)^2  - {n_\text{eq}}(t)^2 \right) \notag \\
&\Leftrightarrow &\qquad 
T(t)^3 \; \frac{d}{dt} \left( \frac{n(t)}{T(t)^3} \right)
&= - \langle \sigma_{\chi \chi} \, v \rangle \; 
  \left( n(t)^2  - {n_\text{eq}}(t)^2 \right) \notag \\
&\Leftrightarrow &\qquad 
\frac{d Y(t)}{dt} 
&= - \langle \sigma_{\chi \chi} \, v \rangle \; T(t)^3 \; 
  \left( Y(t)^2  - {Y_\text{eq}}(t)^2 \right) 
\qquad \text{with} \quad Y(t) := \frac{n(t)}{T^3} \; .
\end{alignat}
Throughout these lecture notes we have always replaced the time by some other
variable describing the history of the Universe. We again switch
variables to $x = m_\chi/T$.  For the Jacobian we assume that most of
the dark matter decoupling happens with $\rho_r \gg \rho_m$; in the
early, radiation-dominated Universe we can link the time and $x$
through the Hubble constant,
\begin{align}
\frac{1}{2t} \eqx{eq:components} H 
\eqx{eq:hubble_temp} \frac{H(x=1)}{x^2}
\quad &\Leftrightarrow& \quad
x &= \sqrt{2 t H(x=1)} \notag \\
\quad &\Leftrightarrow& \quad
\frac{dx}{dt} 
&= \frac{2 H(x=1)}{2 \sqrt{2 t H(x=1)}} 
 = \frac{H(x=1)}{x} \; ,
\end{align}
where $x=1$ means $T = m_\chi$.  With this Jacobian the Boltzmann
equation becomes
\begin{align}
\frac{d Y(x)}{dx} 
&= \frac{x}{H(x=1)} \; \frac{d Y(t)}{dt} \notag \\
&= - \langle \sigma_{\chi \chi} \, v \rangle \; \frac{x}{H(x=1)} \; \frac{m_\chi^3}{x^3} \; 
  \left( Y(x)^2  - {Y_\text{eq}}(x)^2 \right) \notag \\
&= - \frac{\lambda(x)}{x^2} \; \left( Y(x)^2  - {Y_\text{eq}}(x)^2 \right) 
\qquad \text{with} \quad 
\lambda(x) := \frac{m_\chi^3 \, \langle \sigma_{\chi \chi} \, v \rangle}{H(x=1)} 
= \frac{\sqrt{90} \, \mpl m_\chi \, }{\pi \sqrt{\geff}} \, \langle \sigma_{\chi \chi} \, v \rangle(x) \; .
\label{eq:wimp_boltzmann}
\end{align}
%
To analytically solve this Boltzmann equation we make two
approximations: first, according to Eq.\eqref{eq:n_vs_temp} the
equilibrium density drops like $e^{-x}$ towards later times or
increasing $x$. Assuming that the actual number density $n(x)$ drops \index{number density}
more slowly, we can safely approximate the Boltzmann equation by
\begin{align}
\frac{d Y(x)}{dx} 
= - \frac{\lambda(x)}{x^2} \; Y(x)^2 \; .
\label{eq:boltzmann_approx}
\end{align}
Second, we can estimate $\lambda(x)$ by expanding the thermally
averaged annihilation WIMP cross section for small velocities. We use
Eq.\eqref{eq:wimp_ann_approx} as the leading term in the annihilation
cross section and approximate $v$ following
Eq.\eqref{eq:non_relativistic_v}, giving us 
%
\begin{align}
\lambda(x) 
&= \frac{\sqrt{90} \, \mpl m_\chi \, }{\pi \sqrt{\geff}} \, 
 \sigma_{\chi \chi} \, v + \ope (v^2) \notag \\
&\approx \frac{\sqrt{90} \, \mpl m_\chi \, }{\pi \sqrt{\geff}} \, 
 \sqrt{\frac{2}{x}} \; \frac{\pi \alpha^2 m_\chi^2}{c_w^4 m_Z^4} 
\equiv \frac{\bar \lambda}{\sqrt{x}} \; .
\label{eq:boltzmann_def_lambda}
\end{align}
The value $\bar{\lambda}$ depends on $x$ independently through
$\geff$, so we can assume it to be constant as long as $\geff$ does
not change much.  Under this assumption we can then solve the
Boltzmann equation with the simple substitution $\overline{Y} = 1/Y$,
\begin{align}
\frac{d Y(x)}{dx} 
= \frac{d}{dx} \; \frac{1}{\overline Y}
&=  - \frac{1}{\overline Y(x)^2} \frac{d \overline Y(x)}{dx}
\really - \frac{\bar \lambda}{x^{5/2}} \frac{1}{\overline Y(x)^2}
\qquad \Leftrightarrow \qquad 
\frac{d \overline Y(x)}{dx} = \frac{\bar \lambda}{x^{5/2}} \; .
\label{eq:dYdx}
\end{align}
From Eq.\eqref{eq:xdec_wimp} we know that thermal WIMPs have masses
well above $10$~GeV, which corresponds to $\geff \approx 100$. This
value only changes once the temperature reaches the bottom mass and
then drops to $\geff \approx 3.6$ today. This allows us to separate
the leading effects driving the dark matter density into the
decoupling phase described by the Boltzmann equation and an expansion
phase with its drop in $\geff$. For the first phase we can just
integrate the Boltzmann equation for constant $\geff$ starting just
before decoupling ($\xdec$) and to a point $\xdec' \gg \xdec$ after
decoupling but above the bottom mass,
\begin{align}
\frac{1}{Y(\xdec')} - \frac{1}{Y(\xdec)} 
= \overline Y(\xdec') - \overline Y(\xdec) 
= - \frac{\bar \lambda}{\xdec'^{3/2}} + \frac{\bar \lambda}{\xdec^{3/2}}  \; .
\end{align}
From the form of the Boltzmann equation in
Eq.\eqref{eq:boltzmann_approx} we see that $Y(x)$ drops rapidly with
increasing $x$. If we choose $\xdec' \gg \xdec = 23$ it follows
that $Y(\xdec') \ll Y(\xdec)$ and hence
\begin{align}
\frac{1}{Y(\xdec')} = \frac{\bar \lambda}{\xdec^{3/2}}
\qquad \Leftrightarrow \qquad 
Y(\xdec')
= \frac{m_\chi^3 \, \langle \sigma_{\chi \chi} \, v \rangle}{H(x=1)} 
= \frac{\xdec}{\lambda(\xdec)}
\eqx{eq:wimp_boltzmann}
\xdec \frac{\pi \sqrt{\geff}}{\sqrt{90} \, \mpl m_\chi \, } \, 
      \frac{1}{\langle \sigma_{\chi \chi} \, v \rangle} \; .
\end{align}
In this expression $\geff$ is evaluated around the point of
decoupling.  For the second, expansion phase we can just follow
Eq.\eqref{eq:wimp_relic_approx2a} and compute
\begin{align}
\rho_\chi(T_0) 
&= m_\chi n_\chi(T_0) \notag \\
&=m_\chi Y(\xdec')\Tdec'^3\left(\frac{a(\Tdec')}{a(T_0)}\right)^3 
 \eqx{eq:thermodynamics2} m_\chi Y(\xdec') \; T_0^3 \; \frac{\geff(T_0)}{\geff(\Tdec')}
 = m_\chi \frac{Y(\xdec') T_0^3}{28} \; .
\end{align}
For the properly normalized relic density this means
 \begin{align}
 \label{eq:relic_approx}
\Rightarrow \qquad
\Omega_\chi h^2 
&= m_\chi \,  \,\frac{ Y(\xdec')   T_0^3}{28}
  \; \frac{h^2}{3 \mpl^2 H_0^2} \notag \\
&= \frac{h^2 \pi \sqrt{\geff}}{28\sqrt{90} \, \mpl} \, 
      \frac{\xdec}{\langle \sigma_{\chi \chi} \, v \rangle} \; 
   \frac{T_0^3}{3 \mpl^2 H_0^2} \\
&= \frac{h^2 \pi \sqrt{\geff}}{28\sqrt{90} \, \mpl} \, 
      \frac{\xdec}{\langle \sigma_{\chi \chi} \, v \rangle} \; 
   \frac{(2.4 \cdot 10^{-4})^3}{(2.5 \cdot 10^{-3})^4} \frac{1}{\ev} \quad \Rightarrow \quad 
\boxed{
\Omega_\chi h^2 
\approx 0.12 \; \frac{\xdec}{23} \; \frac{\sqrt{\geff}}{10} \; \frac{1.7 \cdot 10^{-9} \, \gev^{-2}}{\langle \sigma_{\chi \chi} \, v \rangle}}\; .\notag
\end{align}
%
%
%
\index{relic!abundance}
We can translate this result into different units. In the cosmology
literature people often use $\ev^{-1} = 2 \cdot 10^{-5}$~cm.  In particle
physics we measure cross sections in barn, where $1~\fb =
10^{-39}~\text{cm}^2$. Our above result is a very good approximation
to the correct value for the relic density in terms of the
annihilation cross section
\begin{align}
\Omega_\chi h^2 
\approx 0.12 \; \frac{\xdec}{23} \; \frac{\sqrt{\geff}}{10} \; \frac{1.7 \cdot 10^{-9} \, \gev^{-2}}{\langle \sigma_{\chi \chi} \, v \rangle}
\approx 0.12 \; \frac{\xdec}{23} \; \frac{\sqrt{\geff}}{10} \; \frac{2.04 \cdot 10^{-26} \text{cm}^3/\text{s}}{\langle \sigma_{\chi \chi} \, v \rangle} \; .
\label{eq:wimp_relic_4}
\end{align}
With this result we can now insert the WIMP annihilation rate given by
Eq.\eqref{eq:wimp_ann_approx} and Eq.\eqref{eq:non_relativistic_v},
\begin{align}
\langle \sigma_{\chi \chi} \, v \rangle 
&= \sigma_{\chi \chi} \, v + \ope (v^2) 
\approx 
 \sqrt{\frac{2}{x}} \; \frac{\pi \alpha^2 m_\chi^2}{c_w^4 m_Z^4} \notag \\
\Rightarrow \quad 
\Omega_\chi h^2 
&= 0.12 \;  \frac{\xdec}{23} \; \frac{\sqrt{\geff}}{10} \; 
\frac{c_w^4 m_Z^4 \sqrt{x}}{\sqrt{2} \pi \alpha^2 m_\chi^2} \; 
\frac{1.7 \cdot 10^{-9}}{\gev^2}
= 0.12 \;\left(\frac{\xdec}{23}\right)^{3/2} \; \frac{\sqrt{\geff}}{10} \; 
\left( \frac{35\, \gev}{m_\chi} \right)^2
\, .
\end{align}
We can compare this result to our earlier estimate in
Eq.\eqref{eq:wimp_relic_approx3} and confirm that these numbers make
sense for a weakly interacting particle with a weak-scale
mass.

Alternatively, we can replace the scaling of the annihilation cross
section given in Eq.\eqref{eq:wimp_ann_approx} by a simpler form, only
including the WIMP mass and certainly valid for heavy dark matter,
$m_\chi > m_Z$. We find
\begin{align}
\boxed{
\langle \sigma_{\chi \chi} \, v \rangle \approx \frac{g^4}{16 \pi m_\chi^2} 
\really \frac{1.7 \cdot 10^{-9}}{\gev^2} }
\qquad \Leftrightarrow \qquad 
g^2 \approx \frac{m_\chi}{3400~\gev} = \frac{m_\chi}{3.4~\tev} \; .
\label{eq:wimp_rate_vs_mass}
\end{align}
This form of the cross section does not assume a weakly interacting
origin, it simply follows for the scaling with the coupling and from
dimensional analysis. Depending on the coupling, its prediction for
the dark matter mass can be significantly higher. Based on this
relation we can estimate an \ul{upper limit on $m_\chi$} from
the unitarity condition for the annihilation cross section
\begin{align}
g^2 \lesssim 4\pi
\qquad \Leftrightarrow \qquad 
m_\chi < 54~\tev \; .
\end{align}
A lower limit does not exist, because we can make a lighter particle
more and more weakly coupled. Eventually, it will be light enough to
be relativistic at the point of decoupling, bringing us back to the
relic neutrinos discussed in Section~\ref{sec:neutrinos}.\bigskip

Let us briefly recapitulate our argument which through the Boltzmann
equation leads us to the WIMP miracle: we start with a $2 \to 2$
scattering process linking dark matter to Standard Model particles
through a so-called mediator,\index{mediator} which can for example be a weak
boson. This allows us to compute the dark matter relic density as a
function of the mediating coupling and the dark matter mass, and it
turns out that a weak-coupling combined with a dark matter mass below
the TeV scale fits perfectly. There are two ways in which we can
modify the assumed dark matter annihilation process given in
Eq.\eqref{eq:wimp_annihilation}: first, in the next
Section~\ref{sec:coannihilation} we will introduce additional
annihilation channels for an extended dark matter sector. Second, in
Section~\ref{sec:models_portal} we will show what happens if the
annihilation process proceeds through an $s$-channel Higgs resonance.

\subsection{Co-annihilation}
\label{sec:coannihilation}
\index{co-annihilation}

In many models the dark matter sector consists of more than one
particle, separated from the Standard Model particles for example
through a specific quantum number. A typical structure are
\ul{two dark matter particles} $\chi_1$ and $\chi_2$ with
$m_{\chi_1} < m_{\chi_2}$. In analogy to
Eq.\eqref{eq:wimp_annihilation} they can annihilate into a pair of
Standard Model particles through the set of processes
\begin{align}
\chi_1 \chi_1 \to f \bar{f}
\qqquad 
\chi_1 \chi_2 \to f \bar{f}
\qqquad 
\chi_2 \chi_2 \to f \bar{f} \; .
\label{eq:proc_coann}
\end{align}
This set of processes can mediate a much more efficient annihilation
of the dark matter state $\chi_1$ together with the second state
$\chi_2$, even in the limit where the actual dark matter process
$\chi_1 \chi_1 \to f \bar{f}$ is not allowed. Two
non-relativistic states will have number densities both given by
Eq.\eqref{eq:n_vs_temp}.  We know from Eq.\eqref{eq:xdec_wimp} that
decoupling of a WIMP \index{WIMP} happens at typical values $\xdec = m_\chi/\Tdec
\approx 28$, so if we for example assume $\Delta m_\chi = m_{\chi_2} -
m_{\chi_1} = 0.2 \; m_{\chi_1}$ and $g_1 = g_2$ we find
\begin{align}
\frac{n_2(\Tdec)}{n_1(\Tdec)}
\eqx{eq:n_vs_temp}
 \frac{g_2}{g_1} \; \left( 1 + \frac{\Delta m_\chi}{m_{\chi_1}}  \right)^{3/2} \; 
  e^{-\Delta m_\chi/\Tdec} 
= 1.31 \;  e^{-0.2 \xdec} 
 \approx \frac{1}{206} \; .
\label{eq:ratio_coann}
\end{align}
Just from statistics the heavier state will already be rare by the
time the lighter, actual dark matter agent annihilates. For a mass difference
around 10\% this suppression is reduced to a factor 1/15, gives us an
estimate that efficient co-annihilation will prefer two states with
mass differences in the 10\% range or closer.\bigskip

Let us assume that there are two particles present at the time of
decoupling. In addition, we assume that the first two processes shown
in Eq.\eqref{eq:proc_coann} contribute to the annihilation of the dark
matter state $\chi_1$.  In this case the Boltzmann equation from
Eq.\eqref{eq:boltzmann} reads
\begin{align}
\dot n_1(t) + 3 H(t) n_1(t) 
&= - \langle \sigma_{\chi_1 \chi_1} v \rangle \; 
  \left( n_1(t)^2  - n_{1,\text{eq}}(t)^2 \right) 
  - \langle \sigma_{\chi_1 \chi_2} v \rangle \; 
  \left( n_1(t) n_2(t)  - n_{1,\text{eq}}(t) n_{2,\text{eq}}(t) \right) \notag \\
&\approx - \langle \sigma_{\chi_1 \chi_1} v \rangle \; 
  \left( n_1(t)^2  - n_{1,\text{eq}}(t)^2 \right) 
  - \langle \sigma_{\chi_1 \chi_2} v \rangle \; 
  \left( n_1^2(t) - n_{1,\text{eq}}(t)^2 \right) \; \frac{n_2}{n_1} \notag \\
&\eqx{eq:ratio_coann}
 - \left[ \langle \sigma_{\chi_1 \chi_1} v \rangle 
                 + \langle \sigma_{\chi_1 \chi_2} v \rangle \; \frac{g_2}{g_1} \; 
                 \left( 1 + \frac{\Delta m_\chi}{m_{\chi_1}} \right)^{3/2}  \; e^{- \Delta m_\chi/T} 
     \right]
  \left( n_1(t)^2  - n_{1,\text{eq}}(t)^2 \right) \; . 
\label{eq:coann_boltzmann}
\end{align}
In the second step we assume that the two particles decouple
simultaneously, such that their number densities track each other
through the entire process, including the assumed equilibrium values.
This means that we can throughout our single-species calculations 
just replace
\begin{align}
\langle \sigma_{\chi \chi} \, v \rangle
&\to \langle \sigma_{\chi_1 \chi_1} v \rangle 
   + \langle \sigma_{\chi_1 \chi_2} v \rangle \; \frac{g_2}{g_1} \; 
     \left( 1 + \frac{\Delta m_\chi}{m_{\chi_1}} \right)^{3/2}  \; e^{- \Delta m_\chi/T} \; .
\end{align}
In the co-annihilation setup it is not required that the direct
annihilation process dominates.  The annihilation of more than one
particle contributing to a dark matter sector can include many other
aspects, for example when the dark matter state only interacts
gravitationally and the annihilation proceeds mostly through a
next-to-lightest, weakly interacting state. The Boltzmann equation
will in this case split into one equation for each state and include
decays of the heavier state into the dark matter state.  Such a system
of Boltzmann equations cannot be solved analytically in general.

What we can assume is that the two co-annihilation partners have very
similar masses, $\Delta m_\chi \ll m_{\chi_1}$, similar couplings,
$g_1 = g_2$, and that the two annihilation processes in
Eq.\eqref{eq:proc_coann} are of similar size, $\langle
\sigma_{\chi_1 \chi_1} v \rangle \approx \langle \sigma_{\chi_1 \chi_2} v
\rangle$. In that limit we simply find $\langle \sigma_{\chi \chi} \, v
\rangle \to 2 \langle \sigma_{\chi_1 \chi_1} v \rangle $ in the Boltzmann
equation. We know from Eq.\eqref{eq:wimp_relic_4} how the correct
relic density depends on the annihilation cross section. Keeping the
relic density constant we absorb the rate increase through
co-annihilation into a shift in the typical WIMP masses of the two
dark matter states. According to Eq.\eqref{eq:wimp_rate_vs_mass} the
WIMP masses should now be
\begin{align}
\langle \sigma_{\chi \chi} \, v \rangle 
\approx \frac{g^4}{16 \pi m_{\chi}^2}
\equiv 2 \; \frac{g^4}{32 \pi \, m_{\chi_1}^2}
\qquad \text{or} \qquad 
m_{\chi_1} \approx m_{\chi_2} \approx \sqrt{2} m_\chi \; .
\label{eq:simple_coannihilaion}
\end{align}
A simple question we can ask for example when we will talk about
collider signatures is how easy it would be to discover a single WIMP
compared to the pair of co-annihilating, slightly heavier
WIMPs.\bigskip

An interesting question is how co-annihilation channels modify the
WIMP mass scale which is required by the observed relic density. From
Eq.\eqref{eq:simple_coannihilaion} we see that an increase in the
total annihilation rate leads to a larger mass scale of the dark
matter particles, as expected from our usual scaling.  On the other
hand, the annihilation cross section really enters for example
Eq.\eqref{eq:wimp_relic_4} in the combination $\langle
\sigma_{\chi_1 \chi_1} v \rangle/\sqrt{\geff}$. If we increase the number
of effective degrees of freedom significantly, while the
co-annihilation channels really have a small effect on the total
annihilation rate, the dark matter mass might also decrease.

\subsection{Velocity dependence}
\label{sec:velocity}
\index{dark matter!annihilation}

While throughout the early estimates we use the dark matter
annihilation rate $\sigma_{\chi \chi}$, we introduce the more
appropriate thermal expectation value of the velocity times the
annihilation rate $\langle \sigma_{\chi \chi} v \rangle$ in
Eq\eqref{eq:def_sigma_v}. This combination has the nice feature that
its leading term can be independent of the velocity $v$. In general,
the velocity-weighted cross section will be of the form
\begin{align}
\langle \sigma_{\chi \chi} v \rangle
= \langle s_0 + s_1 v^2 + \ope(v^4) \rangle
\label{eq:threshold_series}
\end{align}
This pattern follows from the partial wave analysis of relativistic
scattering.  The first term $s_0$ is velocity-independent and arises
from $S$-wave scattering. An example is the scattering of two scalar
dark matter particles with an $s$-channel scalar mediator or two Dirac
fermions with an $s$-channel vector mediator. The second term $s_1$
with a vanishing rate at threshold is generated by $S$-wave and
$P$-wave scattering. It occurs for example for Dirac fermion
scattering through an $s$-channel vector mediator. All $t$-channel processes
have an $S$-wave component and are not suppressed at threshold.

For dark matter phenomenology, the dependence on a potentially small
velocity shown in Eq.\eqref{eq:threshold_series} is the important
aspect.  Different dark matter agents with different interaction patterns lead to distinct threshold dependences. For $s$-channel
and $t$-channel mediators and several kinds of couplings to Standard
Model fermions $f$ in the final state we find

\begin{center}
\begin{tabular}{rllll|llll}
\hline
& \multicolumn{4}{c}{$s$-channel mediator} 
& \multicolumn{4}{c}{$t$-channel mediator} \\
 & $\bar{f} f$ & $\bar{f} \gamma^5 f$ & $\bar{f} \gamma^\mu f$ & $\bar{f} \gamma^\mu \gamma^5 f$
 & $\bar{f} f$ & $\bar{f} \gamma^5 f$ & $\bar{f} \gamma^\mu f$ & $\bar{f} \gamma^\mu \gamma^5 f$ \\ \hline
Dirac fermion       & $v^2$ & $v^0$ & $v^0$ & $v^0$
                    & $v^0$ & $v^0$ & $v^0$ & $v^0$ \\
Majorana fermion    & $v^2$ & $v^0$ & $0$ &  $v^0$  
                    & $v^0$ & $v^0$ & $v^0$ &  $v^0$  \\
real scalar & $v^0$ & $v^0$ & $0 $ &  $0 $ \\
complex scalar & $v^0$ & $v^0$ & $ v^2$ &  $ v^2$ \\
\hline
\end{tabular}
\end{center}
\index{mediator!s-channel}
\index{mediator!t-channel}

Particles who are their own anti-particles, like Majorana fermions and
real scalars, do not annihilate through $s$-channel vector
mediators. The same happens for complex scalars and axial-vector
mediators. In general, $t$-channel annihilation to two Standard Model
fermions is not possible for scalar dark matter.\bigskip

To allow for an efficient dark matter annihilation to today's relic
density, we tend to prefer an un-suppressed contribution $s_0$ to
increase the thermal freeze-out cross section. \index{freeze out} The
problem with such large annihilation rates is that they are strongly
constrained by early-universe physics.  For example, the PLANCK
measurements of the \index{PLANCK} matter power spectrum discussed in
Section~\ref{sec:structure} constrain the light dark matter very
generally, just based on the fact that such light dark matter can
affect the photon background at the time of decoupling.  The problem
arises if dark matter candidates annihilate into Standard Model
particles through non-gravitational interactions,
\begin{align}
\chi \chi \to \text{SM} \,\text{SM} \; . 
\end{align}
As we know from Eq.\eqref{eq:dm_annihilation} this process is the key
ingredient to thermal freeze-out dark matter.  If it happens at the
time of last scattering it injects heat into the intergalactic
medium. This ionizes the hydrogen and helium atoms formed during
recombination. While the ionization energy does not modify the time of
the last scattering, it prolongs the period of recombination or,
alternatively, leads to a broadening of the surface of last
scattering. This leads to a suppression of the temperature
fluctuations and enhance the polarization power spectrum.  The
temperature and polarization data from PLANCK puts an upper limit on
the dark matter annihilation cross section\index{PLANCK}
\begin{align}
f_\text{eff} \frac{\langle \sigma_{\chi\chi} v \rangle}{m_\chi} \lesssim 
\frac{8.5 \cdot 10^{-11}}{\gev^3}
\label{eq:planck}
\end{align}
The factor $f_\text{eff} < 1$ denotes the fraction of the dark matter
rest mass energy injected into the intergalactic medium. It is a
function of the dark matter mass, the dominant annihilation channel,
and the fragmentation patterns of the SM particles the dark matter
agents annihilate into. For example, a $200$~GeV dark matter particle
annihilating to photons or electrons reaches $f_\text{eff} =
0.66~...~0.71$, while an annihilation to muon pairs only gives
$f_\text{eff} = 0.28$.
%
%
As we know from Eq.\eqref{eq:relic_approx} for freeze-out dark matter
an annihilation cross section of the order $\langle
\sigma_{\chi\chi}v\rangle\approx 1.7 \cdot 10^{-9}$/~GeV$^2$ is
needed. This means that the PLANCK constraints of Eq.\eqref{eq:planck}
requires
\begin{align}\label{eq:plancklimit}
m_\chi \gtrsim 10~\gev \; ,
\end{align}
In contrast to limits from searches for dark matter annihilation in
the center of the galaxy or in dwarf galaxies, as we will discuss in
Section \ref{sec:indirect}, this constraint does not suffer from
astrophysical uncertainties, such as the density profile of the dark
matter halo in galaxies.

\subsection{Sommerfeld enhancement}
\label{sec:sommerfeld}
\index{Sommerfeld enhancement}

Radiative corrections can drastically change the threshold behavior
shown in Eq.\eqref{eq:threshold_series}. As an example, we study the
annihilation of two dark matter fermions through an $s$-channel scalar
in the limit of small relative velocity of the two fermions.
The starting point of our discussion is the loop diagram which
describes the exchange of a gauge boson between two incoming (or
outgoing) massive fermions $\chi$:
\vspace{.5cm}
\begin{center}
\begin{fmfgraph*}(100,80)
\fmfset{arrow_len}{2mm}
\fmfright{o1}
\fmfleft{i2,i3}
\fmf{dashes,tension=0.4,width=0.6}{o1,v1}
\fmf{fermion,tension=0.1,label=$\chi(q+k_2)$,lab.side=left,width=0.6}{v1,v2}
\fmf{fermion,tension=0.1,label=$\chi(q+k_1)$,lab.side=left,width=0.6}{v3,v1}
\fmf{photon,tension=0.15,label=$Z(q)$,lab.side=left,width=0.6}{v2,v3}
\fmf{fermion,tension=0.6,width=0.6}{v2,i2}
\fmf{fermion,tension= 0.6,width=0.6}{i3,v3}
\fmflabel{$S$}{o1}
\fmflabel{$\chi(k_2)$}{i2}
\fmflabel{$\chi(k_1)$}{i3}
\end{fmfgraph*}
\end{center}
\vspace{.5cm}

After inserting the Feynman rules we find the expression
\begin{align}
\int d^4 q \; 
\frac{\slashchar{q} + \slashchar{k}_1 + m_\chi}{(q+k_1)^2 - m_\chi^2 }
\gamma_\mu 
\frac{1}{q^2 - m_Z^2}
\gamma^\mu 
\frac{\slashchar{q} + \slashchar{k}_2 + m_\chi}{(q+k_2)^2 - m_\chi^2 } \; .
\label{eq:sommerfeld_int}
\end{align}
The question is where this integral receives large contributions.
Using $k^2 = m_\chi^2$ the denominators of the fermion propagators
read
\begin{align}
\frac{1}{(q+k)^2 - m_\chi^2}
&= \frac{1}{q_0^2 - |\vec q|^2 + 2 q_0 k_0 - 2 \vec q \vec k} \notag \\
&\eqx{eq:scatter_velocity} \frac{1}{q_0^2 - |\vec q|^2 + (2+v^2) m_\chi q_0 - 2 m_\chi v |\vec q| \cos \theta + \ope(q_0 v^2)} \notag \\
&\stackrel{|\vec q|=m_\chi v}{=}
\frac{1}{q_0^2 - m_\chi^2 v^2 (1+2 \cos \theta) + (2+v^2) m_\chi q^0 + \ope(q_0 v^2)} \; .
\end{align}
The particles in the loop are not on their respective mass
shells. Instead, we can identify a particularly dangerous region for
$v \to 0$, namely $q_0=m_\chi v^2$, where
\begin{align}
\frac{1}{(q+k)^2 - m_\chi^2} =
\frac{1}{m_\chi^2 v^2 (1-2 \cos \theta) + \ope(v^4)} \; .
\end{align}
Unless we make an assumption about the angle $\theta$ we cannot make a
stronger statement about the contributions of the fermion propagators.
If we just set $\cos \theta = 0$ we find
\begin{align}
\frac{1}{(q+k)^2 - m_\chi^2} =
\frac{1}{m_\chi^2 v^2 + \ope(v^4)} \; .
\end{align}
In the same phase space region the $Z$ boson propagator in the
integral scales like
\begin{align}
\frac{1}{q^2 - m_Z^2} 
= \frac{1}{m_\chi^2 v^4 - m_\chi^2 v^2  - m_Z^2} 
= - \frac{1}{m_\chi^2 v^2 + m_Z^2 + \ope(v^4)} \; .
\end{align}
In the absence of the gauge boson mass the gauge boson propagator
would diverge for $v \to 0$, just like the fermion propagators.  This
means that we can approximate the loop integral by focussing on the
phase space regime
\begin{align}
q_0 \approx m_\chi v^2 
\qqquad \text{and} \qqquad 
|\vec q| \approx m_\chi v \; .
\end{align}
The complete infrared contribution to the one-loop matrix element of
Eq.\eqref{eq:sommerfeld_int} with a massive gauge boson exchange and
neglecting the Dirac matrix structure is
\begin{align}
\int d^4 q \; 
\frac{m_\chi}{(q + k_1)^2 - m_\chi^2}
\frac{1}{q^2 - m_Z^2}
\frac{m_\chi}{(q - k_2)^2 - m_\chi^2} 
&\approx \Delta q_0 (\Delta |\vec q|)^3 \; 
\frac{1}{m_\chi v^2} \;
\frac{1}{m_\chi^2 v^2 + m_Z^2} \;
\frac{1}{m_\chi v^2} \notag \\
&\approx m_\chi v^2 \; (m_\chi v)^3 \; 
\frac{1}{m_\chi v^2} \;
\frac{1}{m_\chi^2 v^2 + m_Z^2} \;
\frac{1}{m_\chi v^2} \notag \\
&=\frac{v}{v^2 + \dfrac{m_Z^2}{m_\chi^2}}
\stackrel{m_\chi \gg m_Z}{\longrightarrow}
\frac{1}{v} \; .
\end{align}
This means that part of the one-loop correction to the dark matter
annihilation process at threshold scales like $1/v$ in the limit of
massless gauge boson exchange. For massive gauge bosons the divergent
behavior is cut off with a lower limit $v \gtrsim m_Z/m_\chi$. If we
attach an additional gauge boson exchange to form a two-loop integral,
the above considerations apply again, but only to the last, triangular
diagram. The divergence still has the form $1/v$. Eventually, it will
be cut off by the widths of the particles, which is a phrase often used
in the literature and not at all easy to show in detail.

What is more important is the question what the impact of this result
is for our calculations --- it will turn out that while the loop
corrections for slowly moving particles with a massless gauge boson
exchange are divergent, they typically correct a cross section
which vanishes at threshold and only lead to a finite rate at the
production threshold.\bigskip

As long as we limit ourselves to $v \ll 1$ we do not need to use
relativistic quantum field theory for this calculation. We can compute
the same $v$-dependent correction to particle scattering using
non-relativistic quantum mechanics.  We assume two electrically and
weakly charged particles $\chi^\pm$, so their attractive potential has
spherically symmetric Coulomb and Yukawa parts,
\begin{align}
V(r) = - \frac{e^2}{r} - \frac{g_Z^2}{r} e^{-m_Z r} 
\qquad \text{with} \quad r = |\vec r| \; .
\label{eq:pot_sommerfeld}
\end{align}
The coupling $g_Z$ describes an unknown $\chi$-$\chi$-$Z$ interaction.
With such a potential we can compute a two-body scattering process. The
wave function $\psi_k(\vec r)$ will in general be a superposition of
an incoming plane wave in the $z$-direction and a set of spherical
waves with a modulation in terms of the scattering angle $\theta$.
%
%
As in Eq.\eqref{eq:spherical_def} we can expand the wave function in
spherical harmonics, combined with an energy-dependent radial function
$R(r;E)$. We again exploit the symmetry with respect to the
azimuthal angle $\phi$  and obtain
\begin{align}
\psi_k(\vec r) 
&= \sum_{\ell = 0}^\infty \sum_{m=-\ell}^\ell \; 
   a_{\ell m} Y_{\ell m}(\theta, \phi) \; R_{\ell}(r;E) \notag \\
&= \sum_{\ell = 0}^\infty \; (2 \ell +1) \; 
   a_{\ell 0} Y_{\ell 0}(\theta, \phi) \; R_{\ell}(r;E) \notag \\
&\eqx{eq:legendre_def} \sum_{\ell = 0}^\infty \; (2 \ell +1) \; 
   a_{\ell 0} \; \frac{\sqrt{2\ell+1}}{2} P_\ell(\cos \theta) \; R_{\ell}(r;E) 
=: \sum_{\ell = 0}^\infty \; 
   A_\ell \; P_\ell(\cos \theta) \; R_{\ell}(r;E) \; .
\label{def:wavefunction_series}
\end{align}
From the calculation of the hydrogen atom we know that the radial,
time-independent Schr\"odinger equation in terms of the reduced mass
$m$ reads
\begin{align}
\left[
- \frac{1}{2m r^2} \frac{d}{dr} \left( r^2 \frac{d}{dr} \right) 
+ \frac{\ell (\ell +1)}{2m r^2}
+ V(r) - E 
\right] R_\ell(r;E) = 0 \; .
\label{eq:schroedinger}
\end{align}
The reduced mass for a system with two identical masses is given by 
\begin{align}
m = \frac{m_1 m_2}{m_1 + m_2}
  =\frac{m_\chi}{2} \; .
\label{eq:reduced_mass}
\end{align}
As a first step we solve the Schr\"odinger equation\index{Schr\"odinger equation} at large distances,
where we can neglect $V(r)$. We know that the solution will be plane
waves, but to establish our procedure we follow the procedure
starting with Eq.\eqref{eq:schroedinger} step by step,
\begin{alignat}{7}
&& \left[
- \frac{1}{r^2} \frac{d}{dr} \left( r^2 \frac{d}{dr} \right) 
+ \frac{\ell (\ell +1)}{r^2}
- k^2
\right] R_{k \ell}(r) &= 0 \qquad \text{with} & \; k^2 &:= 2mE = m^2 v^2\notag \\
& \Leftrightarrow & \qquad 
\frac{1}{\rho^2} \frac{d}{d\rho} \left( \rho^2 \frac{dR_{k \ell} }{d\rho} \right) 
- \frac{\ell (\ell +1)}{\rho^2} R_{k \ell}
+ R_{k \ell}
 &= 0 \qquad \text{with} & \; \rho &:= kr \notag \\
& \Leftrightarrow & \qquad 
 \frac{1}{\rho^2} \left( 2 \rho \frac{dR_{k \ell} }{d\rho} + \rho^2 \frac{d^2 R_{k \ell}}{d\rho^2} \right) 
- \frac{\ell (\ell +1)}{\rho^2} R_{k \ell}
+ R_{k \ell}
 &= 0  \notag \\
& \Leftrightarrow & \qquad 
 \rho^2 \frac{d^2 R_{k \ell}}{d\rho^2}
+  2 \rho \frac{dR_{k \ell} }{d\rho} 
- \ell (\ell +1) R_{k \ell}
+ \rho^2 R_{k \ell}
 &= 0  
\label{eq:schroedinger_vacuum}
\end{alignat}
This differential equation turns out to be identical to the implicit
definition of the \ul{spherical Bessel functions}
$j_\ell(\rho)$, so we can identify $R_{k \ell}(r) = j_\ell(\rho)$. The
radial wave function can then be expressed in Legendre polynomials,
\begin{align}
R_{k \ell}(r) = j_\ell(\rho) 
&= 
\frac{1}{2 \ell!}  
\left( \frac{\rho}{2} \right)^\ell (-1)^\ell
\int_{-1}^1 dt \; e^{i \rho t} (t^2-1)^\ell \notag \\
&=
\frac{1}{2 \ell!} 
\left( \frac{\rho}{2} \right)^\ell (-1)^\ell
\left[ \frac{1}{i \rho} e^{i \rho t} (t^2-1)^\ell \Bigg|_{-1}^1 
- \int_{-1}^1 \; dt \; \frac{1}{i \rho} e^{i \rho t} \; \frac{d}{dt} (t^2-1)^\ell 
\right] \notag \\
&=
\frac{1}{2 \ell!}  
\left( \frac{\rho}{2} \right)^\ell (-1)^\ell\; \frac{(-1)}{i \rho}
\int_{-1}^1 \; dt \;  e^{i \rho t} \; \frac{d}{dt} (t^2-1)^\ell 
= \cdots \notag \\
&=
\frac{1}{2 \ell!}  
\left( \frac{\rho}{2} \right)^\ell \frac{1}{(i \rho)^\ell}
\int_{-1}^1 \; dt \;  e^{i \rho t} \; \frac{d^\ell}{dt^\ell} (t^2-1)^\ell 
\eqx{eq:legendre_def2}
\frac{(-i)^\ell}{2} \; 
\int_{-1}^1 \; dt \;  e^{i \rho t} \; P_\ell(t) \; .
\label{eq:spherical_bessel}
\end{align}
The integration variable $t$ corresponds to $\cos \theta$ in our
physics problem.  As mentioned above, these solutions to the free
Schr\"odinger equation have to be plane waves. We use
the relation
\begin{align}
\sum_{\ell = 0}^\infty \frac{2\ell + 1}{2} P_\ell(t) P_\ell(t') = \delta(t - t')
\label{eq:sommerfeld_sumrule}
\end{align}
to link the plane wave to this expression in terms of the spherical
Bessel functions and the Legendre polynomials. 
With the correct ansatz we find
\begin{align}
\sum_{\ell = 0}^\infty i^\ell (2\ell + 1) P_\ell(t) j_\ell(\rho)
&\eqx{eq:spherical_bessel} \sum_{\ell = 0}^\infty i^\ell (2\ell + 1) P_\ell(t) 
\frac{(-i)^\ell}{2} \; 
\int_{-1}^1 \; dt' \;  e^{i \rho t'} \; P_\ell(t') \notag \\
&\eqx{eq:sommerfeld_sumrule} 2 i^\ell \; \frac{(-i)^\ell}{2} \; e^{i \rho t} = e^{i k r \cos \theta} \; .
\end{align}
If we know that the series in Eq.\eqref{def:wavefunction_series}
describes such plane waves, we can determine $A_\ell R_{k \ell}$ by comparing the
two sums and find
\begin{align}
A_\ell \; R_{k \ell}(r) 
&= i^\ell (2\ell+1) j_\ell(k r) 
\approx \begin{cases}
i^\ell  (2\ell+1) \; \dfrac{\sin \left(kr  - \dfrac{\ell \pi}{2} \right)}{k r} & \text{for} \; kr \gg \ell^2 \\[4mm]
i^\ell  (2\ell+1) \; \dfrac{(kr)^\ell}{(2\ell +1)!!} & \text{for} \; kr \ll 2 \sqrt{\ell} \; .
\end{cases}
\label{eq:factor_sommerfeld}
\end{align}
We include two limits which can be derived for the spherical Bessel
functions. To describe the interaction with and without a potential
$V(r)$ we are always interested in the wave function at the origin.  The
lower of the above two limits indicates that for small $r$ and hence
small $\rho$ values only the first term $\ell = 0$ will contribute. We
can evaluate $j_0(kr)$ for $kr = 0$ in both forms and find the same
value,
\begin{align}
\left| \psi_k(\vec 0) \right|^2 
\stackrel{\ell=0}{=} \left| A_0 P_0(\cos \theta) R_{k \ell}(0) \right|^2
= \left| A_0 R_{k \ell}(0) \right|^2 
= \lim_{r \to 0} |j_0(kr)|^2
= 1
\label{eq:wavefunction_vacuum}
\end{align}
The argument that only $\ell = 0$ contributes to the wave function at
the origin is not at all trivial to make, and it holds as long as the
potential does not diverge faster than $1/r$ towards the
origin.\bigskip

Next, we add an attractive Coulomb potential to
Eq.\eqref{eq:schroedinger_vacuum}, giving us the radial Schr\"odinger
equation in a slightly re-written form in the first term
\begin{alignat}{7}
&& \left[
- \frac{1}{r} \frac{d^2}{dr^2} r
+ \frac{\ell (\ell +1)}{r^2}
- \frac{2 m e^2}{r}
- k^2
\right] \frac{u_{k \ell}}{r} &= 0 \qquad \text{with} & \; u_{k \ell}(r) &:= r R_{k \ell}(r)  \notag \\
& \Leftrightarrow & \qquad 
  \frac{d^2}{dr^2} u_{k \ell}
- \frac{\ell (\ell +1)}{r^2} u_{k \ell}
+ \frac{2 m e^2}{r} u_{k \ell}
+ k^2 u_{k \ell}
 &= 0  \notag \\ 
& \Leftrightarrow & \qquad 
  \frac{d^2}{d \rho^2} u_{k \ell}
- \frac{\ell (\ell +1)}{\rho^2} u_{k \ell}
+ \frac{2 m e^2}{\rho k} u_{k \ell}
+ u_{k \ell} 
 &= 0  
\label{eq:schroedinger_vacuum_radial}
\end{alignat}
The solution of this equation will lead us to the well-known hydrogen
atom and its energy levels. However, we are not interested in the
energy levels but in the continuum scattering process. Following the
discussion around Eq.\eqref{eq:factor_sommerfeld} and assuming that
the Coulomb potential will not change the fundamental structure of the
solution around the origin we can evaluate the radial wave function
for $\ell =0$,
\begin{align}
  \frac{d^2}{d \rho^2} u_{k 0}
+ \frac{2 m e^2}{\rho k} u_{k 0}
+ u_{k 0} 
 &= 0  
\end{align}
This is the equation we need to solve and then evaluate at the origin,
$\vec r= \vec 0$. We only quote the result,
\begin{align}
\boxed{
\left| \psi_k(\vec 0) \right|^2 
= \frac{2 \pi e^2}{v} \; \frac{1}{1 - e^{- 2 \pi e^2/v}} 
\approx
\begin{cases}
\dfrac{2 \pi e^2}{v} \qquad & \text{for} \; v \to 0 \\[2mm]
\; 1 \qquad & \text{for} \; v \to \infty 
\end{cases}
} \; .
\label{eq:wavefunction_coulomb}
\end{align}
Compared to Eq.\eqref{eq:wavefunction_vacuum} this increased
probability measure is called the Sommerfeld enhancement. It is
divergent at small velocities, just as in the Feynman-diagrammatic
discussion before. For very small velocities, it can lead to an
enhancement of the threshold cross section by several orders of
magnitude.

It can be shown that the calculation based on ladder diagrams in
momentum space and based on the Schr\"odinger equation in position
space are equivalent for simple scattering processes. The resummation
of the ladder diagrams is equivalent to the computation of the wave
function at the origin in the Fourier-transformed position
space.\bigskip

The case of the Yukawa potential shows a similar behavior.  It
involves an amusing trick in the computation of the potential, so we
discuss it in some detail. When we include the Yukawa potential \index{Yukawa potential} in the
Schr\"odinger equation we cannot solve the equation analytically;
however, the \ul{Hulthen potential} \index{Hulthen potential}is an approximation to the
Yukawa potential which does allow us to solve the Schr\"odinger
equation. It is defined as
\begin{align}
V(r) 
= \frac{g_Z^2 \delta e^{- \delta r}}{1 - e^{-\delta r}} \; .
\end{align}
Optimizing the numerical agreement of the Hulthen potential's radial
wave functions with those of the Yukawa potential suggests for the
relevant mass ratio in our calculation
\begin{align}
\delta \approx \frac{\pi^2}{6} \; m_Z \; ,
\label{eq:delta_hulthen}
\end{align}
which we will use later.
Unlike for the Coulomb potential we can now keep the full
$\ell$-dependence of the Schr\"odinger equation. The only additional
approximation we use is for the angular
momentum term
\begin{align}
\frac{\delta^2 e^{-\delta r}}{\left( 1- e^{-\delta r} \right)^2} 
&= \delta^2 \; \dfrac{1 - \delta r + \ope(\delta^2 r^2)}
                      {\left(- \delta r + \dfrac{1}{2} \delta^2 r^2 + \ope(\delta^3 r^3) \right)^2} \notag \\
&= \frac{1}{r^2} \; \dfrac{1 - \delta r + \ope(\delta^2 r^2)}
                      {\left(1 - \dfrac{1}{2} \delta r + \ope(\delta^2 r^2) \right)^2} 
= \frac{1}{r^2} \; \left( 1 + \ope(\delta^2 r^2) \right) \; .
\end{align}
The radial Schr\"odinger equation of
Eq.\eqref{eq:schroedinger_vacuum_radial} with the Hulthen potential
and the above approximation for the angular-momentum-induced potential
term now reads
\begin{align}
&& \left[
- \frac{1}{r} \frac{d^2}{dr^2} r
+ \ell (\ell +1) \; \frac{\delta^2 e^{-\delta r}}{\left( 1- e^{-\delta r} \right)^2} 
+ \frac{g_Z^2 \delta e^{- \delta r}}{1 - e^{-\delta r}} 
- k^2
\right] \frac{u_{k \ell}}{r} &= 0  \qqqquad \notag \\
& \Leftrightarrow & \qquad 
  \frac{d^2}{dr^2} u_{k \ell}
- \ell (\ell +1) \; \frac{\delta^2 e^{-\delta r}}{\left( 1- e^{-\delta r} \right)^2} u_{k \ell}
+ \frac{g_Z^2 \delta e^{- \delta r}}{1 - e^{-\delta r}} u_{k \ell}
+ k^2 u_{k \ell}
 &= 0  \; .
\end{align}
Again, we only quote the result: the leading term for the
corresponding Sommerfeld enhancement factor in the limit $v \ll 1$
arises from
\begin{alignat}{5}
\left| \psi_k(\vec 0) \right|^2 
&= \frac{\pi g_Z^2}{v} \;
   \frac{\sinh \dfrac{2 v m_\chi \pi}{\delta}}
        {\cosh \dfrac{2 v m_\chi \pi}{\delta} - \cos \left( 2\pi \sqrt{\dfrac{g_Z^2 m_\chi}{\delta} - \dfrac{v^2 m_\chi^2}{\delta^2}} \right)} \; .
\label{eq:wavefunction_yukawa}
\end{alignat}
This Sommerfeld enhancement factor will be a combination of a slowly
varying underlying function with a peak structure defined by the
denominator.

\begin{figure}[b!]
\begin{center}
\includegraphics[width=0.55\textwidth]{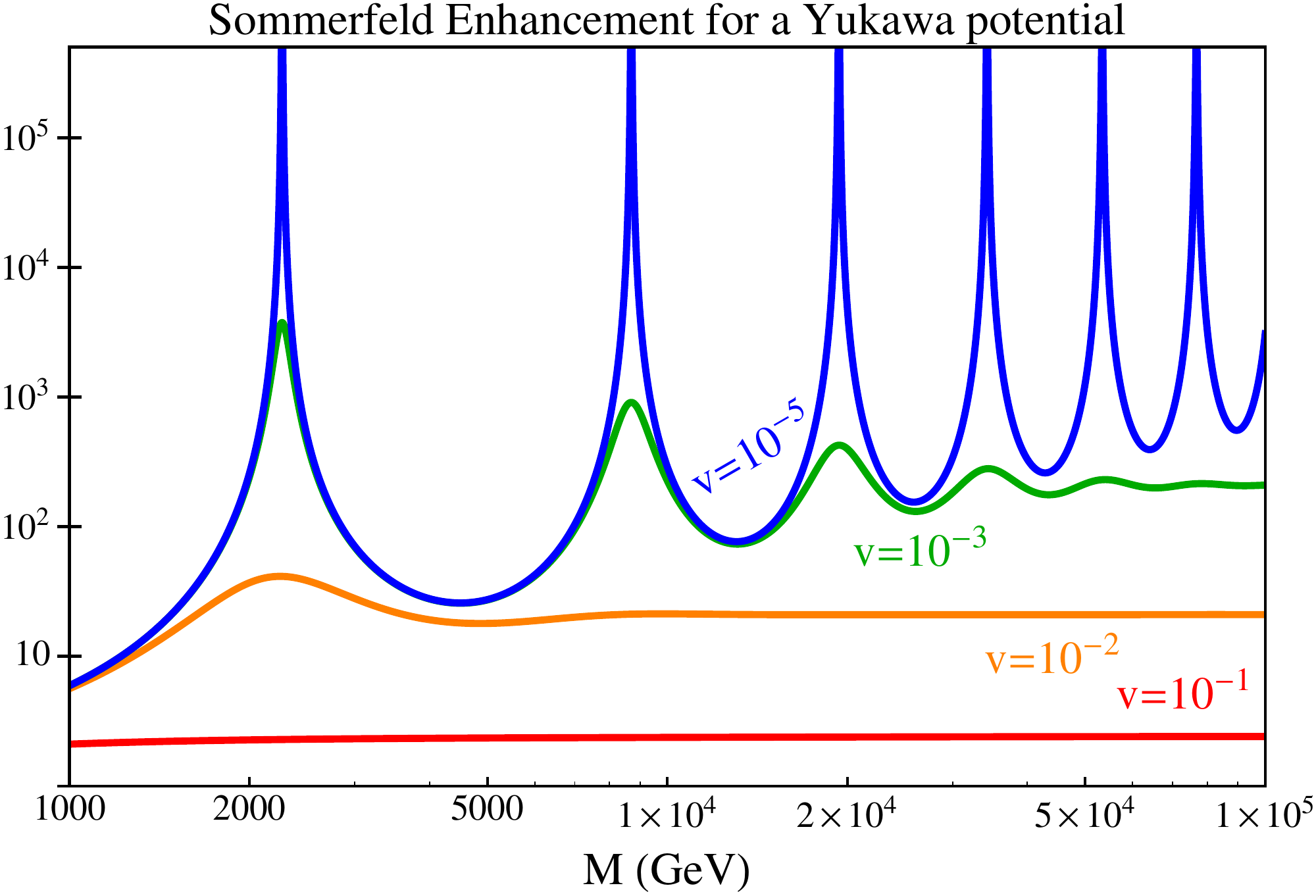}
\end{center}
\caption{Sommerfeld enhancement for a Yukawa potential as a function
  of the dark matter mass ($M \equiv m_\chi$), shown for different
  velocities. It assumes the correct $Z$-mass and a coupling strength
  of $g_Z^2= 1/30$. Figure from Ref.~\cite{tanedo2}, found for example
  in Mariangela Lisanti's lecture notes~\cite{lisanti}.}
\label{fig:sommerfeld}
\end{figure}

We are interested in the position and the height of the first and the
following peaks. We need to two Taylor series 
\begin{align}
\sinh x = x + \ope(x^3)
\qquad \text{and} \qquad 
\cosh x = 1 + \frac{x^2}{2} + \ope(x^4)
\end{align}
The $\cosh$ function is always larger than one and grows rapidly with
increasing argument This means that in the limit $v \ll 1$ the two
terms in the denominator can cancel almost entirely,
\begin{align}
\left| \psi_k(\vec 0) \right|^2 
&= \frac{\pi g_Z^2}{v} \;
   \frac{\dfrac{2 \pi v m_\chi}{\delta} + \ope(v^3)}
        {1 + \ope(v^2) - \cos \left( 2\pi \sqrt{\dfrac{g_Z^2 m_\chi}{\delta} + \ope(v^2)} \right)}
\stackrel{v \to 0}{\longrightarrow}\frac{\dfrac{2 \pi^2 g_Z^2 m_\chi}{\delta}}
        {1 - \cos \sqrt{\dfrac{4 \pi^2 g_Z^2 m_\chi}{\delta}}} \; .
\label{eq:sommerfeld_pole1}
\end{align}
The finite limit for $v \to 0$ is well defined except for mass ratios
$m_\chi/\delta$ or $m_\chi/m_Z$ right on the pole.  The positions of the
peaks in this oscillating function of the mass ratio $m_\chi/m_Z$ is
independent of the velocity in the limit $v \ll 1$. The peak positions
are
\begin{align}
\dfrac{4 \pi^2 g_Z^2 m_\chi}{\delta} = (2 n \pi)^2
\qquad \Leftrightarrow \qquad 
\frac{m_\chi}{\delta} = \frac{n^2}{g_Z^2}
\qquad \stackrel{\text{Eq.\eqref{eq:delta_hulthen}}}{\Leftrightarrow} \qquad 
\frac{m_\chi}{m_Z} = \frac{\pi^2}{6 g_Z^2} \, n^2
\qquad \text{with} \quad n=1,2,... 
\label{eq:sommerfeld_pole2}
\end{align}
For example assuming $g_Z^2 \approx 1/20$ we expect the first peak at
dark matter masses below roughly 3~TeV. For the Sommerfeld enhancement
factor on the first peak we have to include the second term in the
Taylor series in Eq.\eqref{eq:wavefunction_yukawa} and find
\begin{align}
\left| \psi_k(\vec 0) \right|^2 
= \frac{\dfrac{2 \pi^2 g_Z^2 m_\chi}{\delta}}
       {\dfrac{1}{2} \left( \dfrac{2 v m_\chi \pi}{\delta}\right)^2 + \ope(v^4)} 
= \frac{g_Z^2 \delta}{m_\chi v^2} 
\qquad \stackrel{\text{Eq.\eqref{eq:sommerfeld_pole2}}}{\Rightarrow} \qquad 
\boxed{\left| \psi_k(\vec 0) \right|^2 
= \frac{g_Z^4}{v^2} }
\; . 
\label{eq:sommerfeld_pole3}
\end{align}
For $v = 10^{-3}$ and $g_Z^2 \approx 1/20$ we find sizeable Sommerfeld
enhancement on the first peak by a factor around $2500$.
Figure~\ref{fig:sommerfeld} illustrates these peaks in the Sommerfeld
enhancement for different velocities. The slightly different numerical
values arise because the agreement of the Hulthen and Yukawa
potentials is limited.\bigskip

From our calculation and this final result it is clear that a large
ratio of the dark matter mass to the electroweak masses modifies the
pure $v$-dependence of the Coulomb-like Sommerfeld enhancement, but is
not its source. Just like for the Coulomb potential the driving force
behind the Sommerfeld enhancement is the vanishing velocity, leading
to long-lived bound states. The ratio $m_\chi/m_Z$ entering the
Sommerfeld enhancement is simply the effect of the $Z$-mass acting as
a regulator towards small velocities.

\subsection{Freeze-in production}
\label{sec:freezeIn}

\index{freeze-in}
In the previous discussion we have seen that thermal freeze-out offers
an elegant explanation of the observed relic density, requiring only
minimal modifications to the thermal history of the Universe. On the
other hand, for cold dark matter and asymmetric dark matter we have
seen that an alternative production mechanism has a huge effect on
dark matter physics.
A crucial assumption behind freeze-out dark matter is that the coupling
between the Standard Model and dark matter cannot be too small,
otherwise we will never reach thermal equilibrium and cannot apply
Eq.\eqref{eq:dm_equilib}.  For example for the Higgs portal model
discussed in Section~\ref{sec:models_portal} this is the case for a
portal coupling of $\lambda_3\lesssim 10^{-7}$.  For such small
interaction rates the (almost) model-independent lower bound on the
dark matter mass from measurements of the CMB temperature variation
and polarization, discussed in Section \ref{sec:cmb} and giving
$m_\chi \gtrsim 10$~GeV, does not apply. This allows for new kinds of
light dark matter.

\begin{figure}[b!]
\begin{center}
\includegraphics[width=0.9\textwidth]{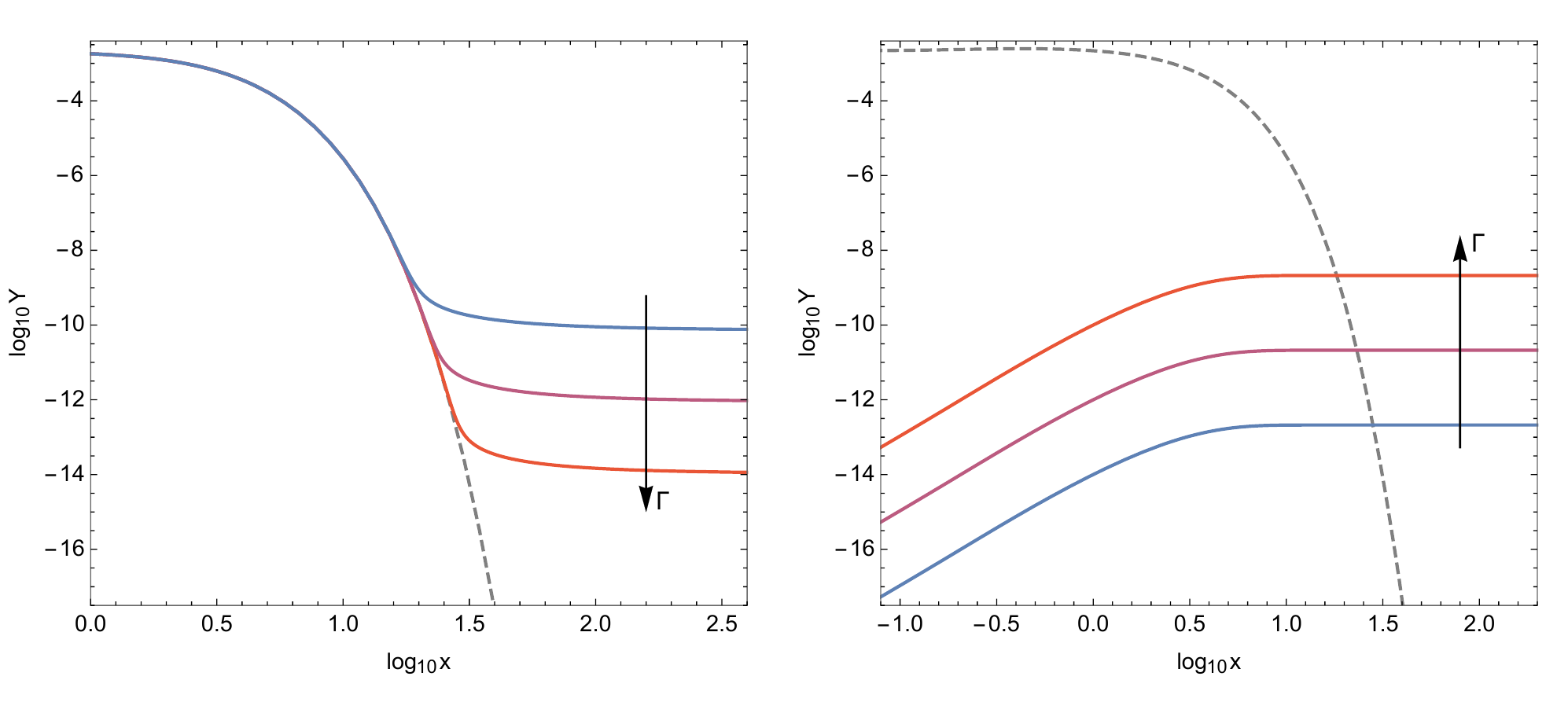}
\end{center}\vspace{-.5cm}
\caption{Scaling of $Y(x)=n_\chi/T^3$ for the freeze-out (left) and
  freeze-in (right) mechanisms for three different interaction rates
  (larger to smaller cross sections along the arrow). In the left
  panel $x=m_\chi/T$ and in the right panel $x=m_B/T$. The dashed
  contours correspond to the equilibrium densities. Figure from
  Ref.~\cite{Bernal:2017kxu}.}
\label{fig:fimp1}
\end{figure}

For such very weakly interacting particles, called feebly interacting
massive particles or \ul{FIMPs}, \index{FIMP} we can invoke the non-thermal,
so-called freeze-in mechanism.  The idea is that the dark matter
sector gets populated through decay or annihilation of SM particles
until the number density of the corresponding SM particles species
becomes Boltzmann-suppressed. For an example SM particle $B$ with an
interaction
\begin{align}\label{eq:exLag}
\mathcal{L}\ni -y m_B \bar \chi B \chi +h.c.
\end{align}
\index{Boltzmann equation}
and $m_B > 2 m_\chi$, the decay $B \to \chi \bar \chi$ allows to populate the dark sector. The Boltzmann
equation in Eq.\eqref{eq:boltzmann} then acquires a source term
\begin{align}
\dot n_\chi(t) + 3 H(t) n_\chi(t) = S(B\to \chi\bar\chi) \; .
\label{eq:FIMPboltzmann}
\end{align}
The condition that the dark matter sector is not in thermal
equilibrium initially translates into a lower bound on the dark matter
mass. Its precise value depends on the model, but for a mediator with\index{mediator}
$m_B\approx 100$~GeV one can estimate $m_\chi\gtrsim 0.1~...~1$~keV from the
fundamental assumptions of the model.

A decay-based source term in terms of the internal number of degrees
of freedom $g_B^*$, the partial width $B \to \chi \bar \chi$, and the
equilibrium distribution $\exp(-E_B/T)$ can be
written as
\begin{align}
S(B\to \chi\bar\chi)
&= g_B^* \,\int \frac{d^3p_B}{(2\pi)^3}\, 
   e^{-E_B/T} \; \frac{m_B}{E_B} \; \Gamma(B \to \chi\bar \chi) \notag \\
&=g_B^* \; \Gamma(B \to \chi\bar \chi) \,\int \frac{|\vec p_B|^2 d|p_B|}{2 \pi^2}\, e^{-E_B/T} \; \frac{m_B}{E_B} \notag \\
&=\frac{g_B^*\,m_B}{2\pi^2} \; \Gamma(B\to\chi\bar\chi) \, \int_{m_B}^\infty dE_B \sqrt{E_B^2-m_B^2} \, e^{-E_B/T} 
 \qquad \text{using} \quad \frac{d|\vec p_B|}{dE_B}=\frac{E_BdE_B}{\sqrt{E_B^2-m_B^2}} \notag\\
&= \frac{g_B^*m_B^2}{2\pi^2} \; \Gamma(B\to\chi\bar\chi)\; T\, K_1(m_B/T)\; ,
\label{eq:FIMPboltzmann1}
\end{align}
where $K_1(z)$ is the the modified Bessel function of the second
kind. For small $z$ it is approximately given by $K_1(z) \approx 1/z$,
while for large $z$ it reproduces the Boltzmann factor, $K_1(z)\propto
e^{-z}/z +\ope(1/z)$. This form suggests that the dark matter density will
increase until $T$ becomes small compared to $m_B$ and the source term becomes suppressed by $e^{-m_B/T}$. The source term is independent of $n_\chi$ and
proportional to the partial decay width. We also expect it to be
proportional to the equilibrium number density of $B$, defined as \index{number density}
\begin{align}
n_B^\text{eq}
&= g_B^* \int \frac{d^3p}{(2\pi)^3} e^{-E_B/T} 
= \frac{g_B^*}{2\pi^2} \int_{m_B}^\infty dE_B \; E_B \; \sqrt{E_B^2-m_B^2}e^{-E_B/T} \notag \\
&= \frac{g_B^*}{2\pi^2} \; m_B^2 \; T \, K_2(m_B/T) \; ,
\end{align}
in analogy to Eq.\eqref{eq:FIMPboltzmann1}, but with the Bessel
function of the first kind $K_1$. We can use this relation to
eliminate the explicit temperature dependence of the source term,
\begin{align}
S(B\to \chi\bar\chi)=\Gamma(B\to \bar\chi\chi) \;
                     \frac{K_1(m_B/T)}{K_2(m_B/T)} \; n_B^\text{eq} \; .
\label{eq:FIMPboltzmann2}
\end{align}
To compute the relic density, we
introduce the notation of Eq.\eqref{eq:wimp_boltzmann}, namely $x =
m_B/T$ and $Y = n_\chi/T^3$.  The Boltzmann equation from
Eq.\eqref{eq:FIMPboltzmann} now reads
\begin{align}
\frac{d Y(x)}{dx}
=\frac{g_B^*}{2\pi^2} \; \frac{\Gamma(B\to \chi\bar \chi)}{H(x_B=1)} x^3 K_1(x)
\qquad \text{with} 
\qquad 
x^3K_1(x) \approx 
\begin{cases} 
x^3 e^{-x} & x \gg 1 \quad \text{or} \; T \ll m_B \\
x^2 \qquad & x \ll 1 \quad \text{or} \; T \gg m_B 
\end{cases}
\; .
\end{align}
Because the function $x^3 K_1(x)$ has a distinct maximum at $x \approx
1$, dark matter production is dominated by temperatures $T \approx
m_B$. We can integrate the dark matter production over the entire
thermal history and find for the final yield $Y(x'_\text{dec})$ with the help of the appropriate integral
table \index{yield}
\begin{align}
Y(x'_\text{dec})
&\equiv\frac{g_B^*}{2\pi^2}\frac{\Gamma(B\to \chi\bar \chi)}{H(x_B=1)}\,\int_0^\infty x^3K_1(x)\,dx=\frac{3g_B^*}{4\pi}\frac{\Gamma(B\to \chi\bar \chi)}{H(x_B=1)}\; .
\end{align}
We can now follow the steps from Eq.\eqref{eq:wimp_relic_approx3} and
Eq.\eqref{eq:relic_approx} and compute the relic density today,
\begin{align}
\Omega_\chi h^2
&= \frac{h^2}{3 \mpl^2 H_0^2} \; \frac{m_\chi}{28} T_0^3  Y(x'_\text{dec})\notag \\
&=\frac{h^2}{112\pi}\frac{m_\chi}{\mpl^2}\frac{T_0^3}{H^2 H(x_B=1)} \; \Gamma(B\to \chi\bar \chi)\notag\\
&\hspace{-.3cm}\eqx{eq:hubble_temp} \frac{\sqrt{90} h^2}{112\pi^2}\frac{g_B^*}{\sqrt{\geff}}\frac{m_\chi}{m_B^2}\frac{T_0^3}{H^2\mpl} \; \Gamma(B\to \chi\bar \chi)\notag\\
&=  \frac{\sqrt{90} h^2}{112\pi^2}\frac{g_B^*}{\sqrt{\geff}}\frac{m_\chi}{m_B^2}\frac{(2.4 \cdot 10^{-4})^3}{(2.5 \cdot 10^{-3})^4} \mpl \; \Gamma(B\to \chi\bar \chi) 
=3.6\cdot 10^{23} \; \frac{g_B^*m_\chi}{m_B^2}\Gamma(B\to \chi\bar \chi)\; .
\end{align}
The calculation up to this point is independent from the details of the interaction between the decaying particle $B$ and the DM candidate $\chi$. For the example interaction Eq.\eqref{eq:exLag}, the partial decay with is given by $\Gamma(B\to \chi\bar\chi)=y^2 m_B/(8\pi)$, and assuming $g_B^*=2$ we find 
\begin{align}
\Omega_\chi h^2=0.12\left(\frac{y}{2\cdot 10^{-12}}\right)^2\frac{m_\chi}{m_B}\; .
\label{eq:freezeInRD}
\end{align}
\index{relic!abundance}
The correct relic density from $B$-decays requires small couplings $y$
and/or dark matter masses $m_\chi$, compatible with the initial
assumption that dark matter was never in thermal equilibrium with the
Standard Model for $T\gtrsim m_B$. Following Eq.\eqref{eq:freezeInRD},
larger interaction rates lead to larger final dark matter
abundances. This is the opposite scaling as for the freeze-out
mechanism of Eq.\eqref{eq:wimp_relic_4}.  In the right panel of
Figure~\ref{fig:fimp1} we show the scaling of $Y(x)$ with $x=m_B/T$,
compared with the scaling of $Y(x)$ with $x=m_\chi/T$ for
freeze-out. Both mechanisms can be understood as the limits of
increasing the interaction strength between the visible and the dark
matter sector (freeze-out) and decreasing this interaction strength
(freeze-in) in a given model.\index{freeze in}

Even though we illustrate the freeze-in mechanism with the example of
the decay of the SM particle $B$ into dark matter, the dark matter
sector could also be populated by an annihilation process $B\bar B\to
\chi\bar \chi$, decays of SM particles into a visible particle and
dark matter $B \to B_2 \chi$, or scenarios where $B$ is not a SM
particle.
If the decay $B \to B_2 \chi$ is responsible for the observed relic
density, it can account for asymmetric dark matter if $\Gamma(B \to
B_2 \chi)\neq\Gamma(\bar B \to \bar B_2\bar \chi)$, as discussed in
Section~\ref{sec:asymmetric}.

\newpage
\section{WIMP models}
\label{sec:models}
\index{WIMP}

If we want to approach the problem of dark matter from a particle
physics perspective, we need to make assumptions about the quantum
numbers of the weakly interacting state which forms dark
matter. During most of these lecture notes we assume that this new particle
has a mass in the GeV to TeV range, and that its density is thermally produced
during the cooling of the Universe. Moreover, we assume that the
entire dark matter density of the Universe is due to one stable
particle.

The first assumption fixes the spin of this particle. From the
Standard Model we know that there exist fundamental scalars, like the
Higgs, fundamental fermions, like quarks and leptons, and fundamental
gauge bosons, like the gluon or the weak gauge bosons. Scalars have
spin zero, fermions have spin 1/2, and gauge bosons have spin
1. Because calculations with gauge bosons are significantly harder, in
particular when they are massive, we limit
ourselves to scalars and fermions.

When we construct particle models of dark matter we are faced with
this wide choice of new, stable particles and their quantum
numbers. Moreover, dark matter has to couple to the Standard Model,
because it has to annihilate to produce the observed relic density  $\Omega_\chi h = 0.12$. 
This means that strictly
speaking we do not only need to postulate a dark matter particle, but
also a way for this state to communicate to the Standard Model along
the line of the table of states in Section~\ref{sec:sommerfeld}. The
second state is usually called a mediator.\index{mediator}

\subsection{Higgs portal}
\label{sec:models_portal}
\index{Higgs!portal}
 \index{Higgs!potential}

An additional scalar particle in the Standard Model can couple to the Higgs sector of the
Standard Model in a unique way. The so-called \ul{Higgs portal
interactions} is renormalizable, which means that the coupling constant
between two Higgs bosons and two new scalars has a mass unit zero and
can be represented by a c-number. All we do in such a model is extend
the renormalizable Higgs potential of the Standard
Model~\cite{lecture}, which has a non-zero vacuum expectation value
(VEV) for $\mu_H^2 < 0$,
\begin{align}
V_\text{SM} 
&= \mu_H^2 \; \phi^\dagger \phi + \lambda_H ( \phi^\dagger \phi )^2 \notag \\
&\supset \mu_H^2 \frac{(H+v_H)^2}{2} + \lambda_H \frac{(H+v_H)^4}{4} 
 \supset - \frac{m_H^2}{2} H^2 + \frac{m_H^2}{2v_H} H^3 + \frac{m_H^2}{8 v_H^2} H^4 \; ,
\label{eq:higgspot_sm}
\end{align}
In the Standard Model this leads to the two observable mass scales
\begin{align}
v_H = \sqrt{ \frac{-\mu_H^2}{\lambda_H}} = 246~\gev
\qquad \text{and} \qquad 
m_H = \sqrt{2 \lambda_H} \, v_H = 2 \sqrt{ -\mu_H^2 } = 125~\gev \approx \frac{v_H}{2} \; .
\label{eq:higgsmass_sm}
\end{align}
The last relation is a numerical accident.  The general Higgs
potential in Eq.\eqref{eq:higgspot_sm} allows us to couple a new
scalar field $S$ to the Standard Model using a renormalizable, dimension-4
term $(\phi^\dagger \phi)(S^\dagger S)$.

For any new scalar field there are two choices we can make. First, we
can give it some kind of multiplicative charge, so we
actually postulate a set of two particles, one with positive and one
with negative charge. This just means that our new scalar field has to
be complex values, such that the two charges are linked by complex
conjugation. In that case the Higgs portal coupling includes the
combination $S^\dagger S$. Alternatively, we can assume that no such
charge exists, in which case our new scalar is real and the Higgs
portal interaction is proportional to $S^2$.

Second, we know from the case of the Higgs boson that a scalar can
have a finite vacuum expectation value. Due to that VEV, the corresponding new state will mix 
with the SM Higgs boson to form two mass eigenstates, and modify the SM Higgs couplings and 
the masses of the $W$ and $Z$
bosons. This is a complication we neither 
want nor need, so we will work with a \ul{dark real
  scalar}. The combined potential reads
\begin{align}
V 
&= \mu_H^2 \; \phi^\dagger \phi + \lambda_H ( \phi^\dagger \phi )^2 
 + \mu_S^2 \; S^2 + \kappa \; S^3 + \lambda_S \; S^4 
 + \kappa_3 \phi^\dagger \phi S
 + \lambda_3 \phi^\dagger \phi S^2 \notag \\
&\supset -\frac{m_H^2}{2} H^2 + \frac{m_H^2}{2v_H} H^3 + \frac{m_H^2}{8 v_H^2} H^4 
 - \mu_S^2 \, S^2 + \kappa \, S^3 + \lambda_S \, S^4 
 + \frac{\kappa_3}{2} (H+v_H)^2 S 
 + \frac{\lambda_3}{2} (H+v_H)^2 S^2 \; .
\label{eq:higgspot_portal1}
\end{align}
A possible linear term in the new, real field is removed by a shift in
the fields. In the above form the new scalar $S$ can couple to two
SM Higgs bosons, which induces a decay either on-shell $S \to HH$ or
off-shell $S \to H^* H^* \to 4b$. To forbid this, we apply the usual
trick, which is behind essentially all WIMP dark matter models; we
require the Lagrangian to obey a \ul{global $\mathbb{Z}_2$
  symmetry} 
\begin{align}
S \to -S,
\qqquad 
H \to +H,  \quad \cdots
\end{align}
This defines an ad-hoc $\mathbb{Z}_2$ parity $+ 1$ for all
SM particles and $-1$ for the dark matter candidate. The combined
potential now reads
\begin{align}
V 
&\supset - \frac{m_H^2}{2} H^2 + \frac{m_H^2}{2v_H} H^3 + \frac{m_H^2}{8 v_H^2} H^4 
 - \mu_S^2 \, S^2 + \lambda_S \, S^4 
 + \frac{\lambda_3}{2} (H+v_H)^2 S^2 \notag \\
&= - \frac{m_H^2}{2} H^2 + \frac{m_H^2}{2v_H} H^3 + \frac{m_H^2}{8 v_H^2} H^4 
 - \left( \mu_S^2 - \lambda_3 \frac{v_H^2}{2} \right) \, S^2 + \lambda_S \, S^4 
 + \lambda_3 v_H \, H S^2  
 + \frac{\lambda_3}{2} H^2 S^2 \; .
\label{eq:higgspot_portal2}
\end{align}
The mass of the dark matter scalar and its phenomenologically relevant
$SSH$ and $SSHH$ couplings read
%
\begin{align}
m_S = \sqrt{ 2 \mu_S^2 - \lambda_3 v_H^2 } 
\qqqquad 
g_{SSH} = - 2 \lambda_3 v_H 
\qqqquad 
g_{SSHH} = - 2 \lambda_3  \; .
\label{eq:portal_coup}
\end{align}
The sign of $\lambda_3$ is a free parameter. Unlike for
singlet models with a second VEV, the dark singlet does not affect the
SM Higgs relations in Eq.\eqref{eq:higgsmass_sm}. However, the $SSH$
coupling mediates $SS$ interactions with pairs of SM particles through
the light Higgs pole, as well as Higgs decays $H \to SS$, provided the
new scalar is light enough. The $SSHH$ coupling can mediate heavy dark
matter annihilation into Higgs pairs. We will discuss more details on
invisible Higgs decays in Section~\ref{sec:coll}.\bigskip

\index{dark matter!annihilation}
For dark matter annihilation, the $SSf \bar{f}$ transition matrix
element based on the Higgs portal is described by the Feynman diagram
\vspace{.5cm}
\begin{center}
\begin{fmfgraph*}(100,60)
\fmfset{arrow_len}{2mm}
\fmfleft{i1,i2}
\fmfright{o1,o2}
\fmf{dashes,tension=0.4,width=0.6}{i2,v1}
\fmf{dashes,tension=0.4,width=0.6}{v1,i1}
\fmf{dashes,tension=0.4,label=$H$,width=0.6}{v1,v2}
\fmf{fermion,tension=0.4,width=0.6}{v2,o2}
\fmf{fermion,tension= 0.4,width=0.6}{o1,v2}
\fmflabel{$S$}{i1}
\fmflabel{$S$}{i2}
\fmflabel{$\bar{b}(k_1)$}{o1}
\fmflabel{$b(k_2)$}{o2}
\end{fmfgraph*}
\end{center}
\vspace{.5cm}

All momenta are defined incoming, giving us for an outgoing fermion
and an outgoing anti-fermion
\begin{align}
\mat 
&= \bar{u}(k_2) \, \frac{-i m_f}{v_H} \, v(k_1) \; 
\frac{-i}{(k_1 + k_2)^2 - m_H^2 + i m_H \Gamma_H} \; 
(- 2 i \lambda_3 v_H ) \; .
\label{eq:matrix_annihilation}
\end{align}
In this expression we see that $v_H$ cancels, but the fermion mass
$m_f$ will appear in the expression for the annihilation rate.  We
have to square this matrix element, paying attention to the spinors
$v$ and $u$, and then sum over the spins of the external fermions,
\begin{align}
\sum_\text{spin} |\mat|^2 
&= 4 \lambda_3^2 m_f^2
   \left( \sum_\text{spin} v(k_1) \bar{v}(k_1) \right) \;
   \left( \sum_\text{spin} u(k_2) \bar{u}(k_2) \right) \;
   \frac{1}{\left| (k_1 + k_2)^2 - m_H^2 + i m_H \Gamma_H \right|^2} \notag \\
&= 4 \lambda_3^2 m_f^2 \;
   \tr \left[ ( \slashchar{k}_1 - m_f \one ) \;
              ( \slashchar{k}_2 + m_f \one ) \right] \;
   \frac{1}{\left[ (k_1 + k_2)^2 - m_H^2 \right]^2 + m_H^2 \Gamma_H^2} \notag \\
&= 4 \lambda_3^2 m_f^2 \;
   4 \left[ k_1 k_2 - m_f^2 \right] \;
   \frac{1}{\left[ (k_1 + k_2)^2 - m_H^2 \right]^2 + m_H^2 \Gamma_H^2} \notag \\
&= 8 \lambda_3^2 m_f^2 \;
   \frac{(k_1 + k_2)^2 - 4 m_f^2}{\left[ (k_1 + k_2)^2 - m_H^2 \right]^2 + m_H^2 \Gamma_H^2} \; .
\label{eq:matrix2_annihilation}
\end{align}
In the sum over spin and color of the external fermions the averaging
is not yet included, because we need to specify which of the external
particles are incoming or outgoing. As an example, we compute the
cross section for the dark matter annihilation process to a pair of
bottom quarks
\begin{align}
S S \to H^* \to b \bar{b} \; .
\end{align}
This $s$-channel annihilation corresponds to the leading on-shell
Higgs decay $H \to b\bar{b}$ with a branching ratio around 60\%.  In
terms of the Mandelstam variable $s = (k_1 + k_2)^2$ it gives us
\begin{align}
\overline{ \sum_\text{spin,color} |\mat|^2 }
&= N_c \;
  8 \lambda_3^2 m_b^2 \;
   \frac{s - 4 m_b^2}{\left( s - m_H^2 \right)^2 + m_H^2 \Gamma_H^2}  \notag \\
\Rightarrow \qquad 
\sigma(SS \to b\bar{b}) 
&= \frac{1}{16 \pi s} \;
   \sqrt{ \frac{1 - 4 m_b^2/s}{1 - 4 m_S^2/s}} \; 
   \overline{ \sum |\mat|^2 } \notag \\
&= \frac{N_c}{2 \pi \sqrt{s}} \;
   \lambda_3^2 m_b^2 \;
   \sqrt{ \frac{1 - 4 m_b^2/s}{s - 4 m_S^2}} \; 
   \frac{s - 4 m_b^2}{\left( s - m_H^2 \right)^2 + m_H^2 \Gamma_H^2}  \; .
\label{eq:portal_annrate1}
\end{align}
To compute the relic density we need the velocity-averaged cross
section. For the contribution of the $b\bar{b}$ final state to the
dark matter annihilation rate we find the leading term in the
non-relativistic limit, $s = 4 m_S^2$
\begin{align}
\langle \sigma v \rangle \Bigg|_{SS \to b\bar{b}}
\equiv \sigma v \Bigg|_{SS \to b\bar{b}}
&\eqx{eq:def_velocity} v \; 
   \frac{N_c \lambda_3^2 m_b^2}{2 \pi \sqrt{s}} \;
   \frac{\sqrt{1 - 4 m_b^2/s}}{m_S v} \; 
   \frac{s - 4 m_b^2}{\left( s - m_H^2 \right)^2 + m_H^2 \Gamma_H^2}  \notag \\
&\stackrel{\text{threshold}}{=}
   \frac{N_c \lambda_3^2 m_b^2}{4 \pi m_S^2} \;
   \sqrt{1 - \frac{m_b^2}{m_S^2}} \; 
   \frac{4 m_S^2 - 4 m_b^2}{\left( 4 m_S^2 - m_H^2 \right)^2 + m_H^2 \Gamma_H^2}  \notag \\
&\stackrel{m_S \gg m_b}{=}
   \frac{N_c \lambda_3^2 m_b^2}{\pi} \;
   \frac{1}{\left( 4 m_S^2 - m_H^2 \right)^2 + m_H^2 \Gamma_H^2}  \; .
\label{eq:portal_annrate2}
\end{align}
This expression holds for all scalar masses $m_S$.  In our estimate we identify the $v$-independent
expression with the thermal average. Obviously, this will become more
complicated once we include the next term in the expansion around $v
\approx 0$. The Breit--Wigner propagator guarantees that the rate
never diverges, even in the case when the annihilating dark matter
hits the Higgs pole in the $s$-channel.\bigskip

\begin{figure}[b!]
\begin{center}
\includegraphics[width=0.45\textwidth]{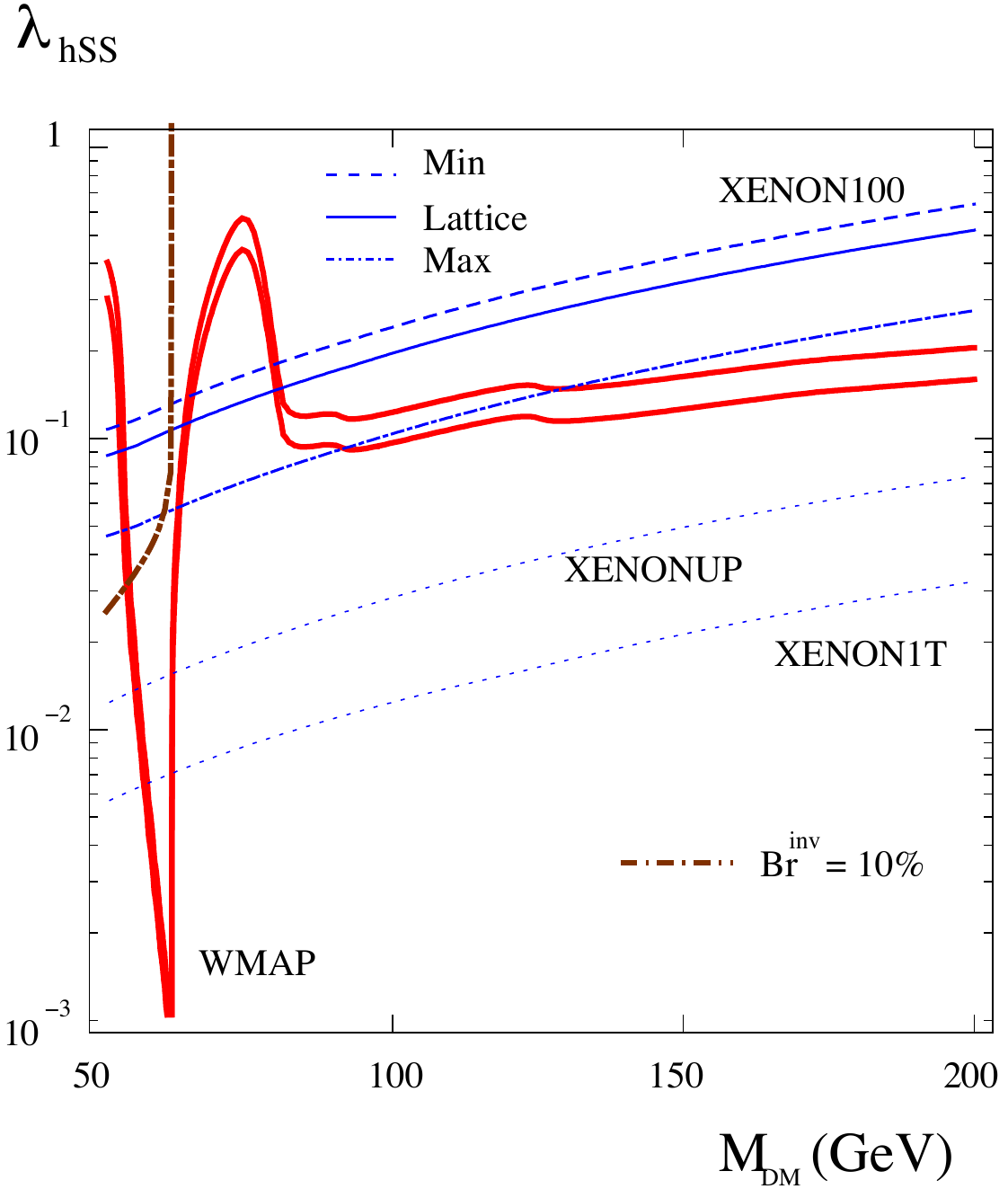}
\end{center}
\caption{Higgs portal parameter space in terms of the self coupling
  $\lambda_{hSS} \sim \lambda_3$ and the dark matter mass $M_\text{DM}
  = m_S$. The red lines indicate the correct relic density
  $\Omega_\chi h^2$. Figure from Ref.~\cite{abdel}.}
\label{fig:portal_mass_coup}
\end{figure}

The simplest parameter point to evaluate this annihilation cross
section is \ul{on the Higgs pole}. This gives us 
\begin{align}
\langle \sigma v \rangle \Bigg|_{SS \to b\bar{b}}
&=  \frac{N_c \lambda_3^2 m_b^2}{\pi} \;
   \frac{1}{\left( 4 m_S^2 - m_H^2 \right)^2 + m_H^2 \Gamma_H^2}  
\stackrel{m_H = 2 m_S}{=}
   \frac{N_c \lambda_3^2 m_b^2}{\pi  m_H^2 \Gamma_H^2} 
\approx \frac{15 \lambda_3^2}{\gev^2} \notag \\ 
\Rightarrow \qquad 
\langle \sigma_{\chi \chi} \, v \rangle 
&= \frac{1}{\br(H \to b\bar{b})} \; \langle \sigma v \rangle \Bigg|_{SS \to b\bar{b}} 
\approx \frac{25 \lambda_3^2}{\gev^2} 
\really 1.7 \cdot 10^{-9} \; \frac{1}{\gev^2} 
\qquad \Leftrightarrow \qquad 
\boxed{ 
\lambda_3 \approx 8\cdot 10^{-6}
} \; ,
\label{eq:portal_annrate_pole}
\end{align}
with $\Gamma_H \approx 4 \cdot 10^{-5} m_H$.  While it is correct that
the self coupling required on the Higgs pole is very small, the full
calculation leads to a slightly larger value $\lambda_3 \approx
10^{-3}$, as shown in Figure~\ref{fig:portal_mass_coup}.\bigskip

\ul{Lighter dark matter scalars} also probe the Higgs mediator
on-shell.  In the Breit-Wigner propagator of the annihilation cross
section, Eq.\eqref{eq:portal_annrate2}, we have to compare to the two
terms
\begin{align}
m_H^2 - 4 m_S^2 = m_H^2 \left( 1 - \frac{4 m_S^2}{m_H^2} \right)
\qquad \Leftrightarrow \qquad 
m_H \Gamma_H \approx 4 \cdot 10^{-5} \; m_H^2 \; .
\end{align}
The two states would have to fulfill exactly the on-shell condition
$m_H = 2 m_S$ for the second term to dominate. We can therefore stick
to the first term for $m_H > 2 m_S$ and find for the dominant decay to
$b\bar{b}$ pairs in the limit $m_H^2 \gg m_S^2 \gg m_b^2$
\begin{align}
\langle \sigma v \rangle \Bigg|_{SS \to b\bar{b}}
= \frac{N_c \lambda_3^2 m_b^2}{\pi m_H^4}  
\approx \frac{\lambda_3^2}{125^2 \; 50^2 \; \gev^2}  
\really 1.7 \cdot 10^{-9} \frac{1}{\gev^2} 
\qquad \Leftrightarrow \qquad 
\boxed{
\lambda_3 
= 0.26
}\; .
\label{eq:ref_portal_2}
\end{align}
\bigskip

\ul{Heavier dark matter scalars} well above the Higgs pole also
include the annihilation channels
\begin{align}
SS \to \tau^+ \tau^-, W^+ W^-, ZZ, HH, t\bar{t}  \; .
\end{align}
Unlike for on-shell Higgs decays, the $b\bar{b}$ final state is not
dominant for dark matter annihilation when it proceeds through a $2 \to 2$
process. Heavier particles couple to the Higgs more strongly, so above
the Higgs pole they will give larger contributions to the dark matter
annihilation rate. For top quarks in the final state this simply means
replacing the Yukawa coupling $m_b^2$ by the much larger $m_t^2$. In
addition, the Breit-Wigner propagator will no longer scale like
$1/m_H^2$, but proportional to $1/m_S^2$. Altogether, this gives us a
contribution to the annihilation rate of the kind
\begin{align}
\langle \sigma v \rangle \Bigg|_{SS \to t\bar{t}}
= \frac{N_c \lambda_3^2 m_t^2}{\pi \left( 4 m_S^2 - m_H^2 \right)^2}  
\stackrel{2 m_S \gg m_H}{=}
 \frac{N_c \lambda_3^2 m_t^2}{16 \pi m_S^4} \; .
\label{eq:portal_annrate3}
\end{align}
The real problem is the annihilation to the weak bosons $W,Z$, because
it leads to a different scaling of the annihilation cross section. In
the limit of large energies we can describe for example the process
$SS \to W^+ W^-$ using spin-0 Nambu-Goldstone bosons in the final state.\index{Nambu-Goldstone boson} These
Nambu-Goldstone modes in the Higgs doublet $\phi$ appear as the longitudinal
degrees of freedom, which means that dark matter annihilation to weak
bosons at large energies follows the same pattern as dark matter
annihilation to Higgs pairs.  Because we are more used to the Higgs
degree of freedom we calculate the annihilation to Higgs pairs,
\begin{align}
SS \to HH \; .
\end{align}
The two Feynman diagrams with the direct four-point interaction and the
Higgs propagator at the threshold $s = 4 m_S^2$ scale like
\begin{align}
\mat_4
&= g_{SSHH} = - 2 \lambda_3 \notag \\
\mat_H 
&= \frac{g_{SSH}}{s - m_H^2} \frac{3 m_H^2}{v_H} 
\stackrel{\text{threshold}}{=}
 - \frac{2 \lambda_3 v_H}{4 m_S^2 - m_H^2} \frac{3 m_H^2}{v_H} 
\stackrel{m_S \gg m_H}{=}
 - \frac{6 \lambda_3 m_H^2}{4 m_S^2} \ll \mat_4 \; .
\end{align}
This means for heavy dark matter we can neglect the $s$-channel Higgs
propagator contribution and focus on the four-scalar interaction.  In
analogy to Eq.\eqref{eq:portal_annrate2} we then compute the
velocity-weighted cross section at threshold,
\begin{align}
\sigma (SS \to HH) 
&= \frac{1}{16 \pi \sqrt{s}} \; 
  \frac{\sqrt{1 - 4 m_H^2/m_S^2}}{\sqrt{s - 4 m_S^2}} \; 
  4 \lambda_3^2 
\eqx{eq:def_velocity} 
  \frac{\lambda_3^2}{4 \pi \sqrt{s}} \sqrt{1 - \frac{4 m_H^2}{m_S^2}} \; 
  \frac{1}{v m_S} \notag \\
\sigma v \Bigg|_{SS \to HH}
&= \frac{\lambda_3^2}{4 \pi m_S \sqrt{s}} \; \sqrt{1 - \frac{4 m_H^2}{m_S^2}} 
\stackrel{\text{threshold}}{=} 
\frac{\lambda_3^2}{8 \pi m_S^2} \; \sqrt{1 - \frac{4 m_H^2}{m_S^2}} 
\stackrel{m_S \gg m_H}{=}
\frac{\lambda_3^2}{8 \pi m_S^2}
\end{align}
For $m_S = 200$~GeV we can derive the coupling $\lambda_3$ which we
need to reproduce the observed relic density,
\begin{align}
1.7 \cdot 10^{-9} \frac{1}{\gev^2} 
&\really  
\frac{\lambda_3^2}{8 \pi m_S^2}
\approx \frac{\lambda_3^2}{10^6 \, \gev^2}
\qquad \Leftrightarrow \qquad 
\boxed{\lambda_3 \approx 0.04} \; .
\end{align}
The curve in Figure~\ref{fig:portal_mass_coup} shows two thresholds
related to four-point annihilation channels, one at $m_S = m_Z$ and
one at $m_S = m_H$.  Starting with $m_S = 200$~GeV and corresponding
values for $\lambda_3$ the annihilation to Higgs and Goldstone boson
pairs dominates the annihilation rate.\bigskip

One lesson to learn from our Higgs portal considerations is the
scaling of the dark matter annihilation cross section with the WIMP \index{WIMP}
mass $m_S$. It does not follow Eq.\eqref{eq:wimp_ann_approx} at all
and only follows Eq.\eqref{eq:wimp_rate_vs_mass} for very heavy dark
matter. For our model, where the annihilation is largely mediated
by a Yukawa coupling $m_b$, we find
\begin{align}
\boxed{
\sigma_{\chi \chi} \propto 
\begin{cases}
\dfrac{\lambda_3^2 m_b^2}{m_H^4} \qquad & m_S \ll \dfrac{m_H}{2} \\[3mm]
\dfrac{\lambda_3^2 m_b^2}{m_H^2 \Gamma_H^2} & m_S = \dfrac{m_H}{2} \\[3mm]
\dfrac{\lambda_3^2}{m_S^2} & m_S > m_Z,m_H \; .
\end{cases}
}
\label{eq:annrate_scaling}
\end{align}
It will turn out that the most interesting scaling is on the Higgs
peak, because the Higgs width is not at all related to the weak
scale.

\subsection{Vector Portal}
\label{sec:vector_portal}
\index{vector portal}

Inspired by the WIMP assumption in Eq.\eqref{eq:wimp_ann_approx} we
can use a new massive gauge boson to mediate thermal freeze-out
production. \index{freeze out}  The combination of a free vector mediator mass and a free
dark matter mass will allow us to study a similar range scenarios as
for the Higgs portal, Eq.\eqref{eq:annrate_scaling}. A physics
argument is given by the fact that the Standard Model has a few
global symmetries which can be extended to anomaly-free gauge
symmetries.

The extension of the Standard Model with its hypercharge symmetry
$U(1)_Y$ by an additional $U(1)$ gauge group defines another
renormalizable portal to dark matter. Since $U(1)$-field strength
tensors are gauge singlets, the kinetic part of the Lagrangian allows
for \ul{kinetic mixing}\index{kinetic mixing},
\begin{align}
\lag_\text{gauge} 
= -\frac{1}{4}\hat B^{\mu\nu}\hat B_{\mu\nu} 
-\frac{s_\chi}{2}\hat V^{\mu\nu} \hat B_{\mu\nu}
-\frac{1}{4}\hat V^{\mu\nu}\hat V_{\mu\nu} 
= -\frac{1}{4}
\begin{pmatrix} \hat{B}_{\mu \nu} & \hat{V}_{\mu \nu} \end{pmatrix}
\begin{pmatrix} 1 & s_\chi \\ s_\chi & 1 \end{pmatrix}
\begin{pmatrix} \hat{B}_{\mu \nu} \\ \hat{V}_{\mu \nu} \end{pmatrix} \; ,
\label{eq:kinmixlag}
\end{align}
where $s_\chi\equiv \sin \chi$ is assumed to be a small mixing
parameter. In principle it does not have to be an angle, but for the
purpose of these lecture notes we assume that it is small, $s_\chi\ll
1$, so we can treat it as a trigonometric function and write
$c_\chi\equiv \sqrt{1-s_\chi}$ and $t_\chi \equiv s_\chi/c_\chi$. Even
if the parameter $s_\chi$ is chosen to be zero at tree-level, loops of
particles charged under both $U(1)_X$ and $U(1)_Y$ introduce a
non-zero value for it. Similar to the Higgs portal, there is no
symmetry that forbids it, so we do not want to assume that all quantum
corrections cancel to a net value $s_\chi=0$.

The notation $\hat B_{\mu\nu}$ indicates that the gauge fields are not
yet canonically normalized, which means that the residue of the
propagator it not one. In addition, the gauge boson propagators
derived from Eq.\eqref{eq:kinmixlag} are not diagonal.  We can
diagonalize the matrix in Eq.\eqref{eq:kinmixlag} and keep the
hypercharge unchanged with a non-orthogonal rotation of the gauge
fields
\begin{align}
\begin{pmatrix} \hat{B}_\mu \\\hat W^3_\mu \\\hat{V}_\mu \end{pmatrix} 
= G(\theta_{V}) \, 
\begin{pmatrix} B_\mu \\ W^3_\mu\\V_\mu \end{pmatrix}  
= \begin{pmatrix} 1  & 0&-s_\chi/c_\chi \\ 0&1&0\\0  &0&1/c_\chi \end{pmatrix}
  \begin{pmatrix} B_\mu \\ W^3_\mu\\V_\mu \end{pmatrix} \; .
\label{eq:gaugefieldrot}
\end{align}
We now include the third component of the $SU(2)_L$ gauge field
triplet $W_\mu=(W^1_\mu, W^2_\mu, W^3_\mu )$ which mixes with the
hypercharge gauge boson through electroweak symmetry breaking to
produce the massive $Z$ boson and the massless photon.  Kinetic mixing
between the $SU(2)_L$ field strength tensor and the $U(1)_X$ field
strength tensor is forbidden because $\hat V^{\mu\nu} \hat
A^a_{\mu\nu}$ is not a gauge singlet.  Assuming a mass $\hat m_V$ for
the $V$-boson we write the combined mass matrix as
\begin{align}
\mathcal{M}^2
\eqx{eq:gaugefieldrot}\frac{v^2}{4}
\begin{pmatrix} g'^2  & -g\,g' &-{g'}^2 s_\chi \\[2mm]
               -g\,g' & g^2    & g\,g' \,s_\chi\\
               -{g'}^2 s_\chi \quad  & \quad g\,g' s_\chi \quad & \quad \dfrac{4\hat m_V^2}{v^2}(1+s_\chi^2)+g^{\prime 2} s_\chi^2
\end{pmatrix} +\ope(s_\chi^3) \; .
\label{eq:matrix_vectors}
\end{align}
This mass matrix can be diagonalized with a combination of two
block-diagonal rotations with the weak mixing matrix and an additional
angle $\xi$,
\begin{align}
R_1(\xi)R_2(\theta_w)=\begin{pmatrix} 1&0 & 0\\
0&c_\xi & s_\xi \\
0&-s_\xi & c_\xi \end{pmatrix}\,
\begin{pmatrix} c_w& s_w & 0\\
-s_w& c_w&0\\
0&0&1\end{pmatrix} \; ,
\end{align}
giving
\begin{align}
R_1(\xi)R_2(\theta_w) \mathcal{M}^2\,R_2(\theta_w)^TR_1(\xi)^T=
\begin{pmatrix} m_\gamma^2&0&0\\ 0&m_Z^2&0\\ 0&0&m_V^2\end{pmatrix}\; ,
\label{eq:physical_v}
\end{align}
provided
\begin{align}\label{eq:tan2xi}
\tan 2 \xi = \frac{2s_\chi s_w}{1- \dfrac{\hat m_V^2}{\hat m_Z^2} }+\ope(s_\chi^2) \; .
\end{align}
For this brief discussion we assume for the mass ratio
\begin{align}
\frac{\hat m_V^2}{\hat m_Z^2} 
= \frac{2 \hat m_V^2}{(g^2+g'^2) v^2}
       \ll 1\,,
\end{align}
and find for the physical masses 
\begin{align}
m_\gamma^2=0\,, \qqquad
m_Z^2 = \hat m_Z^2 \left[ 1+s_\chi^ 2s_w^2 \left( 1+ \frac{\hat m_V^2}{\hat m_Z^2} \right) \right]\,, \qqquad 
m_{V}^2 = \hat m_V^2 \left[ 1+s_\chi^2c_w^2\right] \; .
\end{align}
\bigskip

In addition to the dark matter mediator mass we also need the coupling
of the new gauge boson $V$ to SM matter. Again, we start with the
neutral currents for the not canonically normalized gauge fields and
rotate them to the physical gauge bosons defined in
Eq.\eqref{eq:physical_v},
\begin{align}\label{eq:currentcouplings}
\left(ej_\text{EM} , \frac{e}{\sin \theta_w \cos \theta_w} j_Z, g_{D}j_{D}\right) 
\begin{pmatrix}\hat A\\ \hat Z\\\hat A'\end{pmatrix}
=&\left(ej_\text{EM} , \frac{e}{\sin \theta_w \cos \theta_w} j_Z, g_{D}j_{D}\right) \,K\,\begin{pmatrix}A\\ Z\\  V\end{pmatrix} \,, 
\end{align}
with
\begin{align}
K=\left[ R_1(\xi)R_2(\theta_w)G^{-1}(\theta_\chi)R_2(\theta_w)^{-1}\right]^{-1}
\approx \begin{pmatrix}
1 & 0 & -s_\chi c_w \\
0 & 1& 0 \\
0 & s_\chi s_w&  1
\end{pmatrix} \; .
\label{eq:KK}
\end{align}
The new gauge boson couples to the electromagnetic current with a
coupling strength of $-s_\chi c_w e$, while to leading order in
$s_\chi$ and $\hat m_V/\hat m_Z$ its coupling to the $Z$-current
vanishes. It is therefore referred to as \ul{hidden
  photon}.\index{hidden photon} This behavior changes for larger masses, $\hat m_V/\hat
m_Z \gtrsim 1$, for which the coupling to the $Z-$current can be the
dominating coupling to SM fields. In this case the new gauge boson is
called a \ul{$Z'-$boson}. For the purpose of these lecture
notes we will concentrate on the light $V$-boson, because it will
allow for a light dark matter particle.\bigskip

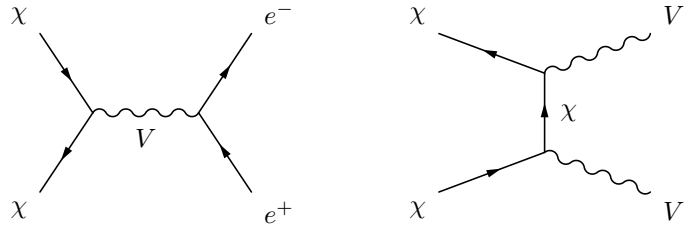
\begin{figure}[b!]
\vspace{.3cm}
\begin{center}
\begin{fmfgraph*}(100,60)
\fmfset{arrow_len}{2mm}
\fmfleft{i1,i2}
\fmfright{o1,o2}
\fmf{fermion, tension=0.4,width=0.6}{i2,v1}
\fmf{fermion,tension=0.4,width=0.6}{v1,i1}
\fmf{wiggly,tension=0.4,label=$V$,width=0.6}{v1,v2}
\fmf{fermion,tension=0.4,width=0.6}{v2,o2}
\fmf{fermion,tension= 0.4,width=0.6}{o1,v2}
\fmflabel{$\chi$}{i1}
\fmflabel{$ \chi$}{i2}
\fmflabel{$e^+$}{o1}
\fmflabel{$e^-$}{o2}
\end{fmfgraph*}
\hspace*{0.1\textwidth}
\begin{fmfgraph*}(100,60)
\fmfset{arrow_len}{2mm}
\fmfleft{i1,i2}
\fmfright{o1,o2}
\fmf{fermion, tension=0.4,width=0.6}{i1,v1}
\fmf{fermion,tension=0.4,width=0.6}{v2,i2}
\fmf{fermion,tension=0.4,label=$\chi$,width=0.6}{v1,v2}
\fmf{wiggly,tension=0.4,width=0.6}{v1,o1}
\fmf{wiggly,tension= 0.4,width=0.6}{o2,v2}
\fmflabel{$\chi$}{i1}
\fmflabel{$ \chi$}{i2}
\fmflabel{$V$}{o1}
\fmflabel{$V$}{o2}
\end{fmfgraph*}
\end{center}
\vspace{.3cm}
\caption{Feynman diagrams contributing to the annihilation of dark
  matter coupled to the visible sector through a hidden photon.}
\label{fig:DVannihilation}
\end{figure}

There are two ways in which the hidden photon could be relevant from a
dark matter perspective. The new gauge boson could be the dark matter
itself, or it could provide a portal to a dark matter sector if the
dark matter candidate is charged under $U(1)_X$. The former case is
problematic, because the hidden photon is not stable and can decay
through the kinetic mixing term.  Even if it is too light to decay
into the lightest charged particles, electrons, it can decay into
neutrinos $V\to \nu\bar \nu$ through the suppressed mixing with the
$Z$-boson and into three photons $V \to 3 \gamma$ through loops of
charged particles. For mixing angles small enough to guarantee
stability on time scales of the order of the age of the universe, the
hidden photon can therefore not be thermal dark matter.

If the hidden photon couples to a new particle
charged under a new $U(1)_X$ gauge group, this particle could be a dark
matter candidate. For a new Dirac fermion with $U(1)_X$ charge $Q_X$,
we add a kinetic term with a covariant derivative to the Lagrangian,
 \begin{align}
\lag_\text{DM}=\bar \chi i\gamma^\mu D_\mu \chi -m_\chi \bar\chi\chi
\qquad \text{with} \quad 
D_\mu=\partial_\mu-ig_D Q_\chi V_\mu \; .
\end{align}
Through the $U(1)_X$-mediator \index{mediator} this dark fermion is in thermal contact
with the Standard Model through the usual annihilation shown in
Eq.\eqref{eq:dm_annihilation}. If the dark matter is lighter than the
hidden photon and the electron $m_V> m_\chi > m_e$, the dominant
s-channel Feynman diagram contributing to the annihilation cross
section is shown on the left of Figure~\ref{fig:DVannihilation}. This
diagram resembles the one shown above
Eq.\eqref{eq:matrix_annihilation} for the case of a Higgs portal and
the cross section can be computed in analogy to
Eq.\eqref{eq:portal_annrate1},
\begin{align}
\sigma(\chi\bar\chi \to e^+e^-)=\frac{1}{12\pi}(s_\chi c_w e g_D Q_\chi)^2 \bigg(1+\frac{2m_e^2}{s}\bigg) \bigg(1+\frac{2m_\chi^2}{s}\bigg) \frac{s}{(s-m_V^2)^2+m_V^2\Gamma_V^2}\,
\frac{\sqrt{1-4\dfrac{m_e^2}{m_V^2}}}{\sqrt{1-4\dfrac{m_\chi^2}{m_V^2}}} \; ,
\end{align}
with $\Gamma_V$ the total width of the hidden photon $V$. For the
annihilation of two dark matter particles $s=4m_\chi^2$, and assuming
$m_V \gg \Gamma_V$, the thermally averaged annihilation cross section
is given by
\begin{align}\label{eq:HPannihilationCS}
\langle \sigma v\rangle=\frac{1}{4\pi}(s_\chi c_w e g_D Q_\chi)^2\sqrt{1-\frac{m_e^2}{m_\chi^2}}\left(1+\frac{m_e^2}{2m_\chi^2}\right)\frac{4m_\chi^2}{(4m_\chi^2-m_V^2)^2}\stackrel{m_\chi\gg m_e}{\approx}\frac{m_\chi^2}{\pi m_V^4}(s_\chi c_w e Q_f g_D Q_\chi)^2\; .
\end{align}
It exhibits the same scaling as in the generic WIMP case of
Eq.\eqref{eq:wimp_ann_approx}. In contrast to the WIMP, however, the
gauge coupling is rescaled by the mixing angle $s_\chi$ and for very
small mixing angles the hidden photon can in principle be very
light. In Eq.\eqref{eq:HPannihilationCS} we assume that the dark
photon decays into electrons. Since the hidden photon branching ratios
into SM final states are induced by mixing with the photon, for masses
$m_V> 2 m_\mu$ the hidden photon also decays into muons and for hidden
photon masses above a few 100 MeV and below $2m_\tau$, the hidden
photon decays mainly into hadronic states. For $m_V > m_\chi$, the
PLANCK bound on the DM mass in Eq.\eqref{eq:planck} implies $m_V > 10$
GeV and Eq.\eqref{eq:HPannihilationCS} would need to be modified by
the branching ratios into the different kinematically accessible final
states. Instead, we illustrate the scaling with the scenario
$Q_\chi=1$, $m_\chi=10$ MeV and $m_V=100\,$MeV that is formally
excluded by the PLANCK bound, but only allows for hidden photon decays
into electrons. In this case, we find the observed relic density given
in Eq.\eqref{eq:relic_approx} for a coupling strength\index{PLANCK}
\begin{align}
\frac{1.7 \cdot 10^{-9}}{\gev^2} \stackrel{!}{=} 0.07\left[\frac{m_\chi}{0.01\,\text{GeV}}\right]^2\left[\frac{0.1 \,\text{GeV}}{m_V}\right]^4(s_\chi g_D)^2\qquad
\Leftrightarrow \qquad \boxed{s_\chi g_D= 0.0015} \; .
\end{align}
In the opposite case of $m_\chi > m_V > m_e$, the annihilation cross
section is dominated by the diagram on the right of
Figure~\ref{fig:DVannihilation}, with subsequent decays of the hidden
photon.  The thermally averaged annihilation cross section then reads
\begin{align}
\langle \sigma v\rangle=\frac{g_D^4 Q_\chi^4}{8\pi}\frac{1}{m_\chi^2}\frac{\left(1-\dfrac{m_V^2}{m_\chi^2}\right)^\frac{3}{2}}{\left(1-\dfrac{m_V^2}{2m_\chi^2}\right)^2} \stackrel{m_\chi \gg m_V}{\approx} \frac{g_D^4 Q_\chi^4}{8\pi m_\chi^2}\; .
\end{align}
\index{dark matter!secluded}
The scaling with the dark matter mass is the same as for a WIMP with
$m_\chi > m_Z$, as shown in Eq.\eqref{eq:wimp_rate_vs_mass}.  The
annihilation cross section is in principle independent of the mixing
angle $s_\chi$, motivating the name \ul{secluded dark matter}
for such models, but the hidden photon needs to eventually decay into
SM particles. Again assuming $Q_\chi=1$, and $m_\chi=10$ GeV, we find
\begin{align}
\frac{1.7 \cdot 10^{-9}}{\gev^{2}} \stackrel{!}{=} \frac{g_D^4}{8\pi m_\chi^2}=\frac{g_D^4}{250~\gev^2}
\qquad \Leftrightarrow \qquad 
\boxed{g_d=0.025} \; .
\end{align}
%


\subsection{Supersymmetric neutralinos}
\label{sec:models_mssm}
\index{neutralino}

Supersymmetry is a (relatively) fashionable model for physics beyond
the Standard Model which provides us with a very general set of dark
matter candidates. Unlike the portal model described in
Section~\ref{sec:models_portal} the lightest supersymmetric partner
(LSP) is typically a fermion, more specifically a Majorana
fermion. 
Majorana fermions are their own anti-particles.  An on-shell Dirac
fermion, like an electron, has four degrees of freedom; for the
particle $e^-$ we have two spin directions, and for the anti-particle
$e^+$ we have another two. The Majorana fermion only has two degrees
of freedom. The reason why the minimal supersymmetric extension of the
Standard Model, the MSSM\index{MSSM}, limits us to Majorana fermions is that the
photon as a massless gauge boson only has two degrees of freedom. This
holds for both, the bino partner of the hypercharge gauge boson $B$
and the wino partner of the still massless $SU(2)_L$ gauge boson
$W^3$. Just like the gauge bosons in the Standard Model mix to the
photon and the $Z$, the bino and wino mix to form so-called
neutralinos. The masses of the physical state can be computed from the
bino mass parameter $M_1$ and the wino mass parameter $M_2$.

For reasons which we do not have to discuss in these lecture notes,
the MSSM includes a non-minimal Higgs sector: the masses of up-type
and down-type fermions are not generated from one Higgs
field. Instead, we have two Higgs doublets with two vacuum expectation
values $v_u$ and $v_d$. Because both contribute to the weak gauge
boson masses, their squares have to add to
\begin{align}
v_u^2 + v_d^2 &= v_H^2 = (246~\gev )^2 \notag \\
\Leftrightarrow \qquad
v_u &= v_H \cos \beta \qquad 
v_d  = v_H \sin \beta 
\qquad \Leftrightarrow \qquad
\tan \beta = \frac{v_u}{v_d} \; .
\end{align}
Two Higgs doublets include eight degrees of freedom, out of which three
Nambu-Goldstone modes are needed to make the weak bosons massive. The five
remaining degrees of freedom form a light scalar $h^0$, a heavy scalar $H^0$,
a pseudo-scalar $A^0$, and a charged Higgs $H^\pm$. Altogether this
gives four neutral and four charged degrees of freedom. In the Standard
Model we know that the one neutral (pseudo-scalar) Nambu-Goldstone mode
forms one particle with the $W^3$ gauge bosons. We can therefore
expect the supersymmetric higgsinos to mix with the bino and wino as
well. Because the neutralinos still are Majorana fermions, the eight
degrees of freedom form four neutralino states $\nni$. Their mass matrix
has the form
\begin{align}
\mat = 
\begin{pmatrix}
 M_1    &   0 &
 -m_Z s_w \cos \beta & \phantom{-}  m_Z s_w \sin \beta  \\
 0                &   M_2 &
 \phantom{-} m_Z c_w \cos \beta &  -m_Z c_w \sin \beta  \\
 -m_Z s_w \cos \beta & \phantom{-}  m_Z c_w \cos \beta &
 \phantom{-} 0               &  -\mu              \\
 \phantom{-} m_Z s_w \sin \beta &  -m_Z c_w \sin \beta &
 -\mu             & \phantom{-} 0
\end{pmatrix} \; .
\label{eq:neutmass}
\end{align}
The mass matrix is real and therefore symmetric.  In the upper left
corner the bino and wino mass parameters appear, without any mixing
terms between them. In the lower right corner we see the two higgsino
states. Their mass parameter is $\mu$, the minus sign is conventional;
by definition of the Higgs potential it links the up-type and
down-type Higgs or higgsino fields, so it has to appear in the
off-diagonal entries. The off-diagonal sub-matrices are proportional
to $m_Z$. In the limit $s_w \to 0$ and $\sin \beta= \cos \beta =
1/\sqrt{2}$ a universal mixing mass term $m_Z/\sqrt{2}$ between the wino and
each of the two higgsinos appears. It is the supersymmetric
counterpart of the combined Goldstone-$W^3$ mass $m_Z$. \index{neutralino!wino} \index{neutralino!higgsino}

As any symmetric matrix, the neutralino mass matrix can be
diagonalized through a real orthogonal rotation,
\begin{align}
  & N \; \mat \; N^{-1} = \text{diag} \left( m_{\nnj} \right)
\qquad j=1,2 
\end{align}
It is possible to extend the MSSM such that the dark matter candidates
become Dirac fermions, but we will not explore this avenue in these
lecture notes.\bigskip

Because the $SU(2)_L$ gauge bosons as well as the Higgs doublet include
charged states, the neutralinos are accompanied by chargino
states. They cannot be Majorana particles, because they carry electric
charge. However, as a remainder of the neutralino Majorana property
they do not have a well-defined fermion number, like electrons or
positrons have. The corresponding chargino mass matrix will not
include a bino-like state, so it reads
\begin{align}
\mat = 
\begin{pmatrix}
 M_2        & \sqrt{2} m_W \sin \beta \\
 \sqrt{2} m_W \cos \beta & \mu 
\end{pmatrix}  
\label{eq:charmass}
\end{align}
It includes the remaining four degrees of freedom from the wino sector
and four degrees of freedom from the higgsino sector. As for the
neutralinos, the wino and higgsino components mix via a weak mass
term.  Because the chargino mass matrix is real and not symmetric, it
can only be diagonalized using two unitary matrices,
\begin{align}
U^* \; \mat \; V^{-1} = \text{diag} \left( m_{\cpmj} \right) \qquad j=1,2 
\end{align}
For the dark matter phenomenology of the neutralino--chargino sector
it will turn out that the mass difference between the lightest
neutralino(s) and the lightest chargino are the relevant
parameters. The reason is a possible co-annihilation process as
described in Section~\ref{sec:coannihilation}\bigskip

We can best understand the MSSM dark matter sector in terms of the
different $SU(2)_L$ representations. The bino state as the partner of
the hypercharge gauge boson is a singlet under $SU(2)_L$. \index{neutralino!bino} The wino
fields with the mass parameter $M_2$ consist of two neutral degrees of
freedom as well as four charged degrees of freedom, one for each
polarization of $W^\pm$. \index{neutralino!wino} Together, the supersymmetric partners of the
$W$ boson vector field also form a triplet under $SU(2)_L$. Finally,
each of the two higgsinos arise as supersymmetric partner of an
$SU(2)_L$ Higgs doublet. The neutralino mass matrix in
Eq.\eqref{eq:neutmass} therefore interpolates between singlet,
doublet, and triplet states under $SU(2)_L$.\bigskip

The most relevant couplings of the neutralinos and charginos we need
to consider for our dark matter calculations are
\begin{align}
g_{Z \nne \nne} 
&= \frac{g}{2 c_w} \; 
   \left( |N_{13}|^2 - |N_{14}|^2 \right) \notag \\ 
g_{h \nne \nne} 
&= \left( g' N_{11} - g N_{12} \right) \; 
   \left( \sin \alpha \; N_{13} + \cos \alpha \; N_{14} \right) \notag \\ 
g_{A \nne \nne}
&= \left( g' N_{11} - g N_{12} \right) \; 
   \left( \sin \beta \; N_{13} - \cos \beta \; N_{14} \right) \notag \\ 
g_{\gamma \cpe \cme}
&= e \notag \\
g_{W \nne \cpe} &=
g \; \left( \frac{1}{\sqrt{2}} N_{14} V_{12}^* - N_{12} V_{11}^* \right) \; ,
\label{eq:coups_neutralinos}
\end{align}
with $e = g s_w$, $s_w^2 \approx 1/4$ and hence
$c_w^2 \approx 3/4$.  The mixing angle $\alpha$ describes the
rotation from the up-type and down-type supersymmetric Higgs bosons
into mass eigenstates. In the limit of only one light Higgs boson with
a mass of $126$~GeV it is given by the decoupling condition $\cos (\beta
- \alpha) \to 0$. The above form means for those couplings which
contribute to the (co-) annihilation of neutralino dark matter
\begin{itemize}
\item[--] neutralinos couple to weak gauge bosons through their higgsino content
\item[--] neutralinos couple to the light Higgs through gaugino--higgsino mixing
\item[--] charginos couple to the photon diagonally, like any other charged particle
\item[--] neutralinos and charginos couple to a $W$-boson diagonally as higgsinos and gauginos
\end{itemize}
Finally, supersymmetry predicts scalar partners of the quarks and
leptons, so-called squarks and sleptons. For the partners of massless
fermions, for example squarks $\tilde{q}$, there exists a $q \tilde{q}
\nnj$ coupling induced through the gaugino content of the
neutralinos. If this kind of coupling should contribute to neutralino
dark matter annihilation, the lightest supersymmetric scalar has to be
almost mass degenerate with the lightest neutralino. Because squarks
are strongly constrained by LHC searches and because of the pattern of
renormalization group running, we usually assume one of the sleptons
to be this lightest state. In addition, the mixing of the scalar
partners of the left-handed and right-handed fermions into mass
eigenstates is driven by the corresponding fermion mass, the most
attractive co-annihilation scenario in the scalar sector is
stau--neutralino co-annihilation. However, in these lecture notes we will
focus on a pure neutralino--chargino dark matter sector and leave the
discussion of the squark--quark--neutralino coupling to
Section~\ref{sec:coll} on LHC searches.\bigskip

Similar to the previous section we now compute the neutralino
annihilation rate, assuming that in the $10~...~1000$~GeV mass range
they are thermally produced. For mostly \ul{bino dark matter} \index{dark matter!supersymmetric} with 
\begin{align}
M_1 \ll M_2, |\mu|
\label{eq:bino_paras}
\end{align}
the annihilation to the observed relic density is a problem. There
simply is no relevant $2 \to 2$ Feynman diagram, unless there is help
from supersymmetric scalars $\tilde{f}$, as shown in
Figure~\ref{fig:neutralinos_relic_feyn}.  If for example light staus
appear in the $t$-channel of the annihilation rate we find \index{neutralino!bino}
\begin{align}
\sigma(\tilde{B} \tilde{B} \to f \bar{f})
\approx \frac{g^4 m_{\nne}^2}{16 \pi m_{\tilde{f}}^4}
\qquad \text{with} \quad m_{\nne} \approx M_1 \ll m_{\tilde{f}} \; .
\end{align}
The problem with pure neutralino annihilation is that in the limit of
relatively heavy sfermions the annihilation cross section drops rapidly, leading to
a too large predicted bino relic density. Usually, this leads us to
rely on stau co-annihilation for a light bino LSP.
Along these lines it is useful to mention that with
gravity-mediated supersymmetry breaking we assume $M_1$ and $M_2$ to
be identical at the Planck scale, which given the beta functions of
the hypercharge and the weak interaction turns into the condition $M_1
\approx M_2/2$ at the weak scale, \ie light bino dark matter would be
a typical feature in these models.\bigskip

\begin{figure}[b!]
\begin{center}
\begin{fmfgraph*}(100,60)
\fmfset{arrow_len}{2mm}
\fmfleft{i1,i2}
\fmfright{o1,o2}
\fmf{plain,tension=0.2,width=0.6}{i1,v1}
\fmf{photon,tension=0.2,width=0.6}{i1,v1}
\fmf{plain,tension=0.2,width=0.6}{v2,i2}
\fmf{photon,tension=0.2,width=0.6}{v2,i2}
\fmf{scalar,tension=0.2,label=$\tilde{f}$,width=0.6}{v2,v1}
\fmf{fermion,tension=0.4,width=0.6}{o2,v2}
\fmf{fermion,tension= 0.4,width=0.6}{v1,o1}
\fmflabel{$f$}{o1}
\fmflabel{$\bar{f}$}{o2}
\fmflabel{$\tilde{B}$}{i1}
\fmflabel{$\tilde{B}$}{i2}
\end{fmfgraph*}
\hspace*{0.1\textwidth}
\begin{fmfgraph*}(100,60)
\fmfset{arrow_len}{2mm}
\fmfleft{i1,i2}
\fmfright{o1,o2}
\fmf{plain,tension=0.2,width=0.6}{i1,v1}
\fmf{photon,tension=0.2,width=0.6}{i1,v1}
\fmf{plain,tension=0.2,width=0.6}{v2,i2}
\fmf{photon,tension=0.2,width=0.6}{v2,i2}
\fmf{plain,tension=0.1,width=0.6}{v1,v2}
\fmf{photon,tension=0.1,label=$\tilde{W}^+$,width=0.6}{v1,v2}
\fmf{photon,tension=0.4,width=0.6}{o2,v2}
\fmf{photon,tension= 0.4,width=0.6}{v1,o1}
\fmflabel{$W^+$}{o1}
\fmflabel{$W^-$}{o2}
\fmflabel{$\tilde{W}$}{i1}
\fmflabel{$\tilde{W}$}{i2}
\end{fmfgraph*}
\hspace*{0.1\textwidth}
\begin{fmfgraph*}(100,60)
\fmfset{arrow_len}{2mm}
\fmfleft{i1,i2}
\fmfright{o1,o2}
\fmf{plain,tension=0.2,width=0.6}{i1,v1}
\fmf{photon,tension=0.2,width=0.6}{i1,v1}
\fmf{plain,tension=0.2,width=0.6}{v1,i2}
\fmf{photon,tension=0.2,width=0.6}{v1,i2}
\fmf{dashes,tension=0.2,label=$A$,width=0.6}{v1,v2}
\fmf{fermion,tension=0.4,width=0.6}{o2,v2}
\fmf{fermion,tension= 0.4,width=0.6}{v2,o1}
\fmflabel{$f$}{o1}
\fmflabel{$\bar{f}$}{o2}
\fmflabel{$\tilde{H}$}{i1}
\fmflabel{$\tilde{H}$}{i2}
\end{fmfgraph*}
\end{center}
\caption{Sample Feynman diagrams for the annihilation of
  supersymmetric binos (left), winos (center), and higgsinos.}
\label{fig:neutralinos_relic_feyn} 
\end{figure}

\begin{figure}[b!]
\begin{center}
\includegraphics[width=0.5\textwidth]{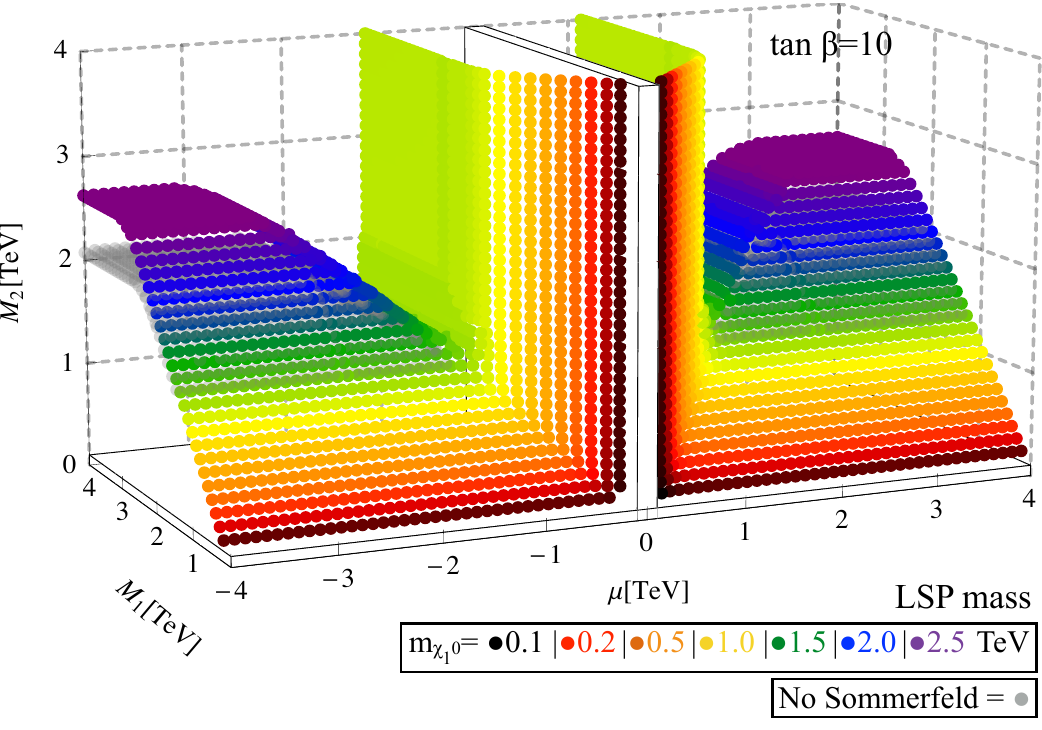}
\end{center}
\caption{Combinations of neutralino mass parameters $M_1, M_2, \mu$
  that produce the correct relic abundance, accounting for
  Sommerfeld-enhancement, along with the LSP mass. The relic surface
  without Sommerfeld enhancement is shown in gray. Figure from
  Ref.~\cite{nimatron2}.}
\label{fig:neutralino_relic1}
\end{figure}

\begin{figure}[b!]
\includegraphics[width=0.5\textwidth]{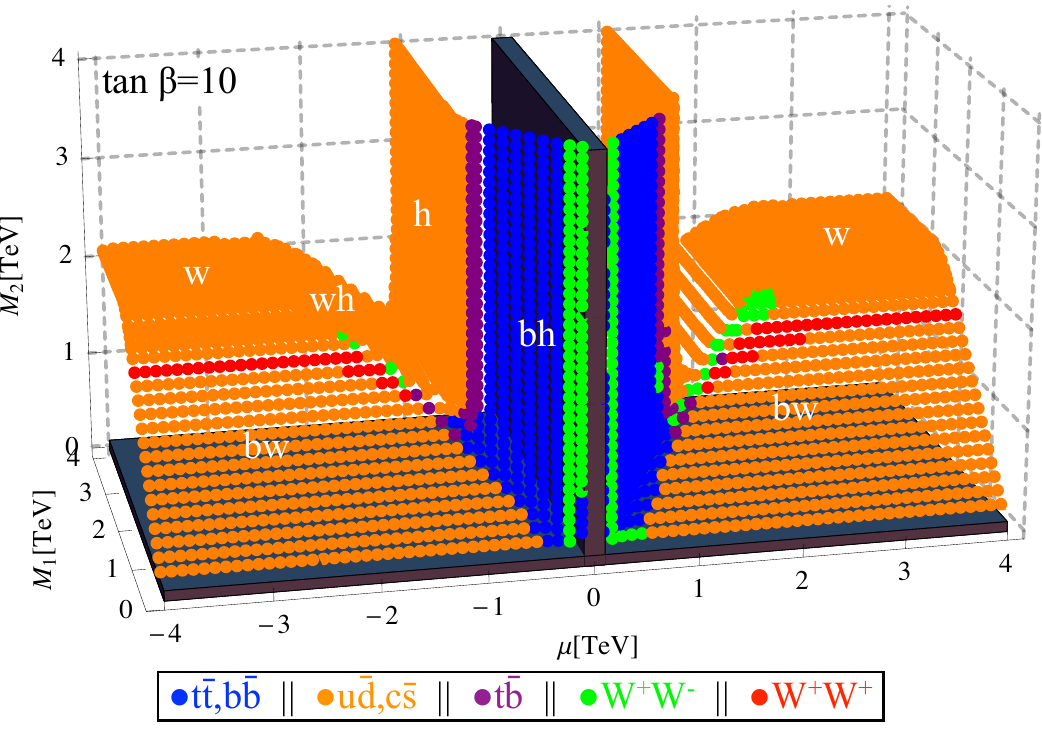}
\includegraphics[width=0.5\textwidth]{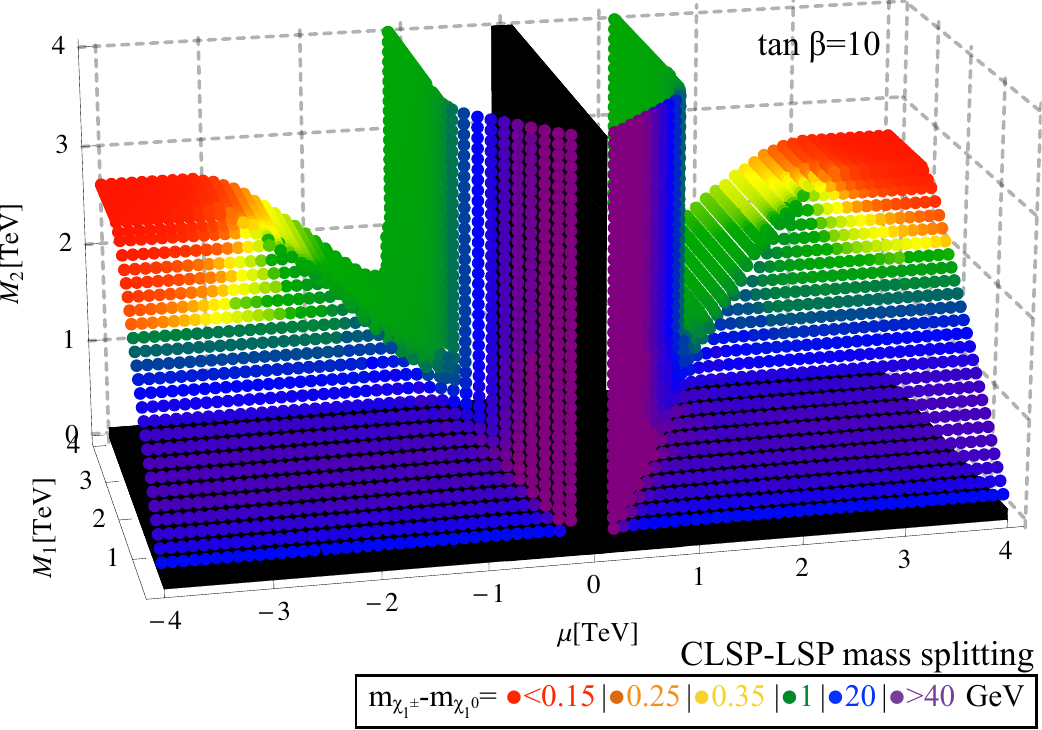}
\caption{Left: combinations of neutralino mass parameters $M_1, M_2,
  \mu$ that produce the correct relic abundance, not accounting for
  Sommerfeld-enhancement, along with the leading annihilation product.
  Parameters excluded by LEP are occluded with a white or black box.
  Right: mass splitting between the lightest chargino and lightest
  neutralino. Parameters excluded by LEP are occluded with a white or
  black box.  Figures from Ref.~\cite{nimatron1}.}
\label{fig:indirect_relic}
\end{figure}

If $M_1$ becomes larger than $M_2$ or $\mu$ we can to a good
approximation consider the limit
\begin{align}
M_2 \ll M_1 \to \infty
\qquad \text{and} \qquad 
|\mu| \ll M_1 \to \infty \; ,
\end{align}
In that case we know that independent of the relation of $M_2$ and
$\mu$ there will be at least the lightest chargino close in mass to
the LSP. It appears in the $t$-channel of the actual annihilation
diagram and as a co-annihilation partner.  To avoid a second
neutralino in the co-annihilation process we first consider
\ul{wino dark matter}, \index{neutralino!wino}
\begin{align}
M_2 \ll \mu, M_1, m_{\tilde{f}} \; .
\end{align}
From the list of neutralino couplings in
Eq.\eqref{eq:coups_neutralinos} we see that in the absence of
additional supersymmetric particles pure wino dark matter can
annihilate through a $t$-channel chargino, as illustrated in
Figure~\ref{fig:neutralinos_relic_feyn}.  Based on the known couplings
and on dimensional arguments the annihilation cross section should
scale like
\begin{align}
\sigma( \tilde{W} \tilde{W} \to W^+ W^-) 
&\approx \frac{1}{16 \pi} \; \sqrt{ \frac{1 - 4 m_W^2/s}{1 - 4 m_{\nne}^2/s} } \;
        \frac{g^4 s_w^2}{c_w^4 m_{\nne}^2} 
\qquad \text{with} \quad
m_{\nne} \approx m_{\cpme} \approx M_2 \gg m_W  \notag \\
&\approx \frac{1}{16 \pi}  \;\frac{1}{ m_{\nne} v}  \;
        \frac{g^4 s_w^4}{c_w^4 m_{\nne}} 
\qquad \Rightarrow \qquad 
\boxed{\sigma_{\chi \chi} \propto \frac{1}{m_{\nne}^2}} \; 
\end{align}
The scaling with the mass of the dark matter agent does not follow our
original postulate for the WIMP miracle in
Eq.\eqref{eq:wimp_ann_approx}, which was $\sigma_{\chi \chi} \propto
m_{\nne}^2/m_W^4$. If we only rely on the direct dark matter
annihilation, the observed relic density translated into a comparably
light neutralino mass,
\begin{align}
\langle \sigma v \rangle \Bigg|_{\tilde{W} \tilde{W} \to W^+ W^-}
=  \frac{g^4 s_w^4}{16 \pi c_w^4 m_{\nne}^2} 
\approx \frac{0.7^4}{450 \, m_{\nne}^2} 
&\eqx{eq:wimp_rate_vs_mass} 1.7 \cdot 10^{-9} \frac{1}{\gev^2} 
\qquad \Leftrightarrow \qquad 
m_{\nne} 
\approx 560~\gev \; .
\end{align}
However, this estimate is numerically poor. The reason is that in
contrast to this lightest neutralino, the co-annihilating chargino can
annihilate through a photon s-channel diagram into charged Standard Model fermions,
\begin{align}
\sigma( \cpe \cme \to \gamma^* \to f\bar{f} )
\approx \sum_f \frac{N_c e^4}{16 \pi m_{\cpme}^2} 
= \sum_f \frac{N_c g^4 s_w^2}{16 \pi m_{\cpme}^2} \; .
\label{eq:chargino_ann}
\end{align}
For light quarks alone the color factor combined with the sum over
flavors adds a factor $5 \times 3 = 15$ to the annihilation rate.  In
addition, for $\cpe \cme$ annihilation we need to take into account
the Sommerfeld enhancement through photon exchange between the slowly
moving incoming charginos, as derived in
Section~\ref{sec:sommerfeld}. This gives us the correct values
\begin{align}
\boxed{
\Omega_{\tilde{W}} h^2 \approx 0.12
\left(\frac{m_{\nne}}{2.1~\tev}\right)^2
\stackrel{\text{Sommerfeld}}{\longrightarrow} 0.12
\left(\frac{m_{\nne}}{2.6~\tev}\right)^2 
} \; .
\label{eq:winorelic}
\end{align}
In Figure~\ref{fig:neutralino_relic1} and in the left panel of Figure~\ref{fig:indirect_relic} this wino LSP
mass range appears as a horizontal plateau in $M_2$, with and without
the Sommerfeld enhancement. In the right panel of
Figure~\ref{fig:indirect_relic} we show the mass difference between
the lightest neutralino and the lighter chargino. Typical values for a
wino-LSP mass splitting are around $\Delta m = 150$~MeV, sensitive to
loop corrections to the mass matrices shown in Eq.\eqref{eq:neutmass}
and Eq.\eqref{eq:charmass}.\bigskip

Finally, we can study \ul{higgsino dark matter} in the limit \index{neutralino!higgsino}
\begin{align}
|\mu| \ll M_1, M_2, m_{\tilde{f}} \; .
\end{align}
\index{Sommerfeld enhancement}
Again from Eq.\eqref{eq:coups_neutralinos} we see that in addition to
the $t$-channel chargino exchange, annihilation through $s$-channel
Higgs states is possible.  Again, the corresponding Feynman diagrams
are shown in Figure~\ref{fig:neutralinos_relic_feyn}.  At least in the
pure higgsino limit with $N_{i3} = N_{i4}$ the two contributions to
the $\tilde{H} \tilde{H} Z^0$ coupling cancel, limiting the impact of
$s$-channel $Z$-mediated annihilation. Still, these channels make the
direct annihilation of higgsino dark matter significantly more
efficient than for wino dark matter. The Sommerfeld enhancement plays
a sub-leading role, because it mostly affects the less relevant
chargino co-annihilation,
\begin{align}
\boxed{
\Omega_{\tilde{H}} h^2 
\approx 0.12 \left( \frac{m_{\nne}}{1.13~\tev} \right)^2
\stackrel{\text{Sommerfeld}}{\longrightarrow}
0.12 \left(\frac{m_{\nne}}{1.14~\tev}\right)^2  
} \; .
\label{eq:higgsinorelic}
\end{align}
The higgsino LSP appears in 
Figure~\ref{fig:neutralino_relic1} as a vertical plateau in
$\mu$. The corresponding mass difference between the lightest
neutralino and chargino is much larger than for the wino LSP; it now
ranges around a GeV.\bigskip

Also in Figure~\ref{fig:neutralino_relic1} we see that a dark matter
neutralino in the MSSM can be much lighter than the pure wino and
higgsino results in Eqs.\eqref{eq:winorelic}\index{Higgs!funnel}
and~\eqref{eq:higgsinorelic} suggest. For a strongly mixed neutralino
the scaling of the annihilation cross section with the neutralino mass
changes, and poles in the $s$-channels appear. In the left panel of
Figure~\ref{fig:indirect_relic} we add the leading Standard Model
final state of the dark matter annihilation process, corresponding to
the distinct parameter regions
\begin{itemize}
\item[--] the light Higgs funnel region with $2 m_{\nne} = m_h$ .  The
  leading contribution to dark matter annihilation is the decay to $b$
  quarks. As a consequence of the tiny Higgs width the neutralino mass
  has to be finely adjusted. According to
  Eq.\eqref{eq:coups_neutralinos} the neutralinos couple to the Higgs
  though gaugino-higgsino mixing.  A small, $\ope(10\%)$ higgsino
  component can then give the correct relic density. This very narrow
  channel with a very light neutralino is not represented in
  Figure~\ref{fig:indirect_relic}. Decays of the Higgs mediator
  to lighter fermions, like tau leptons, are suppressed by their
  smaller Yukawa coupling and a color factor;

\item[--] the $Z$-mediated annihilation with $2 m_{\nne} \approx m_Z$,
  with a final state mostly consisting out of light-flavor jets. The
  corresponding neutralino coupling requires a sizeable higgsino
  content. Again, this finely tuned low-mass channel in not shown in
  Figure~\ref{fig:indirect_relic};
\index{mediator!s-channel}
\item[--] $s$-channel annihilation through the higgsino content with
  some bino admixture also occurs via the heavy Higgs bosons
  $A^0,H^0$, and $H^\pm$ with their large widths.  This region extends
  to large neutralino masses, provided the Higgs masses follows the
  neutralino mass. The main decay channels are $b\bar{b}$, $t\bar{t}$,
  and $t\bar{b}$. The massive gauge bosons typically decouple from the
  heavy Higgs sector;

\item[--] with a small enough mass splitting between the lightest
  neutralino and lightest chargino, co-annihilation in the
  neutralino--chargino sector becomes important.  For a higgsino-bino
  state there appears a large annihilation rate to $\nne \nne \to W^+
  W^-$ with a $t$-channel chargino exchange.  The wino-bino state will
  mostly co-annihilate into $\nne \cpme \to W^\pm \to q\bar{q}'$, but
  also contribute to the $W^+ W^-$ final state.  Finally, as shown in
  Figure~\ref{fig:indirect_relic} the co-annihilation of two charginos
  can be efficient to reach the observed relic density, leading to a
  $W^+ W^+$ final state;

\item[--] one channel which is absent from our discussion of purely
  neutralino and chargino dark matter appears for a mass splitting
  between the scalar partner of the tau lepton, the stau, and the
  lightest neutralino of few per-cent or less the two states can
  efficiently co-annihilate.  In the scalar quark sector the same
  mechanism exists for the lightest top squark, but it leads to issues
  with the predicted light Higgs mass of $126$~GeV.

\end{itemize}
In the right panel of Figure~\ref{fig:indirect_relic} we show the mass
difference between the lightest chargino and the lightest
neutralino. In all regions where chargino co-annihilation is required,
this mass splitting is small. From the form of the mass matrices shown
in Eq.\eqref{eq:neutmass} and Eq.\eqref{eq:charmass} this will be the
case when either $M_2$ or $\mu$ are the lightest mass
parameters. Because of the light higgsino, the two higgsino states in the
neutralino sector lead to an additional level separation between the two
lightest neutralinos, the degeneracy of the lightest chargino and the
lightest neutralino masses will be less precise here. For pure winos
the mass difference between the lightest chargino and the lightest
neutralino can be small enough that loop corrections matter and the
chargino becomes long-lived.

Note that all the above listed channels correspond to ways of
enhancing the dark matter annihilation cross section, to allow for
light dark matter closer to the Standard Model masses. In that sense
they indicate a fine tuning around the generic scaling
$\sigma_{\chi \chi} \propto 1/m_{\nne}^2$ which in the MSSM predicts
TeV-scale higgsinos and even heavier winos.

\subsection{Effective field theory}
\label{sec:models_eft}

As another, final theoretical framework to describe the dark matter
relic density we introduce an effective theory of dark
matter~\cite{tim_eft}.\index{effective field theory} We will start from the MSSM description and
show how the heavy mediator can decouple from the annihilation
process.  This will put us into a situation similar to the description
for example of the muon decay in Fermi's theory. Next, we will
generalize this result to an effective Lagrangian. Finally, we will
show how this effective Lagrangian describes dark matter annihilation
in the early universe.\bigskip

\index{dark matter!annihilation}
Let us start with dark matter annihilation mediated by a heavy
pseudoscalar $A$ in the MSSM, as illustrated in the right panel of
Figure~\ref{fig:neutralinos_relic_feyn}. The $A\nne \nne$ coupling is
defined in Eq.\eqref{eq:coups_neutralinos}. If we assume the heavy
Higgs to decay to two bottom quarks, the $2 \to 2$ annihilation
channel is
\begin{align}
 \nne \nne \to A^* \to b\bar{b} \; ,
\label{eq:nn_annihilation_eft}
\end{align}
This description of dark matter annihilation includes two different
mass scales, the dark matter mass $m_{\nne}$ and a decoupled mediator
mass $m_A \gg m_{\nne}$.  The matrix element for the dark matter
annihilation process includes the $A$-propagator.  From
Section~\ref{sec:boltzmann} we know that for WIMP annihilation the
velocity of the incoming particles is small, $v \ll 1$.  If the energy
of the scattering process, which determines the momentum flowing
through the $A$-propagator is much smaller than the $A$-mass, we can
approximate the intermediate propagator as
\index{degrees of freedom}
\begin{align} 
 \frac{1}{q^2 - m_A^2} \to - \frac{1}{m_A^2}
 \qquad \Leftrightarrow \qquad 
 \sigma( \nne \nne \to b\bar{b}) \propto 
 g_{A\nne \nne}^2 g_{A bb}^2\,\frac{m_b^2}{m_A^4} \; .
\end{align}
The fact that the propagator of the heavy scalar $A$ does not include
a momentum dependence is equivalent of removing the kinetic term of
the $A$-field from the Lagrangian. We remove the heavy scalar field
from the \ul{propagating degrees of freedom} of our theory. The
only actual particles we can use in our description of the
annihilation process of Eq.\eqref{eq:nn_annihilation_eft} are the dark
matter fermions $\nne$ and the bottom quarks. Between them we observe
a four-fermion interaction.\bigskip

On the Lagrangian level, such a four-fermion interactions mediated by
a non-propagating state is given by an operator of the type
\begin{align}
g_\text{ann} \; \overline{\psi}_{\nne} \Gamma^\mu\psi_{\nne}   \overline{\psi}_b \Gamma_\mu\psi_b \; ,
\end{align}
where $\Gamma^\mu=\{1,\gamma_5, \gamma_\mu, \gamma_\mu\gamma_5, [\gamma_\mu,\gamma_\nu]\}$ represents some kind of Lorentz structure.  We know that a
Lagrangian has mass dimension four, and a fermion spinor has mass
dimension 3/2. The four-fermion interaction then has mass dimension
six, and has to be accompanied by a mass-dependent prefactor,
\begin{align}
\lag \supset \frac{g_\text{ann}}{\Lambda^2} \; 
\overline{\psi}_{\nne} \Gamma^\mu \psi_{\nne} \overline{\psi}_b \Gamma_\mu\psi_b \; .
\label{eq:lag_annihilation_eft}
\end{align}
Given this Lagrangian, the question arises if we want to use this
interaction as a simplified description of the MSSM annihilation
process or view it as a more general structure without a known
ultraviolet completion. For example for the muon decay we nowadays
know that the suppression is given by the $W$-mass of the weak
interaction. Using our derivation of
Eq.\eqref{eq:lag_annihilation_eft} we are inspired by the MSSM
annihilation channel through a heavy pseudoscalar.  In that case the
scale $\Lambda$ should be given by the mass of the lightest particle
we integrate out. This defines, modulo order-one factors, the
\ul{matching condition}
\begin{align}
\Lambda = m_A 
\qqquad \text{and} \qqquad 
g_\text{ann} = g_{A \nne \nne} \; g_{Abb} \; . 
\end{align}
\bigskip

From Eq.\eqref{eq:def_eft} we see that all predictions by the
effective Lagrangian are invariant under a simultaneous scaling of the
new physics scale $\Lambda$ and the underlying coupling
$g_\text{ann}$.  Moreover, we know that the annihilation process $\nne
\nne \to f \bar{f}$ can be mediated by a scalar in the $t$-channel.
In the limit $m_f \ll m_{\nne} \ll m_{\tilde{f}}$ this defines
essentially the same four-fermion interaction as given in
Eq.\eqref{eq:lag_annihilation_eft}.

Indeed, the effective Lagrangian is more general than
its interpretation in terms of one half-decoupled model. This suggests
to regard the Lagrangian term of Eq.\eqref{eq:lag_annihilation_eft} as
the fundamental description of dark matter, not as an approximation 
to a full model. For excellent reasons we usually prefer
renormalizable Lagrangians, only including operators with mass dimension
four or less. Nevertheless, we can extend this approach to examples
including all operators up to mass dimension six. This allows to
describe all kinds of four-fermion interactions. From constructing the
Standard Model Lagrangian we know that given a set of particles we
need selection rules to choose which of the possible operators make it
into our Lagrangian.  Those rules are given by the symmetries of the
Lagrangian, local symmetries as well as global symmetries, gauge
symmetries as well as accidental symmetries. This way we define a
general Lagrangian of the kind \index{effective field theory}
\begin{align}
\boxed{
\lag = \lag_\text{SM} + \sum_{j} \frac{c_j}{\Lambda^{n-4}} \; \ope_j 
} \; ,
\label{eq:def_eft}
\end{align}
where the operators $\ope_j$ are organized by their
dimensionality. The $c_j$ are couplings of the kind shown in
Eq.\eqref{eq:lag_annihilation_eft}, called Wilson coefficients, and
$\Lambda$ is the new physics scale.

The one aspect which is crucial for any \ul{effective field
  theory} or EFT analysis is the choice of operators contributing to a
Lagrangian. Like for any respectable theory we have to assume that any
interaction or operator which is not forbidden by a symmetry will be
generated, either at tree level or at the quantum level. In practice,
this means that any analysis in the EFT framework will have to include a
large number of operators. Limits on individual Wilson coefficients
have to be derived by marginalizing over all other Wilson coefficients
using  Bayesian
integration (or a frequentist profile likelihood).

From the structure of the Lagrangian we know that there are several
ways to generate a higher dimensionality for additional operators,
\begin{itemize}
\item[--] external particles with field dimensions adding to more than
  four. The four-fermion interaction in Eq.\eqref{eq:lag_annihilation_eft} is
  one example;
\item[--] an energy scale of the Lagrangian normalized to the suppression
  scale, leading to corrections to lower-dimensional operators of the kind
  $v^2/\Lambda^2$;
\item[--] a derivative in the Lagrangian, which after Fourier
  transformation becomes a four-momentum in the Feynman rule. This gives
  corrections to lower-dimensional operators of the kind
  $p^2/\Lambda^2$.
\end{itemize}
For dark matter annihilation we usually rely on dimension-6 operators
of the first kind. Another example would be a $\nne \nne WW$
interaction, which requires a dimension-5 operator if we couple to the
gauge boson fields and a dimension-7 operator if we couple to the
gauge field strengths. The limitations of an EFT treatment are obvious
when we experimentally observe poles, for example the $A$-resonance in
the annihilation process of Eq.\eqref{eq:nn_annihilation_eft}. In the
presence of such a resonance it does not help to add higher and higher
dimensions --- this is similar to Taylor-expanding a pole at a finite
energy around zero. Whenever there is a new particle which can be
produced on-shell we have to add it to the effective Lagrangian as a
new, propagating degree of freedom.  Another limiting aspect is most
obvious from the third kind of operators: if the correction has the
form $p^2/\Lambda^2$, and the available energy for the process allows
for $p^2 \gtrsim \Lambda^2$, higher-dimensional operators are no
longer suppressed. However, this kind of argument has to be worked out
for specific observables and models to decide whether an EFT approximation
is justified.\bigskip

Finally, we can estimate what kind of effective theory of dark matter
can describe the observed relic density, $\Omega_\chi h^2 \approx
0.12$. As usual, we assume that there is one thermally produced dark
matter candidate $\chi$.  Two mass scales given by the propagating
dark matter agent and by some non-propagating mediator govern our dark
matter model. If a dark matter EFT should ever work we need to require
that the dark matter mass is significantly smaller than the mediator
mass,
\begin{align}
m_\chi \ll \mmed \; .
\end{align}
In terms of one coupling constant $g$ governing the annihilation
process we can use the usual estimate of the WIMP annihilation rate,
similar to Eq.\eqref{eq:wimp_ann_approx},
\begin{align}
 \langle \sigma_{\chi \chi} \, v \rangle 
\approx \frac{g^4 m_\chi^2}{4 \pi \, \mmed^4} 
\eqx{eq:relic_approx} \frac{1.7 \cdot 10^{-9}}{\gev^2} \; .
\label{eq:eft_sigann}
\end{align}
We know that it is crucial for this rate to be large enough to bring
the thermally produced dark matter rate to the observed level. This
gives us a lower limit on the ratio $m_\chi/\mmed^2$ or alternatively an
upper limit on the mediator mass for fixed dark matter mass.  As a
rough relation between the mediator and dark matter masses we find
\begin{alignat}{5}
\frac{\mmed^2}{g^2 m_\chi} = 6.8~\tev
&\qquad \stackrel{m_\chi = 10~\gev}{\Rightarrow} \qquad 
&\frac{\mmed}{g} &=  260~\gev \gg m_\chi  \notag \\
&\qquad \stackrel{m_\chi = \mmed/2}{\Rightarrow} \qquad 
&\frac{\mmed}{g} &=  3.4~\tev = m_\chi \; .
\label{eq:relic_rough}
\end{alignat}
The dark matter agent in the EFT model can be very light, and the
mediator will typically be significantly heavier. An EFT description
of dark matter annihilation seems entirely possible.\bigskip

Going back to our two models, the Higgs portal and the MSSM
neutralino, it is less clear if an EFT description of dark matter
annihilation works well. In part of the allowed parameter space, dark
matter annihilation proceeds through a light Higgs in the $s$-channel
on the pole.  Here the mediator is definitely a propagating degree of
freedom. For neutralino dark matter we discuss $t$-channel
chargino-mediated annihilation, where $m_{\cpme} \approx
m_{\nne}$. Again, the chargino is clearly propagating at the relevant
energies.

Finally, to fully rely on a dark matter EFT we need to make sure that
all relevant processes are correctly described. For our WIMP models
this includes the annihilation predicting the correct relic density,
indirect detection and possibly the Fermi galactic center excess 
introduced in Section~\ref{sec:indirect}, the limits from direct
detection discussed in Section~\ref{sec:direct}, and the collider
searches of Section~\ref{sec:coll}. We will comment on the related
challenges in the corresponding sections.

\newpage

\section{Indirect searches}
\label{sec:indirect}
\index{indirect detection}

There exist several ways of searching for dark matter in earth-bound
or satellite experiments. All of them rely on the interaction of the
dark matter particle with matter, which means they only work if the
dark matter particles interacts more than only gravitationally. This
is the main assumption of these lecture notes, and it is motivated by
the fact that the weak gauge coupling and the weak mass scale happen
to predict roughly the correct relic density, as described in
Section~\ref{sec:miracle}.

The idea behind indirect searches for WIMPS \index{WIMP} is that the generally
small current dark matter density is significantly enhanced wherever
there is a clump of gravitational matter, as for example in the sun or
in the center of the galaxy. In these regions dark matter should
efficiently annihilate even today, giving us either photons or pairs
of particles and anti-particles coming from there. Particles like
electrons or protons are not rare, but anti-particles in the
appropriate energy range should be detectable. The key ingredient to
the calculation of these spectra is the fact that dark matter
particles move only very slowly relative to galactic objects. This
means we need to compute all processes with incoming dark matter
particles essentially at rest. This approximation is even better than at the time of the dark matter freeze-out discussed in 
Section~\ref{sec:boltzmann}.\bigskip
\index{dark matter!annihilation}
\begin{figure}[b!]
\begin{center}
\includegraphics[width=0.55\textwidth]{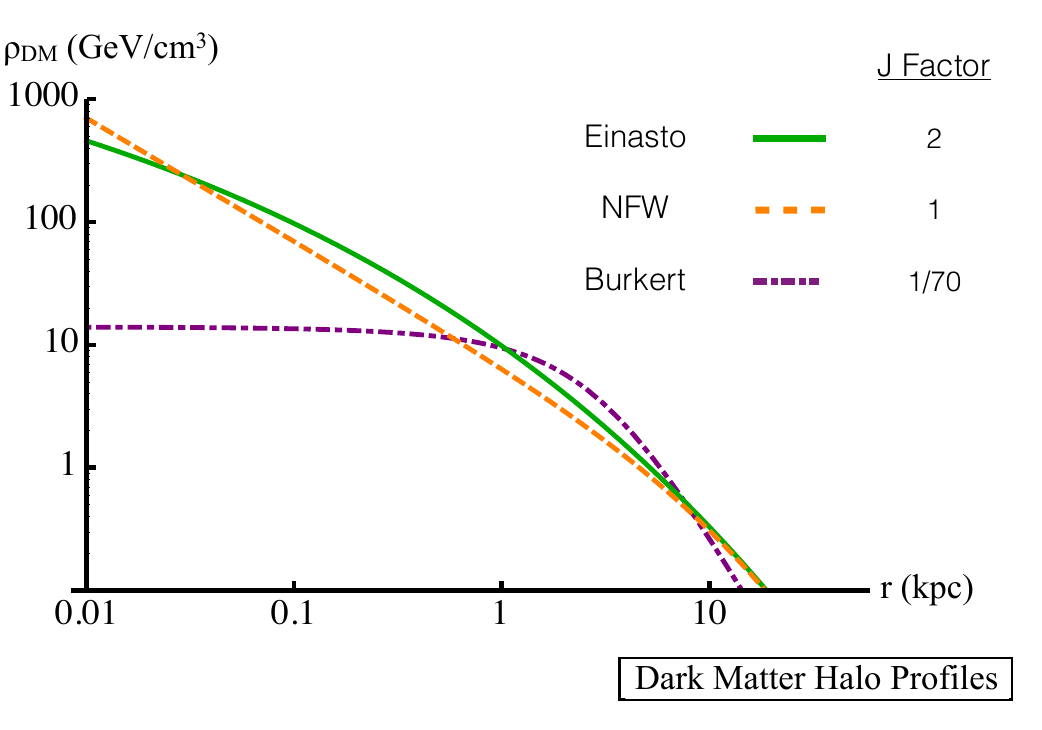}
\end{center}
\vspace*{-7mm}
\caption{Dark matter galactic halo profiles, including standard
  Einasto and NFW profiles along with a Burkert profile with a 3 kpc
  core. $J$ factors are obtained assuming a spherical dark matter distribution
  and integrating over the radius from the galactic center from
  $r\simeq 0.05$ to 0.15~kpc. $J$ factors are normalized so that
  $J(\rho_\text{NFW})=1$. Figure from Ref.\cite{nimatron2} \index{halo profile}}
\label{fig:profiles}
\end{figure}

Indirect detection experiments search for many different particles
which are produced in dark matter annihilation. First, this might be
the particles that dark matter directly annihilated into, for example
in a $2 \to 2$ scattering process. This includes protons and
anti-protons if dark matter annihilates into quarks. Second, we might
see decay products of these particles. An example for such signatures
are neutrinos. Examples for dark matter annihilation processes are
\begin{align}
\nne \nne &\to \ell^+ \ell^- \notag \\
\nne \nne &\to q \bar{q} \to p \bar{p} + X \notag \\
\nne \nne &\to \tau^+ \tau^-, W^+ W^-, b \bar{b} + X \to \ell^+ \ell^-, p\bar{p} + X 
\qquad ...
\end{align}
The final state particles are stable leptons or protons propagating
large distances in the Universe. While the leptons or protons can come
from many sources, the anti-particles appear much less frequently. One
key experimental task in many indirect dark matter searches is
therefore the ability to measure the charge of a lepton, typically
with the help of a magnetic field. For example, we can study the
energy dependence of the antiproton--proton ratio or the
positron--electron ratio as a function of the energy. The dark matter
signature is either a line or a shoulder in the spectrum, with a
cutoff
\begin{align}
E_{e^+} \approx m_{\nne}
\qquad \text{or} \qquad 
E_{e^+} < m_{\nne} \; .
\end{align}
The main astrophysical background is pulsars, which produce for
example electron--positron pairs of a given energy. There exists a
standard tool to simulate the propagation of all kinds of particles
through the Universe, which is called GALPROP. For example Pamela has
seen such a shoulder with a positron flux pointing to a very large
WIMP annihilation rate.  An interpretation in terms of dark matter is
inconclusive, because pulsars could provide an alternative explanation
and the excess is in tension with PLANCK results from CMB
measurements, as discussed in Sec~\ref{sec:velocity}.\bigskip

In these lecture notes we will focus on photons from dark matter
annihilation, which we can search for in gamma ray surveys over a wide
range of energies. They also will follow one of two kinematic
patterns: if they occur in the direct annihilation process, they will
appear as a mono-energetic line in the spectrum
\begin{align}
\chi \chi \to \gamma \gamma 
\qquad \text{with} \quad E_\gamma \approx m_\chi \; ,
\label{eq:indirect_line}
\end{align}
for any weakly interacting dark matter particle $\chi$. This is
because the massive dark matter particles are essentially at rest when
colliding. If the photons are radiated off charged particles or appear
in pion decays $\pi^0 \to \gamma \gamma$
\begin{align}
\chi \chi \to \tau^+ \tau^-, b\bar{b}, W^+ W^- 
          \to \gamma + \cdots \; ,
\label{eq:indirect_frag}
\end{align}
they will follow a fragmentation pattern. We can either compute this
photon spectrum or rely on precise measurements from the LEP
experiments at CERN (see Section~\ref{sec:coll_lepton} for a more
detailed discussion of the LEP experiments). This photon spectrum will
constrain the kind of dark matter annihilation products we should
consider, as well as the mass of the dark matter particle.\bigskip

The energy dependence of the photon flow inside a solid angle $\Delta
\Omega$ is given by
\begin{align}
\frac{d \Phi_\gamma}{d E_\gamma} 
= \frac{\left\langle \sigma v \right\rangle}{8 \pi m_{\nne}^2} \;
  \frac{d N_\gamma}{d E_\gamma} \; 
  \int_{\Delta \Omega} d \Omega \int_l\, d z \; \rho_\chi^2(z) \; ,
\label{eq:photon_flux}
\end{align}
where $E_\gamma$ is the photon energy, $\left\langle \sigma v
\right\rangle$ is the usual velocity-averaged annihilation
cross-section, $N_\gamma$ is the number of photons produced per
annihilation, and $l$ is the distance from the observer to the actual
annihilation event (line of sight). The photon flux depends on the dark matter density
squared because it arises from the annihilation of two dark matter
particles. A steeper dark matter halo profile, \ie the dark matter
density increasing more rapidly towards the center of the galaxy,
results in a more stringent bound on dark matter annihilation. The key
problem in the interpretation of indirect search results in terms of
dark matter is that we cannot measure the dark matter distributions
$\rho_\chi(l)$ for example in our galaxy directly. Instead, we have to
rely on numerical simulations of the dark matter profile, which
introduce a sizeable parametric or theory uncertainty in any
dark-matter related result. Note that the dark matter profile is not
some kind of multi-parameter input which we have the freedom to assume
freely. It is a prediction of numerical dark matter simulations with
associated error bars. Not all papers account for this uncertainty
properly.  In contrast,  the constraints derived from CMB anisotropies discussed in Section~\ref{sec:cmb} are largely free of astrophysical uncertainties.  

There exist three standard density profiles; the steep
\ul{Navarro-Frenk-White (NFW) profile}\index{halo profile!Navarro-Frenk-White} is given by
\begin{align}
\rho_\text{NFW}(r) 
= \frac{\rho_\odot}{\left( \dfrac{r}{R} \right)^\gamma \left( 1+ \dfrac{r}{R} \right)^{3 - \gamma}}
\stackrel{\gamma=1}{=} \frac{\rho_\odot}{\dfrac{r}{R} \left( 1+ \dfrac{r}{R} \right)^2}
\; ,
\label{eq:nfw_profile}
\end{align}
where $r$ is the distance from the galactic center. Typical parameters
are a characteristic scale $R=20~\text{kpc}$ and a solar position dark
matter density $\rho_\odot = 0.4~\gev/\text{cm}^3$ at $r_\odot
=8.5$~kpc. In this form we can easily read off the scaling of the dark
matter density in the center of the galaxy, \ie $r \ll R$; there we
find $\rho_\text{NFW} \propto r^{-\gamma}$. The second steepest is
the exponential \ul{Einasto profile} \index{halo profile!Einasto},
\begin{align}
\rho_\text{Einasto}(r) = \rho_\odot~ \exp \left[-\frac{2}{\alpha} \left(\left( \frac{r}{R}\right)^\alpha -1\right) \right] \; ,
\label{eq:einasto_profile}
\end{align}
with $\alpha=0.17$ and $R=20~\text{kpc}$. It fits micro-lensing and
star velocity data best. Third is the \ul{Burkert profile} \index{halo profile!Burkert} with
a constant density inside a radius $R$,
\begin{align}
\rho_\text{Burkert}(r) =
\frac{\rho_\odot}{\left( 1+ \dfrac{r}{R} \right) \left(1+ \dfrac{r^2}{R^2} \right)} \; ,
\label{eq:burkert_profile}
\end{align}
where we assume $R=3$~kpc. Assuming a large core results in very
diffuse dark matter at the galactic center, and therefore yields the
weakest bound on neutralino self annihilation.  Instead assuming $R =
0.1$~kpc only alters the dark matter annihilation constraints by an
order-one factor.  We show the three profiles in
Figure~\ref{fig:profiles} and observe that the difference between the
Einasto and the NFW parametrizations are marginal, while the Burkert
profile has a very strongly reduced dark matter density in the center
of the galaxy. One sobering result of this comparison is that whatever
theoretical considerations lie behind the NFW and Einasto profiles,
once their parameters are fit to data the possibly different
underlying arguments play hardly any role. The impact on gamma ray
flux of different dark matter halo profiles is conveniently
parameterized by the factor
\begin{align}
J \propto \int_{\Delta \Omega} d \Omega \int_\text{line of sight} d z \; \rho_\chi^2(z) 
\qquad \text{with} \quad 
J(\rho_\text{NFW}) \equiv 1 \; .
\end{align}
Also in Figure~\ref{fig:profiles} we quote the $J$ factors integrated
over the approximate HESS galactic center gamma ray search range, $r =
0.05~...~0.15$~kpc. As expected, the Burkert profile predicts a photon
flow lower by almost two orders of magnitude. In a quantitative
analysis of dark matter signals this difference should be included as
a theory error or a parametric error, similar to for example parton
densities or the strong coupling in LHC searches.\bigskip

\begin{figure}[b!]
\begin{center}
\includegraphics[width=0.5\textwidth]{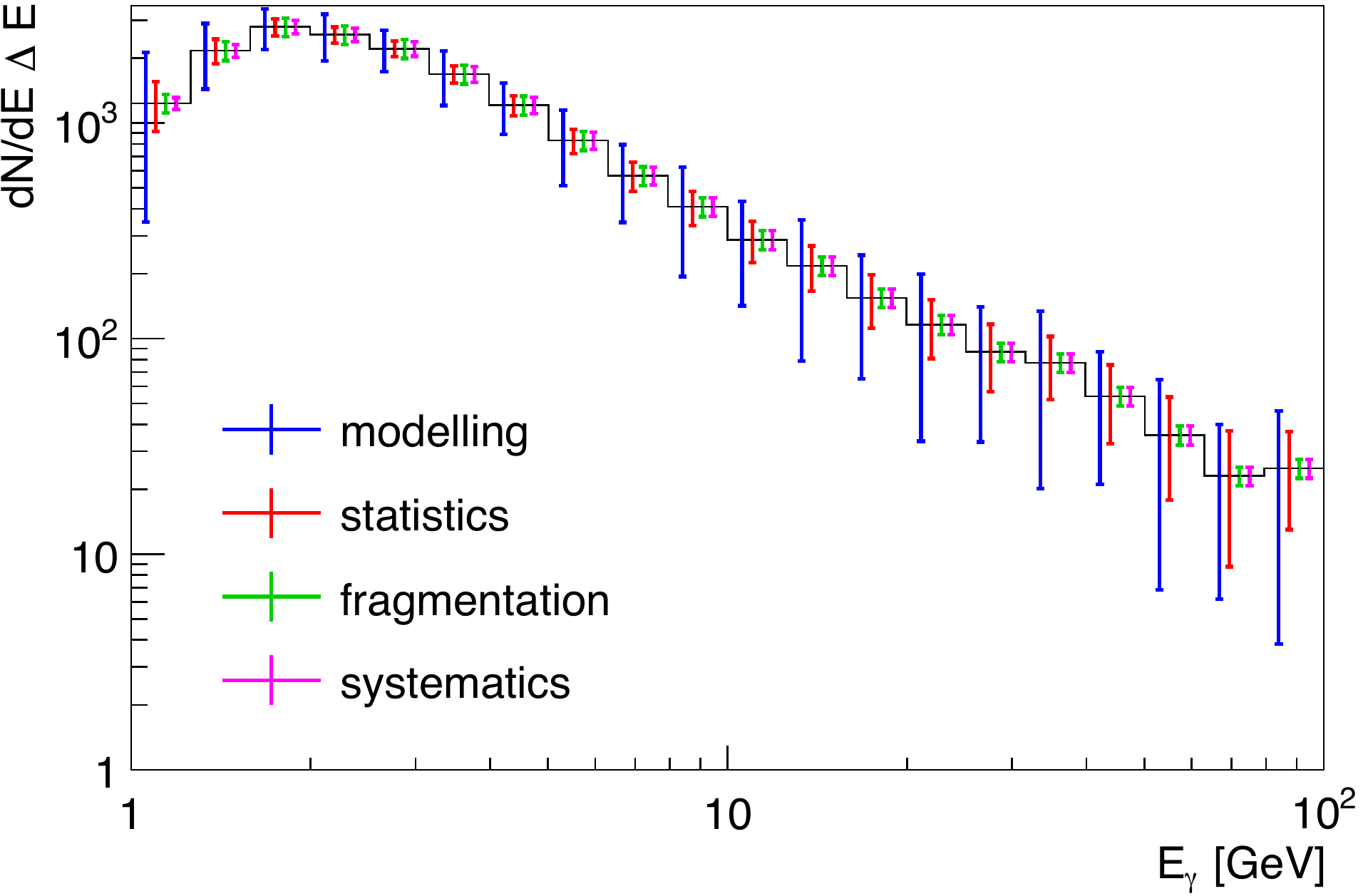}
\end{center}
\vspace*{-7mm}
\caption{Excess photon spectrum of the Fermi galactic center excess.\index{galactic center excess}
  Figure from Ref.~\cite{MSSM_GCE}, including the original data and
  error estimates from Ref.~\cite{fermi}.}
\label{fig:fermi}
\end{figure}

\begin{figure}[b!]
\begin{center}
\includegraphics[width=0.45\textwidth]{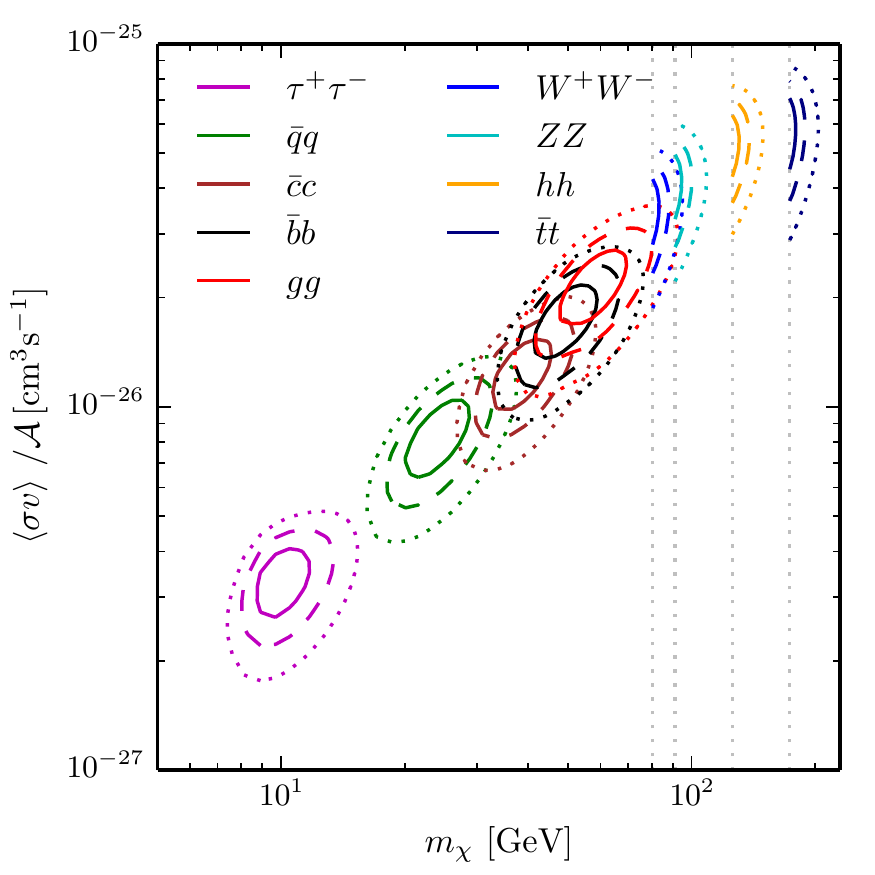}
\end{center}
\vspace*{-7mm}
\caption{Preferred dark matter masses and cross sections for different
  annihilation channels~\cite{Berlin:2014tja}. Figure from
  Ref.\cite{Calore}.}
\label{fig:indirect_channels}
\end{figure}

While at any given time there is usually a sizeable set of
experimental anomalies discussed in the literature, we will focus on
one of them: the \ul{photon excess in the center of our
  galaxy}, observed by Fermi, but discovered in their data by several
non-Fermi groups. The excess is shown in Figure~\ref{fig:fermi} and
covers the wide photon energy range
\begin{align}
E_\gamma = 0.3~...~5~\gev \; ,
\end{align}
and clearly does not form a line. The error bars refer to the
interstellar emission model, statistics, photon fragmentation, and
instrumental systematics.  Note that the statistical uncertainties are
dominated not by the number of signal events, but by the statistical
uncertainty of the subtracted background events. The fact that
uncertainties on photon fragmentation, means photon radiation off
other Standard Model particles are included in the analysis,
indicates, that for an explanation we resort to photon radiation off
dark matter annihilation products, Eq.\eqref{eq:indirect_frag}. This
allows us to link the observed photon spectrum to dark matter
annihilation, where the photon radiation off the final state particles
is known very well from many collider studies. Two aspects of
Figure~\ref{fig:fermi} have to be matched by any explanation. First, the
total photon rate has to correspond to the dark matter annihilation
rate. It turns out that the velocity-averaged annihilation rate has to
be in the same range as the rate required for the observed relic
density,
\begin{align}
\langle \sigma_{\chi \chi} \, v \rangle
= \frac{10^{-8}~...~10^{-9}}{\gev^2} \; ,
\label{eq:indirect_rate}
\end{align}
but with a much lower velocity spectrum now. Second, the energy
spectrum of the photons reflects the mass of the dark matter
annihilation products. Photons radiated off heavier, non-relativistic
states will typically have higher energies. This information is used
to derive the preferred annihilation channels given in
Figure~\ref{fig:indirect_channels}. The official Fermi data confirms
these ranges, but with typically larger error bars. As an example, we
quote the fit information under the assumption of two dark matter
Majorana fermions decaying into a pair of Standard Model
states~\cite{Calore}:

\begin{center}
\begin{tabular}{lrr}
\hline
Channel & $\langle \sigma_{\chi \chi} \, v \rangle$~[fb] & $m_\chi$~[GeV] \\
\hline
$q\bar{q}$     & $275 \pm 45$  & $24 \pm 3$ \\
$b\bar{b}$     & $580 \pm 85$  & $49 \pm 6$ \\
$\tau^+ \tau^-$ & $110 \pm 17 $ & $10 \pm 1$ \\
$W^+ W^-$       & $1172 \pm 160$ & $80 \pm 1$ \\
\hline
\end{tabular}
\end{center}

For each of these annihilation channels the question arises if we can
also generate a sizeable dark matter annihilation rate at the center
of the galaxy today, while also predicting the correct relic density
$\Omega_\chi h^2$.

\subsection{Higgs portal}
\label{sec:indirect_portal}
\index{Higgs!portal}

Similar to our calculation of the relic density, we will first show
what range of annihilation cross sections from the galactic center can
be explained by Higgs portal dark matter. Because the Fermi data
prefers a light dark matter particle we will focus on the two
velocity-weighted cross sections accounting for the observed relic
density and for the galactic center excess around the Higgs pole
$m_S/2 = m_H$. First, we determine \ul{how large an
  annihilation cross section} in the galactic center we can
achieve. The typical cross sections given in
Eq.\eqref{eq:indirect_rate} can be explained by $m_S = 220$~GeV and
$\lambda_3 = 1/10$ as well as a more finely tuned $m_S = m_H/2 =
63$~GeV with $\lambda_3 \approx 10^{-3}$, as shown in
Figure~\ref{fig:portal_mass_coup}.

We can for example assume that the Fermi excess is due to on-shell
Higgs-mediated annihilation, while the observed relic density does not
probe the Higgs pole.  The reason we can separate these two
annihilation signals based on the same Feynman diagram this way is
that the Higgs width is smaller than the typical velocities,
$\Gamma_H/m_H \ll v$.  We start with the general annihilation rate of
a dark matter scalar, Eq.\eqref{eq:portal_annrate1} and express it
including the leading relative velocity dependence from
Eq.\eqref{eq:def_velocity},
\begin{align} 
s = 4 m_S^2 + m_S^2 v^2 = 4 m_S^2 \, \left( 1 + \frac{v^2}{4} \right) \, .
\label{eq:indirect_velocity}
\end{align}
The WIMP velocity at the point of dark matter decoupling in the early
universe we find roughly
\begin{align}
\xdec := \frac{m_S}{\Tdec} 
\eqx{eq:xdec_wimp} 28
\qquad \Leftrightarrow \qquad 
\Tdec \approx \frac{m_S}{28} 
= \frac{m_S}{2} v_\text{ann}^2
\qquad \Leftrightarrow \qquad 
v_\text{ann}^2 = \frac{1}{14} \; .
\label{eq:indirect_velocity2}
\end{align}
Today the Universe is colder, and the WIMP velocity is strongly
red-shifted.  Typical galactic velocities today are 
\begin{align}
v_0 \approx 2.3 \cdot 10^{5} \, \frac{\text{m}}{\text{s}} \; \frac{1}{c}
\approx \frac{1}{1300} \ll v_\text{ann} \; ,
\end{align}
This hierarchy in typical velocities between the era of thermal dark
matter production and annihilation and dark matter annihilation today
is what will drive our arguments below.\bigskip

Only assuming $m_b \ll s$ the general form of the scalar dark matter
annihilation rate is
\begin{align}
 \sigma v \Bigg|_{SS \to b\bar{b}}
&= \frac{N_c}{2 \pi} \;
   \lambda_3^2 m_b^2 \;
   \frac{1}{m_S \, \sqrt{s}} \; 
   \frac{s}{\left( s - m_H^2 \right)^2 + m_H^2 \Gamma_H^2}  \notag \\
&= \left( 1 + \dfrac{v^2}{8} + \ope (v^4) \right) \;
   \frac{N_c}{2 \pi} \;
   \lambda_3^2 m_b^2 \;
   \frac{1}{2 m_S^2} \; 
   \frac{4 m_S^2}{\left( 4 m_S^2 - m_H^2 + m_S^2 v^2 \right)^2 + m_H^2 \Gamma_H^2}  \notag \\
&= \left( 1 + \dfrac{v^2}{8} \right) \;
   \frac{N_c}{2 \pi} \;
   \lambda_3^2 m_b^2 \;
   \frac{1}{2 m_S^2} \; 
   \frac{4 m_S^2}{\left( 4 m_S^2 - m_H^2 \right)^2 + 2 ( 4 m_S^2 - m_H^2 ) m_S^2 v^2 + m_H^2 \Gamma_H^2}  + \ope (v^4) \; .
\label{eq:indirect_intermediate}
\end{align}
The typical velocity of the dark matter states only gives a small
correction for scalar, $s$-wave annihilation. It includes two aspects:
first, an over-all reduction of the annihilation cross section for
finite velocity $v > 0$, and second a combined cutoff of the
Breit-Wigner propagator,
\begin{align}
\max \left[ 2 (4 m_S^2 - m_H^2 ) m_S^2 v^2, m_H^2 \Gamma_H^2 \right]
= m_S^4 \; \max \left[ 8 v^2 \left( 1 - \frac{m_H^2}{4 m_S^2} \right), 16 \cdot 10^{-10} \right] \; .
\label{eq:indirect_scales}
\end{align}
Close to but not on the on-shell pole $m_H = m_S/2$ the modification
of the Breit-Wigner propagator can be large even for small velocities,
while the rate reduction can clearly not account for a large boost
factor describing the galactic center excess. We therefore ignore the
correction factor $(1 + v^2/8)$ when averaging the velocity-weighted
cross section over the velocity spectrum. If, for no good reason, we
assume a narrow Gaussian velocity distribution centered around $\bar{v}$ we
can approximate Eq.\eqref{eq:indirect_intermediate} as~\cite{hitoshi}
\begin{align}
\boxed{
\langle \sigma v \rangle \Bigg|_{SS \to b\bar{b}}
\approx
   \frac{N_c}{2 \pi} \;
   \lambda_3^2 m_b^2 \;
   \frac{1}{2 m_S^2} \; 
   \frac{4 m_S^2}{\left( 4 m_S^2 - m_H^2 + \xi \, m_S^2 \bar{v}^2 \right)^2 + 4 m_S^2 \Gamma_H^2} 
}
\qquad \text{with a fitted} \quad \xi \approx 2 \sqrt{2} \; .
\label{eq:hitoshi}
\end{align}
This modified on-shell pole condition shifts the required dark matter
mass slightly below the Higgs mass $2 m_S \lesssim m_H$. The size of
this shift depends on the slowly dropping velocity, first at the time
of dark matter decoupling, $\bar{v} \equiv v_\text{ann}$, and then
today, $\bar{v} \equiv v_0 \ll v_\text{ann}$. This means that during
the evolution of the Universe the Breit-Wigner propagator in
Eq.\eqref{eq:hitoshi} is always probed above its pole, probing the
actual pole only today.

We first compute the Breit--Wigner suppression of $\langle v \sigma
\rangle$ in the early universe, starting with today's on-shell
condition responsible for the galactic center excess,
\begin{align}
m_S \really \frac{m_H}{2 \sqrt{1 + \dfrac{v_0^2}{\sqrt{2}}}}
\approx \frac{m_H}{2} 
\qquad \Rightarrow \qquad 
4 m_S^2 - m_H^2 + \xi \, m_S^2 v_\text{ann}^2 
&= 4 m_S^2 \left( 1 + \frac{v_\text{ann}^2}{\sqrt{2}} \right) - m_H^2 \notag \\[-4mm]
&= 4 m_S^2 \left( \dfrac{v_\text{ann}^2}{\sqrt{2}} - \dfrac{v_0^2}{\sqrt{2}} \right)
\stackrel{v_\text{ann} \gg v_0}{\approx} 
\frac{m_S^2}{5} \; .
\label{eq:indirect_pole1}
\end{align}
This means that the dark matter particle has a mass just slightly
below the Higgs pole. Using Eq.\eqref{eq:hitoshi} the ratio of
the two annihilation rates, for all other parameters constant, then
becomes
\begin{align}
\frac{\langle  \sigma_0\,v \rangle}{\langle  \sigma_\text{ann} v \rangle}
= \frac{8 m_S^4 v_\text{ann}^4}{4 m_S^2 \Gamma_H^2} 
= \frac{2 v_\text{ann}^4}{16 \cdot 10^{-10}} 
= \frac{1}{8} \, \; \frac{1}{14^2} \; 10^{10} 
\gtrsim 10^6 \; .
\label{eq:indirect_max}
\end{align}
This is the maximum enhancement we can generate to explain Fermi's
galactic center excess.  The corresponding Higgs coupling $\lambda_3$
is given in Figure~\ref{fig:portal_mass_coup}.\bigskip

We can turn the question around and compute the \ul{smallest
  annihilation cross section} in the galactic center consistent with the observed relic abundance in the Higgs portal model. For this purpose we assume that
unlike in Eq.\eqref{eq:indirect_pole1} the pole condition is fulfilled
in the early universe, leading to a Breit-Wigner suppression today of
\begin{align}
m_S \really \frac{m_H}{2 \sqrt{1 + \dfrac{v_\text{ann}^2}{\sqrt{2}}}}
\qquad \Rightarrow \qquad 
4 m_S^2 - m_H^2 + \xi \, m_S^2 v_0^2 
&= 4 m_S^2 \left( 1 + \frac{v_0^2}{\sqrt{2}} \right) - m_H^2 
\stackrel{v_\text{ann} \gg v_0}{\approx} 
- \frac{m_S^2}{5}  \; .
\label{eq:indirect_pole2}
\end{align}
This gives us a ratio of the two velocity-mediated annihilation rates 
\begin{align}
\frac{\langle  \sigma_0\,v \rangle}{\langle  \sigma_\text{ann}\, v \rangle}
= \frac{4 m_S^2 \Gamma_H^2}{8 m_S^4 v_\text{ann}^4}
\stackrel{\text{Eq.\eqref{eq:indirect_max}}}{\lesssim} 10^{-6}
\qqquad \text{for} \qqquad 
m_S 
\approx 
\frac{m_H}{2}  \left( 1 - \dfrac{v_\text{ann}^2}{2 \sqrt{2}} \right)
\eqx{eq:indirect_velocity2} 62.91~\gev \; .
\end{align}
The dark matter particle now has a mass further below the pole.  This
means that we can interpolate between the two extreme ratios of
velocity-averaged annihilation rates using a very small range of $m_S
< m_H/2$. If we are willing to tune this mass relation we can
accommodate essentially any dark matter annihilation rate today with
the Higgs portal model, close to on-shell Higgs pole annihilation. The
key to this result is that following Eq.\eqref{eq:indirect_scales} the
Higgs width-to-mass ratio is small compared to $v_\text{ann}$, so we
can decide to assign the on-shell condition to each of the two
relevant annihilation processes. In between, none of the two processes
will proceed through the on-shell Higgs propagator, which indeed gives
$\langle  \sigma_0 v\rangle \approx \langle 
\sigma_\text{ann} v\rangle$.  The corresponding coupling $\lambda_3$ we
can read off Figure~\ref{fig:portal_mass_coup}. Through this argument
it becomes clear that the success of the Higgs portal model rests on
the the wide choice of scalings of the dark matter annihilation rate,
as shown in Eq.\eqref{eq:annrate_scaling}.

\subsection{Supersymmetric neutralinos}
\label{sec:indirect_mssm}

An explanation of the galactic center excess has to be based on the
neutralino mass matrix given in Eq.\eqref{eq:neutmass}, defining a
dark matter Majorana fermion as a mixture of the bino singlet, the
wino triplet, and two higgsino doublets.  Some of its relevant
couplings are given in
Eq.\eqref{eq:coups_neutralinos}. Correspondingly, some annihilation
processes leading to the observed relic density and underlying our
interpretation of the Fermi galactic center excess are illustrated in
Figure~\ref{fig:neutralinos_relic_feyn}. One practical advantage of
the MSSM is that it offers many neutralino parameter regions to play
with.  We know that pure wino or higgsino dark matter particles
reproducing the observed relic density are much heavier than the Fermi
data suggests. Instead of these pure states we will rely on mixed
states.  A major obstacle of all MSSM interpretations are the mass
ranges shown in Figure~\ref{fig:indirect_channels}, indicating a clear
preference of the galactic center excess for neutralino masses
$m_{\nne} \lesssim 60$~GeV. This does not correspond to the typical MSSM
parameter ranges giving us the correct relic density. This means that
in an MSSM analysis of the galactic center excess the proper error
estimate for the photon spectrum is essential.\bigskip

\begin{figure}[b!]
\centering
\includegraphics[width=0.495\textwidth]{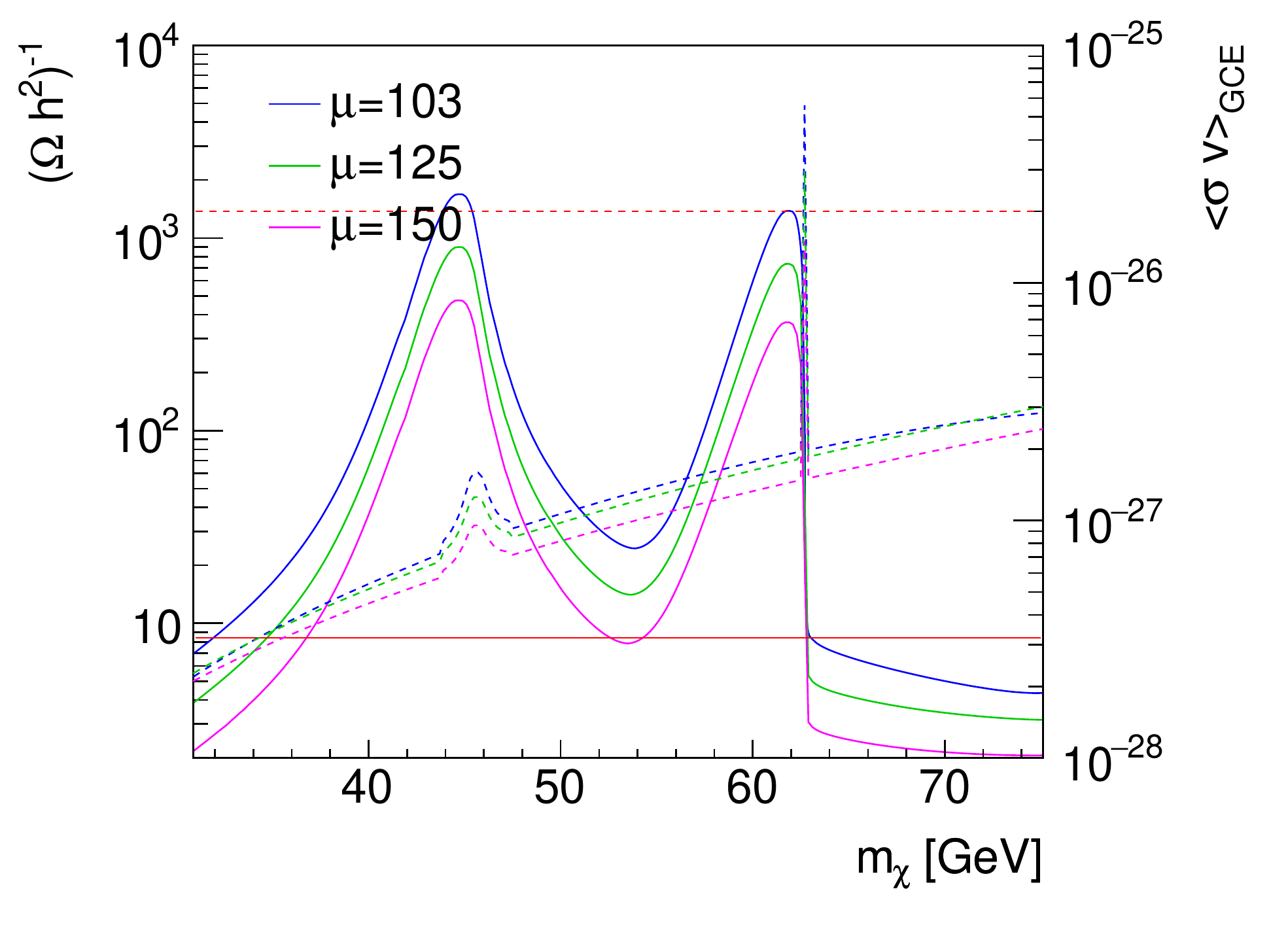}
\vspace*{-3mm}
\caption{Inverse relic density (solid, left axis) and annihilation
  rate in the galactic center (dashed, right axis) for an MSSM
  parameter point where the annihilation is dominated by $\nne \nne
  \to b\bar{b}$. Figure from Ref.~\cite{MSSM_GCE}.}
\label{fig:bb}
\end{figure}

We start our discussion with the finely tuned annihilation through a
SM-like light Higgs or through a $Z$-boson, \ie $\nne \nne \to h^*,Z^*
\to b\bar{b}$. The properties of this channel are very similar to
those of the Higgs portal. On the left $y$-axes of Figure~\ref{fig:bb}
we show the (inverse) relic density for a bino-higgsino LSP, both for
a wide range of neutralino masses and zoomed into the Higgs pole
region. \index{neutralino!bino} We decouple the wino to $M_2 = 700$~GeV and vary $M_1$ to give
the correct relic density for three fixed, small higgsino mass
values. We see that the $b\bar{b}$ annihilation channel only predicts
the correct relic density in the two pole regions of the MSSM
parameter space, with $m_{\nne} = 46$~GeV and $m_{\nne} = 63$~GeV.
The width of both peaks is given by the momentum smearing through
velocity spectrum rather than physical Higgs width and $Z$-width.  The
enhancement of the two peaks over the continuum is comparable, with
the $Z$-funnel coupled to the velocity-suppressed axial-vector current
and the Higgs funnel suppressed by the small bottom Yukawa coupling.

On the right $y$-axis of Figure~\ref{fig:bb}, accompanied by dashed
curves, we show the annihilation rate in the galactic center.  The
rough range needed to explain the Fermi excess is indicated by the
horizontal line. As discussed for the Higgs portal, the difference to
the relic density is that the velocities are much smaller, so the widths
of the peaks are now given by the physical widths of the two
mediators. The scalar Higgs resonance now leads to a much higher peak
than the velocity-suppressed axial-vector coupling to the
$Z$-mediator. This implies that continuum annihilation as well as
$Z$-pole annihilation would not explain the galactic center excess,
while the Higgs pole region could.

This is why in the right panel of Figure~\ref{fig:bb} we zoom into the
Higgs peak regime. A valid explanation of the galactic center excess
requires the solid relic density curves to cross the solid horizontal
line and at the same time the dashed galactic center excess lines
to cross the dashed horizontal line. We see that there exist finely
tuned regions around the Higgs pole which allow for an explanation of
the galactic center excess via a thermal relic through the process
$\nne \nne \to b\bar{b}$. The physics of this channel is very similar
to scalar Higgs portal dark matter.\bigskip

\begin{figure}[b!]
\includegraphics[width=0.495\textwidth]{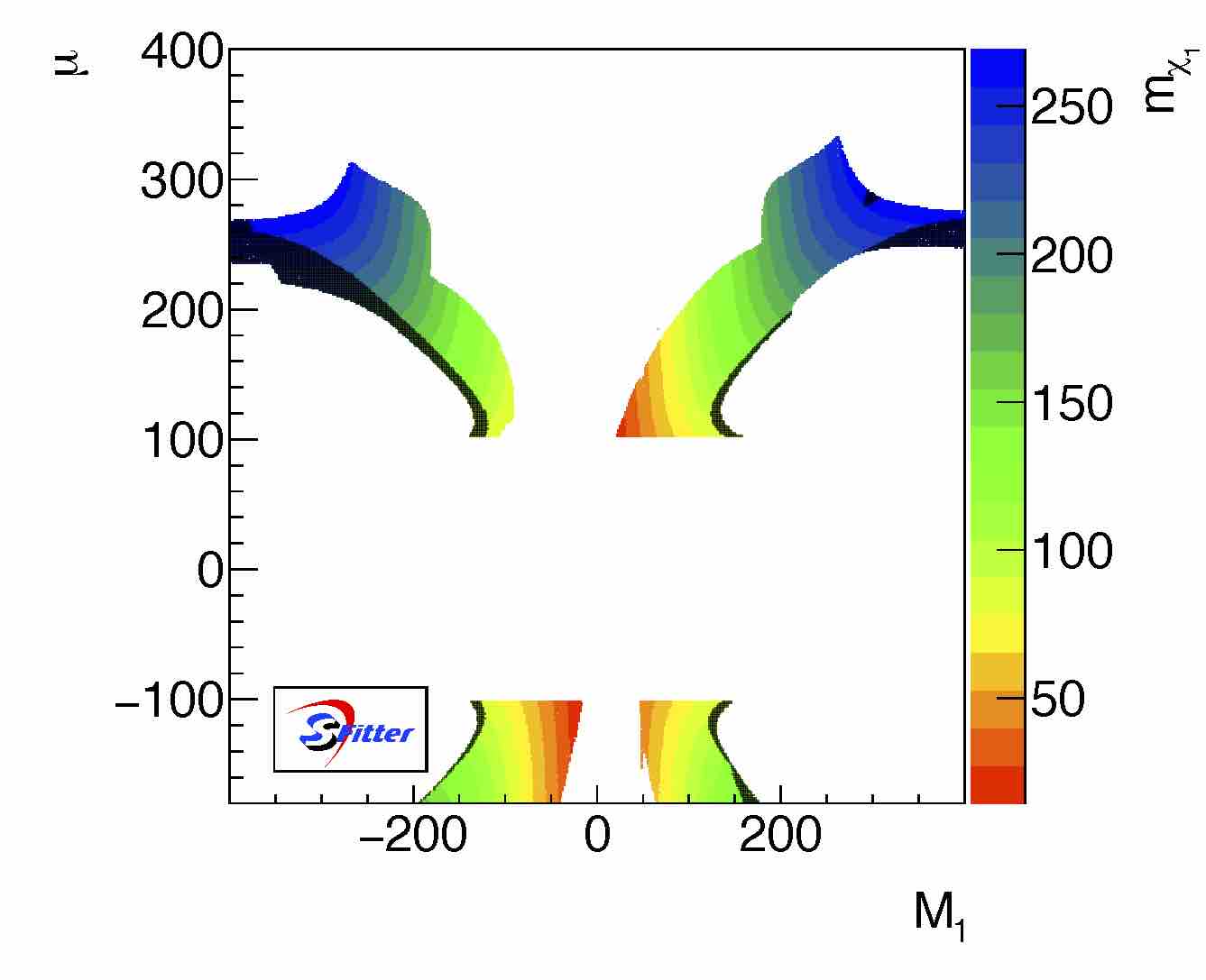}
\hspace*{0.02\textwidth}
\includegraphics[width=0.495\textwidth]{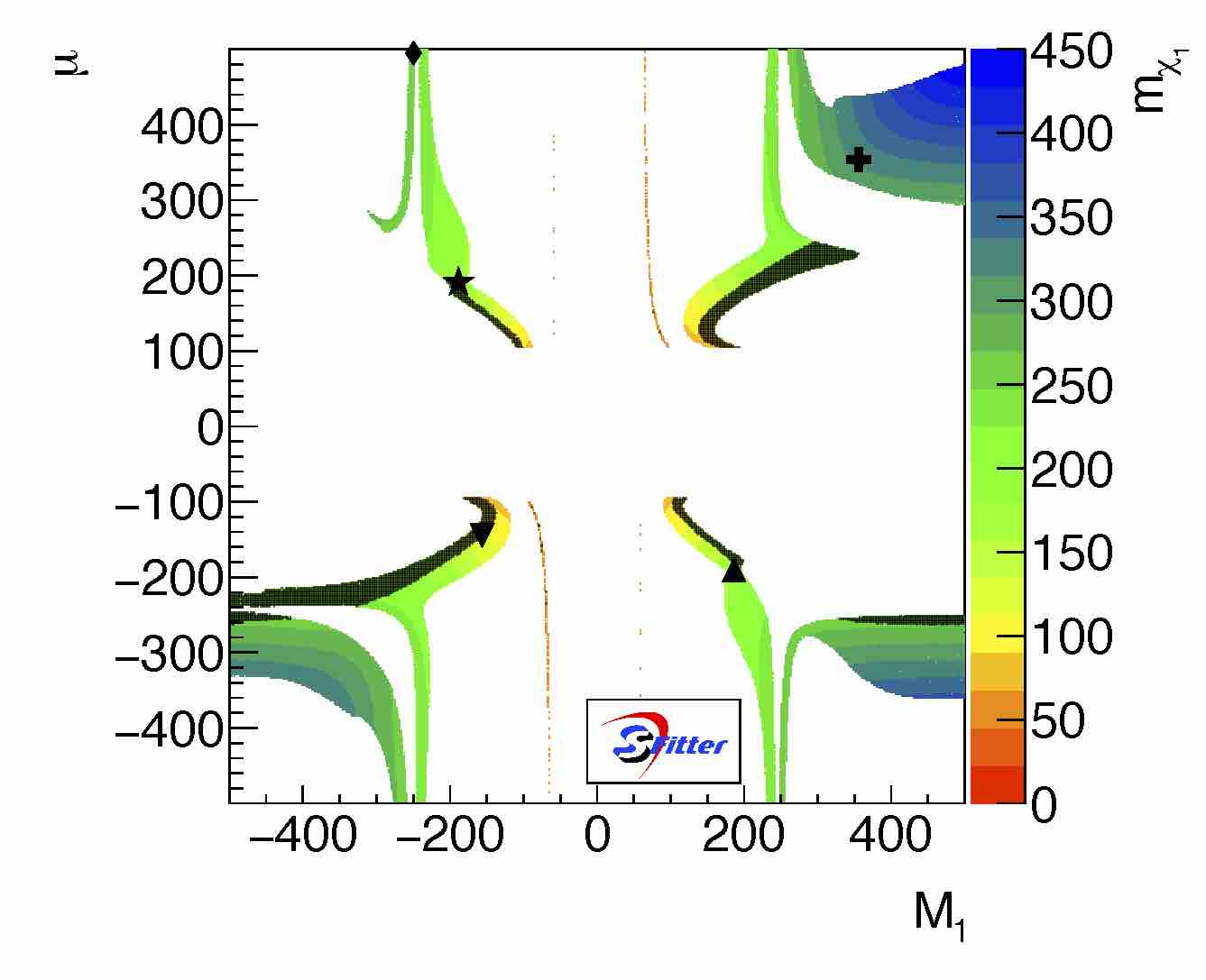}
\vspace*{-3mm}
\caption{Left: lightest neutralino mass based on the Fermi photon
  where $\nne \nne \to WW$ is a dominant annihilation channel. Right:
  lightest neutralino mass based on the Fermi photon spectrum for $m_A
  = 500$~GeV, where we also observe $\nne \nne \to t \bar{t}$.  The
  five symbols indicate local best-fitting parameter points. The
  black shaded regions are excluded by the Fermi limits from dwarf spheroidal
  galaxies. }
\label{fig:mum1_ww}
\end{figure}

For slightly larger neutralino masses, the dominant annihilation
becomes $\nne \nne \to WW$, mediated by a light $t$-channel chargino
combined with chargino-neutralino co-annihilation for the relic
density.  Equation~\eqref{eq:coups_neutralinos} indicates that in this
parameter region the lightest neutralino requires either a wino
content or a higgsino content. In the left panel of
Figure~\ref{fig:mum1_ww} we show the bino--higgsino mass plane
indicating the preferred regions from the galactic center excess. The
lightest neutralino mass varies from $m_{\nne} \approx 50$~GeV to more
than $250$~GeV. Again, we decouple the wino to $M_2 = 700$~GeV, so the
LSP is a mixture of higgsino, coupling to electroweak bosons, and
bino. For this slice in parameter space an increase in $|\mu|$
compensates any increase in $M_1$, balancing the bino and higgsino
contents.  The MSSM parameter regions which allow for efficient dark
matter annihilation into gauge bosons are strongly correlated in $M_1$
and $\mu$, but not as tuned as the light Higgs funnel region with its
underlying pole condition. Around $M_1 = |\mu| = 200$~GeV a change in
shape occurs. It is caused by the on-set of neutralino annihilation to
top pairs, in spite of a heavy Higgs mass scale of 1~TeV.\bigskip

To trigger a large annihilation rate for $\nne \nne \to t \bar{t}$ we
lower the heavy pseudoscalar Higgs mass to $m_A = 500$~GeV. In the
right panel of Figure~\ref{fig:mum1_ww} we show the preferred
parameter range again in the bino-higgsino mass plane and for heavy
winos, $M_2 = 700$~GeV.  As expected, for $m_{\nne} > 175$~GeV the
annihilation into top pairs follows the $WW$ annihilation region in
the mass plane.  The main difference between the $WW$ and $t\bar{t}$
channels is the smaller $M_1$ values around $|\mu| = 200$~GeV. The
reason is that an increased bino fraction compensates for the much
larger top Yukawa coupling. The allowed LSP mass range extends to
$m_{\nne} \gtrsim 200$~GeV.

The only distinctive feature for $m_A = 500$~GeV in the $M_1$ vs $\mu$
plane is the set of peaks around $M_1 \approx 300$~GeV. Here the
lightest neutralino mass is around $250$~GeV, just missing the $A$-pole
condition. Because on the pole dark matter annihilation through a $2
\to 1$ process becomes too efficient, the underlying coupling is
reduced by a smaller higgsino fraction of the LSP. The large-$|M_1|$
regime does not appear in the upper left corner of
Figure~\ref{fig:mum1_ww} because at tree level this parameter region
features $m_{\cpe} < m_{\nne}$ and we have to include loop corrections
to revive it.\bigskip

In principle, for $m_{\nne} > 126$~GeV we should also observe
neutralino annihilation into a pair of SM-like Higgs bosons. However,
the $t$-channel neutralino diagram which describes this process will
typically be overwhelmed by the annihilation to weak bosons with the
same $t$-channel mediator, shown in
Figure~\ref{fig:neutralinos_relic_feyn}. From the annihilation into
top pairs we know that $s$-channel mediators with $m_{A,H} \approx 2
m_h$ are in principle available, and depending on the MSSM parameter
point the heavy scalar Higgs can have a sizeable branching ratio into
two SM-like Higgses.  For comparably large velocities in the early
universe both $s$-channel mediators indeed work fine to predict the
observed relic density.  For the smaller velocities associated with
the galactic center excess the CP-odd mediator $A$ completely
dominates, while the CP-even $H$ is strongly velocity-suppressed.  On
the other hand, only the latter couples to two light Higgs bosons, so
an annihilation into Higgs pairs responsible for the galactic center
excess is difficult to realize in the MSSM.\bigskip

Altogether we see that the annihilation channels 
\begin{align}
\nne \nne \to b\bar{b}, WW, t\bar{t} 
\qquad \text{with} \quad m_{\nne} = 63~...~250~\gev
\end{align}
can explain the Fermi galactic center excess and the observed relic
density in the MSSM. Because none of them correspond to the central
values of a combined fit to the galactic center excess, it is crucial
that we take into account all sources of (sizeable) uncertainties. An
additional issue which we will only come to in
Section~\ref{sec:direct} is that direct detection constraints in
addition to requiring the correct relic density and the correct
galactic center annihilation rate is a serious challenge to the MSSM
explanations.

\subsection{Next-to-minimal neutralino sector}
\label{sec:indirect_nmssm}
\index{NMSSM}
An obvious way out of the MSSM limitations is to postulate an addition
particle in the $s$-channel of the neutralino annihilation process and
a new, lighter dark matter fermion. This leads us to the
next-to-minimal supersymmetric extension, the NMSSM. It introduces an
additional singlet under all Standard Model gauge transformations, with
its singlino partner. The singlet state forms a second scalar $H_1$
and a second pseudo-scalar $A_1$, which will appear in the dark matter
annihilation process. As singlets they will only couple to gauge
bosons through mixing with the Higgs fields, which guarantees that
they are hardly constrained by many searches.  The singlino will add a
fifth Majorana state to the neutralino mass matrix in
Eq.\eqref{eq:neutmass},
\begin{align}
\mat =
\begin{pmatrix}
M_1 & 0 & -m_Z \cos \beta s_w & \phantom{-}m_Z \sin \beta s_w & 0 \\
0 & M_2 & \phantom{-}m_Z \cos \beta c_w & -m_Z \sin \beta c_w & 0 \\ 
-m_Z \cos \beta s_w & \phantom{-}m_Z \cos \beta c_w & \phantom{-}0 & -\mu & -m_Z \sin \beta \tilde{\lambda} \\
\phantom{-}m_Z \sin \beta s_w & -m_Z \sin \beta c_w & -\mu & \phantom{-}0 & -m_Z \cos \beta \tilde{\lambda} \\
0 & 0 & -m_Z \sin \beta \tilde{\lambda} & -m_Z \cos \beta \tilde{\lambda} & 2 \tilde{\kappa} \mu
\end{pmatrix}\;.
\label{eq:neutmass_nmssm}
\end{align}
The singlet/singlino sector \index{neutralino!singlino}can be described by two parameters, the
mass parameter $\tilde \kappa$ and the coupling for example to the
other neutralinos $\tilde \lambda$~\cite{anja}.  First, we need to
include the singlino in our description of the neutralino
sector. While the wino and the two higgsinos form a triplet or two
doublets under $SU(2)_L$, the singlino just adds a second singlet
under $SU(2)_L$. The only difference to the bino is that the singlino
is also a singlet under $U(1)_Y$, which makes no difference unless we
consider co-annihilation driven by hypercharge interaction. A singlet
neutralino will therefore interact and annihilate to the observed
relic density through its mixing with the wino or with the higgsinos,
just like the usual bino.

What is crucial for the explanation of the galactic center excess is
the $s$-channel dark matter annihilation through the new pseudoscalar,
\begin{align}
\nne \nne \to A_1 \to b\bar{b}
\qquad \text{with} \quad 
m_{\nne} &= \frac{m_{A_1}}{2} \approx 50~\gev \notag \\
g_{A_1^0 \nne \nne} &= \sqrt{2} \, g \tilde{\lambda} \; \left( N_{13} N_{14} - \tilde{\kappa} N_{15}^2 \right) \; .
\end{align}
We can search for these additional singlet and singlino states at
colliders. One interesting aspect is the link between the neutralino
and the Higgs sector, which can be probed by looking for anomalous Higgs
decays, for example into a pair of dark matter particles. Because an
explanation of the galactic center excess requires the singlet and the
singlino to be light and to mix with their MSSM counterparts, the
resulting invisible branching ratio of the Standard-Model-like Higgs
boson can be large.

\subsection{Simplified models and vector mediator}
\label{sec:indirect_simp}
\index{vector portal}

The discussion of the dark matter annihilation processes responsible
for today's dark matter density as well as a possible galactic center
excess nicely illustrates the limitations of the effective theory
approach introduced in Section~\ref{sec:models_eft}. To achieve the
currently observed density with light WIMPs we have to rely on an
efficient annihilation mechanism, which can be most clearly seen in
the MSSM. For example, we invoke $s$-channel annihilation or
co-annihilation, both of which are not well captured by an effective
theory description with a light dark matter state and a heavy,
non-propagating mediator. In the effective theory language of
Section~\ref{sec:models_eft} this means the mediators are not light
compared to the dark matter agent,
\begin{align}
m_\chi \lesssim \mmed \; .
\end{align}

In addition, the MSSM and the NMSSM calculations illustrate how one
full model extending the Standard Model towards large energy scales
can offer several distinct explanations, only loosely linked to each
other.  In this situation we can collect all the necessary degrees of
freedom in our model, but ignore additional states for example
predicted by an underlying supersymmetry of the Lagrangian. This
approach is called \ul{simplified models}. \index{simplified model}It typically
describes the dark matter sector, including co-annihilating particles,
and a mediator coupling the dark matter sector to the Standard
Model. In that language we have come across a sizeable set of
simplified models in our explanation of the Fermi galactic center
excess:
\begin{itemize}
\item[--] dark singlet scalar with SM Higgs mediator (Higgs
  portal, $SS \to b\bar{b}$);
\item[--] dark fermion with SM $Z$ mediator (MSSM, $\nne \nne \to f\bar{f}$, not good for galactic center excess);
\item[--] dark fermion with SM Higgs mediator (MSSM, $\nne \nne \to b\bar{b}$);
\item[--] dark fermion with $t$-channel fermion mediator (MSSM, $\nne \nne \to WW$);
\item[--] dark fermion with heavy $s$-channel pseudo-scalar mediator (MSSM, $\nne \nne \to t\bar{t}$);
\item[--] dark fermion with light $s$-channel pseudo-scalar mediator (NMSSM, $\nne \nne \to b\bar{b}$). 
\end{itemize}
In addition, we encountered a set of models in our discussion of the
relic density in the MSSM in Section~\ref{sec:models_mssm}:
\begin{itemize}
\item[--] dark fermion with fermionic co-annihilation partner and charged
  $s$-channel mediator (MSSM, $\nne \cme \to \bar t b$);
\item[--] dark fermion with fermionic co-annihilation partner and SM $W$-mediator (MSSM, $\nne \cme \to \bar u d$);
\item[--] dark fermion with scalar $t$-channel mediator (MSSM, $\nne \nne \to \tau \tau$); 
\item[--] dark fermion with scalar co-annihilation partner (MSSM, $\nne \tilde{\tau} \to \tau^*$)
\end{itemize}
Strictly speaking, all the MSSM scenarios require a Majorana fermion
as the dark matter candidate, but we can replace it with a Dirac
neutralino in an extended supersymmetric setup.\bigskip

One mediator
which is obviously missing in the above list is a new, heavy vector $V$ or
axial-vector. Heavy gauge bosons are ubiquitous in models for physics
beyond the Standard Model, and the only question is how we would link
or couple them to a dark matter candidate. In principle, there exist 
different mass regimes in the $m_\chi - m_V$ mass plane,
\begin{alignat}{5}
m_V & > 2 m_\chi  \qqqquad && \text{possible effective theory} \notag \\
m_V &\approx 2 m_\chi && \text{on-shell, simplified model} \notag \\
m_V & < 2 m_\chi && \text{light mediator, simplified model}\; .
\label{eq:s_regimes}
\end{alignat}
To allow for a global analysis including direct detection as well as
LHC searches, we couple the vector mediator to a dark matter fermion
$\chi$ and the light up-quarks,
\begin{align}
\lag \supset g_u\ \bar{u}\ \gamma^\mu V_\mu \ u
            + g_\chi\ \bar{\chi}\ \gamma^\mu V_\mu \ \chi \; .
\label{eq:s_model}
\end{align}
The typical mediator width for $m_\chi \ll m_V$ is
\begin{align}
\frac{\Gamma_V}{m_V} \lesssim 0.4~...~10\% 
\qquad \text{for} \quad 
g_u = g_\chi=0.2~...~1 \; .
\end{align}
\bigskip

\begin{figure}[b!]
\begin{center}
\includegraphics[width=0.40\textwidth]{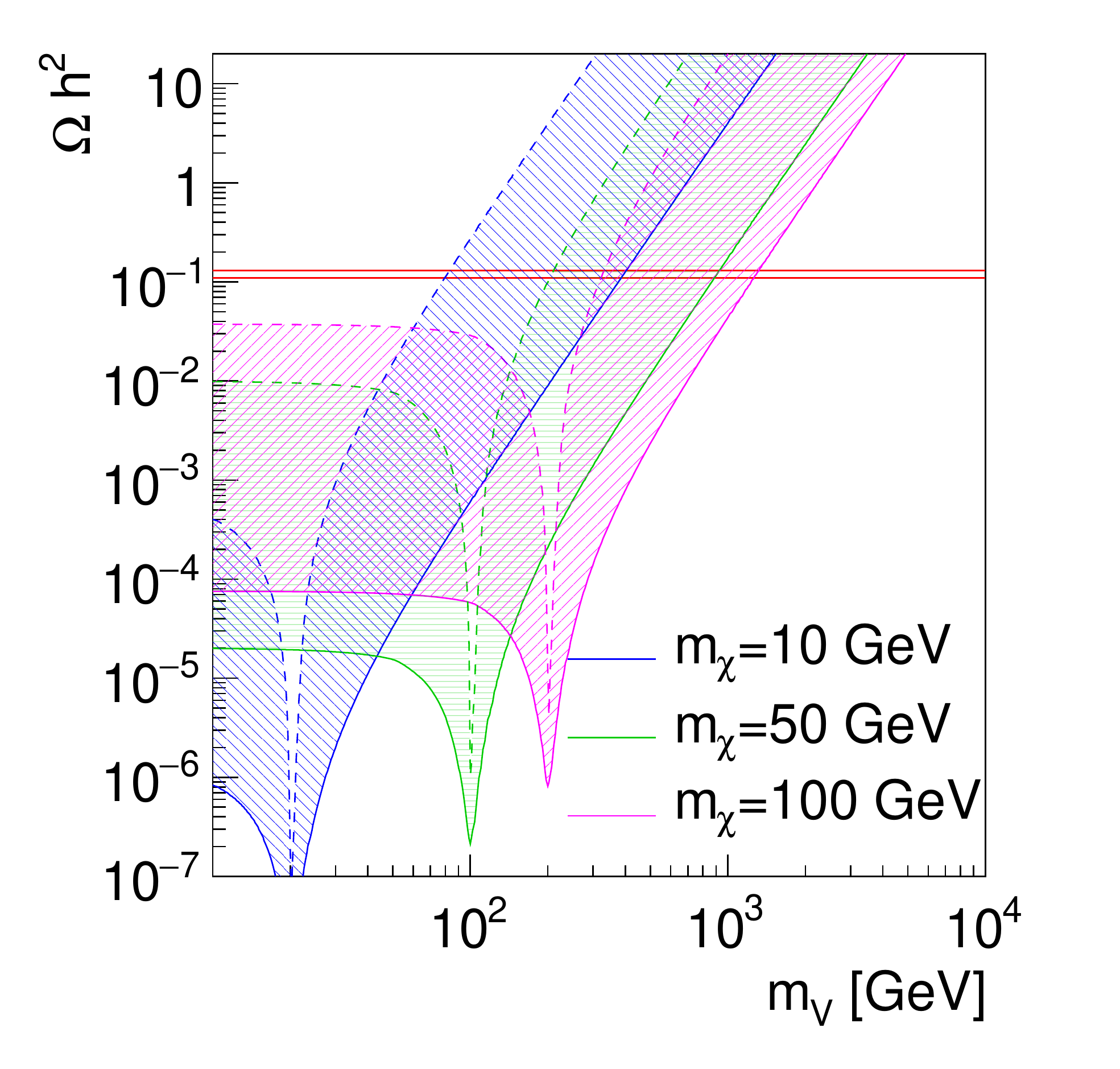}\hspace*{0.05\textwidth}
\includegraphics[width=0.40\textwidth]{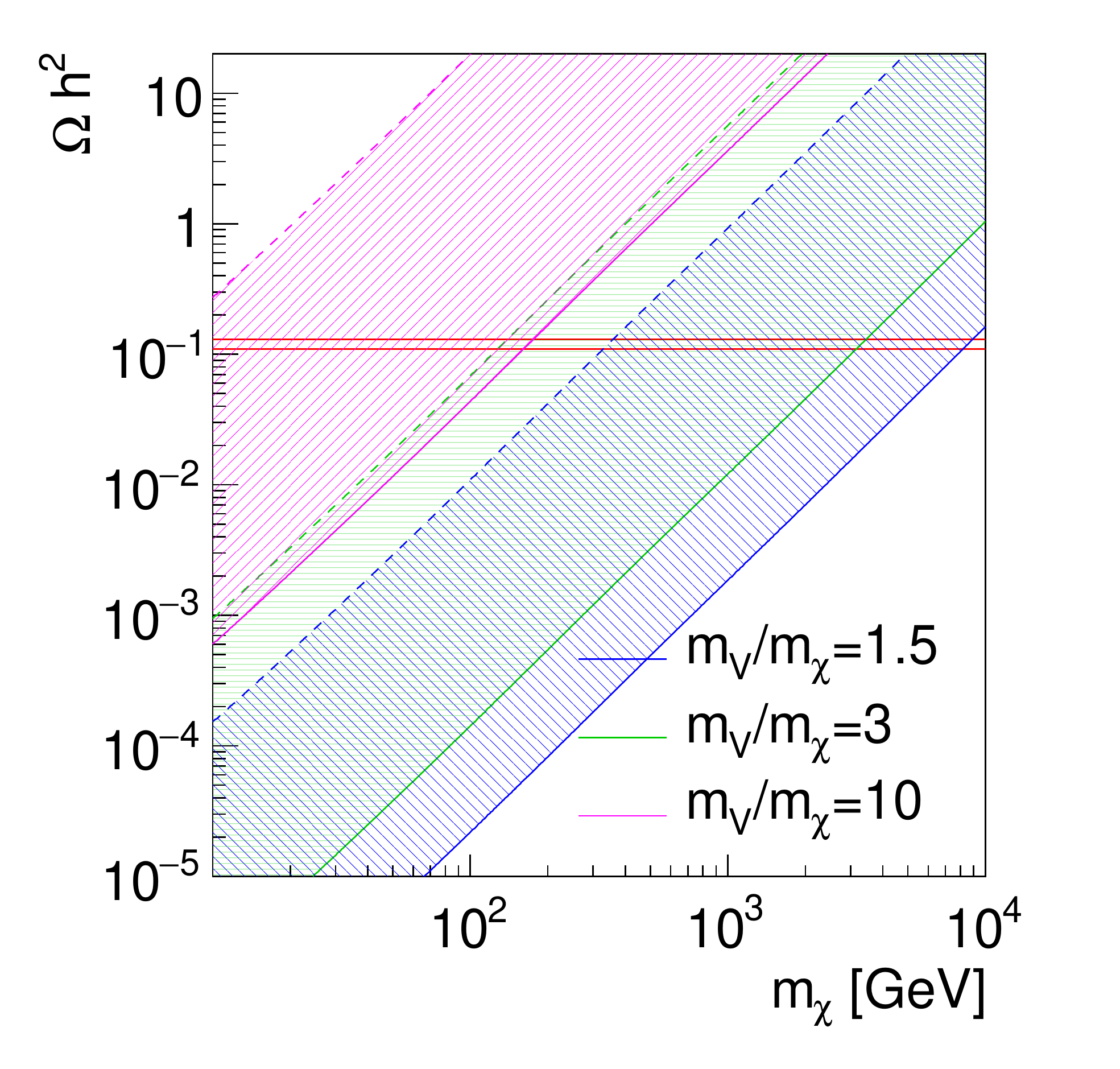}
\end{center}
\vspace*{-7mm}
\caption{Relic density for the simplified vector mediator model of
  Eq.\eqref{eq:s_model} as a function of the mediator mass for
  constant dark matter mass (left) and as a function of the dark
  matter mass for a constant ratio of mediator to dark matter
  mass (right). Over the shaded bands we vary the couplings
  $g_u=g_\chi=0.2~...~1$. Figure from Ref.~\cite{dm_eff}.}
\label{fig:relic_schannel}
\end{figure}

Based on the annihilation process 
\begin{align}
\chi \chi \to V^* \to u \bar{u} 
\end{align}
we can compute the predicted relic density or the indirect detection
prospects.  While the $\chi-\chi-V$ interaction also induces a
$t$-channel process $\chi \chi \to V^* V^*$, its contribution to the
total dark matter annihilation rate is always strongly suppressed by
its 4-body phase space. The on-shell annihilation channel
\begin{align}
\chi \chi \to VV
\end{align}
becomes important for $m_V < m_\chi$, with a subsequent decay of the
mediator for example to two Standard Model fermions. In that case the
dark matter annihilation rate becomes independent of the mediator
coupling to the Standard Model, giving much more freedom to avoid
experimental constraints.

In Figure~\ref{fig:relic_schannel} we observe that for a light
mediator the predicted relic density is smaller than the observed
values, implying that the annihilation rate is large.  In the left
panel we see the three kinematic regimes defined in
Eq.\eqref{eq:s_regimes}. First, for small mediator masses the $2 \to
2$ annihilation process is $\chi \chi \to u\bar{u}$. The
dependence on the light mediator mass is small because the mediator is
always off-shell and the position of its pole is far away from the
available energy of the incoming dark matter particles.  Around the
pole condition $2 m_\chi \approx m_V \pm \Gamma_V$ the model predicts
the correct relic density with very small couplings.
For heavy mediators the $2 \to 2$ annihilation process rapidly
decouples with large mediator masses, as follows for example from
Eq.\eqref{eq:wimp_ann_approx}. In the right panel of
Figure~\ref{fig:relic_schannel} we assume a constant mass ratio
$m_V/m_\chi \gtrsim 1$, finding that our simplified vector model has
no problems predicting the correct relic density over a wide range of
model parameters.\bigskip

One issue we can illustrate with this non-MSSM simplified model \index{simplified model}
is a
strong dependence of our predictions on the assumed model
features. The Lagrangian of Eq.\eqref{eq:s_model} postulates a
coupling to up-quarks, entirely driven by our goal to link dark matter
annihilation with direct detection and LHC observables. From a pure
annihilation perspective we can also define the mediator coupling to
the Standard Model through muons, without changing any of the results
shown in Figure~\ref{fig:relic_schannel}.  Coupling to many
SM fermions simultaneously, as we expect from an extra gauge group,
will increase the predicted annihilation rate easily by an order of
magnitude.  Moreover, it is not clear how the new gauge group is
related to the $U(1)_Y \times SU(2)_L$ structure of the electroweak
Standard Model. All this reflects the fact that unlike the Higgs
portal model or supersymmetric extensions a simplified model is hardly
more than a single tree-level or loop-level Feynman diagram describing
dark matter annihilation. It describes the leading effects for example
in dark matter annihilation based on $2 \to 2$ or $2 \to 1$ kinematics
or the velocity dependence at threshold.  However, because simplified
models are usually not defined on the full quantum level, they leave a
long list of open questions. For new gauge bosons, also discussed in
Section~\ref{sec:vector_portal}, they include fundamental properties
like gauge invariance, unitarity, or freedom from anomalies.

\newpage
\section{Direct searches}
\label{sec:direct}
\index{direct detection}
\index{WIMP}

The experimental strategy for direct dark matter detection is based on
measuring a recoil of a nucleus after scattering with WIMP dark
matter.  For this process we can choose the optimal nuclear target
based on the largest possible recoil energy.  We start with the
non-relativistic relation between the momenta in relative coordinates
between the nucleus and the WIMP, assuming a nucleus composed
of $A$ nucleons and with charge $Z$.  The relative WIMP velocity $v_0/2$
is defined in Eq.\eqref{eq:def_velocity}, so in terms of the reduced
mass $m_A m_\chi/(m_A + m_\chi)$ we find
\begin{alignat}{7}
&&2 m_A E_A 
&= |\vec{p}_A|^2  
\approx \left( \frac{m_A m_\chi}{m_A + m_\chi} \right)^2  
  \frac{v_0^2}{4} 
\qquad \Leftrightarrow \qquad 
E_A 
= \frac{m_A}{(m_A + m_\chi)^2} 
  m_\chi^2 \frac{v_0^2}{8} \notag \\
&\Rightarrow& \qquad 
\frac{d E_A}{d m_A} 
&= \left[ \frac{1}{(m_A + m_\chi)^2} + \frac{(-2) m_A }{(m_A + m_\chi)^3} \right]
   m_\chi^2 \frac{v_0^2}{8}  \really 0 
\quad \Leftrightarrow \quad \boxed{m_A = m_\chi} \notag \\
&\Rightarrow& \; \qquad
E_A
&= \frac{m_\chi^2}{4} \; \frac{1}{2 m_\chi} \; \frac{v_0^2}{4}
= \frac{m_\chi}{32} v_0^2 
\approx 10^{4}~\ev \; ,
\label{eq:direct_recoil}
\end{alignat}
with $v_0 \approx 1/1300$ and for a dark matter around 1~TeV. Because
of the above relation, an experimental threshold from the lowest
observable recoil can be directly translated into a lower limit on
dark matter masses we can probe with such experiments. This also tells
us that for direct detection all momentum transfers are very small
compared to the electroweak or WIMP mass scale.  Similar masses of
WIMP and nuclear targets produce the largest recoil in the 10~keV
range.  Remembering that the Higgs mass in the Standard Model is
roughly the same as the mass of the gold atom we know that it should
be possible to find appropriate nuclei, for example Xenon with a
nucleus including $A = 131$ nucleons, of which $Z = 54$ are
protons.\bigskip

Strictly speaking, the dark matter velocity relevant for direct
detection is a combination of the thermal, un-directional velocity $v_0
\approx 1/1300$ and the earth's movement around the sun,
\begin{align}
v_\text{earth-sun} \, c
&= 15000 \cos \left( 2\pi \, \frac{t - 152.5~\text{d}}{365.25~\text{d}} \right) \frac{\text{m}}{\text{s}} \notag \\
\Leftrightarrow \qquad 
v_\text{earth-sun} &= 5 \cdot 10^{-5} \cos \left( 2\pi \, \frac{t - 152.5~\text{d}}{365.25~\text{d}} \right) 
\approx \frac{v_0}{15} \;  \cos \left( 2\pi \, \frac{t - 152.5~\text{d}}{365.25~\text{d}} \right) \; .
\end{align}
If we had full control over all annual modulations in a direct
detection scattering experiment we could use this modulation to
confirm that events are indeed due to dark matter scattering.\bigskip

Given that a dark matter particle will (typically) not be charged
under $SU(3)_c$, the interaction of the WIMP with the partons inside
the nucleons bound in the nucleus will have to be mediated by
electroweak bosons or the Higgs. We expect a WIMP charged under
$SU(2)_L$ to couple to a nucleus by directly coupling to the partons
in the nucleons through $Z$-exchange. This means with increased
resolution we have to compute the scattering processes for the
nucleus, the nucleons, and the partons:\\[8mm]

\begin{center}
\begin{fmfgraph*}(80,50)
\fmfset{arrow_len}{2mm}
\fmfleft{i1,i2}
\fmfright{o1,o2}
\fmf{fermion,tension=0.4,width=0.6}{i2,v2}
\fmf{fermion,tension=0.4,width=0.6}{v2,o2}
\fmf{photon,tension=0.4,label=$Z$,width=0.6}{v1,v2}
\fmf{fermion,tension=0.4,width=0.6}{i1,v1}
\fmf{fermion,tension= 0.4,width=0.6}{v1,o1}
\fmflabel{$(A,Z)$}{i1}
\fmflabel{$(A,Z)$}{o1}
\fmflabel{$\chi$}{i2}
\fmflabel{$\chi$}{o2}
\end{fmfgraph*}
\hspace*{0.15\textwidth}
\begin{fmfgraph*}(80,50)
\fmfset{arrow_len}{2mm}
\fmfleft{i1,i2}
\fmfright{o1,o2}
\fmf{fermion,tension=0.4,width=0.6}{i2,v2}
\fmf{fermion,tension=0.4,width=0.6}{v2,o2}
\fmf{photon,tension=0.4,label=$Z$,width=0.6}{v1,v2}
\fmf{fermion,tension=0.4,width=0.6}{i1,v1}
\fmf{fermion,tension= 0.4,width=0.6}{v1,o1}
\fmflabel{$N=p,n$}{i1}
\fmflabel{$N=p,n$}{o1}
\fmflabel{$\chi$}{i2}
\fmflabel{$\chi$}{o2}
\end{fmfgraph*}
\hspace*{0.15\textwidth}
\begin{fmfgraph*}(80,50)
\fmfset{arrow_len}{2mm}
\fmfleft{i1,i2}
\fmfright{o1,o2}
\fmf{fermion,tension=0.4,width=0.6}{i2,v2}
\fmf{fermion,tension=0.4,width=0.6}{v2,o2}
\fmf{photon,tension=0.4,label=$Z$,width=0.6}{v1,v2}
\fmf{fermion,tension=0.4,width=0.6}{i1,v1}
\fmf{fermion,tension= 0.4,width=0.6}{v1,o1}
\fmflabel{$q=u,d$}{i1}
\fmflabel{$q=u,d$}{o1}
\fmflabel{$\chi$}{i2}
\fmflabel{$\chi$}{o2}
\end{fmfgraph*}
\end{center}\bigskip

This gauge boson exchange will be dominated by the valence
quarks in the combinations $p \approx (uud)$ and $n \approx (udd)$.
Based on the interaction of individual nucleons, which we will
calculate below, we can express the dark matter interaction with a
heavy nucleus as
\begin{align}
\sigma^\text{SI} (\chi A \to \chi A) = 
\frac{1}{16 \pi s} 
   \overline{ \sum_N | Z \mat_p + (A-Z) \mat_n|^2 } 
= \begin{cases}
 \dfrac{A^2}{64 \pi s} \; \overline{ \sum_N | \mat_p + \mat_n|^2 } 
& \text{for} \quad Z = A/2 \\[4mm]
 \dfrac{A^2}{16 \pi s} \; \overline{ \sum_N | \mat_n|^2 } 
& \text{for} \quad \mat_p = \mat_n \; .
\end{cases} 
\end{align}
We refer to this coherent interaction as \ul{spin-independent} \index{spin-independent cross section}
scattering with the cross section $\sigma^\text{SI}$.  The scaling
with $A^2$ appears as long as the exchange particle probes all nuclei
in the heavy nucleon coherently, which means we have to square the sum of the
individual matrix elements. The condition for coherent
scattering can be formulated in terms of the size of the nucleus and
the wavelength of the momentum transfer $\sqrt{2 m_A E_A}$,
\begin{align}
\sqrt{2 m_A E_A} \approx \sqrt{2 A m_p E_A}
&< \frac{1}{A^{1/3} r_p} \approx \frac{m_p}{A^{1/3}} 
\qquad \Leftrightarrow \qquad
E_A < \frac{10^9~\ev}{2 A^{5/3}} \; .
\end{align}
for $m_p \approx 1$~GeV. This is clearly true for the typical recoils
given in Eq.\eqref{eq:direct_recoil}.  In the next step, we need to
compute the interaction to the individual $A$ nucleons in terms of
their partons. Because there are very different types of partons,
valence quarks, sea quarks, and gluons, with different quantum
numbers, this calculation is best described in a specific
model.\bigskip

One of the most interesting theoretical questions in direct detection
is how different dark matter candidates couple to the
non-relativistic nuclei.  The general trick is to link the
\ul{nucleon mass (operator)} to the nucleon-WIMP interaction
(operator).  We know that three quarks can form a color singlet state;
in addition, there will be a gluon and a sea quark content in the
nucleons, but in a first attempt we assume that those will play a
sub-leading role for the nucleon mass or its interaction to dark
matter, as long as the mediator couples to the leading valence quarks.
We start with the nucleon mass operator evaluated between
two nucleon states and write it in terms of the partonic quark
constituents,
\begin{align}
\langle N | m_N \one | N \rangle 
= m_N \langle N | N \rangle
= \sum_q \langle N |  m_q \, \bar{q} q| N \rangle 
= \sum_q m_q \langle N | \bar{q} q| N \rangle \; ,
\label{eq:direct_weak_f}
\end{align}
assuming an appropriate definition of the constituent masses. Based on the
same formalism we can write the \ul{nucleon--WIMP interaction
  operator} in terms of the quark parton content,
\begin{align}
\langle N | \sum_q \chi \chi \bar{q} q | N \rangle
= \sum_q \chi \chi \; \langle N | \bar{q} q | N \rangle \; .
\end{align}
These two estimates suggest that we can link the nucleon interaction
operator to the nucleon mass operator in the naive quark parton
model. Based on the nucleon mass we define a non-relativistic quark
density inside the nucleon as
\begin{align}
f_N := \langle N | N \rangle
\eqx{eq:direct_weak_f}
  \sum_q \frac{m_q}{m_N}   \langle N | \bar{q} q | N \rangle
= \sum_q f_q 
\qquad \Leftrightarrow \qquad 
f_q :=& \frac{m_q}{m_N}   \langle N | \bar{q} q | N \rangle \notag \\
\qquad \Rightarrow \qquad 
\langle N | \sum_q \chi \chi \bar{q} q | N \rangle
=& \sum_q \chi \chi \; \frac{m_N}{m_q} f_q \; .
\label{eq:formfactors1}
\end{align}
The form factors $f_q$ describe the probability of finding a (valence)
quark inside the proton or neutron at a momentum transfer well below
the nucleon mass.  They can for example be computed using lattice
gauge theory.  

\subsection{Higgs portal}
\label{sec:direct_portal}
\index{Higgs!portal}

The issue with Eq.\eqref{eq:formfactors1} is that it neither includes
gluons nor any quantum effects. Things become more interesting with a
\ul{Higgs-mediated WIMP-nucleon interaction}, as we encounter
it in our Higgs portal models.  To cover this case we need to compute
both, the nucleon mass and the WIMP--nucleon interaction operators
beyond the quark parton level.  From LHC we know that at least for
relativistic protons the dominant Higgs coupling is through the gluon
content. In the Standard Model the Higgs coupling to gluons is
mediated by a top loop, which \ul{does not decouple} for large
top masses. The fact that, in contrast, the top quark does decouple
from the nucleon mass will give us a non-trivial form factor for
gluons.\bigskip

Defining our quantum field theory framework, in \ul{proper QCD}
two terms contribute to the nucleon mass: the valence quark masses
accounted for in Eq.\eqref{eq:direct_weak_f} and the strong
interaction, or gluons, leading to a binding energy.  This view is
supported by the fact that pions, consisting of two quarks, are almost
an order of magnitude lighter than protons and neutrons, with three
quarks.  We can describe both sources of the nucleon mass using the
energy--momentum tensor $T^{\mu \nu}$ as it appears for example in the
Einstein--Hilbert action in Eq.\eqref{eq:einstein},
\begin{align}
\boxed{ m_N \langle N | N \rangle 
= \langle N | T_\mu^\mu | N \rangle } \; .
\label{eq:direct_traceanomaly}
\end{align}
Scale invariance, or the lack of fundamental mass scales in our theory
implies that the energy--momentum tensor is traceless.  A non-zero
trace of the energy--momentum tensor indicates a change in the
Lagrangian with respect to a scale variation, where in our units a
variation of the length scale and a variation of the energy scale are
equivalent. Lagrangians which are symmetric under such a scale
variation cannot include explicit mass terms, because those correspond
to a fixed energy scale. 

In addition to the quark masses, for the general form of the nucleon mass
given in Eq.\eqref{eq:direct_traceanomaly} we need to consider contributions
from the running strong coupling to the trace of the energy--momentum
tensor. At one-loop order the running of $\alpha_s$ with the
underlying energy scale $p^2$ is given by
\begin{align}
\alpha_s(p^2) =  \frac{1}{b_0 \log \dfrac{p^2}{\Lambda_\text{QCD}^2}} 
\qquad \text{with} \quad 
b_0 = - \frac{1}{4 \pi} \; \left( \frac{2 n_q}{3} - \frac{11}{3} N_c \right) 
\; , 
\label{eq:running_alphas}
\end{align}
and an appropriate reference value of $\Lambda_\text{QCD} \approx
200$~MeV. Mathematically, such a reference mass scale has to appear in
any problem which involves a logarithmic running, \ie which would
otherwise force us to take the logarithm of a dimensionful scale
$p^2$. Physically, this scale is defined by the point at which the
strong coupling explodes and we need to switch degrees of
freedom. That occurs at positive energy scales as long as $b_0 > 0$,
or as long as the gluons dominate the running of $\alpha_s$. Because
the running of the strong coupling turns the dimensionless parameter
$\alpha_s$ into the dimensionful parameter $\Lambda_\text{QCD}$, this
mechanism is called \ul{dimensional transmutation}\index{dimensional transmutation}.

The contribution of the running strong coupling to the nucleon mass is
given through the kinetic gluon term in the Lagrangian, combined with
the momentum variation of the strong coupling. Altogether we find
\begin{align}
m_N \langle N | N \rangle 
&= \sum_q m_q \, \langle N | \bar{q} q | N \rangle
+ \frac{2}{\alpha_s} \; \frac{d\alpha_s}{d \log p^2} \; \langle N | G_{\mu \nu}^a G^{a \, \mu \nu} | N \rangle  \notag \\
&= \sum_q m_q \, \langle N | \bar{q} q | N \rangle 
- \frac{\alpha_s b_0}{2}  \; \langle N | G_{\mu \nu}^a G^{a \, \mu \nu} | N \rangle\notag \\ 
&= \sum_q m_q \, \langle N | \bar{q} q | N \rangle 
+ \frac{\alpha_s}{8 \pi} \;
   \left( \frac{2 n_q}{3} - \frac{11}{3} N_c \right) \; \langle N | G_{\mu \nu}^a G^{a \, \mu \nu} | N \rangle \; ,
\end{align}
again written at one loop and neglecting the anomalous dimension of
the quark fields. One complication in this formula is the appearance
of all six quark fields in the sum, suggesting that all quarks
contribute to the nucleon mass. While this is true for the up
and down valence masses, and possibly for the strange mass, the three
heavier quarks hardly appear in the nucleon. Instead, they contribute
to the nucleon mass through gluon splitting or self energy diagrams in
the gluon propagator.  We can compute this contribution in terms of a
heavy quark effective theory, giving us the leading contribution per
heavy quark\index{heavy quark effective theory}
\begin{align}
\langle N | \bar{q} q | N \rangle \Bigg|_{c,b,t}
= - \frac{\alpha_s}{12 \pi m_q} 
  \; \langle N | G_{\mu \nu}^a G^{a \, \mu \nu} | N \rangle
+ \ope \left( \frac{1}{m_q^3} \right) \; .
\end{align}
We can insert this result in the above expression and find the
complete expression for the nucleon mass operator
\begin{align}
m_N \langle N | N \rangle 
&= \sum_{u,d,s} m_q \, \langle N | \bar{q} q | N \rangle 
- \sum_{c,b,t} \frac{\alpha_s}{12 \pi} \langle N | G_{\mu \nu}^a G^{a \, \mu \nu}  | N \rangle 
+ \frac{\alpha_s}{8 \pi} \left( \frac{2 \times 6}{3} - \frac{11}{3} N_c \right) \langle N | G_{\mu \nu}^a G^{a \, \mu \nu} | N \rangle \notag \\
&= \sum_{u,d,s} m_q \, \langle N | \bar{q} q | N \rangle 
+ \frac{\alpha_s}{8 \pi} \left( \frac{2 \times 3}{3} - \frac{11}{3} N_c \right) \langle N | G_{\mu \nu}^a G^{a \, \mu \nu} | N \rangle \; .
\end{align}
Starting from the full beta function of the strong coupling this
result implies that we only need to consider the running due to the
three light-flavor quarks and the gluon itself for the nucleon mass
prediction,
\begin{align}
\boxed{ 
m_N \langle N | N \rangle 
= \sum_{u,d,s} m_q \, \langle N | \bar{q} q | N \rangle 
- \frac{\alpha_s b_0^{(u,d,s)}}{2}  \; \langle N | G_{\mu \nu}^a G^{a \, \mu \nu} | N \rangle }
\qquad \text{with} \quad 
b_0^{(u,d,s)} = - \frac{1}{4 \pi} \; \left( \frac{2 n_\text{light}}{3} - \frac{11}{3} N_c \right) \; .
\label{eq:nucleon_mass}
\end{align}
This reflects a \ul{full decoupling of the heavy quarks} in their
contribution to the nucleon mass. From the derivation it is clear that
the same structure appears for any number of light quarks defining our
theory.\bigskip

Exactly in the same way we now describe the WIMP--nucleon interaction in
terms of six quark flavors. The light quarks, including the strange
quark, form the actual quark content of the nucleon. Virtual heavy
quarks occur through gluon splitting at the one-loop level.  In
addition to the small Yukawa couplings of the light quarks we know
from LHC physics that we can translate the Higgs-top interaction into
an effective Higgs--gluon interaction.  In the limit of large quark
masses the loop-induced coupling defined by the Feynman diagram\\[12mm]

\begin{center}
\begin{fmfgraph*}(80,70)
\fmfset{arrow_len}{2mm}
\fmfleft{i1}
\fmfright{o1,o2}
\fmf{dashes,tension=0.6,width=0.6}{i1,v1}
\fmf{fermion,tension=0.2,width=0.6}{v1,v2}
\fmf{fermion,tension=0.15,label=$t$,width=0.6}{v2,v3}
\fmf{fermion,tension=0.2,width=0.6}{v3,v1}
\fmf{gluon,tension=0.5,width=0.6}{v2,o1}
\fmf{gluon,tension=0.5,width=0.6}{o2,v3}
\fmflabel{$H$}{i1}
\fmflabel{$g$}{o1}
\fmflabel{$g$}{o2}
\end{fmfgraph*}
\end{center}
is given by
\begin{align}
\lag_{ggH} \supset \; \frac{1}{v_H} \;
                          g_{ggH} \; H \, G^{\mu\nu}G_{\mu\nu} 
\qqquad \text{with} \qquad
 g_{ggH}  = \frac{\alpha_s}{12 \pi} \; .
\label{eq:higgs_eff1}
\end{align}
In terms of an effective field theory the dimension-5 operator scales
like $1/v$ and not $1/m_t$. The reason is that the dependence on the
top mass in the loop and on the Yukawa coupling in the numerator
cancel exactly in the limit of small momentum transfer through the
Higgs propagator. Unlike for the nucleon mass operators this means
that in the Higgs interaction the Yukawa coupling induces a
non-decoupling feature in our theory.  Using this effective field
theory level we can successively compute the $ggH^{n+1}$ coupling from
the $ggH^n$ coupling via
\begin{alignat}{5}
g_{ggH^{n+1}} = m_q^{n+1} \; 
              \frac{\p}{\p m_q} \left( \frac{1}{m_q^n}  g_{ggH^n} \right) \; .
\label{eq:higgs_eff3}
\end{alignat}
This relation also holds for $n=0$, which means it formally links the
Higgs--nucleon coupling operator to the nucleon mass operator in
Eq.\eqref{eq:nucleon_mass}. The only difference between the effective
Higgs-gluon interaction at LHC energies and at direct detection
energies is that in direct detection all three quarks $c,b,t$
contribute to the effective interaction defined in
Eq.\eqref{eq:higgs_eff1}.\bigskip

Keeping this link in mind we see that the Higgs-mediated WIMP
interaction operator again consists of two terms
\begin{align}
\langle N | \sum_{u,d,s} m_q \, H \bar{q} q | N \rangle
- \langle N | \sum_{c,b,t} \frac{\alpha_s}{12 \pi} \;  H G_{\mu \nu}^a G^{a \, \mu \nu} | N \rangle \; .
\label{eq:higgs_eff2}
\end{align}
The corresponding Feynman diagrams at the parton level are:\medskip

\begin{center}
\begin{fmfgraph*}(80,50)
\fmfset{arrow_len}{2mm}
\fmfleft{i1,i2}
\fmfright{o1,o2}
\fmf{dashes,tension=0.4,width=0.6}{i2,v2}
\fmf{dashes,tension=0.4,width=0.6}{v2,o2}
\fmf{dashes,tension=0.4,label=$H$,width=0.6}{v1,v2}
\fmf{fermion,tension=0.4,width=0.6}{i1,v1}
\fmf{fermion,tension= 0.4,width=0.6}{v1,o1}
\fmflabel{$u,d,s$}{i1}
\fmflabel{$u,d,c$}{o1}
\fmflabel{$S$}{i2}
\fmflabel{$S$}{o2}
\end{fmfgraph*}
\hspace*{0.15\textwidth}
\begin{fmfgraph*}(80,50)
\fmfset{arrow_len}{2mm}
\fmfleft{i1,i2}
\fmfright{o1,o2}
\fmf{dashes,tension=0.4,width=0.6}{i2,v2}
\fmf{dashes,tension=0.4,width=0.6}{v2,o2}
\fmf{dashes,tension=0.4,label=$H$,width=0.6}{v1,v2}
\fmf{gluon,tension=0.4,width=0.6}{i1,v1}
\fmf{gluon,tension= 0.4,width=0.6}{v1,o1}
\fmflabel{$g$}{i1}
\fmflabel{$g$}{o1}
\fmflabel{$S$}{i2}
\fmflabel{$S$}{o2}
\end{fmfgraph*}
\end{center}\medskip

The Yukawa interaction, described by the first terms in
Eq.\eqref{eq:higgs_eff2} has a form similar to the nucleon mass in
Eq.\eqref{eq:nucleon_mass}.  Comparing the two formulas for \ul{light
quarks only} we indeed find
\begin{align}
\langle N | \sum_q m_q H \bar{q} q | N \rangle \Bigg|_{u,d,s}
&= H \sum_{u,d,s} \; m_q \, \langle N | \bar{q} q | N \rangle
\eqx{eq:nucleon_mass} H m_N \langle N | N \rangle \Bigg|_{u,d,s} \; .
\end{align}
This reproduces the simple recipe for computing the
light-quark-induced WIMP-nucleon interaction as proportional to the
nucleon mass. The remaining, numerically dominant gluonic terms is
defined in the so-called chiral limit $m_{u,d,s} = 0$. Because of the
non-decoupling behavior this contribution is independent of the heavy
quark mass, so we find for $n_\text{heavy}$ heavy quarks
\begin{alignat}{7}
\langle N | \sum_q  m_q H \bar{q} q | N \rangle \Bigg|_{c,b,t}
&\stackrel{\text{Eq.\eqref{eq:higgs_eff2}}}{=} 
- \frac{2n_\text{heavy} }{3} \; \frac{\alpha_s}{8 \pi} \; H \; \langle N | G_{\mu \nu}^a G^{a \, \mu \nu} | N \rangle \notag \\
&\quad \neq 
- \frac{\alpha_s}{8 \pi} \left( \frac{11}{3} N_c - \frac{2 n_\text{light}}{3} \right) \; H \; \langle N | G_{\mu \nu}^a G^{a \, \mu \nu} | N \rangle 
\eqx{eq:nucleon_mass} H m_N \langle N | N \rangle \Bigg|_{c,b,t,g}  \; .
\end{alignat}
The contribution to the nucleon mass comes from the gluon and
$n_\text{light}$ light quark loops, while the gluonic contribution to
the nucleon--Higgs coupling is driven by the $n_\text{heavy}$ heavy
quark loops. The boundary condition is $n_\text{light} +
n_\text{heavy} = 6$. At the energy scale of direct detection we can
compensate for this mismatch in the Higgs--nucleon coupling of the
naive scaling between the nucleon mass and nucleon Yukawa interaction
shown in Eq.\eqref{eq:formfactors1}.  We simply include an additional
factor
\begin{alignat}{7}
\boxed{
\sum_q \frac{m_q}{m_N} \langle N | H \bar{q} q | N \rangle \Bigg|_{c,b,t}
= \frac{\dfrac{2n_\text{heavy}}{3}}{\dfrac{11}{3} N_c - \dfrac{2 n_\text{light}}{3}}
\; H \; \langle N | N \rangle \Bigg|_{c,b,t,g}  
}\; ,
\end{alignat}
which we can estimate at leading order and at energy scales relevant
for direct dark matter searches to be
\begin{align}
\frac{\dfrac{2n_\text{heavy}}{3}}{\dfrac{11}{3} N_c - \dfrac{2 n_\text{light}}{3}}
\stackrel{n_\text{light} = 3}{=} \dfrac{3 \times \dfrac{2}{3}}{11 - \dfrac{2\times 3}{3}}
= \frac{2}{9} \; .
\end{align}
%
%
This effect leads to a suppression of the already small Higgs--nucleon
interaction at low momentum transfer.  The exact size of the
suppression depends on the number of active light quarks in our effective theory,
which in turn depends on the momentum transfer.\bigskip

At the parton level, the weakly interacting part of the calculation of
the nucleon--WIMP scattering rate closely follows the calculation of
WIMP annihilation in
Eq.\eqref{eq:matrix_annihilation}. In the case of direct detection the
valence quarks in the nucleons couple through a $t$-channel Higgs to
the dark matter scalar $S$. We account for the parton nature of the
three relevant heavy quarks by writing the nucleon Yukawa coupling as
$f_N m_N \times 2/9$,
\begin{align}
\mat 
&= \bar{u}(k_2) \, \frac{-2 i f_N m_N}{9 v_H} \, u(k_1) \; 
\frac{-i}{(k_1 - k_2)^2 - m_H^2} \; 
(- 2 i \lambda_3 v_H ) \; .
\label{eq:matrix_direct}
\end{align}
For an incoming and outgoing fermion the two spinors are $\bar{u}$ and $u$.
As long as the Yukawa coupling is dominated by the heavy quarks, it
will be the same for neutrons and protons, \ie $\mat_p = \mat_n$.  We
have to square this matrix element, paying attention to the spinors
$v$ and $u$, and then sum over the spins of the external fermions. In
this case we already know that we are only interested in scattering in
the low-energy limit, \ie $|(k_1 - k_2)^2| \ll m_N^2 \ll m_H^2$,
\begin{align}
\sum_\text{spin} |\mat|^2 
&= \frac{16}{81} \; \lambda_3^2 \, f_N^2 m_N^2
   \left( \sum_\text{spin} u(k_2) \bar{u}(k_2) \right) \;
   \left( \sum_\text{spin} u(k_1) \bar{u}(k_1) \right) \;
   \frac{1}{\left[ (k_1 - k_2)^2 - m_H^2 \right]^2} \notag \\
&= \frac{16}{81} \; \lambda_3^2 \, f_N^2 m_N^2 \;
   \tr \left[ ( \slashchar{k}_2 + m_N \one ) \;
              ( \slashchar{k}_1 + m_N \one ) \right] \;
   \frac{1}{\left[ (k_1 - k_2)^2 - m_H^2 \right]^2} \notag \\
 &= \frac{32}{81} \; \lambda_3^2 \, f_N^2 m_N^2 \;
   \left[2 k_1\cdot k_2 + 2 m_N^2 \right] \;
   \frac{1}{\left[ (k_1 - k_2)^2 - 2 m_H^2 \right]^2} \notag \\
 &= \frac{32}{81} \; \lambda_3^2 \, f_N^2 m_N^2 \;
  \left[ - (k_1 - k_2)^2 + 4 m_N^2 \right] \;
   \frac{1}{\left[ (k_1 - k_2)^2 - m_H^2 \right]^2} \notag \\
&\approx 
\frac{128}{81} \; \lambda_3^2 \, f_N^2 \, \frac{m_N^4}{m_H^4} 
\qqquad \Rightarrow \qqquad 
\overline{ \sum_\text{spin,color} |\mat|^2 }
= \frac{64}{81} \; \lambda_3^2 \, f_N^2 \, \frac{m_N^4}{m_H^4} 
\label{eq:matrix2_direct}
\end{align}
The cross section in the low-energy limit is by definition
spin-independent and becomes
\begin{align}
\sigma^\text{SI} (SN \to SN) 
&= \frac{1}{16 \pi s} \;
   \overline{ \sum |\mat|^2 } \notag \\
&= \frac{1}{16 \pi (m_S + m_N)^2} \;
   \frac{64}{81} \; \lambda_3^2 \, f_N^2 \, \frac{m_N^4}{m_H^4} 
\approx \frac{4 \lambda_3^2 \, f_N^2}{81 \pi} \;  \frac{m_N^4}{m_H^4} \; 
 \frac{1}{m_S^2} \; ,
\end{align}
where in the last step we assume $m_S \gg m_N$. For WIMP--Xenon
scattering this gives us \index{Xenon}
\begin{align}
\boxed{
\sigma^\text{SI} (SA \to SA) 
= \frac{4 \lambda_3^2 \, f_N^2\, A^2}{81 \pi} \;  \frac{m_N^4}{m_H^4} \; \frac{1}{m_S^2} 
= 6 \cdot 10^{-7} \; \frac{\lambda_3^2}{m_S^2} } 
\; .
\label{eq:portal_directrate}
\end{align}
The two key ingredients to this expression can be easily understood:
the suppression $1/m_H^4$ appears after we effectively integrate out
the Higgs in the $t$-channel, and the high power of $m_N^4$ occurs
because in the low-energy limit the Higgs coupling to fermions involve
a chirality flip and hence one power of $m_N$ for each coupling.  The
angle-independent matrix element in the low-energy limit can easily be
translated into a spectrum of the scattering angle, which will then
give us the recoil spectrum, if desired. We limit ourselves to the
total rate, assuming that the appropriate WIMP mass range ensures that
the total cross section gets converted into measurable recoil. This
approach reflects the fact that we consider the
kinematics of scattering processes and hence the existence of phase
space a topic for experimental lectures.

Next, we can ask which range of Higgs portal parameters with the correct
relic density, as shown in Figure~\ref{fig:portal_mass_coup}, is
accessible to direct detection experiments. According to
Eq.\eqref{eq:portal_directrate} the corresponding cross section first
becomes small when $\lambda_3 \ll 1$, which means $m_S \lesssim m_H/2$
with the possibility of explaining the Fermi galactic center excess.
Second, the direct detection cross section is suppressed for heavy
dark matter and leads to a scaling $\lambda_3 \propto m_S$.

From Eq.\eqref{eq:annrate_scaling} we know that a constant
annihilation rate leading to the correct relic density also
corresponds to $\lambda_3 \propto m_S$. However, while the direct
detection rate features an additional suppression through the nucleon
mass $m_N$, the annihilation rate benefits from several subleading
annihilation channels, like for example the annihilation to two gauge
bosons or two top quarks. This suggests that for large $m_S$ the two
lines of constant cross sections in the $\lambda_3$-$m_S$ plane run
almost in parallel, with a slightly smaller slope for the annihilation
rate. This is exactly what we observe in
Figure~\ref{fig:portal_mass_coup}, leaving heavy Higgs portal dark
matter with $m_S \gtrsim 300$~GeV a viable model for all observations
related to cold dark matter. This minimal dark matter mass constraint
rapidly increases with new direct detection experiments coming
online. On the other hand, from our discussion of the threshold
behavior in Section~\ref{sec:sommerfeld} is should be clear that we
can effectively switch off all direct detection constraints by making
the scalar Higgs mediator a pseudo-scalar.\bigskip

Finally, we can modify our model and the quantitative link between the
relic density and direct detection, as illustrated in
Figure~\ref{fig:portal_mass_coup}.  The typical renormalizable Higgs
portal includes a scalar dark matter candidate. However, if we are
willing to include higher-dimensional terms in the Lagrangian we can
combine the Higgs portal with fermionic and vector dark matter. This
is interesting in view of the velocity dependence discussed in
Section~\ref{sec:velocity}. The annihilation of dark matter fermions is
velocity-suppressed at threshold, so larger dark matter couplings
predict  the observed relic density. Because direct detection is not
sensitive to the annihilation threshold, it will be able to rule out
even the mass peak region for fermionic dark matter. 

\begin{figure}[b!]
\begin{center}
\includegraphics[width=0.45\textwidth]{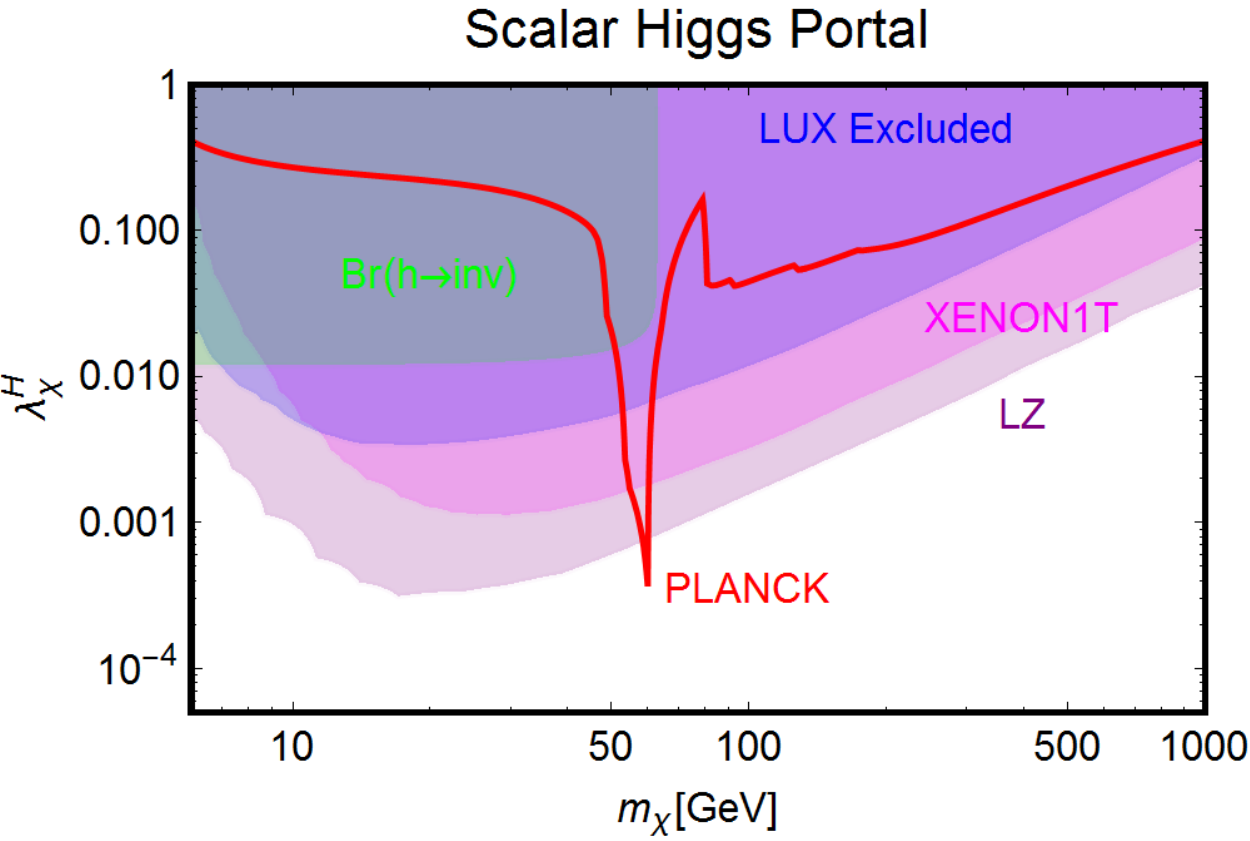}
\hspace*{0.05\textwidth}
\includegraphics[width=0.45\textwidth]{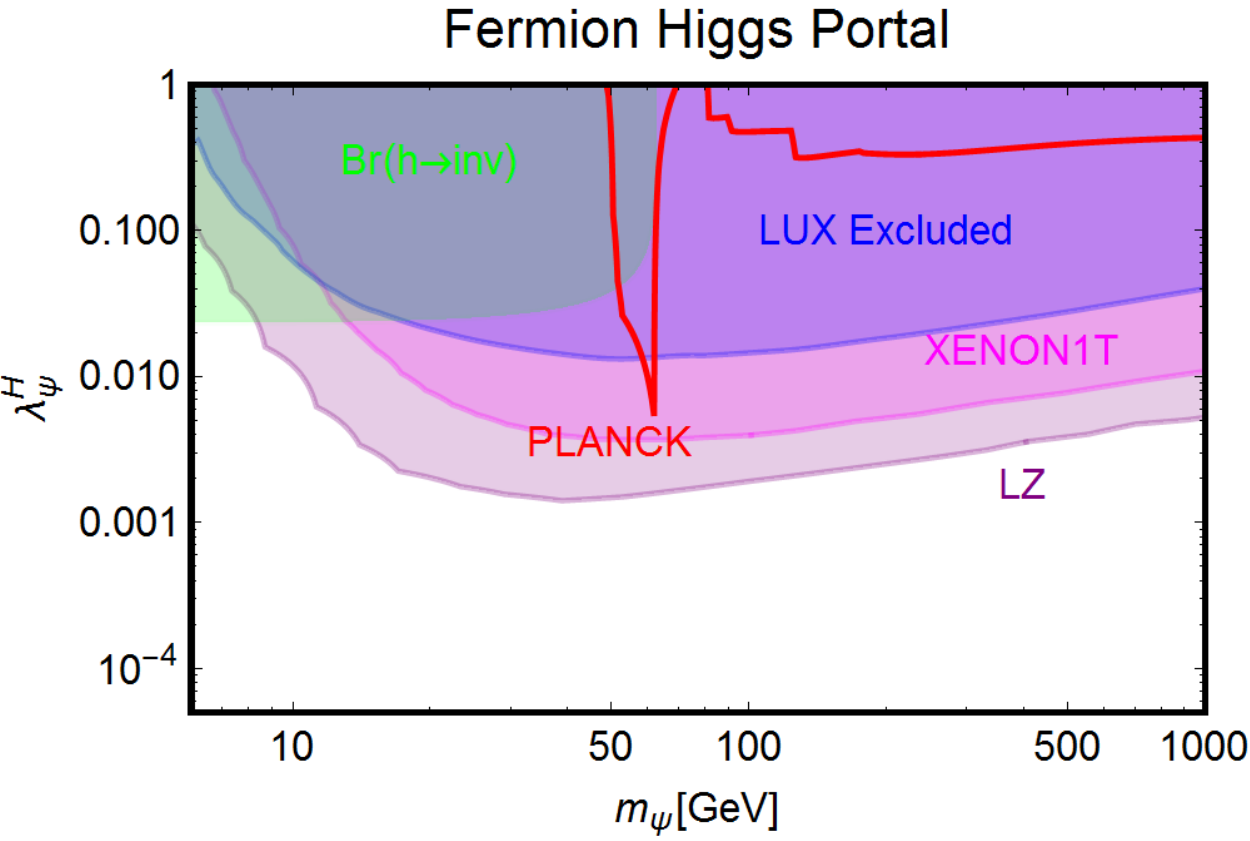}
\end{center}
\caption{Relic density (labelled PLANCK) vs direct dark matter
  detection constraints. The dark matter agent is switched from a real
  scalar (left) to a fermion (right).  Figure from
  Ref.~\cite{Arcadi:2017kky}.}
\label{fig:neutralino_relic}
\end{figure}

\subsection{Supersymmetric neutralinos}
\label{sec:direct_mssm}

\index{neutralino}
In supersymmetry with its fermionic dark matter candidate,
nucleon--neutralino scattering is described by four-fermion operators,
just like in Fermi's theory. The reason is that all intermediate
particles coupling the two neutralinos to two quarks are far below
their mass shell.  Accounting for the mass dimension through a scalar
mediator mass scale $\Lambda \approx m_{h^0}$, the matrix element
reads
\begin{align}
\mat &= \frac{g_{NN \nne \nne}}{\Lambda^2} \; \bar{v}_{\nne} v_{\nne} \; \bar{u}_N u_N \notag \\ 
\sum_\text{spins} |\mat|^2 & =
 \frac{g_{NN \nne \nne}^2}{\Lambda^4} \; 
\tr \left[ ( \slashchar{p}_2 - m_{\nne} \one ) \;
           ( \slashchar{p}_1 - m_{\nne} \one ) \right]
\tr \Big[ ( \slashchar{k}_2 + m_N \one ) \;
           ( \slashchar{k}_1 + m_N  \one ) \Big] \notag \\
&\approx 64  g_{NN \nne \nne}^2 \; \frac{m_{\nne}^2 m_N^2}{\Lambda^4}
\qqquad \Rightarrow \qqquad 
\overline{ \sum_\text{spin,color} |\mat|^2 }
= 16 \; g_{NN \nne \nne}^2 \, \; \frac{m_{\nne}^2 m_N^2}{\Lambda^4} \; .
\end{align}
The corresponding spin-independent cross section mediated by the
Standard Model Higgs in the low-energy limit is then
\begin{align}
\boxed{
\sigma^\text{SI} (\nne N \to \nne N) 
\approx \frac{g_{NN \nne \nne}^2}{\pi} \; \frac{m_N^2}{m_{h^0}^4}  \;} .
\label{eq:susy_directrate1}
\end{align}
As for the Higgs portal case in Eq.\eqref{eq:portal_directrate} the
rate is suppressed by the mediator mass to the fourth power. The lower
power of $m_N^2$ appears only because we absorb the Yukawa coupling in
$g_{NN \nne \nne} = 2 m_N f_N/9$,
\begin{align}
g_{NN \nne \nne} 
= \frac{g_{h^0 \nne \nne} \; 2 f_N m_N}{9 }
\propto \frac{2 f_N m_N}{9 } \;
   \left( g' N_{11} - g N_{12} \right) \; 
   \left( \sin \alpha \; N_{13} + \cos \alpha \; N_{14} \right) 
\; ,
\end{align}
following Eq.\eqref{eq:coups_neutralinos}. We see that this scaling is
identical to the Higgs portal case in Eq.\eqref{eq:portal_directrate},
but with an additional suppression through the difference in mixing
angles in the neutralino and Higgs sectors.\bigskip

\begin{figure}[b!]
\includegraphics[scale=.75]{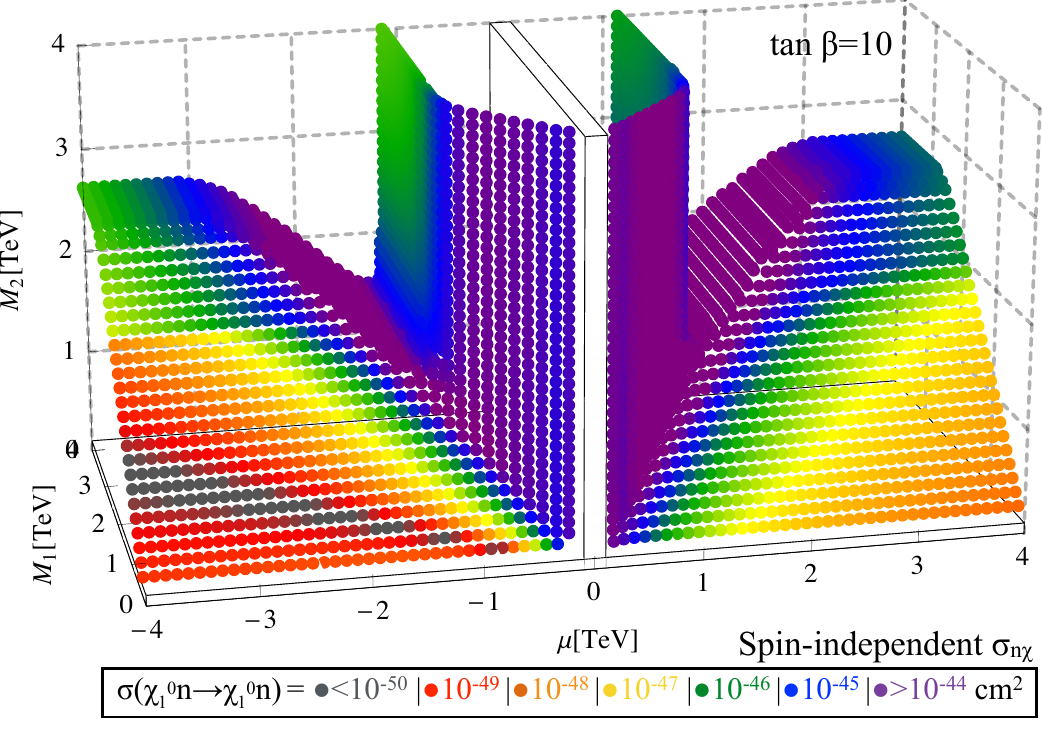}
\includegraphics[scale=.75]{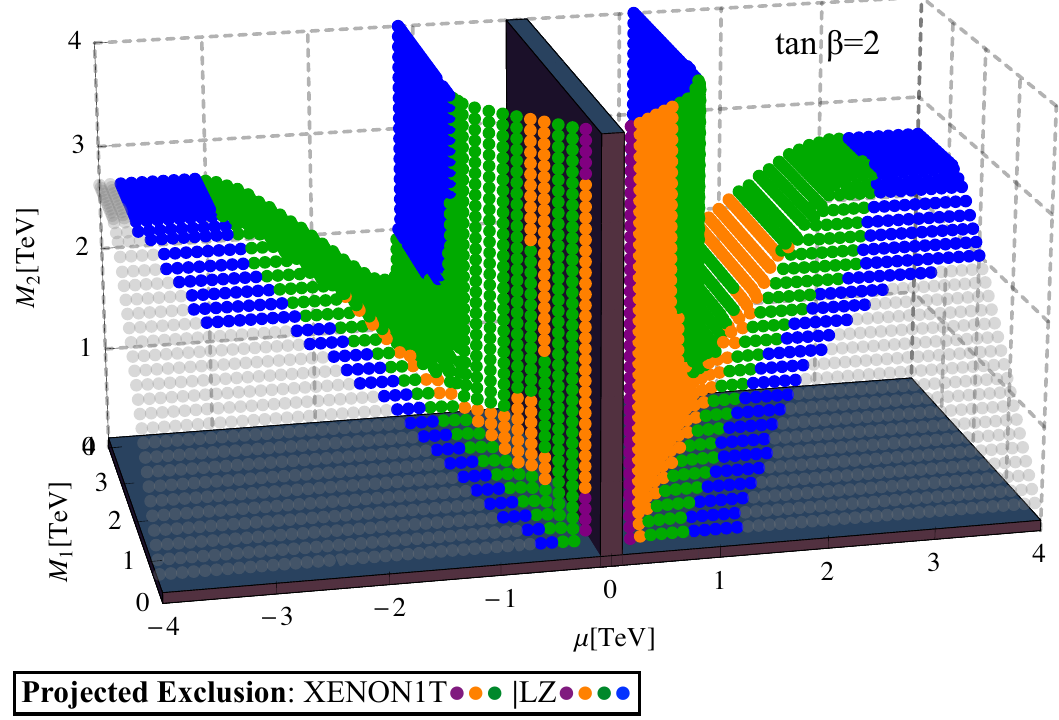}
\caption{Left: spin-independent nucleon-scattering cross-section for
  relic neutralinos. Right: relic neutralino exclusions from XENON100
  and LUX and prospects from XENON1T and LZ.  The boxed out area
  denotes the LEP exclusion. Figure from Ref.~\cite{nimatron2}.}
\label{fig:spininde}
\end{figure}

\begin{figure}[b!]
\begin{center}
\includegraphics[width=0.6\textwidth]{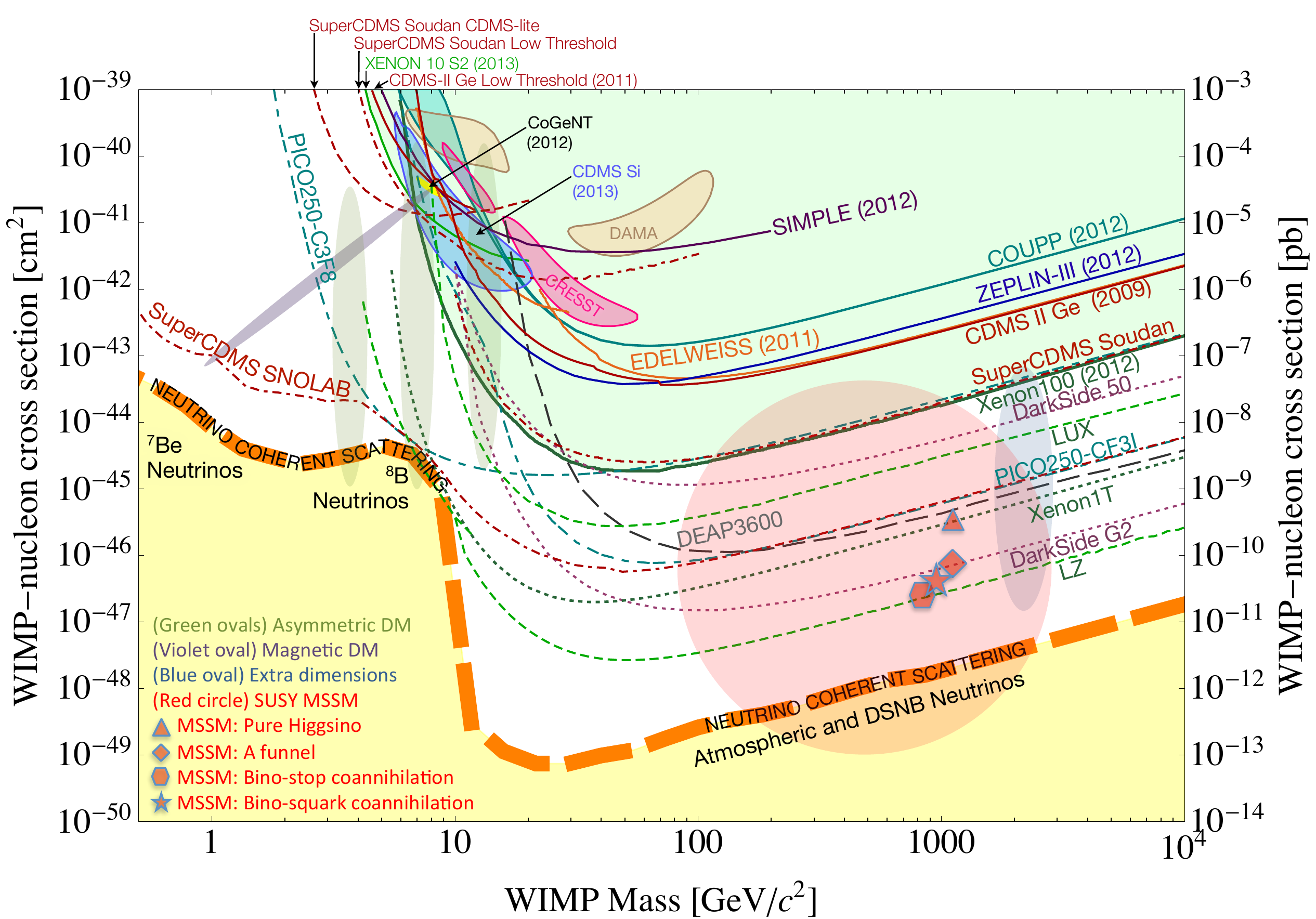}
\end{center}
\caption{Spin-independent WIMP--nucleon cross section limits and
  projections (solid, dotted, dashed curves) and hints for WIMP
  signals (shaded contours) and projections (dot and dot-dashed
  curves) for direct detection experiments. The yellow region
  indicates dangerous backgrounds from solar, atmospheric, and diffuse
  supernovae neutrinos. Figure from Ref.~\cite{snowmass}.}
\label{fig:snowmass}
\end{figure}

However, in supersymmetric models the dark matter mediator will often
be the $Z$-boson, because the interaction $g_{NN \nne \nne}$ is not
suppressed by a factor of the kind $m_N/v$.  In this case we need to
describe a (transverse) vector coupling between the WIMP and the
nucleon in our four-fermion interaction. Following exactly the argument
as for the scalar exchange we can look at a $Z$-mediated interaction
between a (Dirac) fermion $\chi$ and the nucleons,
\begin{align}
\mat &= \frac{g_{NN \chi \chi}}{\Lambda^2} \; \bar{v}_{\chi} \gamma_\mu v_{\chi} \; \bar{u}_N \gamma^\mu u_N \notag \\ 
\sum_\text{spins} |\mat|^2 & =
 \frac{g_{NN \chi \chi}^2}{\Lambda^4} \; 
\tr \Big[ ( \slashchar{p}_2 - m_{\chi} \one ) \; \gamma_\mu \; 
           ( \slashchar{p}_1 - m_{\chi} \one ) \; \gamma_\nu \Big]
\tr \Big[ ( \slashchar{k}_2 + m_N \one ) \; \gamma^\mu \;
           ( \slashchar{k}_1 + m_N  \one ) \; \gamma^\nu \Big] \notag \\
& \approx \frac{g_{NN \chi \chi}^2}{\Lambda^4} \; ( 8 m_{\chi}^2)(8 m_N^2) \notag \\
&\approx 64 g_{NN \chi \chi}^2 \; \frac{m_{\chi}^2 m_N^2}{\Lambda^4}
\qqquad \Rightarrow \qqquad 
\boxed{ 
\sigma^\text{SI} (\chi N \to \chi N) 
\approx \frac{4 g_{NN \chi \chi}^2}{\pi} \; \frac{m_N^2}{\Lambda^4} } \; .
\label{eq:susy_directrate2}
\end{align}
The spin-independent cross section mediated by a gauge boson is
typically several orders of magnitude larger than the cross section
mediated by Higgs exchange. This means that models with dark matter
fermions coupling to the $Z$-boson will be in conflict with direct
detection constraints. For the entire relic neutralino surface with
pure and mixed states the spin-independent cross sections are shown in
Figure~\ref{fig:spininde}.  The corresponding current and future
exclusion limits are indicated in Figure~\ref{fig:snowmass}. The
so-called neutrino floor, which can be reached within the next decade,
is the parameter region where the expected neutrino background will
make standard direct detection searches more challenging.\bigskip

The problem with this result in Eq.\eqref{eq:susy_directrate2} is that
it does not hold for \ul{Majorana fermions}, like the
neutralino $\nne$.  From the discussion in
Section~\ref{sec:sommerfeld} we know that a vector mediator cannot
couple Majorana fermions to the nucleus, so we are left with the
corresponding axial vector exchange,
\begin{align}
\mat &= \frac{g_{NN \nne \nne}}{\Lambda^2} \; \bar{v}_{\nne} \gamma_\mu \gamma_5 v_{\nne} \; \bar{u}_N \gamma^\mu \gamma_5 u_N \notag;, \\ 
\sum_\text{spins} |\mat|^2 & =
 \frac{g_{NN \nne \nne}^2}{\Lambda^4} \; 
\tr \left[ ( \slashchar{p}_2 - m_{\nne} \one ) \; \gamma_5 \gamma_\mu \; 
           ( \slashchar{p}_1 - m_{\nne} \one ) \; \gamma_5 \gamma_\nu \right]
\tr \Big[ ( \slashchar{k}_2 + m_N \one ) \; \gamma_5 \gamma^\mu \;
           ( \slashchar{k}_1 + m_N  \one ) \; \gamma_5 \gamma^\nu \Big] \; .
\end{align}
\index{spin-dependent cross section}
For axial vector couplings the current is defined by $\gamma_\mu
\gamma_5$. This means it depends on the chirality or the helicity of
the fermions.  The spin operator is defined in terms of the Dirac
matrices as $\vec{s} = \gamma_5 \gamma^0 \vec{\gamma}$. This indicates
that the axial vector coupling actually is a coupling to the spin of
the nucleon.  This is why the result is called a
\ul{spin-dependent} cross section, which for each nucleon reads
\begin{align}
\sum_\text{spins} |\mat|^2 
&\approx 16 \times 4  g_{NN \nne \nne}^2 \; \frac{m_{\nne}^2 m_N^2}{\Lambda^4}
\qqquad \Rightarrow \qqquad 
\boxed{ 
\sigma^\text{SD} (\nne N \to \nne N) 
\approx \frac{4 g_{NN \nne \nne}^2}{\pi} \; \frac{m_N^2}{\Lambda^4} } \; .
\label{eq:susy_directrate3}
\end{align}
Again, we can read off Eq.\eqref{eq:coups_neutralinos} that for the
light quarks $q = u,d,s$ the effective coupling should have the form
\begin{align}
g_{NN \nne \nne} 
= g_{Z \nne \nne} \, g_{Zqq}
\propto g^2 \, \left( |N_{13}|^2 - |N_{14}|^2 \right) \; .
\end{align}
The main difference between the spin-independent and spin-dependent
scattering is that for the coupling to the nucleon spin we cannot
assume that the couplings to all nucleons inside a nucleus add
coherently. Instead, we need to link the spin representations of the
nucleus to spin representations of each nucleon. Instead of finding a
coherent enhancement with $Z^2$ or $A^2$ the result scales like $A$,
weighted by Clebsch-Gordan coefficients which appear from reducing out
the combination of the spin-1/2 nucleons.  Nevertheless, direct
detection strongly constrains the higgsino content of the relic
neutralino, with the exception of a pure higgsino, where the two terms
in $g_{Z \nne \nne}$ cancel each other.

\newpage 
\section{Collider searches}
\label{sec:coll}
\index{collider}

Collider searches for dark matter rely on two properties of the dark
matter particle: first, the new particles have to couple to the
Standard Model. This can be either a direct coupling for example to
the colliding leptons and quarks, or an indirect coupling through an
\ul{mediator}.\index{mediator} Second, we need to measure traces of particles
which interact with the detectors as weakly as for example neutrinos
do. And unlike dedicated neutrino detectors their collider counter
parts do not include hundreds of cubic meters of interaction
material. Under those boundary conditions collider searches for dark matter
particles will benefit from several advantages:
\begin{enumerate}
\item we know the kinematic configuration of the dark matter
  production process. This is linked to the fact that most collider
  detectors are so-called multi-purpose detectors which can measure a
  great number of observables;
\item the large number of collisions (parametrized by the luminosity $\lumi$) 
  can give us a large number of dark matter particles to analyze. This
  allows us to for example measure kinematic distributions which
  reflect the properties of the dark matter particle;
\item all background processes and all systematic uncertainties can be
  studied, understood, and simulated in detail. Once an observation of
  a dark matter particle passes all conditions the collider
  experiments require for a discovery, we will know that we discovered
  such a new particle. Otherwise, if an anomaly turns out to not pass
  these conditions we have at least in my life time always been able
  to identify what the problem was.
\end{enumerate}
One weakness we should always keep in mind is that a particle which
does not decay while crossing the detector and which interacts
weakly enough to not leave a trace does not have to be stable on
cosmological time scales. To make this statement we need to measure
enough properties of the dark matter particle to for example predict
its relic density the way we discuss it in
Section~\ref{sec:relic}.

\subsection{Lepton colliders}
\label{sec:coll_lepton}

The key observable we can compute and analyze at colliders is the
number of events expected for a certain production and decay process
in a given time interval. The number of events is the product of the
\ul{luminosity} $\lumi$ measured for example in inverse
femtobarns, the total \ul{production cross section} measured in
femtobarns, and the detection efficiency measured in
per-cent,\footnote{Cross sections and luminosities are two of the few
  observables which we do not measure in eV.}
\begin{align}
N_\text{events} = \sigma_\text{tot} \; \lag \; \Pi_j \epsilon_j \; .
\label{eq:n_events}
\end{align}
This way the event rate is split into a collider--specific number
describing the initial state, a process--specific number describing
the physical process, and a detector--specific efficiency for each
final state particle. The efficiency includes for example phase-space
dependent cuts defining the regions of sensitivity of a given
experiment, as well as the so-called trigger requirements defining
which events are saved and looked at.  This structure holds for every
collider.\bigskip

When it comes to particles with electroweak interactions the most
influential experiments were ALEPH, OPAL, DELPHI, and L3 at the Large
Electron-Positron Collider (LEP) at CERN. It ran from 1989 until 2000,
first with a $e^+e^-$ energy right on the $Z$ pole, and then with
energies up to $209$~GeV. Its life-time integrated luminosity is
$1~\ifb$. The results form running on the $Z$ pole are easily
summarized: the $SU(2)_L$ gauge sector shows no hints for deviations
from the Standard Model predictions. Most of these results are based
on an analysis of the Breit--Wigner propagator of the $Z$ boson which
we introduce in Eq.\eqref{eq:portal_annrate2},
\begin{align}
\sigma(e^+ e^- \to Z) \propto \frac{E_{e^+e^-}^2}{(E_{e^+e^-}^2-m_Z^2)^2 + m_Z^2 \Gamma_Z^2} \; .
\end{align}
If we know what the energy of the incoming $e^+ e^-$ system is we can
plot the cross section as a function of $E_{e^+e^-}$ and measure the
$Z$ mass and the $Z$ width,
\begin{align}
m_Z = ( 91.19 \pm 0.003 )~\gev\;, \qqqquad 
\Gamma_Z = ( 2.49 \pm 0.004 )~\gev \; .
\end{align}
From this $Z$ mass measurement in relation to the $W$ mass and the
vacuum expectation value $v_H = 246$~GeV we can extract the top quark
and Higgs masses, because these particles contribute to quantum
corrections of the $Z$ properties.  The total $Z$ width includes a
partial width from the decay $Z \to \nu \bar{\nu}$, with a branching
ratio around 20\%. It comes from three generations of light neutrinos
and is much larger than for example the 3.4\% branching ratio of the
decay $Z \to e^+ e^-$. Under the assumption that only neutrinos
contribute to the invisible $Z$ decays we can translate the
measurement of the partial width into a measurement of the number of
light neutrinos, giving $2.98 \pm 0.008$. Alternatively, we can assume
that there are three light neutrinos and use this measurement to
constrain light dark matter with couplings to the $Z$ that would lead to
an on-shell decay, for example $Z \to \nne \nne$ in our supersymmetric
model. If a dark matter candidate relies on its electroweak couplings
to annihilate to the observed relic density, this limit means that any
WIMP has be heavier than
\begin{align}
m_{\nne, S} > \frac{m_Z}{2} = 45~\gev \; .
\end{align}
The results from the higher-energy runs are equally simple: there is
no sign of new particles which could be singly or pair-produced in
$e^+ e^-$ collisions. The Feynman diagram for the production of a pair
of new particles, which could be dark matter particles, is \\[10mm]

\begin{center}
\begin{fmfgraph*}(80,50)
\fmfset{arrow_len}{2mm}
\fmfleft{i1,i2}
\fmfright{o1,o2}
\fmf{fermion,tension=0.4,width=0.6}{i1,v1}
\fmf{fermion,tension=0.4,width=0.6}{v1,i2}
\fmf{photon,tension=0.4,label=$Z$,width=0.6}{v1,v2}
\fmf{fermion,tension=0.4,width=0.6}{v2,o2}
\fmf{fermion,tension=0.4,width=0.6}{o1,v2}
\fmflabel{$e^-$}{i1}
\fmflabel{$e^+$}{i2}
\fmflabel{$\chi$}{o1}
\fmflabel{$\chi$}{o2}
\end{fmfgraph*}
\end{center}

The experimental results mean that it is very
hard to postulate new particles which couple to the $Z$ boson or to
the photon. The Feynman rules for the corresponding $f\bar{f} Z$ and
$f\bar{f} \gamma$ couplings are
\begin{alignat}{5}
- i \gamma^\mu \left( \ell \prol + r \pror \right)
 \qqquad \text{with} \quad 
 \ell &= \frac{e}{s_w c_w} \; \left( T_3 - 2 Q s_w^2 \right)
 \qquad 
 r = \ell \Big|_{T_3=0}
 \qqquad &&(Zf \bar{f})
\notag \\
 \ell &= r = Q e
 &&(\gamma f \bar{f}) \; ,
\label{eq:dy_feyn}
\end{alignat}
with the isospin quantum number $T_3 = \pm 1/2$ and $s_w^2 \approx
1/4$. Obviously, a pair of charged fermions will always be produced
through an $s$-channel photon. If a particle has $SU(2)_L$ quantum
numbers, the $Z$-coupling can be cancelled with the help of the
electric charge, which leads to photon-induced pair production.  Dark
matter particles cannot be charged electrically, so for WIMPs there
will exist a production process with a $Z$-boson in the
$s$-channel. This result is important for co-annihilation in a more
complex dark matter sector. For example in our supersymmetric model
the charginos couple to photons, which means that they have to be
heavier than
\begin{align}
m_{\cpme} > \frac{E_{e^+ e^-}^\text{max}}{2} = 104.5~\gev;,
\label{eq:lep_charlimit}
\end{align}
in order to escape LEP constraints.
The problem of producing and detecting a pair of dark matter particles
at any collider is that if we do not produce anything else those
events with `nothing visible happening' are hard to identify. Lepton
colliders have one big advantage over hadron colliders, as we will see
later: we know the kinematics of the initial state. This means that
if, for example, we produce one invisibly decaying particle we can
reconstruct its four-momentum from the initial state momenta and the
final-state recoil momenta. We can then check whether for the majority of
events the on-shell condition $p^2 = m^2$ with a certain mass is
fulfilled. This is how OPAL managed to extract limits on Higgs
production in the process $e^+ e^- \to ZH$ without making any
assumptions about the Higgs decay, notably including a decay to two
invisible states. Unfortunately, because this analysis did not reach
the observed Higgs mass of $126$~GeV, it does not constrain our dark
matter candidates in Higgs decays.\bigskip

The pair production process 
\begin{align}
e^+ e^- \to \gamma^* Z^* \to \chi \chi
\end{align}
is hard to extract experimentally, because we cannot distinguish it
from an electron and positron just missing each other.
The way out is to produce another particle in association with the
dark matter particles, for example a photon with sufficiently large
transverse momentum $p_T$
\index{missing energy}
\begin{align}
e^+ e^- \to \nne \nne \gamma , \, SS \gamma\;,
\label{eq:radiative_return}
\end{align}
Experimentally, this photon recoils against the two dark matter
candidates, defining the signature as a \ul{photon plus missing
  momentum}. A Feynman diagram for the production of a pair of dark
matter particles and a photon through a $Z$-mediator is\\[8mm]

\begin{center}
\begin{fmfgraph*}(80,50)
\fmfset{arrow_len}{2mm}
\fmfleft{i1,i2}
\fmfright{o1,o2,o3}
\fmf{fermion,tension=0.8,width=0.6}{i1,v1}
\fmf{fermion,tension=0.6,width=0.6}{v1,v2}
\fmf{fermion,tension=0.4,width=0.6}{v2,i2}
\fmf{photon,tension=0.7,label=$Z$,width=0.6}{v2,v3}
\fmf{photon,tension=0.4,width=0.6}{v1,o1}
\fmf{fermion,tension=0.5,width=0.6}{v3,o3}
\fmf{fermion,tension=0.7,width=0.6}{o2,v3}
\fmflabel{$e^-$}{i1}
\fmflabel{$e^+$}{i2}
\fmflabel{$\gamma$}{o1}
\fmflabel{$\chi$}{o2}
\fmflabel{$\chi$}{o3}
\end{fmfgraph*}
\end{center}

Because the photon can only be radiated off the incoming electrons,
this process is often referred to as \ul{initial state
  radiation} (ISR). Reconstructing the four-momentum of the photon allows
us to also reconstruct the four-momentum of the pair of dark matter
particles. The disadvantage is that a hard photon is only present in a
small fraction of all $e^+e^-$ collisions for example at LEP. This is one of the few
instances where the luminosity or the size of the cross section makes
a difference at LEP. Normally, the relatively clean $e^+ e^-$
environment allows us to build very efficient and very precise
detectors, which altogether allows us to separate a signal from a
usually small background cleanly. For example, the chargino mass limit
in Eq.\eqref{eq:lep_charlimit} applies to a wide set of new particles
which decay into leptons and missing energy and is hard to
avoid.\bigskip

We should mention that for a long time people have discussed building
another $e^+ e^-$ collider. Searching for new particles with
electroweak interactions is one of the main motivations. Proposals range
from a circular Higgs factory with limited energy due to energy loss in synchrotron radiation(FCC-ee/CERN or
CEPC/China) to a linear collider with an energy up to 1~TeV
(ILC/Japan), to a multi-TeV linear collider with a driving beam
technology (CLIC/CERN).

\subsection{Hadron colliders and mono-X}
\label{sec:coll_hadron}
\index{mono-X}

Historically, hadron colliders have had great success in discovering
new, massive particles. This included UA1 and UA2 at SPS/CERN
discovering the $W$ and $Z$ bosons, CDF and D0 at the
Tevatron/Fermilab discovering the top quark, and most recently ATLAS
and CMS at the LHC with their Higgs discovery. The simple reason is
that protons are much heavier than electrons, which makes it easier to
store large amounts of kinetic energy and release them in a
collision. On the other hand, hadron collider physics is much harder
than lepton collider physics, because the experimental environment is
more complicated, there is hardly any process with negligible
backgrounds, and calculations are generically less precise.

This means that at the LHC we need to consider two kinds of
processes. The first involves all known particles, like electrons or
$W$ and $Z$ bosons, or the top quark, or even the Higgs boson.  These
processes we call \ul{backgrounds}, and they are described by
QCD.  The Higgs boson is in the middle of a transition to a
background, only a few years ago is was the most famous example for a
\ul{signal}. By definition, signals are very rare compared to
backgrounds.  As an example, Figure~\ref{fig:higgs_lhcall} shows that
at the LHC the production cross section for a pair of bottom quarks is
larger than $10^5$~nb or $10^{11}$~fb, the typical production rate for
$W$ or $Z$ bosons ranges around 200~nb or $2 \cdot 10^8$~fb, the rate
for a pair of $500$~GeV supersymmetric gluinos would have been $4 \cdot
10^4$~fb.

\begin{figure}[b!]
\begin{center}
\includegraphics[width=0.45\textwidth]{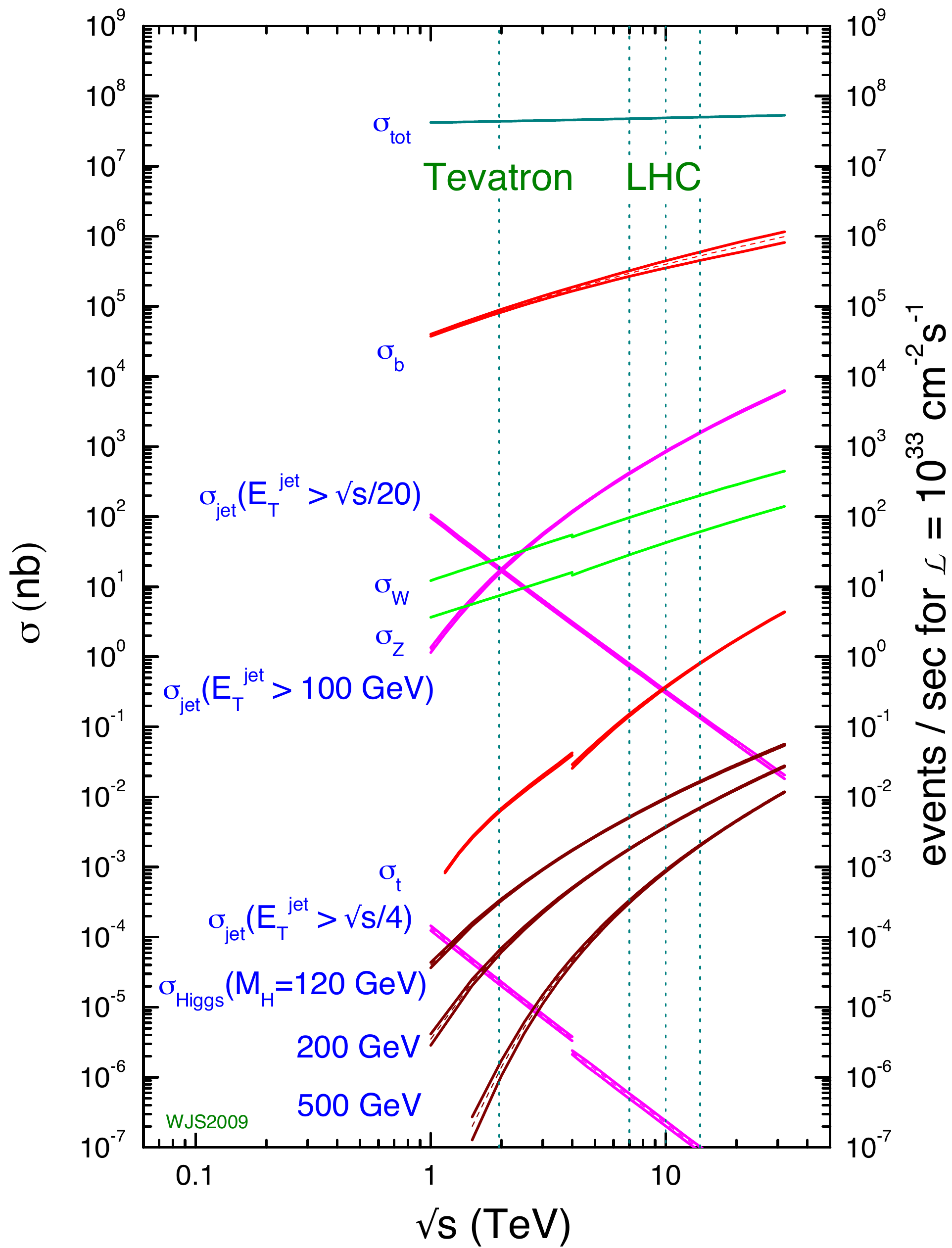}
\end{center}
\vspace*{-4mm}
\caption{Production rates for signal and background processes at
  hadron colliders. The discontinuity is due to the Tevatron being a
  proton--antiproton collider while the LHC is a proton--proton
  collider. The two colliders correspond to the $x$--axis values of
  2~TeV and something between 7~TeV and 14~TeV. Figure from
  Ref.~\cite{Campbell:2006wx}.}
\label{fig:higgs_lhcall}
\end{figure}

One LHC aspect we have to mention in the context of dark matter searches
is the \ul{trigger}. At the LHC we can only save and study a
small number of all events.  This means that we have to decide very
fast if an event has the potential of being interesting in the light
of the physics questions we are asking at the LHC; only these events
we keep. For now we can safely assume that above an energy threshold
we will keep all events with leptons or photons, plus, if at all
possible, events with missing energy, like neutrinos in the Standard
Model and dark matter particles in new physics models and jets with
high energy coming from resonance decays.\bigskip

When we search for dark matter particles at hadron colliders like the
LHC, these analyses cannot rely on our knowledge of the initial state
kinematics.  What we know is that in the transverse plane the incoming
partons add to zero three-momentum. In contrast, we are missing the
necessary kinematic information in the beam direction. This means that
dark matter searches always rely on production with another particle,
leading to un-balanced three-momenta in the plane transverse to the
beam direction.  This defines an an observable \ul{missing
  transverse momentum} three-vector with two relevant dimensions. The
missing transverse energy is the absolute value of this two-dimensional
three-vector. The big problem with missing transverse momentum is that it
relies on reconstructing the entire recoil. This causes several
experimental problems:
\begin{enumerate}
\item there will always be particles in events which are not observed
  in the calorimeters. For example, a particle can hit a support
  structure of the detector, generating fake missing energy;
\item in particular hadronic jets might not be fully reconstructed,
  leading to fake missing energy in the direction of this jet. This is
  the reason why we usually require the missing momentum vector to not
  be aligned with any hard object in an event;
\item slight mis-measurements of the momenta of each of the particles
  in an event add, approximately, in quadrature to a mis-measurement
  of the missing energy vector;
\item QCD activity from the underlying event or from pile-up which
  gets subtracted before we analyze anything. This subtraction adds to
  the error on the measured missing momentum;
\item non-functional parts of the detector automatically lead to a
  systematic bias in the missing momentum distribution.
\end{enumerate}
Altogether these effects imply that missing transverse energy below
$30~...~50$~GeV at the LHC could as well be zero. Only cuts on
$E_{T,miss} \gtrsim 100$~GeV can guarantee a significant background
rejection.\bigskip
\index{parton density function}
Next, we want to compute the production rates for dark matter
particles at the LHC. To do that we need to follow the same path as
for direct detection in Section~\ref{sec:direct}, namely link the
calculable partonic cross section to the observable hadronic cross
section. We cannot compute the energy distributions of the incoming
partons inside the colliding protons from first principles, but we can
start with the assumption that all partons move collinearly with the
surrounding proton. In that case the parton kinematics is described by
a one-dimensional probability distribution for finding a parton just
depending on the respective fraction of the proton's momentum, the
\ul{parton density function} (pdf) $f_i(x)$ with $x = 0~...~1$
and $i= u,d,c,s,g$.  This parton density itself is not an observable;
it is a distribution in the mathematical sense, which means it is only
defined when we integrate it together with a partonic cross
section. Different parton densities have very different behavior ---
for the valence quarks ($uud$) they peak somewhere around $x \lesssim
1/3$, while the gluon pdf is negligible at $x \sim 1$ and grows very
rapidly towards small $x$, $f_g(x) \propto x^{-2}$. Towards $x <
10^{-3}$ it becomes even steeper.

In addition, we can make some arguments based on symmetries and
properties of the hadrons.  For example the parton distributions
inside an anti-proton are linked to those inside a proton through the
CP symmetry, which is an exact symmetry of QCD,
\begin{align}
f^{\bar{p}}_q(x) = f_{\bar{q}}(x)\,, \qqqquad 
f^{\bar{p}}_{\bar{q}}(x) = f_q(x)\,, \qqqquad 
f^{\bar{p}}_g(x) = f_g(x)\,, 
\end{align}
for all $x$.  The proton consists of $uud$ quarks, plus quantum
fluctuations either involving gluons or quark--antiquark pairs. The
expectation values for up- and down-quarks have
to fulfill
\begin{alignat}{5}
\int_0^1 \, dx \; \left( f_u(x) - f_{\bar{u}}(x) \right) = 2
\qqqquad
\int_0^1 \, dx \; \left( f_d(x) - f_{\bar{d}}(x) \right) = 1 \; .
\end{alignat}
Finally, the proton momentum has to be the sum of
all parton momenta, defining the QCD sum rule 
\begin{align}
\langle \sum x_i \rangle = 
\int_0^1 \, dx \; x \; \left( \sum_q f_q(x) 
                            + \sum_{\bar{q}} f_{\bar{q}}(x) 
                            + f_g(x) \right) 
= 1\;.
\end{align}
We can compute this sum accounting for quarks and antiquarks.  The sum
comes out to 1/2, which means that half of the proton momentum is
carried by gluons.

Using the parton densities we can compute the \ul{hadronic
  cross section},
\begin{align}
\boxed{
\sigma_\text{tot} = \int_0^1 dx_1 \int_0^1 dx_2 \;
                   \sum_{ij} f_i(x_1) \, f_j(x_2) \;
                   \hat{\sigma}_{ij}(x_1 x_2 S)
} \; ,
\label{eq:qcd_sigtot}
\end{align}
where $i,j$ are the incoming partons. The partonic energy of the
scattering process is $s=x_1 x_2 S$ with the LHC proton energy of
currently around $\sqrt{S}=13$~TeV. The partonic cross section
includes energy--momentum conservation.\bigskip

On the parton level, the analogy to photon radiation in $e^+ e^-$
production will be dark matter production together with a quark or a
gluon. Two Feynman diagrams for this \ul{mono-jet signature}
with an unspecified mediator are\\[8mm]

\begin{center}
\begin{fmfgraph*}(80,50)
\fmfset{arrow_len}{2mm}
\fmfleft{i1,i2}
\fmfright{o1,o2,o3}
\fmf{fermion,tension=0.8,width=0.6}{i1,v1}
\fmf{fermion,tension=0.6,width=0.6}{v1,v2}
\fmf{fermion,tension=0.4,width=0.6}{v2,i2}
\fmf{gluon,tension=0.4,width=0.6}{v1,o1}
\fmf{fermion,tension=0.5,width=0.6}{v2,o3}
\fmf{fermion,tension=0.7,width=0.6}{o2,v2}
\fmflabel{$q(k_1)$}{i1}
\fmflabel{$\bar{q}$}{i2}
\fmflabel{$g(k_2)$}{o1}
\fmflabel{$\chi$}{o2}
\fmflabel{$\chi$}{o3}
\fmfblob{.2w}{v2} 
\end{fmfgraph*}
\hspace*{0.2\textwidth}
\begin{fmfgraph*}(80,50)
\fmfset{arrow_len}{2mm}
\fmfleft{i1,i2}
\fmfright{o1,o2,o3}
\fmf{gluon,tension=0.8,width=0.6}{i1,v1}
\fmf{fermion,tension=0.6,width=0.6}{v1,v2}
\fmf{fermion,tension=0.4,width=0.6}{v2,i2}
\fmf{fermion,tension=0.4,width=0.6}{o1,v1}
\fmf{fermion,tension=0.5,width=0.6}{v2,o3}
\fmf{fermion,tension=0.7,width=0.6}{o2,v2}
\fmflabel{$g(k_1)$}{i1}
\fmflabel{$\bar{q}$}{i2}
\fmflabel{$\bar{q}(k_2)$}{o1}
\fmflabel{$\chi$}{o2}
\fmflabel{$\chi$}{o3}
\fmfblob{.2w}{v2} 
\end{fmfgraph*}
\end{center} 
\vspace*{5mm}

In addition to this experimental argument there is a theoretical, QCD
argument which suggests to look for initial state radiation of a quark
or a gluon. Both of the above diagrams include an intermediate quark
or gluon propagator with the denominator
\begin{align}
\frac{1}{(k_1 - k_2)^2}
&= \frac{1}{k_1^2 - 2 k_1^0 k_2^0 + 2 (\vec{k}_1 \vec{k}_2) + k_2^2} \notag \\
&= \frac{1}{2} \; \frac{1}{|\vec{k}_1| |\vec{k}_2| \cos \theta_{12} - k_1^0 k_2^0 }
= \frac{1}{2 k_1^0 k_2^0} \; \frac{1}{\cos \theta_{12} - 1} \; .
\end{align}
This propagator diverges when the radiated parton is soft ($k_2^0 \to
0$) or collinear with the incoming parton ($\theta_{12} \to 0$).
Phenomenologically, the soft divergence is less dangerous, because the
LHC experiments can only detect any kind of particle above a certain
momentum or transverse momentum threshold. The actual pole in the
\ul{collinear divergence} \index{collinear divergence}gets absorbed into a re-definition of
the parton densities $f_{q,g}(x)$, as they appear for example in the
hadronic cross section of Eq.\eqref{eq:qcd_sigtot}. This so-called
\ul{mass factorization} is technically similar to a
renormalization procedure for example of the strong coupling, except
that renormalization absorbs ultraviolet divergences and works on the
fundamental Lagrangian level~\cite{lecture}. One effect of this
re-definition of the parton densities is that relative to the original
definition the quark and gluon densities mix, which means that the two
Feynman diagrams shown above cannot actually be separated on a
consistent quantum level.\bigskip
\index{Mandelstam variables}

Experimentally, the scattering or polar angle $\theta_{12}$ is not the
variable we actually measure. The reason is that it is not boost
invariant and that we do not know the partonic rest frame in the beam
direction. Instead, we can use two standard kinematic variables,
\begin{alignat}{5}
t &= -s \; \left( 1- \frac{m_{\chi \chi}^2}{s} \right) \; \frac{1-\cos \theta_{12}}{2} 
&\qqquad& \text{(Mandelstam variable)} \notag \\
p_T^2 &=  s \; \left( 1- \frac{m_{\chi \chi}^2}{s} \right)^2 
          \; \frac{1-\cos \theta_{12}}{2} \; \frac{1+\cos \theta_{12}}{2} 
&\qqquad& \text{(transverse momentum)} \; .
\label{eq:coll_div}
\end{alignat}
Comparing the two forms we see that the transverse momentum is
symmetric under the switch $\cos \theta_{12} \leftrightarrow - \cos
\theta_{12}$, which in terms of the Mandelstam variables corresponds
to $t \leftrightarrow u$. From Eq.\eqref{eq:coll_div} we see that the
collinear divergence appears as a divergence of the partonic
transverse momentum distribution,
\begin{align}
\frac{d \sigma_{\chi \chi j}}{d p_{T,j}} 
\propto | \mat_{\chi \chi j} |^2
\propto \frac{1}{t} \propto \frac{1}{p_{T,j}^2} \; .
\end{align}
An obvious question is whether this divergence is integrable, \ie if it leads
to a finite cross section $\sigma_{\chi \chi j}$. We can approximate
the phase space integration in the collinear regime using an
appropriate constant $C$ to write
\begin{align}
\sigma_{\chi \chi j} 
\approx \int_{p_{T,j}^\text{min}}^{p_{T,j}^\text{max}} d p_{T,j}^2 \frac{C}{p_{T,j}^2}
= 2 \int_{p_{T,j}^\text{min}}^{p_{T,j}^\text{max}} d p_{T,j} \frac{C}{p_{T,j}}
= 2 C \; \log \frac{p_{T,j}^\text{max}}{p_{T,j}^\text{min}} \; .
\label{eq:coll_log}
\end{align}
For an integration of the full phase space including a lower limit
$p_{T,j}^\text{min} = 0$, this logarithm is divergent. When we apply
an experimental cut to generate for example a value of
$p_{T,j}^\text{min} = 10$~GeV, the logarithm gets large, because
$p_{T,j}^\text{max} \gtrsim 2 m_\chi$ is given by the typical energy
scales of the scattering process.  When we absorb the collinear
divergence into re-defined parton densities and use the parton shower
to enforce and simulate the correct behavior
\begin{align}
\frac{d \sigma_{\chi \chi j}}{d p_{T,j}} \stackrel{p_{T,j} \to 0}{\longrightarrow} 0 \; ,
\end{align}
the large collinear logarithm in Eq.\eqref{eq:coll_log} gets re-summed
to all orders in perturbation theory. However, over a wide range of
values the transverse momentum distribution inherits the collinearly
divergent behavior. This means that most jets radiated from incoming
partons appear at small transverse momenta, and even after including
the parton shower regulator the collinear logarithm significantly
enhances the probability to radiate such collinear jets. The same is
(obviously) true for the initial state radiation of photons. The main
difference is that for the photon process we can neglect 
the amplitude with an initial state photon due to the small photon parton
density.\bigskip

Once we know that at the LHC we can generally look for the production
of dark matter particles with an initial state radiation object, we
can study different \ul{mono-$X$} channels. \index{mono-X} Some example
Feynman diagrams for mono-jet, mono-photon, and mono-$Z$ production
are\\[6mm]

\begin{center}
\vspace{.5cm}
\begin{fmfgraph*}(80,50)
\fmfset{arrow_len}{2mm}
\fmfleft{i1,i2}
\fmfright{o1,o2,o3}
\fmf{fermion,tension=0.8,width=0.6}{i1,v1}
\fmf{fermion,tension=0.6,width=0.6}{v1,v2}
\fmf{fermion,tension=0.4,width=0.6}{v2,i2}
\fmf{gluon,tension=0.4,width=0.6}{v1,o1}
\fmf{fermion,tension=0.5,width=0.6}{v2,o3}
\fmf{fermion,tension=0.7,width=0.6}{o2,v2}
\fmflabel{$q$}{i1}
\fmflabel{$\bar{q}$}{i2}
\fmflabel{$g$}{o1}
\fmflabel{$\chi$}{o2}
\fmflabel{$\chi$}{o3}
\fmfblob{.2w}{v2} 
\end{fmfgraph*}
\hspace*{0.1\textwidth}
\begin{fmfgraph*}(80,50)
\fmfset{arrow_len}{2mm}
\fmfleft{i1,i2}
\fmfright{o1,o2,o3}
\fmf{fermion,tension=0.8,width=0.6}{i1,v1}
\fmf{fermion,tension=0.6,width=0.6}{v1,v2}
\fmf{fermion,tension=0.4,width=0.6}{v2,i2}
\fmf{photon,tension=0.4,width=0.6}{v1,o1}
\fmf{fermion,tension=0.5,width=0.6}{v2,o3}
\fmf{fermion,tension=0.7,width=0.6}{o2,v2}
\fmflabel{$q$}{i1}
\fmflabel{$\bar{q}$}{i2}
\fmflabel{$\gamma$}{o1}
\fmflabel{$\chi$}{o2}
\fmflabel{$\chi$}{o3}
\fmfblob{.2w}{v2} 
\end{fmfgraph*}
\hspace*{0.1\textwidth}
\begin{fmfgraph*}(80,50)
\fmfset{arrow_len}{2mm}
\fmfleft{i1,i2}
\fmfright{o0,o1,o2,o3}
\fmf{fermion,tension=0.8,width=0.6}{i1,v1}
\fmf{fermion,tension=0.6,width=0.6}{v1,v2}
\fmf{fermion,tension=0.4,width=0.6}{v2,i2}
\fmf{photon,tension=0.4,width=0.6,label=$Z$}{v1,v3}
\fmf{fermion,tension=0.5,width=0.6}{v2,o3}
\fmf{fermion,tension=0.7,width=0.6}{o2,v2}
\fmf{fermion,tension=0.8,width=0.6}{o0,v3}
\fmf{fermion,tension=0.6,width=0.6}{v3,o1}
\fmflabel{$q$}{i1}
\fmflabel{$\bar{q}$}{i2}
\fmflabel{$\bar{f}$}{o0}
\fmflabel{$f$}{o1}
\fmflabel{$\chi$}{o2}
\fmflabel{$\chi$}{o3}
\fmfblob{.2w}{v2} 
\end{fmfgraph*}
\end{center} 
\medskip

For the radiated $Z$-boson we need to specify a decay. While hadronic
decays $Z \to q\bar{q}$ come with a large branching ratio, we need to
ask what they add to the universal mono-jet signature. Leptonic decays
like $Z \to \mu \mu$ can help in difficult experimental environments,
but are suppressed by a branching ratio of 3.4\% per lepton
generation. Mono-$W$ events can occur through initial state radiation
when we use a $q \bar{q}'$ initial state to generate a hard $q\bar{q}$
scattering. Finally, mono-Higgs signatures obviously make no sense for
initial state radiation. From the similarity of the above Feynman
diagrams we can first assume that at least in the limit $m_Z \to 0$
the total rates for the different mono-$X$ processes relative to the
mono-jet rate scale like
\begin{align}
\frac{\sigma_{\chi \chi \gamma}}{\sigma_{\chi \chi j}} 
&\approx \frac{\alpha}{\alpha_s} \frac{Q_q^2}{C_F}
\approx \frac{1}{40} \notag \\
\frac{\sigma_{\chi \chi \mu \mu}}{\sigma_{\chi \chi j}} 
&\approx \frac{\alpha}{\alpha_s} \; \frac{Q_q^2\,s_w^2}{C_F} \; \br(Z \to \mu \mu)
\approx \frac{1}{4000} \; .
\label{eq:monox_scaling}
\end{align}
The actual suppression of the mono-$Z$ channel is closer to $10^{-4}$,
once we include the $Z$-mass suppression through the available phase
space. In addition, the similar Feynman diagrams also suggest that any
kinematic $x$-distribution scales like
\begin{align}
\frac{1}{\sigma_{\chi \chi j}} \; \frac{d \sigma_{\chi \chi g}}{d x}
\approx \frac{1}{\sigma_{\chi \chi \gamma}} \; \frac{d \sigma_{\chi \chi \gamma}}{d x}
\approx \frac{1}{\sigma_{\chi \chi ff}} \; \frac{d \sigma_{\chi \chi ff}}{d x}
\; .
\end{align}
Here, the suppression of the mono-photon is stronger, because the
rapidity coverage of the detector for jets extends to $|\eta| < 4.5$,
while photons rely on an efficient electromagnetic calorimeter with
$|\eta| <2.5$. On the other hand, photons can be detected to
significantly smaller transverse momenta than jets.

Note that the same scaling as in Eq.\eqref{eq:monox_scaling} applies
to the leading mono-$X$ backgrounds, namely
\begin{align}
pp \to Z_{\nu \nu} X 
\qqquad \text{with} \quad X=j,\gamma,Z \; ,
\end{align}
possibly with the exception of mono-$Z$ production, where the hard
process and the collinear radiation are now both described by
$Z$-production. This means that the signal scaling of
Eq.\eqref{eq:monox_scaling} also applies to backgrounds,
\begin{align}
\frac{\sigma_{\nu \nu \gamma}}{\sigma_{\nu \nu j}} 
&\approx \frac{\alpha}{\alpha_s}  \frac{Q_q^2}{C_F}
\approx \frac{1}{40} \notag \\
\frac{\sigma_{\nu \nu \mu \mu}}{\sigma_{\nu \nu j}} 
&\approx \frac{\alpha}{\alpha_s} \;  \frac{Q_q^2\,s_w^2}{C_F} \; \br(Z \to \mu \mu)
\approx \frac{1}{4000} \; .
\label{eq:monox_scaling2}
\end{align}
If our discovery channel is \ul{statistics limited}, the
significances $n_\sigma$ for the different channels are given in terms
of the luminosity, efficiencies, and the cross sections
\begin{align}
n_{\sigma, j} 
=  \sqrt{\epsilon_j \lumi} \; 
  \frac{\sigma_{\chi \chi j} }{\sqrt{\sigma_{\nu \nu j}}} 
\qquad \Rightarrow \qquad 
n_{\sigma, \gamma} 
=&  \sqrt{\epsilon_\gamma \lumi} \; 
  \frac{\sigma_{\chi \chi \gamma} }{\sqrt{\sigma_{\nu \nu \gamma}}} \notag \\
\approx&  \sqrt{\epsilon_j \lumi} \; \frac{1}{\sqrt{40}} \; 
  \sqrt{ \frac{\epsilon_\gamma}{\epsilon_j}} \;
  \frac{\sigma_{\chi \chi j} }{\sqrt{\sigma_{\nu \nu j}}}
= \frac{1}{6.3} \; \sqrt{ \frac{\epsilon_\gamma}{\epsilon_j}} \;
  n_{\sigma, j} \; .
\label{eq:monox_stat}
\end{align}
\index{systematics}
Unless the efficiency correction factors, including acceptance cuts
and cuts rejecting other backgrounds, point towards a very significant
advantage if the mono-photon channel, the mono-jet channel will be the
most promising search strategy. Using the same argument, the factor in
the expected mono-jet and mono-$Z$ significances will be around
$\sqrt{6000} = 77$.

This estimate might change if the uncertainties are
dominated by \ul{systematics or a theory uncertainty}. These
errors scale proportional to the number of background events in the
signal region, again with a signature-dependent proportionality factor
$\epsilon$ describing how well we know the background
distributions. This means for the significances
\begin{align}
n_{\sigma, \gamma} 
= \epsilon_\gamma \frac{\sigma_{\chi \chi \gamma}}{\sigma_{\nu \nu \gamma}}
= \frac{\epsilon_\gamma}{\epsilon_j} \; 
  \epsilon_j \frac{\sigma_{\chi \chi j}}{\sigma_{\nu \nu j}} 
= \frac{\epsilon_\gamma}{\epsilon_j} \; n_{\sigma, j} \; .
\label{eq:monox_syst}
\end{align}
Typically, we understand photons better than jets, both experimentally
and theoretically. On the other hand, systematic and theory
uncertainties at the LHC are usually limited by the availability and
the statistics in \ul{control regions}, regions which we can
safely assume to be described by the Standard Model.\bigskip

\begin{figure}[b!]
\begin{center}
  \includegraphics[width=0.50\textwidth]{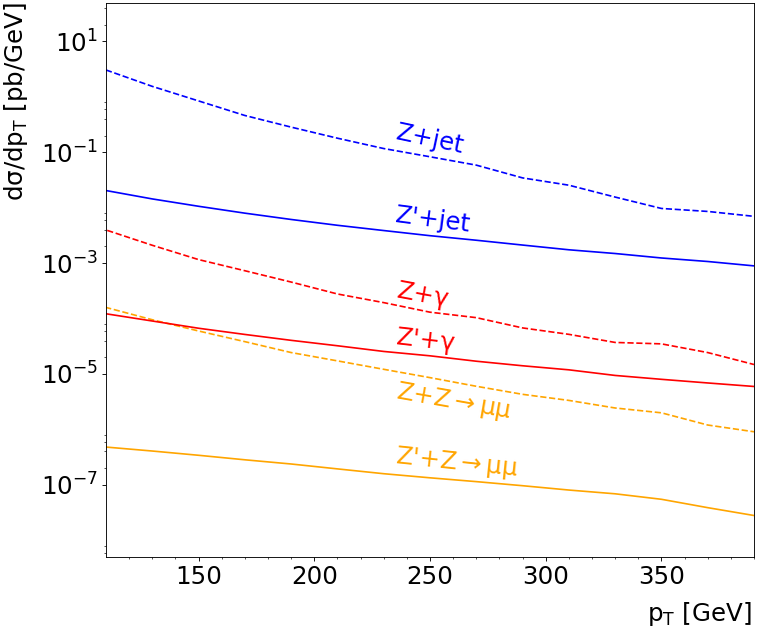}
\end{center}
\vspace*{-3mm}
\caption{Transverse momentum spectrum for signals and backgrounds in
  the different mono-$X$ channels for a heavy vector mediator with
  $m_{Z'} = 1$~TeV.  Figure from Ref.~\cite{Bernreuther:2018nat}.}
\label{fig:jan}
\end{figure}

We can simulate mono-$X$ signatures for \ul{vector mediators},
described in Section~\ref{sec:indirect_simp}. In that case the three
mono-$X$ signatures are indeed induced by initial state radiation.
The backgrounds are dominated by $Z$-decays to neutrinos.  The
corresponding LHC searches are based on the missing transverse
momentum distribution and the transverse momentum $p_{T,X}$ of the
mono-$X$ object. There are (at least) two strategies to control for
example the mono-jet background: first, we can measure it for example
using $Z \to \mu^+ \mu^-$ decays or hard photons produced in
association with a hard jet. Second, if the dark matter signal is
governed by a harder energy scale, like the mass of a heavy mediator,
we can use the low-$p_T$ region as a control region and only
extrapolate the $p_T$ distributions.

Figure~\ref{fig:jan} gives an impression of the transverse momentum
spectra in the mono-jet, mono-photon, and mono-$Z$ channels. Comparing
the mono-jet and mono-photon rates we see that the shapes of the
transverse momentum spectra of the jet or photon, recoiling against
the dark matter states, are essentially the same in both cases, for
the respective signals as well as for the backgrounds. The signal and
background rates follow the hierarchy derived above.  Indeed, the
mono-photon hardly adds anything to the much larger mono-jet channel,
except for cases where in spite of advanced experimental strategies
the mono-jet channel is limited by systematics. The mono-$Z$ channel
with a leptonic $Z$-decay is kinematically almost identical to the
other two channels, but with a strongly reduced rate. This means that for
mono-$X$ signatures induced by initial state radiation the leading
mono-jet channel can be expected to be the most useful, while other
mono-$X$ analyses will only become interesting when the production
mechanism is not initial state radiation.\bigskip

Finally, one of the main challenges of mono-$X$ signatures is that by
definition the mediator has to couple to the Standard Model and to
dark matter. This means for example in the case of the simple model of
Eq.\eqref{eq:s_model}
\begin{align}
\br(V \to q\bar{q}) + \br(V \to \chi \chi) = 100\% \; .
\end{align}
The relative size of the branching ratios is given by the ratio of
couplings $g_\chi^2/g_u^2$. Instead of the mono-$X$ signature we can
constrain part of the model parameter space through resonance searches
with the same topology as the mono-$X$ search and without requiring
a hard jet,

\begin{center}
\begin{fmfgraph*}(80,50)
\fmfset{arrow_len}{2mm}
\fmfleft{i1,i2}
\fmfright{o1,o2,o3}
\fmf{fermion,tension=0.8,width=0.6}{i1,v1}
\fmf{fermion,tension=0.6,width=0.6}{v1,v2}
\fmf{fermion,tension=0.4,width=0.6}{v2,i2}
\fmf{photon,tension=0.7,lab.side=left,label=$V$,width=0.6}{v2,v3}
\fmf{gluon,tension=0.4,width=0.6}{v1,o1}
\fmf{fermion,tension=0.5,width=0.6}{v3,o3}
\fmf{fermion,tension=0.7,width=0.6}{o2,v3}
\fmflabel{$q$}{i1}
\fmflabel{$\bar{q}$}{i2}
\fmflabel{$g$}{o1}
\fmflabel{$\bar{q}$}{o2}
\fmflabel{$q$}{o3}
\end{fmfgraph*}
\hspace*{0.1\textwidth}
\begin{fmfgraph*}(80,50)
\fmfset{arrow_len}{2mm}
\fmfleft{i1,i2}
\fmfright{o1,o2,o3}
\fmf{fermion,tension=0.8,width=0.6}{i1,v2}
\fmf{fermion,tension=0.4,width=0.6}{v2,i2}
\fmf{photon,tension=0.7,lab.side=left,label=$V$,width=0.6}{v2,v3}
\fmf{fermion,tension=0.5,width=0.6}{v3,o3}
\fmf{fermion,tension=0.7,width=0.6}{o2,v3}
\fmflabel{$q$}{i1}
\fmflabel{$\bar{q}$}{i2}
\fmflabel{$\bar{q}$}{o2}
\fmflabel{$q$}{o3}
\end{fmfgraph*}
\end{center}
\index{mediator!s-channel}
On the other hand, for the parameter space $g_u \ll g_\chi$ but constant $g_u g_\chi$ and mediator mass, the impact of resonance searches is reduced, whereas mono-X searches remain relevant. 
\subsection{Higgs portal}
\label{sec:coll_portal}
\index{Higgs!portal}
In addition to the very general mono-jet searches for dark matter, we
will again look at our two specific models.  The Higgs portal model
only introduces one more particle, a heavy scalar with $m_S \gg m_H$
and only coupling to the Higgs. This means that the Higgs has to act
as an \ul{$s$-channel mediator} not only for dark matter
annihilation, but also for LHC production,
\begin{align}
pp \to H^* \to SS +\text{jets} \; .
\end{align}
The Higgs couples to gluons in the incoming protons through a top
loop, which implies that its production rate is very small. The
Standard Model predicts an on-shell Higgs rate of 50~pb for gluon
fusion production at a 14~TeV LHC. Alternatively, we can look for
weak-boson-fusion off-shell Higgs production, \ie production in
association with two forward jets. The corresponding Feynman diagram
is\\[12mm]

\begin{center}
\begin{fmfgraph*}(120,70)
\fmfset{arrow_len}{2mm}
\fmfleft{i1,i2}
\fmfright{o1,o2,o3,o4}
\fmf{fermion,tension=0.6,width=0.6}{i1,v1}
\fmf{fermion,tension=0.6,width=0.6}{v1,o1}
\fmf{fermion,tension=0.6,width=0.6}{i2,v2}
\fmf{fermion,tension=0.6,width=0.6}{v2,o4}
\fmf{photon,tension=0.6,label=$W$,width=0.6}{v1,v3}
\fmf{photon,tension=0.6,label=$W$,width=0.6}{v2,v3}
\fmf{dashes,tension=0.6,label=$H$,width=0.6}{v3,v4}
\fmf{dashes,tension=0.6,width=0.6}{v4,o2}
\fmf{dashes,tension=0.6,width=0.6}{v4,o3}
\fmflabel{$q$}{i1}
\fmflabel{$q$}{i2}
\fmflabel{$q$}{o1}
\fmflabel{$S$}{o2}
\fmflabel{$S$}{o3}
\fmflabel{$q$}{o4}
\end{fmfgraph*}
\end{center}

These so-called tagging jets will allow us to trigger the events. For
an on-shell Higgs boson the weak boson fusion cross section at the LHC
is roughly a factor 1/10 below gluon fusion, and its advantages are
discussed in detail in Ref.~\cite{lecture}.

In particular in this \ul{weak-boson-fusion channel} ATLAS and
CMS are conducting searches for invisibly decaying Higgs bosons. The
main backgrounds are invisible $Z$-decays into a pair of neutrinos,
and $W$-decays where we miss the lepton and are only left with one
neutrino. For high luminosities around $3000~\ifb$ and assuming an
essentially unchanged Standard Model Higgs production rate, the LHC
will be sensitive to invisible branching ratios around
\begin{align}
\br(H \to \text{invisible}) = (2-3)\% \; .
\end{align}
The key to this analysis is to understand not only the tagging jet
kinematics, but also the central jet radiation between the two forward
tagging jets.\bigskip

Following the discussion
in Section~\ref{sec:models_portal} the partial width for the SM Higgs
boson decays into light dark matter is
\begin{align}
\Gamma (H \to SS) 
= \frac{\lambda_3^2 v_H^2}{32 \pi M_H} \; \sqrt{1 - \frac{4 m_S^2}{m_H^2} }
\qquad \Leftrightarrow \qquad 
\frac{\Gamma (H \to SS)}{m_H} 
\approx \frac{\lambda_3^2}{8 \pi} \; 
\left( 1 - \frac{2 m_S^2}{m_H^2} \right)
< \frac{\lambda_3^2}{8 \pi} \; .
\end{align}
This value has to be compared to the Standard Model prediction
$\Gamma_H / m_H = 4 \cdot 10^{-5}$. For example, a 10\% invisible
branching ratio $\br(H\to SS)$ into very light scalars $m_S \ll
m_H/2$ corresponds to a portal coupling
\begin{align}
\frac{\lambda_3^2}{8 \pi}  = 4 \cdot 10^{-6}
\qquad \Leftrightarrow \qquad
\lambda_3 = \sqrt{32 \pi} \cdot 10^{-3} \approx 10^{-2} \; .
\label{eq:inv_higgs_portal}
\end{align}
The light scalar reference point in agreement with the observed relic
density Eq.\eqref{eq:ref_portal_2} has $\lambda_3= 0.3$ and roughly
assuming $m_S \lesssim 50$~GeV.  This is well above the approximate
final reach for the invisible Higgs branching ratio at the
high-luminosity LHC.

For \ul{larger dark matter masses} above $m_S = 200$~GeV the
LHC cross section for pair production in weak boson fusion is tiny,
namely
\begin{align}
\sigma(SSjj) 
\approx \frac{\lambda_3^2}{10}~\fb
\stackrel{\lambda_3 = 0.1}{=} 10^{-3}~\fb
\end{align}
Without going into much detail this means that heavy scalar dark
matter is unlikely to be discovered at the LHC any time soon,
because the final state is heavy and the coupling to the Standard
Model is strongly constrained through the observed relic density.

\subsection{Supersymmetric neutralinos}
\label{sec:coll_mssm}

\index{neutralino}

The main feature of supersymmetry is that it is not just a theory
predicting a dark matter particle, it is a complete, renormalizable 
\ul{ultraviolet completion} of the Standard Model valid to the
Planck scale. From Section~\ref{sec:models_mssm} we know that the MSSM
and the NMSSM offer a wide variety of particles, including messengers
linking the visible matter and dark matter sectors. Obviously, the
usual mono-$X$ signatures from Section~\ref{sec:coll_hadron} or the
invisible Higgs decays from Section~\ref{sec:coll_portal} will also
appear in supersymmetric models. For example SM-like Higgs decays into
a pair of light neutralinos can occur for a mixed gaugino-higgsino LSP
with $M_1 \lesssim |\mu| \lesssim 100$~GeV. An efficient annihilation
towards the observed relic density goes through an $s$-channel
$Z$-mediator coupling to the higgsino fraction. Here we can find
\begin{align}
\br(h \to \nne \nne) = (10~...~50)\%
\qqquad 
m_{\nne} = (35~...~40)~\gev 
\quad \text{and} \quad  
(50~...~55)~\gev \; ,
\end{align}
mostly constrained by direct detection.  On the other hand,
supersymmetric models offer many more dark matter signatures and provide a UV completion to a number of different simplified models. \index{simplified model}
They are often linked
to generic features of heavy new particle production, which is what we
will focus on below.

If our signature consists of a flexible number of visible and
invisible particles we rely on global observables. The
\ul{visible mass} is based on the assumption that we are
looking for the decay of two heavy new states, where the parton
densities will ensure that these two particles are produced close to
threshold. We can then approximate the partonic energy $\sqrt{\hat{s}}
\sim m_1 + m_2$ by some kind of visible energy.  Without taking into
account missing energy and just adding leptons $\ell$ and jets $j$ the
visible mass looks like
\begin{align}
m^2_\text{visible} = \left[ \sum_{\ell,j} \; E \right]^2 
                   - \left[ \sum_{\ell,j} \; \vec{p} \right]^2 \; .
\end{align}
Similarly, Tevatron and LHC experiments have for a long time used an
effective transverse mass scale which is usually evaluated for jets
only, but can trivially be extended to leptons,
\begin{align}
H_T = \sum_{\ell,j} \; E_T = \sum_{\ell,j} \; p_T  \; ,
\end{align}
assuming massless final state particles.  In an alternative definition
of $H_T$ we sum over a number of jets plus the missing energy and skip
the hardest jet in this sum. Obviously, we can add the missing
transverse momentum to this sum, giving us
\begin{align}
 m_\text{eff} = \sum_{\ell,j,\text{miss}} \; E_T 
             = \sum_{\ell,j,\text{miss}} \; p_T  \; .
\end{align}
This effective mass is known to trace the mass of the heavy new
particles decaying for example to jets and missing energy. This
interpretation relies on the non--relativistic nature of the production
process and our confidence that all jets included are really decay
jets.\bigskip

In the Standard Model the neutrino produces such missing transverse
energy, typically through the decays $W \to \ell^+ \nu$ and $Z \to \nu
\bar{\nu}$.  In $W+$~jets events we can learn how to reconstruct the
$W$ mass from one observed and one missing particle. We construct a
\ul{transverse mass} in analogy to an invariant mass, but
neglecting the longitudinal momenta of the decay products
\begin{align}
m_T^2 &= \left( E_{T,\text{miss}} + E_{T, \ell} \right)^2
       - \left( \vec{p}_{T,\text{miss}} + \vec{p}_{T,\ell} \right)^2 
\notag \\
      &= m_{\ell}^2 + m_\text{miss}^2
       + 2 \left( E_{T,\ell} E_{T,\text{miss}} 
       - \vec{p}_{T,\ell} \cdot \vec{p}_{T,\text{miss}} \right) \; ,
\label{eq:sig_mt}
\end{align}
in terms of a transverse energy $E_T^2 = \vec{p}_T^2 + m^2$.  By
definition, it is invariant under --- or better independent of ---
longitudinal boosts.  Moreover, as the projection of the invariant
mass onto the transverse plane it is also invariant under transverse
boosts. The transverse mass is always smaller than the actual mass and
reaches this limit for a purely transverse momentum direction, which
means that we can extract $m_W$ from the upper endpoint in the $m_{T,W}$
distribution. To reject Standard Model backgrounds we can simply
require $m_T > m_W$.\bigskip

The first supersymmetric signature we discuss makes use of the fact
that already the neutralino-chargino sector involves six particles,
four neutral and two charged. Two LHC processes reflecting this
structure are
\begin{alignat}{5}
pp &\to \nnz \nne 
   &&\to ( \ell^+ \ell^- \nne ) \; \nne \notag \\
pp &\to \cpe \cme
   &&\to ( \ell^+ \nu_\ell \nne ) \; ( \ell^- \bar{\nu}_\ell \nne) \; .
\label{eq:lhc_channels1}
\end{alignat}
The leptons in the decay of the heavier neutralinos and charginos can
be replaced by other fermions. Kinematically, the main question is if
the fermions arise from on-shell gauge bosons or from intermediate
supersymmetric scalar partners of the leptons. The corresponding Feynman
diagrams for the first of the two above processes are\\[8mm]

\begin{center}
\begin{fmfgraph*}(120,70)
\fmfset{arrow_len}{2mm}
\fmfleft{i1,i2}
\fmfright{o1,o2,o3,o4}
\fmf{fermion,tension=0.6,width=0.6}{i1,v1}
\fmf{fermion,tension=0.6,width=0.6}{v1,i2}
\fmf{photon,tension=0.6,width=0.6}{v1,v2}
\fmf{plain,tension=0.2,width=0.6}{v2,o4}
\fmf{photon,tension=0.2,width=0.6}{v2,o4}
\fmf{plain,tension=0.2,lab.side=left,label=$\nnz$,width=0.6}{v3,v2}
\fmf{photon,tension=0.2,width=0.6}{v3,v2}
\fmf{photon,tension=0.2,lab.side=right,label=$Z$,width=0.6}{v3,v4}
\fmf{plain,tension=0.1,width=0.6}{v3,o3}
\fmf{photon,tension=0.1,width=0.6}{v3,o3}
\fmf{fermion,tension=0.2,width=0.6}{v4,o1}
\fmf{fermion,tension=0.2,width=0.6}{o2,v4}
\fmflabel{$q$}{i1}
\fmflabel{$\bar{q}$}{i2}
\fmflabel{$\ell^-$}{o1}
\fmflabel{$\ell^+$}{o2}
\fmflabel{$\nne$}{o3}
\fmflabel{$\nne$}{o4}
\end{fmfgraph*}
\hspace*{0.1\textwidth}
\begin{fmfgraph*}(120,70)
\fmfset{arrow_len}{2mm}
\fmfleft{i1,i2}
\fmfright{o1,o2,o3,o4}
\fmf{fermion,tension=0.6,width=0.6}{i1,v1}
\fmf{fermion,tension=0.6,width=0.6}{v1,i2}
\fmf{photon,tension=0.6,width=0.6}{v1,v2}
\fmf{plain,tension=0.2,width=0.6}{v2,o4}
\fmf{photon,tension=0.2,width=0.6}{v2,o4}
\fmf{plain,tension=0.2,lab.side=left,label=$\nnz$,width=0.6}{v3,v2}
\fmf{photon,tension=0.2,width=0.6}{v3,v2}
\fmf{scalar,tension=0.2,lab.side=left,label=$\tilde{\ell}$,width=0.6}{v4,v3}
\fmf{fermion,tension=0.1,width=0.6}{v3,o3}
\fmf{fermion,tension=0.1,width=0.6}{v4,o1}
\fmf{photon,tension=0.1,width=0.6}{v4,o1}
\fmf{fermion,tension=0.2,width=0.6}{o2,v4}
\fmflabel{$q$}{i1}
\fmflabel{$\bar{q}$}{i2}
\fmflabel{$\ell^-$}{o2}
\fmflabel{$\ell^+$}{o3}
\fmflabel{$\nne$}{o4}
\fmflabel{$\nne$}{o1}
\end{fmfgraph*}
\end{center} 
\medskip

The question which decay topologies of the heavier
neutralino dominate, depends on the point in parameter space. The
first of the two diagrams predicts dark matter production in
association with a $Z$-boson. This is the same signature as found to
be irrelevant for initial state radiation in \index{endpoint methods}
Section~\ref{sec:coll_hadron}, namely \ul{mono-$Z$ production}.

The second topology brings us the question how many masses we can
extract from two observed external momenta. \ul{Endpoint
  methods} rely on lower (threshold) and upper (edge) kinematic
endpoints of observed invariant mass distributions.  The art is to
identify distributions where the endpoint is probed by realistic phase
space configurations. The most prominent example is $m_{\ell \ell}$ in
the heavy neutralino decay in Eq.\eqref{eq:lhc_channels1}, proceeding
through an on-shell slepton.  In the rest frame of the intermediate
slepton the $2\to 2$ process corresponding to the decay of the heavy
neutralino,
\begin{align}
\nnz \ell^- \to \tilde{\ell} \to \nne \ell^-
\end{align}
resembles the Drell--Yan process. Because of the scalar in the
$s$-channel, angular correlations do not influence the $m_{\ell \ell}$
distribution, so it will have a triangular shape. Its upper limit or
edge can be computed in the slepton rest frame.  The incoming and
outgoing three-momenta have the absolute values
\begin{align}
|\vec{p}| 
= \frac{|m_{\tilde{\chi}^0_{1,2}}^2 - m_{\tilde{\ell}}^2|}{2 m_{\tilde{\ell}}} \; ,
\end{align}
assuming $m_\ell = 0$. The invariant mass of the two leptons reaches
its maximum if the two leptons are back--to--back and the scattering
angle is $\cos \theta = -1$
\begin{align}
m_{\ell \ell}^2 
&= (p_{\ell^+} + p_{\ell^-})^2 \notag \\
&= 2 \, \left( E_{\ell^+} E_{\ell^-}
             - |\vec{p}_{\ell^+}| |\vec{p}_{\ell^-}| \cos \theta \right) \notag \\
&< 2 \, \left( E_{\ell^+} E_{\ell^-}
             + |\vec{p}_{\ell^+}| |\vec{p}_{\ell^-}| \right) \notag \\
&= 4 \, \frac{m_{\tilde{\chi}^0_2}^2 - m_{\tilde{\ell}}^2}{2 m_{\tilde{\ell}}}
     \; \frac{m_{\tilde{\ell}}^2 - m_{\tilde{\chi}^0_1}^2}{2 m_{\tilde{\ell}}} 
\qqqquad \text{using} \quad E_{\ell^\pm}^2 = \vec{p}_{\ell^\pm}^2 \; .
\end{align}
The kinematic endpoint is then given by
\begin{align}
0 < m^2_{\ell\ell} 
  < \frac{(m_{\tilde{\chi}^0_2}^2-m_{\tilde{\ell}}^2)
          (m_{\tilde{\ell}}^2-m_{\tilde{\chi}^0_1}^2)}{m_{\tilde{\ell}}^2} \; .
\label{eq:sig_mll}
\end{align}
A generic feature or all methods relying on decay kinematics is that
it is easier to constrain the differences of squared masses than the
absolute mass scale. This is because of the form of the endpoint
formulas, which involve the difference of mass squares $m_1^2 - m_2^2
= (m_1+m_2)(m_1-m_2)$. This combination is much more sensitive to
$(m_1-m_2)$ than it is to $(m_1+m_2)$. The common lore that kinematics
only constrain mass differences is not true for two body decays, but
mass differences are indeed easier.\bigskip

The second set of supersymmetric dark matter signatures involves the
same extended dark matter sector with its neutralino and chargino
spectra or a slepton. Because the slepton and the chargino are
electrically charged, they can be produced through a photon mediator,
\begin{alignat}{5}
pp &\to \tilde{\ell} \tilde{\ell}^* 
   &&\to ( \ell^- \nne ) \; ( \ell^+ \nne) \notag \\
pp &\to \cpe \cme
   &&\to ( \pi^+ \nne ) \; ( \pi^- \nne) \; .
\label{eq:lhc_channels2}
\end{alignat}
For the slepton case one of the corresponding Feynman diagrams is\\[8mm]

\begin{center}
\begin{fmfgraph*}(120,70)
\fmfset{arrow_len}{2mm}
\fmfleft{i1,i2}
\fmfright{o1,o2,o3,o4}
\fmf{fermion,tension=0.6,width=0.6}{i1,v1}
\fmf{fermion,tension=0.6,width=0.6}{v1,i2}
\fmf{photon,tension=0.6,label=$\gamma$,width=0.6}{v1,v2}
\fmf{scalar,tension=0.4,label=$\tilde{\ell}$,width=0.6}{v2,v4}
\fmf{scalar,tension=0.4,lab.side=right,label=$\tilde{\ell}^*$,width=0.6}{v3,v2}
\fmf{fermion,tension=0.6,width=0.6}{v4,o1}
\fmf{photon,tension=0.2,width=0.6}{v4,o2}
\fmf{plain,tension=0.2,width=0.6}{v4,o2}
\fmf{photon,tension=0.2,width=0.6}{v3,o3}
\fmf{plain,tension=0.2,width=0.6}{v3,o3}
\fmf{fermion,tension=0.6,width=0.6}{o4,v3}
\fmflabel{$q$}{i1}
\fmflabel{$\bar{q}$}{i2}
\fmflabel{$\ell^-$}{o1}
\fmflabel{$\nne$}{o2}
\fmflabel{$\nne$}{o3}
\fmflabel{$\ell^+$}{o4}
\end{fmfgraph*}
\end{center}

Again, the question arises how many masses we can extract from the
measured external momenta. For this topology the \ul{variable
  $m_{T2}$} generalizes the transverse mass known from $W$ decays to
the case of two massive invisible particles, one from each leg of the
event. First, we divide the observed missing energy in the event into
two scalar fractions $p_{T,\text{miss}} = q_1 + q_2$. Then, we
construct the transverse mass for each side of the event, assuming
that we know the invisible particle's mass or scanning over
hypothetical values $\hat{m}_\text{miss}$.

Inspired by the transverse mass in Eq.\eqref{eq:sig_mt} we are
interested in a mass variable with a well--defined upper endpoint. For
this purpose we construct some kind of minimum of $m_{T,j}$ as a
function of the fractions $q_j$.  We know that maximizing the
transverse mass on one side of the event will minimize it on the other
side, so we define
\begin{align}
m_{T2}(\hat{m}_\text{miss})
     = \min_{p_{T,\text{miss}} = q_1 + q_2}
       \left[ \max_j \; m_{T,j}(q_j;\hat{m}_\text{miss}) \right] \; .
\end{align} 
We can show that by construction the transverse mass fulfills
\begin{align}
m_\text{light} + \hat{m}_\text{miss} &< m_{T2}(\hat{m}_\text{miss})  
\notag \\
m_\text{light} + m_\text{miss} &< m_{T2}(m_\text{miss}) 
                                  < m_\text{heavy} \; .
\end{align}
For the correct value of $m_\text{miss}$ the $m_{T2}$ distribution has
a sharp edge at the mass of the decaying particle. In favorable cases
$m_{T2}$ allows the measurement of both, the decaying particle and the
invisible particle masses. These two aspects for the
correct value $\hat{m}_\text{miss} = m_\text{miss}$ we can see in
Figure~\ref{fig:sim_mt2}: the lower threshold is indeed given by
$m_{T2} - m_{\tilde{\chi}^0_1} = m_\pi$, while the upper edge of
$m_{T2} - m_{\tilde{\chi}^0_1}$ coincides with the dashed line for
$m_{\tilde{\chi}^+_1} - m_{\tilde{\chi}^0_1}$.

An interesting aspect of $m_{T2}$ is that it is boost invariant if and
only if $\hat{m}_\text{miss} = m_\text{miss}$. For a wrong assignment
of $m_\text{miss}$ the value of $m_{T2}$ has nothing to do with the
actual kinematics and hence with any kind of invariant (and house
numbers are not boost invariant). We can exploit this aspect by
scanning over $m_\text{miss}$ and looking for so-called kinks, defined
as points where different events kinematics all return the same value
for $m_{T2}$.\bigskip

\begin{figure}[b!]
\begin{center}
  \includegraphics[width=0.30\textwidth]{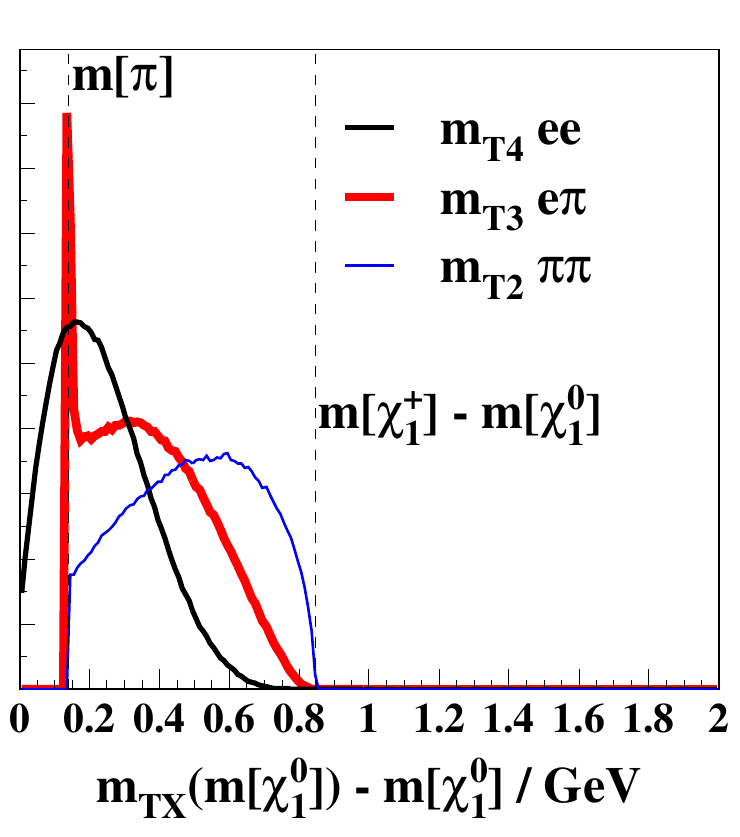}
\end{center}
\vspace*{-5mm}
\caption{Simulations for the decay $\tilde{\chi}^+_1 \to
  \tilde{\chi}^0_1 \pi$ or $\tilde{\chi}^+_1 \to \tilde{\chi}^0_1 e^+
  \nu$. The blue $m_{T2}$ line applies to the two-body decay. Figure
  from Ref.~\cite{Barr:2003rg}.}
\label{fig:sim_mt2}
\end{figure}

Finally, we can account for the fact that supersymmetry predicts new
strongly interacting particles. These are the scalar partners of the
quarks and the fermionic partner of the gluon. For dark matter physics
the squarks are more interesting, because their
quark-squark-neutralino coupling makes them dark matter mediators to
the strongly interacting visible matter sector. The same coupling also
allows the squarks to decay into dark matter and a jet, leading to the
dark matter signature
\begin{alignat}{5}
pp &\to \tilde{q} \tilde{q}^* 
   &&\to ( q \nne ) \; ( \bar{q} \nne) 
\label{eq:lhc_channels3}
\end{alignat}
Example Feynman diagrams showing the role of squarks as
\ul{$t$-channel colored mediators} and as heavy particles
decaying to dark matter are\\[8mm]

\begin{center}
\begin{fmfgraph*}(100,50)
\fmfset{arrow_len}{2mm}
\fmfleft{i1,i2}
\fmfright{o1,o2}
\fmf{fermion,tension=0.6,width=0.6}{i1,v1}
\fmf{fermion,tension=0.6,width=0.6}{v2,i2}
\fmf{scalar,tension=0.4,label=$\tilde{q}$,width=0.6}{v2,v1}
\fmf{plain,tension=0.3,width=0.6}{v1,o1}
\fmf{photon,tension=0.3,width=0.6}{v1,o1}
\fmf{plain,tension=0.3,width=0.6}{v2,o2}
\fmf{photon,tension=0.3,width=0.6}{v2,o2}
\fmflabel{$q$}{i1}
\fmflabel{$\bar{q}$}{i2}
\fmflabel{$\nne$}{o1}
\fmflabel{$\nne$}{o2}
\end{fmfgraph*}
\hspace*{0.2\textwidth}
\begin{fmfgraph*}(120,70)
\fmfset{arrow_len}{2mm}
\fmfleft{i1,i2}
\fmfright{o1,o2,o3,o4}
\fmf{fermion,tension=0.6,width=0.6}{i1,v1}
\fmf{fermion,tension=0.6,width=0.6}{v1,i2}
\fmf{gluon,tension=0.6,width=0.6}{v1,v2}
\fmf{scalar,tension=0.4,label=$\tilde{q}$,width=0.6}{v2,v4}
\fmf{scalar,tension=0.4,lab.side=right,label=$\tilde{q}^*$,width=0.6}{v3,v2}
\fmf{fermion,tension=0.6,width=0.6}{v4,o1}
\fmf{photon,tension=0.2,width=0.6}{v4,o2}
\fmf{plain,tension=0.2,width=0.6}{v4,o2}
\fmf{photon,tension=0.2,width=0.6}{v3,o3}
\fmf{plain,tension=0.2,width=0.6}{v3,o3}
\fmf{fermion,tension=0.6,width=0.6}{o4,v3}
\fmflabel{$q$}{i1}
\fmflabel{$\bar{q}$}{i2}
\fmflabel{$q$}{o1}
\fmflabel{$\nne$}{o2}
\fmflabel{$\nne$}{o3}
\fmflabel{$\bar{q}$}{o4}
\end{fmfgraph*}
\end{center}
\bigskip

Note that these two squark-induced signatures cannot be separated,
because they rely on the same two couplings, the
quark-squark-neutralino coupling and the QCD-induced squark coupling
to a gluon. Kinematically, they add nothing new to the above
arguments: the first diagram will contribute to the mono-jet
signature, with the additional possibility to radiate a gluon off the
$t$-channel mediator, and to pair-production of neutralinos and
charginos; the second diagram asks for a classic $m_{T2}$ analysis.
Moreover, the production process of Eq.\eqref{eq:lhc_channels3} is
QCD-mediated and the 100\% branching fraction gives us no information
about the mediator interaction to dark matter. In other words, for
this pair-production process there exists no link between LHC
observables and dark matter properties.

The non-negligible effect of the $t$-channel squark mediator adding to
the $s$-channel $Z$-mediator for processes of the kind
\begin{align}
pp \to \nni \nnj 
\end{align}
has to do with the couplings.  From Eq.\eqref{eq:coups_neutralinos} we
know that for neutralinos the higgsino content couples to the
$Z$-mediator while the gaugino content couples to light-flavor
squarks. In addition, the $s$-channel and $t$-channel diagrams
typically interfere destructively, so we can tune the squark mass to
significantly reduce the neutralino pair production cross section.
The largest cross section for direct neutralino-chargino production is
usually
\index{cascade decay}
\begin{align}
pp \to \cpe \nnz
   \to (\ell^+ \nu \nne) \; (\ell^+ \ell^- \nne) 
\qquad \text{with} \qquad 
\sigma(\cpme \nnz) \lesssim 1~\pb \; ,
\end{align}
for $m_\chi > 200$~GeV. This decay leads to a tri-lepton signature
with off-shell gauge bosons in the decay.  The backgrounds are pair
production of weak bosons and hence small. Just as a comparison, squark
pair production, $pp \to \tilde{q} \tilde{q}^*$, can reach cross
sections in the pico-barn range even for squark mass above
1~TeV.\bigskip

Before the LHC started running, studies of decay chains with dark
matter states at their end were in fashion. Here, squark decays had
large impact through the stereotypical \ul{cascade decay}
\begin{align}
  \tilde{q} 
  \to \tilde{\chi}_2^0 \; q 
  \to \tilde{\ell}^\pm \ell \; q 
  \to \tilde{\chi}_1^0 \ell^+\ell^- \; q \; .
\label{eq:sim_squarkchain}
\end{align}
First, we need to remove top pair production as the main background
for such signatures. The key observation is that in cascade decays the
leptons are flavor--locked, which means the combination $e^+e^- +
\mu^+\mu^- - e^-\mu^+- e^+\mu^-$ is roughly twice $\mu^+\mu^-$ for the
signal, while it cancels for top pairs. In addition, such cascade
decays are an opportunity to search for kinematic endpoints in many
distributions, like $m_{\ell \ell}$, $m_{q\ell}$, or three-body
combinations. Unfortunately, the general interest in the kinematics of
supersymmetric cascade decays is for now postponed.\bigskip

\begin{figure}[b!]
\begin{center}
  \includegraphics[width=0.40\textwidth]{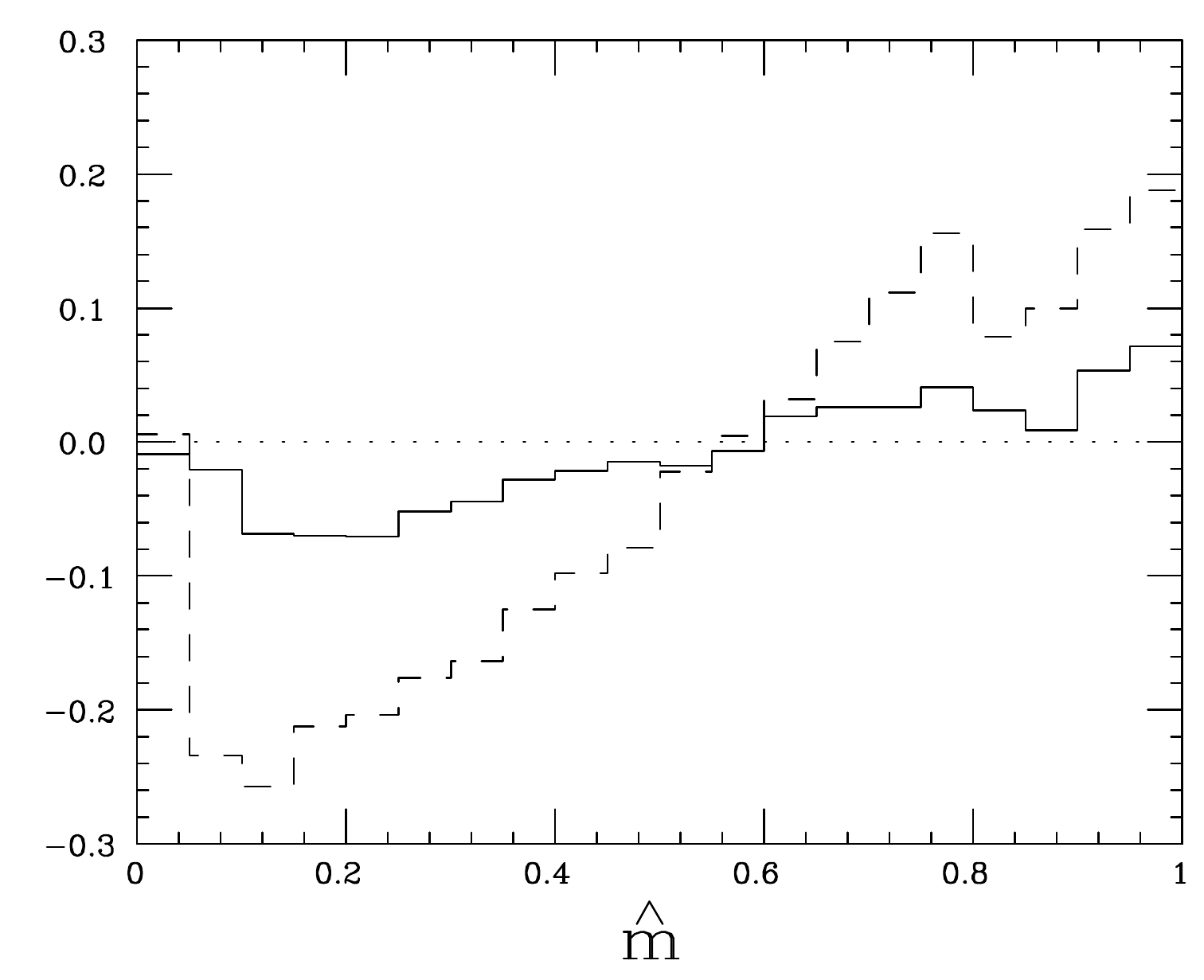}
\end{center}
\vspace*{-7mm}
\caption{Asymmetry in $m_{j \ell}/m_{j \ell}^\text{max}$ for
  supersymmetry (dashed) and universal extra dimensions (solid). The
  spectrum is assumed to be hierarchical, which is typical for supersymmetric
  theories. Figure from Ref.~\cite{Smillie:2005ar}.}
\label{fig:sim_sig_bryan}
\end{figure}

One thing we know for example from the di-lepton edge is that
invariant masses can just be an invariant way of writing angular
correlations between outgoing particles. Those depend on the
\ul{spin and quantum numbers} of all particles involved.  While
measuring for example the spin of new particles is hard in the absence
of fully reconstructed events, we can try to probe it in the
kinematics of cascade decays.  The squark decay chain was the first
case where such a strategy was worked out~\cite{Smillie:2005ar}:
\begin{enumerate}
\item Instead of measuring individual spins in a cascade decay we
  assume that cascade decays radiate particles with known spins. For
  radiated quarks and leptons the spins inside the decay chain
  alternate between fermions and bosons. Therefore, we contrast
  supersymmetry with another hypothesis, where the spins in the decay
  chain follow the Standard Model assignments. An example for such a model are
  Universal Extra Dimensions, where each Standard Model particle
  acquires a Kaluza--Klein partner from the propagation in the bulk of
  the additional dimensions;

\item The kinematical endpoints are completely determined by the masses
  and cannot be used to distinguish between the spin assignments.  In contrast,
  the distributions between endpoints reflect angular correlations. For
  example, the $m_{j \ell}$ distribution in principle allows us to
  analyze spin correlations in squark decays in a Lorentz-invariant
  way. The only problem is the link between $\ell^\pm$ and their
  ordering in the decay chain;

\item As a proton--proton collider the LHC produces considerably more
  squarks than anti-squarks in the squark--gluino production process.
  A decaying squark radiates a quark while an antisquark radiates an
  antiquark, which means that we can define a non-zero production-side
  asymmetry between $m_{j \ell^+}$ and $m_{j \ell^-}$. Such an
  asymmetry we show in Figure~\ref{fig:sim_sig_bryan}, for the SUSY
  and for the UED hypotheses. Provided the masses in the decay chain
  are not too degenerate we can indeed distinguish the two hypotheses.
\end{enumerate}
%

\subsection{Effective field theory}
\label{sec:coll_eft}
\index{effective field theory}

In Section~\ref{sec:models_eft} we introduced an effective field
theory of dark matter to describe dark matter annihilation in the
early universe. If the annihilation process is the usual $2 \to 2$
WIMP scattering process it is formulated in terms of a dark matter
mass $m_\chi$ and a mediator mass $\mmed$, where the latter does not
correspond to a propagating degree of freedom. It can hence be
identified with a general suppression scale $\Lambda$ in an effective
Lagrangian, like the one illustrated in Eq.\eqref{eq:def_eft}. All
experimental environments discussed in the previous sections,
including the relic density, indirect detection, and direct detection,
rely on non-relativistic dark matter scattering. This means they can
be described by a \ul{dark matter EFT} if the mediator is much
heavier than the dark matter agent,
\begin{align}
\boxed{
m_\chi \ll \mmed 
} \; .
\end{align}
In contrast, LHC physics is entirely relativistic and neither the
incoming partons nor the outgoing dark matter particles in the
schematic diagram shown in Section~\ref{sec:photons} are at low
velocities. This means we have to add the partonic energy of the
scattering process to the \ul{relevant energy scales},
\begin{align}
\{ \, m_\chi, \mmed, \sqrt{s} \, \} \; .
\end{align}
In the case of mono-jet production, described in
Section~\ref{sec:coll_hadron}, the key observables here are the $\met$
and $p_{T,j}$ distributions. For simple hard processes the two
transverse momentum distributions are rapidly dropping and strongly
correlated. This defines the relevant energy scales as
\begin{align}
\{ \, m_\chi, \mmed, \met^\text{min} \, \} \; .
\end{align}
The experimentally relevant $\met$ or $p_{T,j}$ regime is given by a
combination of the signal mass scale, the kinematics of the
dominant $Z_{\nu \nu}$+jets background, and triggering. Our effective
theory then has to reproduce two key observables,
\begin{align}
\sigma_\text{tot}(m_\chi,\mmed) \Bigg|_\text{acceptance}
\qquad \text{and} \qquad
\frac{d \, \sigma(m_\chi,\mmed)}{d \, \met} 
\sim \frac{d \, \sigma(m_\chi,\mmed)}{d \, p_{T,j}} \; .
\end{align}
For the total rate, different phase space regions which individually
agree poorly between the effective theory and some underlying model,
might combine to a decent rate. For the main distributions this is no
longer possible.

Finally, the hadronic LHC energy of 13~TeV, combined with reasonable
parton momentum fractions defines an absolute upper limit, above which
for example a particle in the $s$-channel cannot be produced as a
propagating state,
\begin{align}
\boxed{
\{ \, m_\chi, \mmed, \met^\text{min}, \sqrt{s}_\text{max} \, \} 
}\; .
\label{eq:lhc_allscales}
\end{align}
This fourth scale is not the hadronic collision energy 13~TeV. From
the typical LHC reach for heavy resonances in the $s$-channel we
expect it to be in the range $\sqrt{s}_\text{max} = 5~...~8$~TeV,
depending on the details of the mediator.\bigskip

From what we know from these lecture notes, establishing a completely general
dark matter EFT approach at the LHC is not going to work. The Higgs
portal results of Section~\ref{sec:coll_portal} indicate that the only
way to systematically search for its dark matter scalar is through
invisible Higgs decays. By definition, those will be entirely dominated
by on-shell Higgs production, not described by an effective field
theory with a non-propagating mediator. Similarly, in the MSSM a
sizeable fraction of the mediators are either light SM particles or
$s$-channel particles within the reach of the LHC. Moreover, we need to
add propagating degrees of co-annihilation partners, more or less
close to the dark matter sector. 

On the other hand, the fact that some of our favorite dark matter
models are not described well by an effective Lagrangian does not mean
that we cannot use such an effective Lagrangian for other
\ul{classes of dark matter models}. One appropriate way to test
the EFT approach at the LHC is to rely on specific simplified models,
as introduced in Section~\ref{sec:indirect_simp}. Three simplified
models come to mind for a fermion dark matter agent~\cite{dm_eff}:
\begin{enumerate}
\item tree-level $s$-channel vector mediator, as discussed in
  Section~\ref{sec:indirect_simp};
\item tree-level $t$-channel scalar mediator, realized as light-flavor
  scalar quarks in the MSSM, Section~\ref{sec:coll_mssm};
\item loop-mediated $s$-channel scalar mediator, realized as heavy
  Higgses in the MSSM, Section~\ref{sec:coll_mssm}.
\end{enumerate}
\bigskip
\index{mediator!t-channel}
For the \ul{tree-level vector} the situation at the LHC already
becomes obvious in Section~\ref{sec:indirect_simp}. The EFT approach
is only applicable when also at the LHC the vector mediator is
produced away from its mass shell, requiring roughly $m_V >
5$~TeV. The problem in this parameter range is that the dark matter
annihilation cross section will be typically too small to provide the
observed relic density. This makes the parameter region where this
mediator can be described by global EFT analyses very small.\bigskip

We start our more quantitative discussion with a \ul{tree-level
  $t$-channel scalar} $\tilde{u}$.  Unlike for the vector mediator,
the $t$-channel mediator model only makes sense in the half plane with
$m_\chi < \msu$; otherwise the dark matter agent would decay. At the
LHC we have to consider different production processes. Beyond the
unobservable process $u \bar{u} \to \chi \chi$ the two relevant
topologies leading to \ul{mono-jet production} are
\begin{align}
u \bar{u} \to \chi \bar \chi g 
\qquad \text{and} \qquad 
u g \to \chi \bar \chi u \; .
\label{eq:t_monojet}
\end{align}
\index{pair production}
They are of the same order in perturbation theory and experimentally
indistinguishable. The second process can be dominated
by on-shell mediator production, $u g \to \chi \tilde{u} \to \chi \;
(\bar \chi u)$. We can cross its amplitude to describe the co-annihilation
process $\chi \tilde{u} \to u g$. The difference between the (co-)
annihilation and LHC interpretations of the same amplitude is that it
only contributes to the relic density for $\msu < m_\chi + 10\%$,
while it dominates mono-jet production for a wide range of mediator
masses.

Following Eq.\eqref{eq:lhc_channels3} we can also
\ul{pair-produce} the necessarily strongly interacting
mediators with a subsequent decay to two jets plus missing energy,
\begin{align} 
q\bar{q}/gg 
\quad \stackrel{\text{QCD}}{\longrightarrow} \quad \tilde{u} \tilde{u}^* 
\stackrel{\text{dark matter}}{\longrightarrow} (\bar \chi u) \, (\chi \bar u) \; . 
\label{eq:t_double_pole}
\end{align}
The partonic initial state of this process can be quarks or
gluons. For a wide range of dark matter and mediator masses this
process completely dominates the $\chi \chi$+jets process.

When the $t$-channel mediator becomes heavy, for example mono-jet
production with the partonic processes given in
Eq.\eqref{eq:t_monojet} can be described by an effective four-fermion
operator,
\begin{align}
\lag \supset \frac{c}{\Lambda^2}  \; 
                       \left(\bar{u}_R \chi \right) \; 
                       \left( \bar{\chi} u_R\right) \; .
\label{eq:t_eft}
\end{align}
The natural matching scale will be around $\Lambda=\msu$. Note that
this operator mediates the $t$-channel as well as the single-resonant
mediator production topologies and the pair production process induced
by quarks. In contrast, pair production from two gluons requires a
higher-dimensional operator involving the gluon field strength, like
for example
\begin{align}
\lag \supset 
 \frac{c}{\Lambda^3}(\bar\chi\chi) \, G_{\mu\nu}G^{\mu\nu} \; .
\end{align}
This leads to a much faster decoupling pattern of the pair production
process for a heavy mediator.

Because the $t$-channel mediator carries color charge, LHC constraints
typically force us into the regime $\msu \gtrsim 1$~TeV, where an EFT
approach can be viable. In addition, we again need to generate a large
dark matter annihilation rate, which based on the usual scaling can be
achieved by requiring $\msu \gtrsim m_\chi$.  For heavy mediators,
pair production decouples rapidly and leads to a parameter region
where single-resonant production plays an important role.  It is
described by the same effective Lagrangian as the generic $t$-channel
process, and decouples more rapidly than the $t$-channel diagram for
$\msu \gtrsim 5$~TeV. These actual mass values unfortunately imply
that the remaining parameter regions suitable for an EFT description
typically predict very small LHC rates.\bigskip
\index{mediator!s-channel}
The third simplified model we discuss is a \ul{scalar
  $s$-channel mediator}. \index{simplified model}
 To generate a sizeable LHC rate we do not
rely on its Yukawa couplings to light quarks, but on a loop-induced
\ul{coupling to gluons}, in complete analogy to SM-like light
Higgs production at the LHC. The situation is slightly different for
most of the supersymmetric parameter space for heavy Higgses, which
have reduced top Yukawa couplings and are therefore much harder to
produce at the LHC. Two relevant Feynman diagrams for mono-jet
production are\\[8mm]

\begin{center}
\begin{fmfgraph*}(100,70)
\fmfset{arrow_len}{2mm}
\fmfleft{i1,i2}
\fmfright{o1,o2,o3}
\fmf{fermion,width=0.6,tension=0.6}{o1,v1}
\fmf{fermion,width=0.6,tension=0.6}{v1,o2}
\fmf{dashes,width=0.6,tension=0.6,label=$S$}{v1,v2}
\fmf{gluon,width=0.6,tension=0.6}{i2,v5}
\fmf{gluon,width=0.6,tension=0.4}{v5,v4}
\fmf{gluon,width=0.6,tension=0.6}{v3,i1}
\fmf{gluon,width=0.6,tension=0.6}{v5,o3}
\fmf{fermion,width=0.6,tension=0.2,label=$t$}{v3,v4}
\fmf{fermion,width=0.6,tension=0.2}{v2,v3}
\fmf{fermion,width=0.6,tension=0.2}{v4,v2}
\fmflabel{$\bar \chi$}{o1}
\fmflabel{$\chi$}{o2}
\fmflabel{$g$}{i1}
\fmflabel{$g$}{i2}
\fmflabel{$g$}{o3}
\end{fmfgraph*}
\hspace*{0.2\textwidth}
\begin{fmfgraph*}(100,70)
\fmfset{arrow_len}{2mm}
\fmfleft{i1,i2}
\fmfright{o1,o2,o3}
\fmf{fermion,width=0.6,tension=0.6}{o1,v1}
\fmf{fermion,width=0.6,tension=0.6}{v1,o2}
\fmf{dashes,width=0.6,tension=0.6,label=$S$}{v1,v2}
\fmf{gluon,width=0.6,tension=0.6}{i2,v4}
\fmf{gluon,width=0.6,tension=0.6}{v3,i1}
\fmf{gluon,width=0.6,tension=0.6}{v5,o3}
\fmf{fermion,width=0.6,tension=0.4}{v4,v5}
\fmf{fermion,width=0.6,tension=0.2,label=$t$}{v3,v4}
\fmf{fermion,width=0.6,tension=0.4}{v2,v3}
\fmf{fermion,width=0.6,tension=0.2}{v5,v2}
\fmflabel{$\bar \chi$}{o1}
\fmflabel{$\chi$}{o2}
\fmflabel{$g$}{i1}
\fmflabel{$g$}{i2}
\fmflabel{$g$}{o3}
\end{fmfgraph*}
\end{center}

Coupling the scalar only to the top quark, we define the Lagrangian
for the simplified scalar mediator model as
\begin{align}
\lag \supset 
- \frac{y_t m_t}{v}\,S\,\bar t t 
+ g_\chi S\,\bar \chi \chi 
\label{eq:s_loop_model}
\end{align}
The factor $m_t/v$ in the top Yukawa coupling is conventional, to
allow for an easier comparison to the Higgs case.  The scalar coupling
to the dark matter fermions can be linked to $m_\chi$, but does not
have to.  We know that the SM Higgs is a very narrow resonance, while
in this case the total width is bounded by the partial width from
scalar decays to the top quark,
\begin{align}
\frac{\Gamma_S}{m_S} 
> \frac{3 G_F m_t^2 y_t^2}{4 \sqrt{2} \pi} \; \left( 1 - \frac{4m_t^2}{m_S^2} \right)^{3/2} 
\stackrel{m_S \gg m_t}{=} \frac{3 G_F m_t^2 y_t^2}{4 \sqrt{2} \pi}  
\approx 5\% \; ,
\end{align}
assuming $y_t \approx 1$. Again, this is different from the case of
supersymmetric, heavy Higgses, which can be broad.

To get a rough idea what kind of parameter space might be interesting,
we can look at the relic density. The problem in this prediction is
that for $m_\chi < m_t$ the annihilation channel $\chi \chi \to
t\bar{t}$ is kinematically closed.  Going through the same amplitude
as the one for LHC production, very light dark matter will annihilate
to two gluons through a top loop. If we allow for that coupling, the
tree-level process $\chi \chi \to c\bar{c}$ will dominate for slightly
heavier dark matter. If there also exists a Yukawa coupling of the
mediator to bottom quarks, the annihilation channel $\chi \chi
\to b\bar{b}$ will then take over for slightly heavier dark matter. An
even heavier mediator will annihilate into off-shell top quarks, $\chi
\chi \to (W^+ b)(W^- \bar{b})$, and for $m_\chi > m_t$ the
tree-level $2 \to 2$ annihilation process $\chi \chi \to
t\bar{t}$ will provide very efficient annihilation. None of the
aspects determining the correct annihilation channels are well-defined
within the simplified model. Moreover, in the Lagrangian of
Eq.\eqref{eq:s_loop_model} we can easily replace the scalar $S$ with a
pseudo-scalar, which will affect all non-relativistic
processes.

For our global EFT picture this means that if a scalar $s$-channel
mediator is predominantly coupled to up-quarks, the link between the
LHC production rate and the predicted relic density essentially
vanishes. The two observables are only related if the mediator is very
light and decays through the one-loop diagram to a pair of
gluons. This is exactly where the usual dark matter EFT will not be
applicable.\bigskip

If we only look at the LHC, the situation becomes much simpler. The
dominant production process
\begin{align}
gg \to S +\text{jets} \to \chi \chi + \text{jets}
\end{align}
defines the mono-jet signature through initial-state radiation and
through gluon radiation off the top loop.  The mono-jet rate will
factorize into $\sigma_{S+j} \times \br_{\chi\chi}$.  The production
process is well known from \ul{Higgs physics}, including the
phase space region with a large jet and the logarithmic top mass
dependence of the transverse momentum distribution,
\begin{align}
\frac{d \sigma_{Sj}}{d p_{T,j}} = \frac{d \sigma_{Sj}}{d p_{T,S}}
\propto \log^4 \frac{p_{T,j}^2}{m_t^2} \; .
\label{eq:log_pts}
\end{align}
Based on the Lagrangian given in Eq.\eqref{eq:s_loop_model} and the
transverse momentum dependence given in Eq.\eqref{eq:log_pts}, the
mono-jet signal at the LHC depends on the four energy scales,
\begin{align}
\{ \, m_\chi, m_S, m_t, \met=p_{T,j} \, \} \; ,
\label{eq:lhc_morescales}
\end{align}
which have to be organized in an effective field theory. If we focus
on total rates, we are still left with three mass scales with
different possible hierarchies:
\begin{enumerate}
\item The dark matter agent obviously has to remain a propagating
  degree of freedom, so in analogy to the SM Higgs case we can first
  assume a \ul{non-propagating top quark}
\begin{align}
m_t> m_S > 2m_\chi \; .
\end{align}
This defines the effective Lagrangian
\begin{align}
\lag^{(1)} \supset 
  \frac{c}{\Lambda}S \, G_{\mu\nu}G^{\mu\nu}
- g_\chi \; S\,\bar \chi \chi  \; .
\label{eq:eff_lag1}
\end{align}
It is similar to the effective Higgs--gluon coupling in direct
detection, defined in Eq.\eqref{eq:higgs_eff1}.  The Wilson
coefficient can be determined at the matching scale $\Lambda = m_t$
and assume the simple form
\begin{alignat}{7}
\frac{c}{\Lambda}   
\stackrel{m_t \gg m_S}{=}
\frac{\alpha_s}{12\pi}\frac{y_t}{v} \; ,
\end{alignat}
In this effective theory the transverse momentum spectra will fail to
reproduce large logarithms of the type $\log (p_T/m_t)$, limiting the
agreement between the simplified model and its EFT approximation.

\item Alternatively, we can \ul{decouple the mediator},
\begin{align}
m_S> m_t, 2m_\chi \; ,
\end{align}
leading to the usual dimension-6 four-fermion operators coupling dark
matter to the resolved top loop,
\begin{align}
\lag^{(2)} \supset 
 \frac{c}{\Lambda^2}(\bar t t) \; (\bar\chi\chi) \; .
\label{eq:eff_lag2}
\end{align}
The Wilson coefficients we obtain from matching at $\Lambda = m_S$ are 
\begin{align}
\frac{c}{\Lambda^2}=\frac{y_t g_\chi}{m_S^2}\frac{m_t}{v}\; .
\end{align}
This effective theory will retain all top mass effects in the
distributions.

\item Finally, we can \ul{decouple the top as well as the
  mediator},
\begin{align}
m_S,  m_t > 2m_\chi \; .
\end{align}
The effective Lagrangian reads
\begin{align}
\lag^{(3)} \supset 
 \frac{c}{\Lambda^3}(\bar\chi\chi) \; G_{\mu\nu}G^{\mu\nu} \; .
\label{eq:eff_lag3}
\end{align}
This dimension-seven operators is further suppressed by two equal
heavy mass scales. Matching at $\Lambda = m_S \approx m_t$ gives us
\begin{alignat}{9}
\frac{c}{\Lambda^3}
\stackrel{m_S \approx m_t \gg m_\chi}{=}
\frac{\alpha_s}{12\pi}\frac{y_t g_\chi}{m_S^2}\frac{1}{v} \; ,
\end{alignat}
assuming only the top quark runs in the loop.

\end{enumerate}
%

\begin{figure}[b!]
\begin{center}
\includegraphics[width=0.7\textwidth]{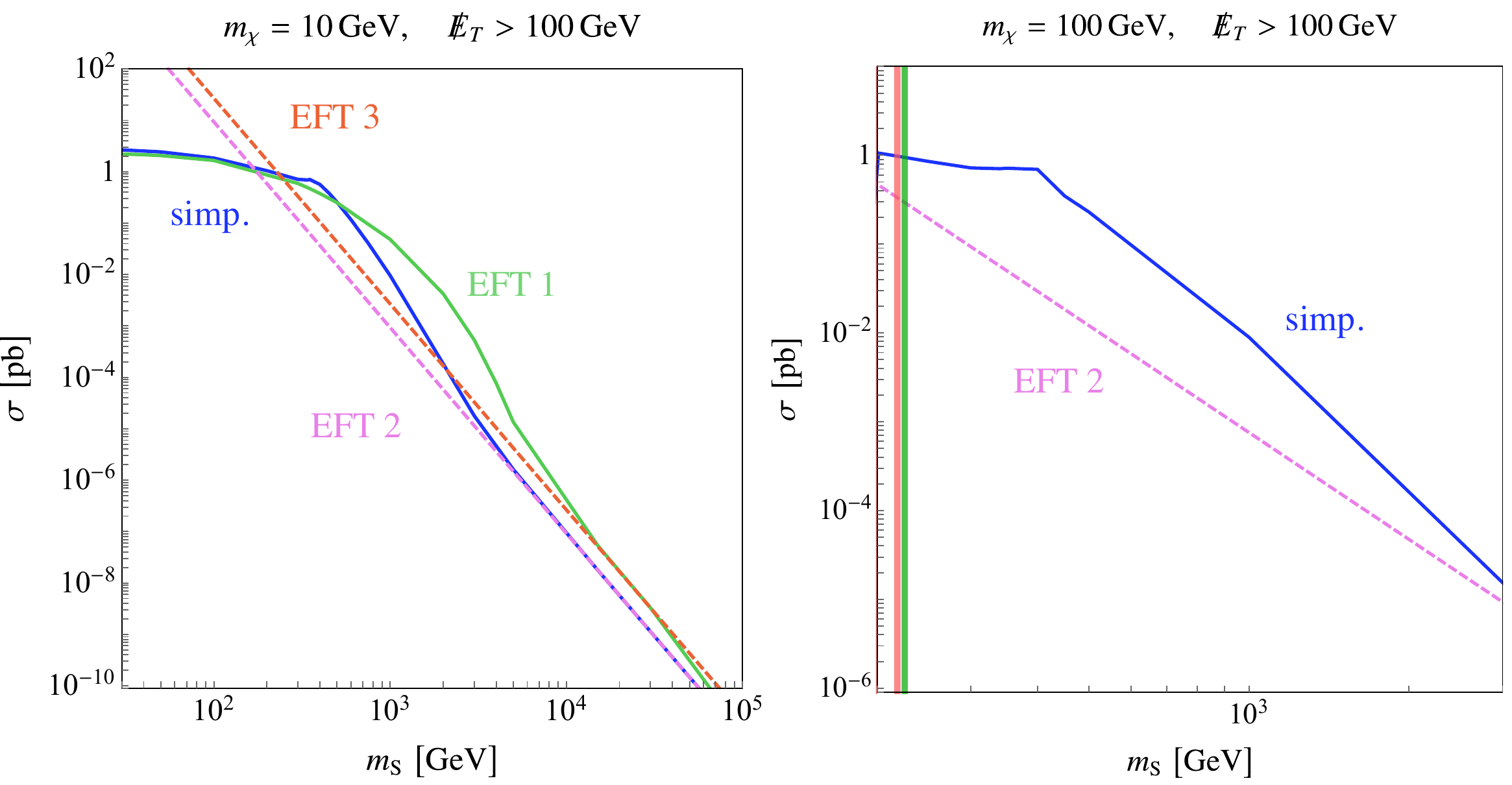}
\end{center}
\vspace{-7mm}
\caption{Total mono-jet rate in the loop-mediated $s$-channel scalar
  model as function of the mediator mass for . We show all three
  different $m_\chi=10$~GeV (left) and $m_\chi=100$~GeV (right).  For
  the shaded regions the annihilation cross section reproduces the
  observed relic density within $\Omega_\chi^\text{obs}/3$ and
  $\Omega_\chi^\text{obs}+10\%$ for a mediator coupling only to up-type
  quarks (red) or to both types of quarks (green). Figure from
  Ref.~\cite{dm_eff}.}
\label{fig:sjet1}
\end{figure}
  
We show the predictions for the total LHC rate based on these three
effective theories and the simplified model in the left panel of
Figure~\ref{fig:sjet1}. The decoupled top ansatz $\lag^{(1)}$ of
Eq.\eqref{eq:eff_lag1} indeed reproduces the correct total rate for
$m_S < 2 m_t$. Above that threshold it systematically overestimates
the cross section. The effective Lagrangian $\lag^{(2)}$ with a
decoupled mediator, Eq.\eqref{eq:eff_lag2}, reproduces the simplified
model for $m_S \gtrsim 5$~TeV. Beyond this value the LHC energy is not
sufficient to produce the mediator on-shell. Finally, the effective
Lagrangian $\lag^{(3)}$ with a simultaneously decoupled top quark and
mediator, Eq.\eqref{eq:eff_lag3}, does not reproduce the total
production rate anywhere.

In the right panel of Figure~\ref{fig:sjet1} we show the mono-jet
rate for heavier dark matter and the parameter regions where the
simplified model predicts a roughly correct relic density. In this
range only the EFT with the decoupled mediator, defined in
Eq.\eqref{eq:eff_lag2}, makes sense.  Because the model gives us this
freedom, we also test what happens to the combination with the relic density
when we couple the mediator to all quarks, rather than up-quarks only.
Altogether, we find that in the region of heavy mediators the EFT is valid for LHC
observables if 
\begin{align}
m_S > 5~\tev \; . 
\end{align}
This is similar to the range of EFT validity for the $s$-channel
vector model.

\newpage 
\section{Further reading}
\label{sec:FR}

First, we would like to emphasize that our list of references is
limited to the, legally required, sources of Figures and to slightly
more advanced material providing more details about the topics
discussed in these lecture notes.

Our discussion on the general relativity background and cosmology is a
very brief summary. Dedicated textbooks include the classics by Kolb \&
Turner~\cite{KolbTurner}, Bergstr\"om \& Goobar~\cite{Bergstrom},
Weinberg~\cite{Weinberg:2008zzc}, as well as the more modern books by
Scott Dodelson~\cite{Dodelson} and Max
Tegmark~\cite{Tegmark:2002dg}. More details on the role of dark matter
in the history of the universe is given in the book by Gianfranco
Bertone and Dan Hooper~\cite{history} and in the notes by Flip
Tanedo~\cite{{Tanedo}} and Yann Mambrini~\cite{Mambrini}.  Jim Cline's
TASI lectures~\cite{Cline:2018fuq} serve as an up-to-date discussion
on the role of dark matter in the history of the Universe.  Further
details on the cosmic microwave background and structure formation are
also in the lecture notes on cosmological perturbation theory by Hannu
Kurki-Suonio that are available online~\cite{Kurki}, as well as in the
lecture notes on Cosmology by Joao Rosa~\cite{Rosa} and Daniel
Baumann~\cite{Baumann}.

For models of particle dark matter, Ref.~\cite{ten_points} provides a
list of consistency tests. For further reading on WIMP dark matter we
recommend the didactic review article Ref.~\cite{Arcadi:2017kky}.
Reference~\cite{Slatyer:2009yq} addresses details on WIMP annihilation
and the resulting constraints from the comic microwave background
radiation. A more detailed treatment of the calculation of the relic
density for a WIMP is given in Ref.~\cite{Steigman:2012nb}. Felix
Kahlh\"ofer has written a nice review article on LHC searches for
WIMPs~\cite{felix}. For further reading on the effect of the
Sommerfeld enhancement, we recommend Ref.~\cite{ArkaniHamed:2008qn}.

Extensions of the WIMP paradigm can result in a modified freeze-out
mechanism, as is the case of the co-annihilation scenario. These
exceptions to the most straightforward dark matter freeze-out have
originally been discussed by Griest and Seckel in
Ref.~\cite{Griest:1990kh}. A nice systematic discussion of recent
research aiming can be found in Ref.~\cite{DAgnolo:2017dbv}.

For models of non-WIMP dark matter, the review article
Ref.~\cite{Baer:2014eja} provides many details.  A very good review of
axions is given in Roberto Peccei's notes~\cite{Peccei:2006as}. while
axions as dark matter candidates are discussed in
Ref.~\cite{Arias:2012az}.  Mariangela Lisanti's TASI
lectures~\cite{lisanti} provide a pedagogical over these different
dark matter candidates. Details on light dark matter, in particular
hidden photons, can be found in Tongyan Lin's notes for her 2018 TASI
lecture~\cite{Tongyan}.

Details on calculations for the direct search for dark matter can be
found in the review by Lewin and Smith~\cite{Lewin:1995rx}. Gondolo
and Silk provide details for dark matter annihilation in the galactic
center~\cite{Gondolo:1999ef}, as do the TASI lecture notes of Dan
Hooper~\cite{Hooper:2009zm}. For many more details on indirect
detection of dark matter we refer to Tracy Slatyer's TASI
lectures~\cite{Slatyer:2017sev}.

Note the one aspect these lecture notes are still missing is the
chapter on the discovery of WIMPs. We plan to add an in-depth
discussion of the WIMP discovery to an updated version of these notes.\bigskip

\begin{center}
{\bf Acknowledgments}
\end{center}

TP would like to thank many friends who have taught him dark matter,
starting with the always-inspiring Dan Hooper. Dan also introduced him
to deep-fried cheese curds and to the best ribs in Chicago. Tim Tait
was of great help in at least two ways: for years he showed us that it
is fun to work on dark matter even as a trained collider physicist;
and then he answered every single email during the preparation of
these notes. Our experimental co-lecturer Teresa Marrodan Undagoitia
showed not only our students, but also us how inspiring dark matter
physics can be. As co-authors of Refs~\cite{nimatron2,nimatron1} Joe
Bramante, Adam Martin, and Paddy Fox gave us a great course on dark
matter physics while we were writing these papers.  Pedro Ruiz-Femenia
was extremely helpful explaining the field theory behind the
Sommerfeld enhancement to us. J\"org J\"ackel for a long time and over
many coffees tried to convince everybody that the axion is a great
dark matter candidate. Teaching and discussing with Bj\"orn-Malte
Sch\"afer was an excellent course on how thermodynamics is actually
useful. Finally, there are many people who helped us with valuable
advice while we prepared this course, like John Beacom, Martin
Schmaltz and Felix Kahlh\"ofer, and people who commented on the notes,
like Elias Bernreuther, Johann Brehmer, Michael Baker, Anja Butter,
Bj\"orn Eichmann, Ayres Freitas, Jan Horak, Michael Ratz, or Michael
Schmidt.

\newpage

\printindex

\end{fmffile}

\end{document}